\documentclass[PAPER, cernpreprint=true, coverpage=false, texlive=2023, UKenglish, texmf, orcidlogo]{atlasdoc}
\usepackage{atlaspackage}
\usepackage{atlasbiblatex}

\usepackage{atlasphysics}
\DeclareUnicodeCharacter{2212}{-}

\graphicspath{{logos/}{figures/}}

\usepackage{ANA-STDM-2023-20-PAPER-defs}

\AtlasTitle{Electroweak, QCD and flavour physics studies  with  ATLAS data from Run~2 of the LHC }

\AtlasAbstract{%
A summary of precision measurements sensitive to electroweak, QCD and quark-flavour effects performed by the ATLAS Collaboration at the Large Hadron Collider is reported.
The measurements  are  predominantly performed  on  proton--proton ($pp$) collision data recorded at a centre-of-mass energy of 13~TeV taken from 2015 to 2018, with  an integrated luminosity of up to 140 fb$^{-1}$, with some results  based on  $pp$ and Pb+Pb data recorded at lower nucleon centre-of-mass energies.
The results cover a wide range of topics, from strong production of particles at low energies and  the spectroscopy of hadrons to perturbative QCD with hadronic jets and electroweak and strong production of single and multiple vector bosons. They provide precise measurements of fundamental constants and stringent tests of the Standard Model with unprecedented precision and in energy ranges never explored before.
They are also used to explore the proton structure and to perform model-independent searches for new physics.
}

\AtlasRefCode{STDM-2023-20}

\PreprintIdNumber{CERN-EP-2024-093}

\AtlasJournalRef{Phys. Rep. 1116 (2025) 57-126}
\AtlasDOI{10.1016/j.physrep.2024.12.003}

\hypersetup{pdftitle={ATLAS document},pdfauthor={The ATLAS Collaboration}}

\begin{document}

\maketitle

\tableofcontents

\section{Introduction}
%

Collisions of protons at the Large Hadron Collider (LHC) at the energy and luminosity frontier, provide a unique opportunity
to study the strong, quantum chromodynamics (QCD), and electroweak (EW) interactions in unprecedented detail.
Run~1 of the LHC left a legacy of pioneering Standard Model (SM) measurements, covering the inclusive production of photons, jets, massive single gauge bosons and pairs of gauge bosons and the first electroweak production processes of gauge bosons, including vector-boson-fusion (VBF) and vector-boson scattering (VBS),  were established. Hadronic event shapes and the substructure of jets were studied, as well as processes at energy scales below where perturbative QCD (pQCD) is applicable. Precision studies of $b$-hadrons and first measurements of fundamental SM parameters ($W$ mass, Weak Mixing angle and the strong coupling constant) were performed.
With the start of the LHC Run~2 in 2015, partial data samples were used to probe the energy dependence of the cross-sections of basic SM processes and establish some rarer processes for the first time. From 2018 onwards, with the full Run~2 data sample available, ATLAS is able to probe new kinematic regions previously inaccessible to measurements and perform more differential measurements. New rare processes, especially in the electroweak sector, are accessible and the measurement of their differential distributions has allowed
ATLAS to perform model-independent searches for new physics.

For Run~2, the LHC increased the centre-of-mass energy ($\sqrt{s}$) in $pp$ collisions from 8 to 13~\TeV. A significant increase in the beam intensity allowed more luminosity to be collected, but also led to a significant increase in the mean number of $pp$ interactions per bunch crossing (pile-up), with correspondingly higher particle multiplicities and trigger rates.
This effect was only partially offset by the reduction of the bunch spacing from 50~ns to 25~ns and required the development of many new techniques to mitigate the adverse effects of these conditions on the measurements.

In parallel with the increased statistical and systematic precision and the increased energy reach, the accuracy of theoretical predictions have substantially advanced for both Monte Carlo (MC) simulations and fixed-order calculations (see e.g.~\cite{LesHouches_2021,LesHouches_2023,Campbell:2022qmc} and references therein). Regarding the former, the combination of next-to-leading order (NLO) in pQCD with a parton shower (PS) is now considered the standard for most analyses and the first predictions at next-to-next-leading order (NNLO) QCD merged with a PS (NNLO+PS) are available.
Substantial modelling and computational progress has also been achieved in multi-leg MC generators that combine matrix elements (ME) of various orders in pQCD with a PS~\cite{PMGR-2021-01,Hoeche13,Bothmann:2019yzt,Alwall:2014hca, Sjostrand:2014zea,Frederix:2012ps,Frederix_2020}.
New developments aim for next-to-leading-logarithmic (NLL) PS accuracy~\cite{Dasgupta_2018}.
Higher-order EW corrections are increasingly included in MC generators and fixed-order predictions~\cite{Denner_2020,Kallweit_2016,Grazzini_2020,Denner_2022}. For fixed-order calculations of $2\to2$ processes, NNLO QCD is now the standard and often combined with next-to-next-to-leading logarithmic (NNLL) order in QCD as NLL/NNLL QCD resummation. First NNLO QCD calculations for $2\to3$ processes~\cite{mitov2022,Alvarez_2023} and N$^3$LO QCD and N$^3$LL QCD calculations for $2\to2$ processes~\cite{Camarda_2021} have been published.
In addition, different settings of model parameters optimised to reproduce experimental results based on the LHC Run 1 data are used in the simulation of QCD phenomena at low energy scales~\cite{ATL-PHYS-PUB-2011-014, Pierog:2013ria}.
While infrared-safe algorithms are routinely used for inclusive jets at the LHC, a variety of such algorithms has now also been developed for flavoured jets~\cite{Gauld:2022lem,Czakon_2023,Caletti_2022}.

The measurements described in this paper, unless stated otherwise, are based on the $pp$ data sample recorded at $\sqrt{s}=13$~\TeV corresponding to an integrated luminosity of $140.1\pm 1.2$~fb$^{-1}$~\cite{DAPR-2021-01}.
Some results are based on $pp$ and lead--lead (Pb+Pb) data recorded at lower nucleon centre-of-mass energies.
This review covers measurements published until spring 2024.

This paper is organised as follows.
The ATLAS detector and its performance
are described in Sections~\ref{sec:atlas} and~\ref{sec:perf}, respectively.
Section~\ref{sec:totalxs} describes the total, elastic and inelastic $pp$ cross-section measurements.
Measurements of inclusive production of charged particles down to low energies are discussed in Section~\ref{sec:soft}.
Sections~\ref{sec:jets} and \ref{sec:photons} summarise measurements with inclusive jets and isolated photons, respectively.
Section~\ref{sec:singlebos} presents the measurements of single gauge ($W$ or $Z$) bosons, while the measurements involving the production of two and three gauge bosons are summarised in Sections~\ref{sec:EWK_measurements} and \ref{sec:vvv}, respectively.
Measurements involving photon--photon interactions are summarised in Section~\ref{sec:gammagamma}.
These photon-induced measurements utilise $pp$ collisions, but also Pb+Pb collision data recorded in 2015 and 2018.
Measurements of fundamental parameters of the SM are presented in Section~\ref{sec:params}.
Sections~\ref{sec:bphys} and \ref{sec:hf} discuss the studies of heavy-flavour hadrons, including charmonium and exotic states.
Finally, Section~\ref{sec:conclusion} summarises the conclusions of the paper.


\section{ATLAS detector in Run 2 of the LHC}
\label{sec:atlas}

%
\newcommand{\AtlasCoordFootnote}{%
ATLAS uses a right-handed coordinate system with its origin at the nominal interaction point (IP)
in the centre of the detector and the \(z\)-axis along the beam pipe.
The \(x\)-axis points from the IP to the centre of the LHC ring,
and the \(y\)-axis points upwards. Cylindrical coordinates \((r,\phi)\) are used in the transverse plane,
where \(\phi\) being the azimuthal angle around the \(z\)-axis and $r$ is the distance from the IP in the transverse plane.
The pseudorapidity is defined in terms of the polar angle \(\theta\) as \(\eta = -\ln \tan(\theta/2)\).
Angular distance is measured in units of \(\Delta R \equiv \sqrt{(\Delta\eta)^{2} + (\Delta\phi)^{2}}\).}

The ATLAS detector~\cite{PERF-2007-01} at the LHC covers nearly the entire solid angle around the collision point.\footnote{\AtlasCoordFootnote}
It consists of an inner tracking detector surrounded by a thin superconducting solenoid, electromagnetic (EM) and hadron calorimeters,
and a muon spectrometer incorporating three large superconducting air-core toroidal magnets.
ATLAS is also equipped with several forward detectors that monitor collision conditions, provide instantaneous luminosity estimates and measure particles scattered at small angles.

The inner-detector system (ID) is immersed in a \qty{2}{\tesla} axial magnetic field
and provides charged-particle tracking in the range of \(|\eta| < 2.5\).
The high-granularity silicon pixel detector covers the vertex region and typically provides four measurements per track, the first hit normally being in the insertable B-layer (IBL) installed before Run~2~\cite{ATLAS-TDR-19,PIX-2018-001}.
It is followed by the silicon microstrip tracker, which usually provides eight measurements per track.
These silicon detectors are complemented by the transition radiation tracker (TRT),
which enables radially extended track reconstruction up to \(|\eta| = 2.0\).
The TRT also provides electron identification information
based on the fraction of hits (typically 30 in total) above a higher energy-deposit threshold corresponding to transition radiation.

The calorimeter system covers the pseudorapidity range of \(|\eta| < 4.9\).
Within the region \(|\eta|< 3.2\), EM calorimetry is provided by barrel and
endcap high-granularity lead/liquid-argon (LAr) calorimeters,
with an additional thin LAr presampler covering \(|\eta| < 1.8\)
to correct for energy loss in material upstream of the calorimeters.
Hadron calorimetry is provided by the steel/scintillator-tile calorimeter,
segmented into three barrel structures within \(|\eta| = 1.7\), and two copper/LAr hadron endcap calorimeters.
The solid angle coverage is completed with forward copper/LAr and tungsten/LAr calorimeter modules
optimised for EM and hadronic energy measurements, respectively.

The muon spectrometer (MS) comprises separate trigger and
high-precision tracking chambers measuring the deflection of muons in a magnetic field generated by the superconducting air-core toroidal magnets.
The field integral of the toroids ranges between \num{2.0} and \qty{6.0}{\tesla\metre}
across most of the detector.
Three layers of precision chambers cover the region \(|\eta| < 2.7\). They consist of layers of monitored drift tubes,
complemented by cathode-strip chambers in the forward region, where the background is highest.
The muon trigger system covers the range of \(|\eta| < 2.4\) with resistive-plate chambers in the barrel, and thin-gap chambers in the endcap regions.

The ALFA detector~\cite{ALFA} is a specific part of the ATLAS experiment designed to measure the trajectory of elastically scattered protons during dedicated runs with special LHC optics.
Because the elastic scattering typically leads to deviations in the proton trajectory by very small angles, these detectors are placed close to the beam and far from the IP\@.
Two stations with scintillating fibre detectors are placed on either side of the central ATLAS detector,
located at distances of $\pm$237 m (inner stations) and $\pm$245 m (outer stations) from the IP\@.
The detectors are housed in `Roman pots' (RPs), an upper one and a lower one, which are movable and can approach the circulating beam in the vertical direction to within 1~mm.

The ATLAS forward proton (AFP) spectrometer~\cite{ATLAS-TDR-24} is designed to measure protons emerging intact from the interactions with significant energy loss, for example, from photon-induced $pp$ interactions.
The AFP system consists of four tracking units located along the beam pipe at $\pm$205 m and $\pm$217 m from the IP, referred to as near and far stations, respectively.
Each station houses a silicon tracker comprising four planes of edgeless silicon pixel sensors.
Movable RPs at each station insert the tracker along the $x$ direction in the beam pipe.
Data taking with the AFP commences once the trackers are at a position where the innermost silicon edge is within 2~mm of the beam centre during stable beams.

The ATLAS zero-degree calorimeters (ZDC) consist of four longitudinal compartments on each side of the IP, each with one nuclear interaction length of tungsten absorber, with the Cherenkov light read out by 1.5~mm quartz rods.
The detectors are located 140 m from the IP in both directions, covering $|\eta|>8.3$.
They detect neutral particles such as neutrons emitted from interacting nuclei.

The ATLAS minimum-bias trigger scintillators (MBTS) consist of scintillator slats positioned between the ID and the endcap calorimeters, with each side having an outer ring of four slats segmented in azimuthal angle, covering $2.07 < |\eta| < 2.76$, and
an inner ring of eight slats, covering $2.76 < |\eta| < 3.86$.

The ATLAS LUCID-2 detector~\cite{Avoni:2018iuv} consists of 32 photomultiplier tubes
for luminosity measurements and luminosity monitoring.
Its two modules are placed symmetrically at about $\pm$17~m from the IP.

Interesting events are selected by the first-level trigger system (L1) implemented in custom hardware,
followed by selections made by algorithms implemented in software in the high-level trigger (HLT)~\cite{TRIG-2016-01}.
The first-level trigger reduces the rate of accepted events from the \qty{40}{\MHz} bunch crossing rate to below \qty{100}{\kHz},
which the high-level trigger further reduces to record events to disk at about \qty{1}{\kHz}.

\begin{figure}[!b]
\begin{center}
\subfloat[]{\includegraphics[width=0.52\textwidth]{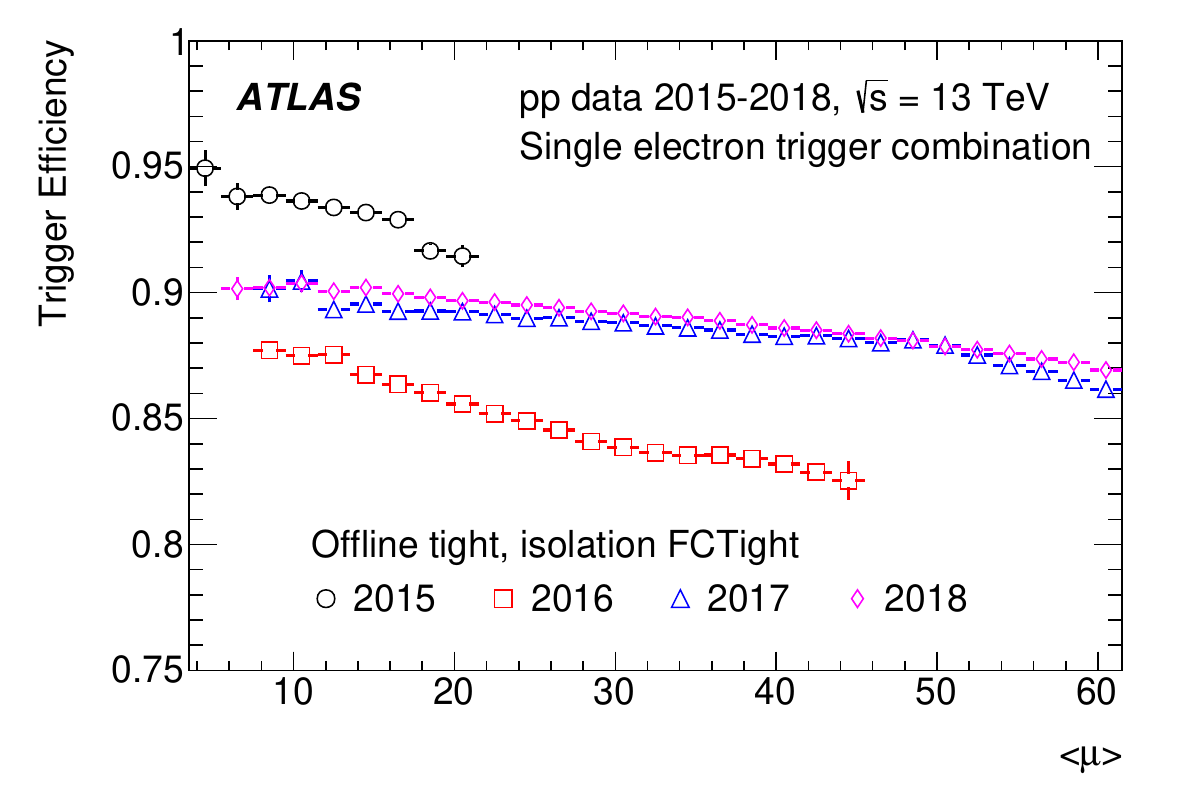}}
\subfloat[]{\includegraphics[width=0.48\textwidth]{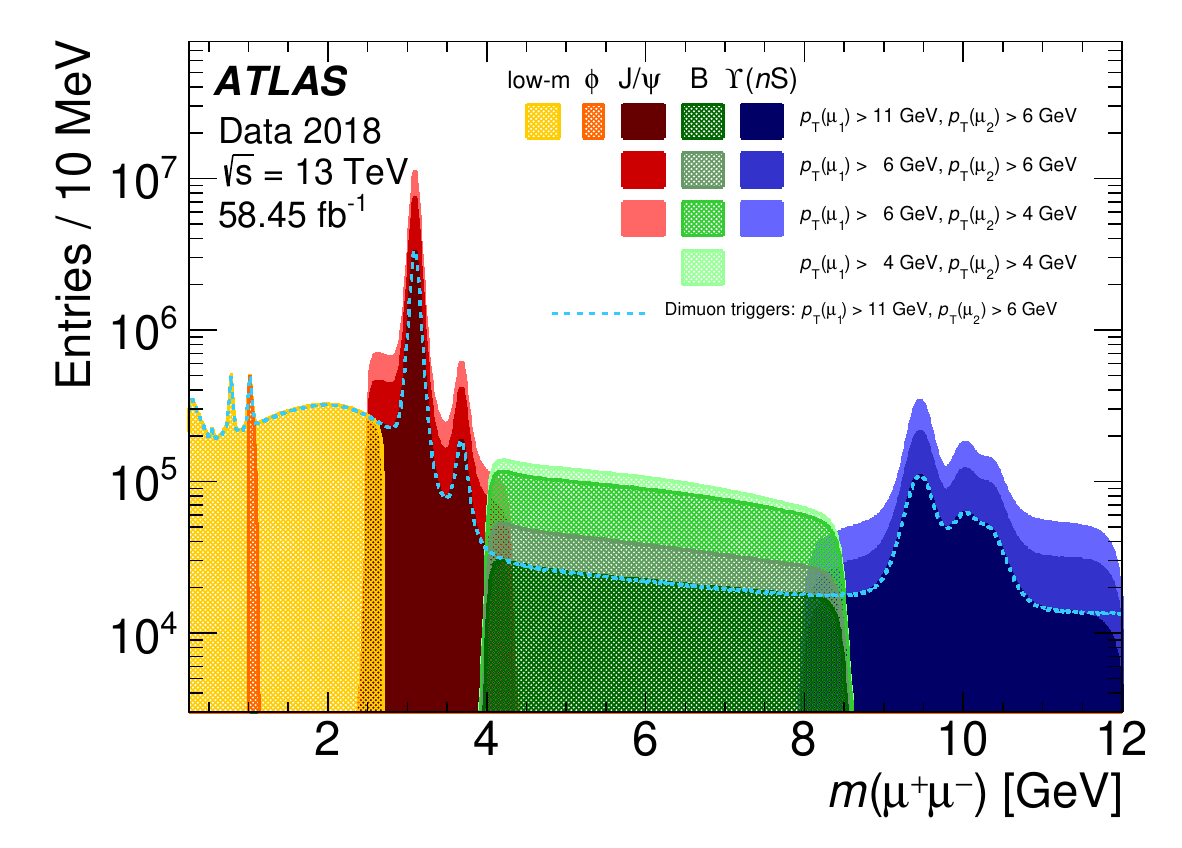}}
\end{center}
\caption{
(a) Evolution of the single-electron trigger efficiency as a function of the pile-up during Run~2~\cite{TRIG-2018-05}.
(b) Distribution of the offline dimuon invariant mass for events collected by various dimuon triggers corresponding to different mass ranges (shown in different colours) and different muon \pT thresholds (different shades) in 2018 data taking~\cite{TRIG-2018-01}. The dashed line represents the events collected by the lowest unprescaled dimuon trigger that is inclusive of the full mass range of interest.}
\label{fig:BphysTriggerDimuMass}
\end{figure}

Most of the analyses described in this report use events recorded with single-lepton (electron or muon), single-photon or single-jet triggers~\cite{TRIG-2018-05, TRIG-2018-01, TRIG-2019-04, TRIG-2019-03}.
Figure~\ref{fig:BphysTriggerDimuMass}(a) shows the evolution of the single-electron trigger efficiency as a function of pile-up during Run~2. The trigger efficiency was almost independent of the pile-up towards the end of Run~2.

Some measurements make use of dilepton and diphoton trigger configurations, benefiting from lower \pT thresholds compared to the corresponding thresholds of single-object triggers.
In particular, the $B$~hadron physics programme of ATLAS is mostly based on events triggered by the presence of two muons at L1 that are subsequently reconstructed in the HLT and successfully fit to a common vertex.
Starting from late 2016, a new topological processor was introduced, allowing for a selection based on
various kinematic properties of L1 objects to be applied.
To reduce the L1 dimuon trigger rates, the two triggering L1 muon objects were required to satisfy both $\Delta R$ and invariant mass criteria.
With those improvements the \pT thresholds on muons in such triggers
were maintained mostly at the level of 4--6\,\GeV during Run~2 (with the lowest-threshold running typically at the end of LHC fills when the instantaneous luminosity drops sufficiently).
Certain analyses still gain much of their sensitivity from earlier data where most of events were collected with the
triggers having 4\,\GeV threshold for both the muons.
Figure~\ref{fig:BphysTriggerDimuMass}(b)
shows the dimuon invariant mass distribution for events collected by various triggers of this type.
To further reduce the dimuon trigger rate at HLT and to achieve as low a muon \pT as possible,
some triggers used the information about other ID tracks to reconstruct the
full final states of particular $B$ hadron decays, such as $B^0_s\to\Jpsi(\mu^+\mu^-)\phi(K^+K^-)$.

An extensive software suite~\cite{ATL-SOFT-PUB-2021-001} is used in data simulation, in the reconstruction
and analysis of real and simulated data, in detector operations, and in the trigger and data acquisition
systems of the experiment.
To cope with a fourfold increase of the peak LHC luminosity from 2015 to 2018, and a similar increase in the number of interactions per beam-crossing to about 60, trigger and offline reconstruction algorithms were optimised to control the rates and retain a high efficiency for physics analyses.

\section{Run 2 detector performance}
\label{sec:perf}
\begin{figure}[!b]
\begin{center}
\subfloat[]{\includegraphics[width=0.48\textwidth]{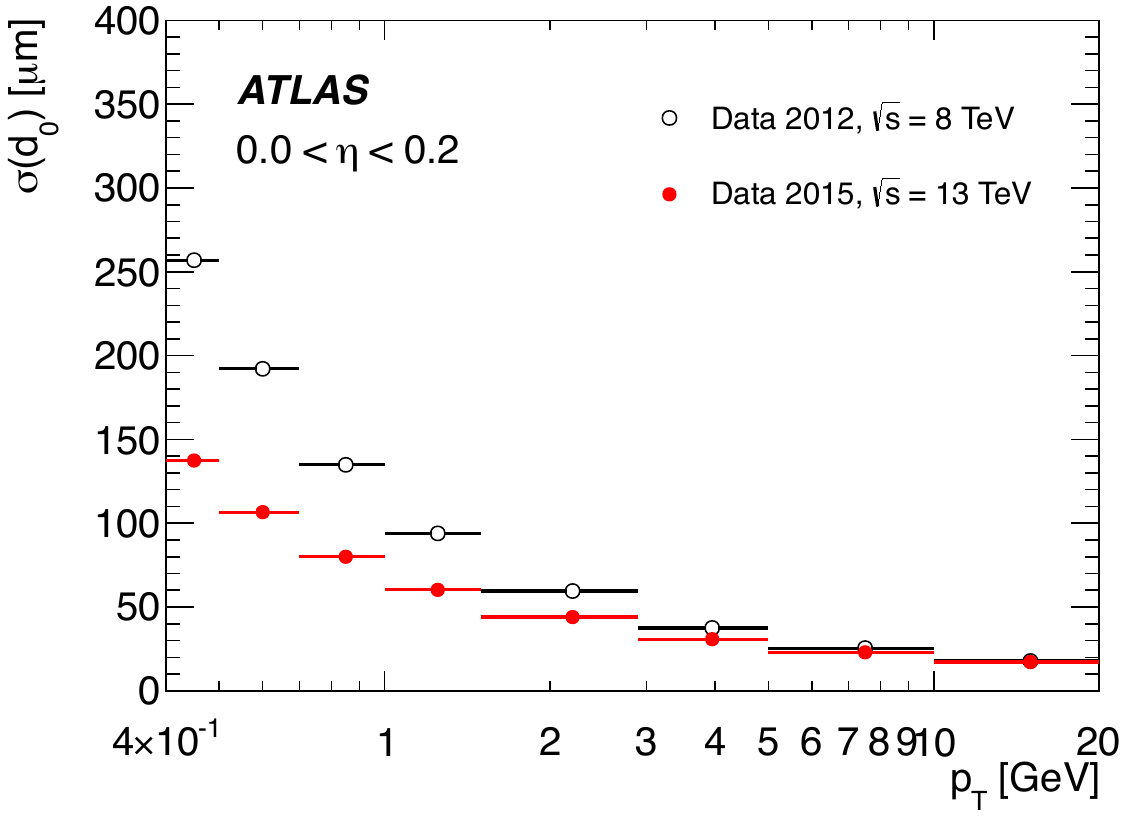}}
\subfloat[]{\includegraphics[width=0.48\textwidth]{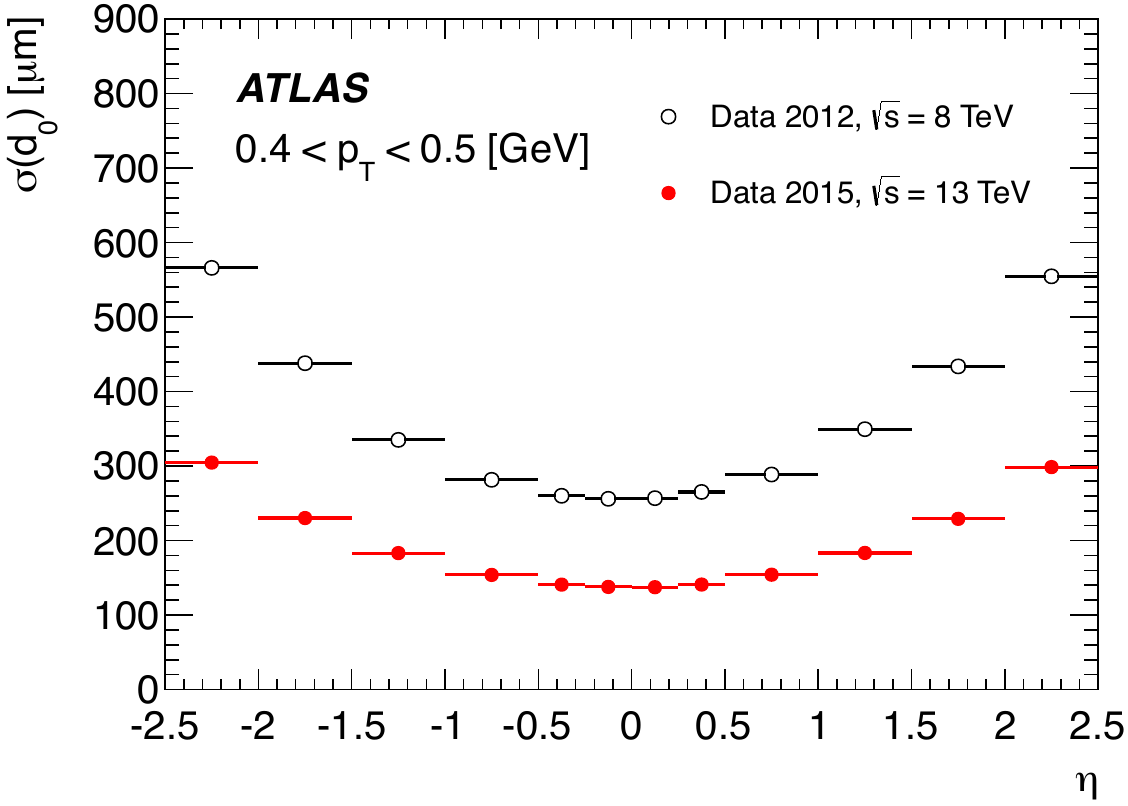}}
\end{center}
\caption{Transverse impact parameter resolution for reconstructed charged-particle tracks measured  in 2015 and 2012 data as a function of (a) track \pT and (b) track $\eta$~\cite{ATLAS:2023dns}.  }
\label{fig:IDTranResolution}
\end{figure}

Several upgrades were made to the ATLAS detector between Run~1 and Run~2. A major improvement of the ID system was the installation of a fourth pixel layer, the IBL~\cite{ATLAS-TDR-19,PIX-2018-001}, together with a new beam pipe in 2014.
The IBL provides a hit measurement at an average radius of 33.3 mm, significantly closer to the interaction point than the closest pixel layer in Run 1 (radius of 50.5 mm).
It improves significantly the track and vertex reconstruction performance at higher instantaneous luminosities during Run~2 and mitigates the impact of radiation damage to the previous innermost layer of the pixel detector, resulting in improved tagging of jets containing $b$-hadrons ($b$-tagging), $\tau$-lepton identification,  and reconstruction of inclusive and exclusive $b$- and $c$-hadron decays.
The improvement in reconstructing the transverse impact parameter of charged-particle tracks, defined as the shortest distance between a track and the beam line in the transverse plane, is shown in Figure~\ref{fig:IDTranResolution}.

In addition, the reconstruction and calibration of physics objects in ATLAS benefited from several improvements made prior to or during Run~2.
Electrons and photons  are reconstructed in ATLAS from clusters of energy deposits in the EM calorimeter cells~\cite{EGAM-2018-01}. Electrons are additionally required to have a matching track reconstructed in the ID\@.
The identification of electrons and photons was revisited in Run~2 to capitalise on the improved cell clustering procedure.
Muons are identified using information from various parts of the detector, the ID, the MS, and the
calorimeters~\cite{MUON-2018-03}.
The performance of the electron, photon and muon reconstruction and identification algorithms was improved to be almost insensitive to the harshening data-taking conditions with increasing pile-up.

\begin{figure}[!t]
\begin{center}
\subfloat[]{\includegraphics[width=0.52\textwidth]{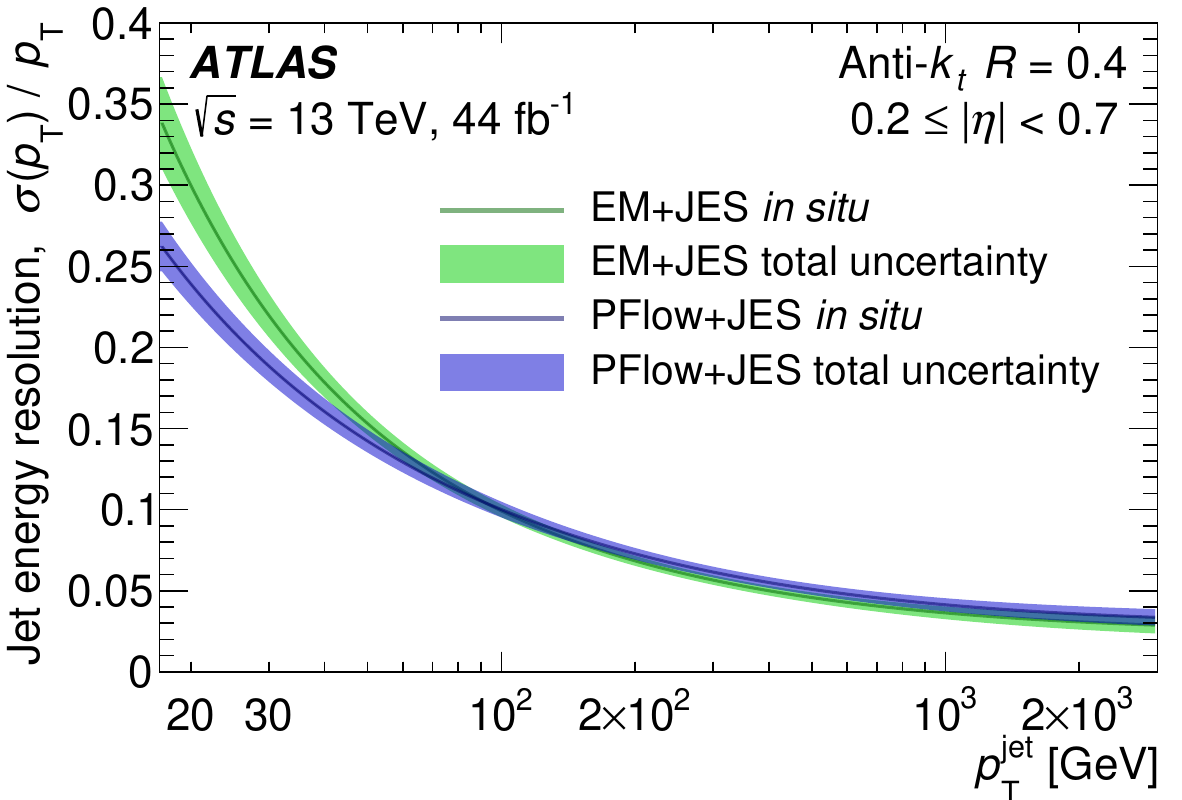}}
\subfloat[]{\includegraphics[width=0.47\textwidth]{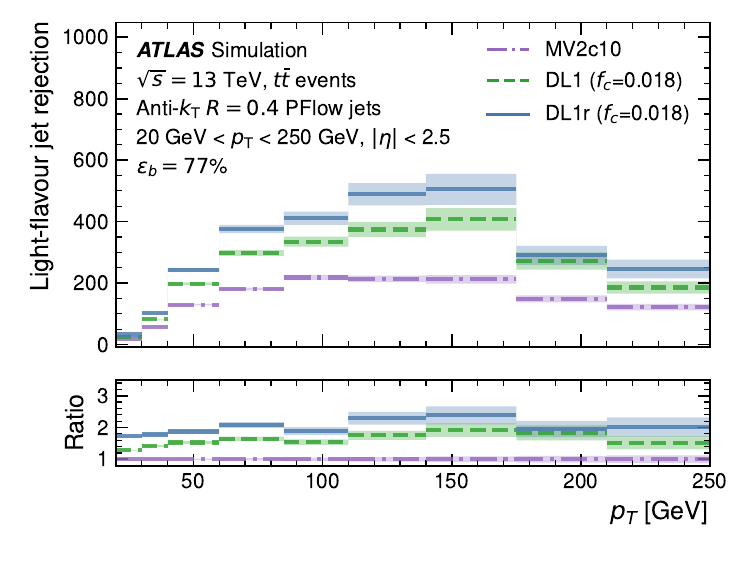}}
\end{center}
\caption{
(a) Comparison of the relative jet energy resolution between fully calibrated particle-flow jets (PFlow+JES) and jets reconstructed using only calorimeter-based energy information (EM+JES) as a function of jet \pT~\cite{JETM-2018-05}.
(b) The factor of light-flavour jet rejection for a given $b$-tagging algorithm at a fixed $b$-tagging efficiency of 77\% as a function of jet \pT for several high-level taggers: DL1r, DL1 (based on recurrent and deep neural networks), and MV2c10 (based on boosted decision trees)~\cite{FTAG-2019-07}. The lower panel shows the ratios of the taggers to MV2c10.}
\label{fig:jet_perf}
\end{figure}

Jets in ATLAS are reconstructed using two different input types: topo-clusters formed from energy deposits in calorimeter cells~\cite{PERF-2014-07}, and an algorithmic combination of charged-particle tracks with those topo-clusters, referred to as the ATLAS particle-flow reconstruction method~\cite{PERF-2015-09}.
Figure~\ref{fig:jet_perf}(a) provides a comparison of the relative jet energy resolution for particle-flow jets and jets reconstructed using only calorimeter-based energy information. The latter was the primary jet definition used in ATLAS physics results by the end of Run~2. The resulting improvement in the jet energy resolution at low \pT is clearly visible.
Similarly, systematic uncertainties in the jet energy scale (JES) can reach a sub-percent level for a large range of high-\pT jets~\cite{JETM-2018-05}.

The $b$-jet identification combines the results of several low-level algorithms with multivariate classifiers into high-level algorithms. The low-level algorithms either exploit the large impact parameters of the tracks originating from the $b$-hadron decay products or attempt to directly reconstruct heavy-flavour hadron decay vertices.
The analysis of the data from Run~2 of the LHC is marked by improvements and retuning of the low-level algorithms, first introduced during Run~1, but also by the introduction of new low- and high-level algorithms respectively based on recurrent and deep neural networks (NNs)~\cite{FTAG-2019-07}.
This yields considerable improvements over previous work~\cite{PERF-2012-04}, which was based on boosted decision trees or likelihood discriminants, as shown in Figure~\ref{fig:jet_perf}(b).

The luminosity determination for the ATLAS detector uses the absolute luminosity scale determined using van der Meer beam separation scans during dedicated LHC fills for each data-taking period, which is extrapolated to the physics data-taking regime using complementary measurements from several luminosity-sensitive detectors~\cite{DAPR-2021-01}.
The resulting total uncertainties in the integrated luminosities for each data-taking period of LHC Run 2, including dedicated runs with reduced instantaneous luminosity and Pb+Pb runs, range from 0.9\% to 2.1\%.

%
%
%
%
%
%
%
%


%


\section{Total, elastic and inelastic $pp$ cross-section measurements}
\label{sec:totalxs}
The total cross-section for $pp$ interactions ($\sigma_{\textrm{tot}}$) characterises a fundamental process of the strong interaction. Its energy evolution has been studied at each new range of centre-of-mass energies available.
Measurements of $\sigma_{\textrm{tot}}$ give unique experimental access to non-perturbative
dynamics, which cannot be calculated from first principles.

The total cross-section at the LHC is measured via elastic scattering using the optical theorem~\cite{Pancheri:2016yel}:
\begin{equation}
\sigma_{\textrm{tot}} = 4\pi \, \textrm{Im} \, [ f_{\textrm{el}}\left(t)]\right|_{t\rightarrow 0} ,
\label{eq:OpticalTheorem}
\end{equation}
which relates $\sigma_{\textrm{tot}}$ to the elastic-scattering amplitude extrapolated to the forward direction $f_{\textrm{el}}(t)$, with $t$ being the four-momentum transfer squared.
The total cross-section can be extracted in different ways using the optical theorem.
ATLAS uses the luminosity-dependent method that requires a measurement of the luminosity to normalise the elastic cross-section, $\sigma_{\textrm{el}}$. With this method, $\sigma_{\textrm{tot}}$ is given by the formula:
\begin{equation}
\sigma_{\textrm{tot}} = \frac{16\pi}{1+\rho^2} \left. \frac{\textrm{d}\sigma_{\textrm{el}}}{\textrm{d}t} \right|_{t\rightarrow 0}~,
\label{eq:totalcrosssect}
\end{equation}
where $\rho$ represents a small correction arising from the ratio of the real to the imaginary part of the elastic-scattering amplitude in the forward direction.
The $\rho$-parameter is sensitive not only to the high-energy evolution of the total hadronic cross-section but also to the
fundamental structure of the elastic-scattering amplitude. Traditionally, the elastic-scattering amplitude at
energies well above 100~\GeV is believed to be dominated by the $t$-channel Pomeron exchange (see e.g.\ Ref.~\cite{compete}). In QCD the Pomeron is
represented by a two-gluon colourless state with spin--parity--charge quantum numbers J$^{\textrm{PC}} = 0^{++}$. The additional possible presence of a
three-gluon colourless state with J$^{\textrm{PC}} = 1^{--}$, the `Odderon'~\cite{Lukaszuk:1973nt}, can also influence the value of
the $\rho$-parameter. Thus, measurements of the $\rho$-parameter at the highest energy of
the LHC are essential.

ATLAS previously reported a measurement of $\sigma_{\textrm{el}}$ and consequently $\sigma_{\textrm{tot}}$ at 7 and $8~\TeV$~\cite{STDM-2013-10,STDM-2015-22}.
The measurements were performed with the ALFA sub-detector of ATLAS\@.
However, those measurements did not extend to the region of very small $|t|$-values where the differential cross-section is sensitive to the  $\rho$-parameter. Such  small $|t|$-values require measurements of angles in the microradian range, which in turn need
even smaller divergence  of the beam at the IP.

A new ATLAS measurement of $\sigma_{\textrm{tot}}$ using $pp$ collision data at $\sqrt{s}= 13$~\TeV, corresponding to an integrated luminosity of 340~$\upmu$b$^{-1}$~\cite{STDM-2018-08} extends $|t|$ by an order of magnitude lower compared to previous ATLAS results. For the first time, the ATLAS measurement reaches the region of small scattering angles where the Coulomb interaction plays an important role.
The necessarily small divergence of the beam at the IP is achieved by using very high-$\beta^*$ optics\footnote{The $\beta$-function determines the variation of
the beam envelope around the LHC ring and depends on the focusing properties of the magnetic lattice.} ($\beta^*=2.5$~km), producing a large beam
spot size but very small beam divergence.
From a fit to the differential elastic cross-section,
the total cross-section and $\rho$-parameter are determined to be:
\begin{equation*}
\sigma_{\textrm{tot}}(pp\rightarrow X) =  \mbox{104.7} \pm 1.1 \; \mbox{mb} , \; \; \;
\rho =  \mbox{0.098} \pm 0.011 .
\end{equation*}
The new ATLAS measurement of $\rho$ is compatible within uncertainties with the recent TOTEM measurement~\cite{TOTEM_2p5km}, but the TOTEM value of the total cross-section is about 5\% higher,
which corresponds to approximately two standard deviation ($\sigma$) tension assuming uncorrelated uncertainties. A similar difference was already
observed at 7 and $8~\TeV$~\cite{STDM-2013-10,STDM-2015-22}. The difference has been traced back to the normalisation of the differential elastic cross-section as measured by ATLAS and TOTEM.

\begin{figure}[b!]
\begin{center}
\subfloat[]{\includegraphics[width=0.52\textwidth]{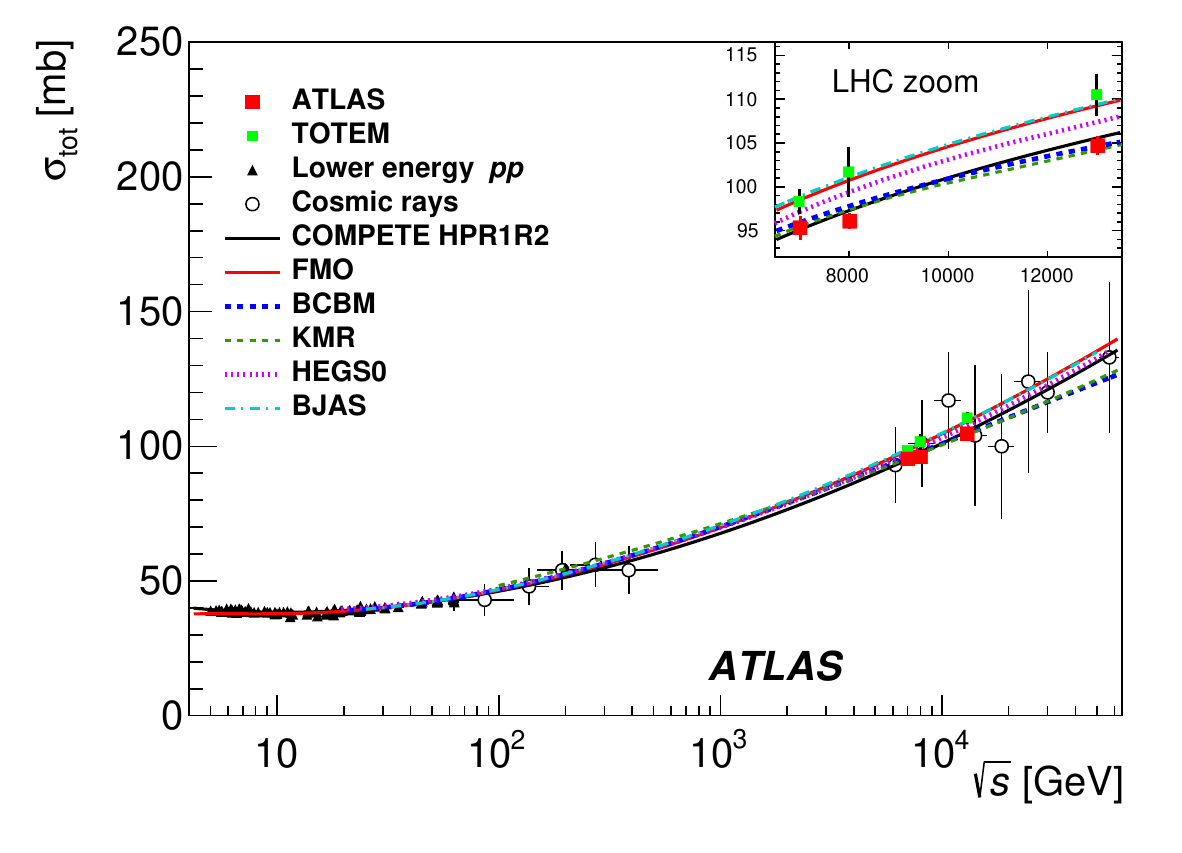}}
\subfloat[]{\includegraphics[width=0.52\textwidth]{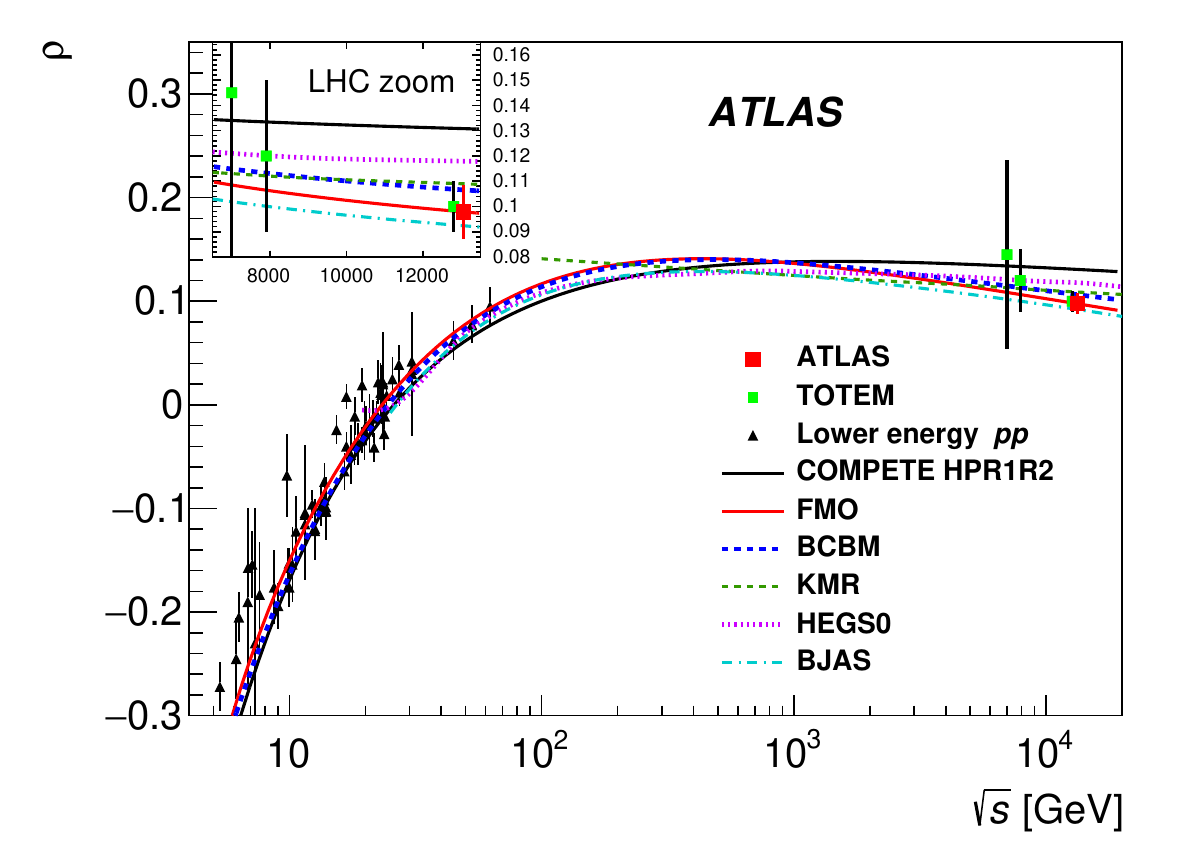}}
\end{center}
\caption{Measurements of (a) $\sigma_{\textrm{tot}}$ and (b) the $\rho$-parameter at different centre-of-mass energies compared with different model predictions~\cite{STDM-2018-08}.}
\label{fig:sigma_tot}
\end{figure}

The new data for $\sigma_{\textrm{tot}}$ and $\rho$ are compared with previous measurements (including lower-energy data), and the energy evolution of these data is analysed
in the context of model studies of the evolution in Figure~\ref{fig:sigma_tot}.
This study shows that the commonly accepted energy evolution as implemented in the COMPETE model~\cite{compete} is in tension with
the $13~\TeV$ elastic-scattering data.
Further research is needed to understand whether the low value of $\rho$ can be attributed to the Odderon, as suggested by the TOTEM+D0 observations~\cite{TOTEM_D0}, or other effects in strong interactions.

The ATLAS analysis of $\sigma_{\textrm{tot}}$ at 13~\TeV also measures the inelastic cross-section, using the relation $\sigma_{\textrm{inel}} = \sigma_{\textrm{tot}} - \sigma_{\textrm{el}}$.
The result is $\sigma_{\textrm{inel}}^{\textrm{ALFA}} = 77.4 \pm 1.1$~mb.
This result using ALFA proton spectrometers can be compared with the ATLAS measurement of the inelastic cross-section using two sets of scintillation counters in a data sample corresponding to an integrated luminosity of 60~$\upmu$b$^{-1}$ collected in 2015~\cite{STDM-2015-05}. In inelastic interactions, one or both protons dissociate as a result of coloured (non-diffractive) or colourless (diffractive) exchange. The counters are insensitive to elastic $pp$ scattering and diffractive dissociation processes in which neither proton dissociates into a system, $X$, of mass $m_X > 13$~\GeV.
The measurement is performed in such a fiducial region, and the result is extrapolated to the total inelastic cross-section using models of inelastic interactions: $\sigma_{\textrm{inel}}^{\textrm{MBTS}} = 78.1 \pm 2.8$~mb.
The two ATLAS measurements of $\sigma_{\textrm{inel}}$ and other LHC measurements at 13~\TeV~\cite{TOTEM:2017asr,LHCb:2018ehw} are compatible within uncertainties, while the ALFA measurement is the most precise of the four available LHC measurements.

\section{Production of charged particles in $pp$ interactions}
\label{sec:soft}
Measurements of charged-particle distributions in $pp$ collisions probe the strong interaction in the non-perturbative regime of QCD characterised by small momentum transfers.
In this region, charged-particle interactions are typically described by QCD-inspired models implemented in MC event generators and measurements are used to constrain the free parameters of these models.
An accurate description of low-energy strong interaction processes is, for example, essential for simulating single $pp$ interactions to estimate the effects of
pile-up at high instantaneous luminosity in hadron colliders.

\subsection{Charged-particle distributions}

Inclusive measurements of primary charged particles with $\pt > 500$~\MeV in $pp$ collisions at $\sqrt{s} = 13$~\TeV, using data  corresponding to an integrated luminosity of approximately 170~$\upmu$b$^{-1}$ are performed by ATLAS~\cite{STDM-2015-02}. A follow-up ATLAS analysis~\cite{STDM-2015-17} extends the measurements to particles with $\pt > 100$~\MeV. While this nearly doubles the overall number of particles in the kinematic acceptance, the measurements are rendered more difficult due to multiple scatterings and imprecise knowledge of the material in the detector.
The results are defined only by the final state and include all processes in $pp$ interactions and no attempt is made to correct for certain types of process such as diffraction.
Corrections for detector effects are made to present these measurements as distributions of primary charged particles in a well-defined fiducial phase space region:
events are required to have at least one primary charged particle with $\pt > 500$~\MeV, or two with $\pt > 100$~\MeV, and absolute pseudorapidity $|\eta| < 2.5$ to be within the geometrical acceptance of the tracking detector.

\begin{figure}[!b]
\begin{center}
\subfloat[]{\includegraphics[width=0.5\textwidth]{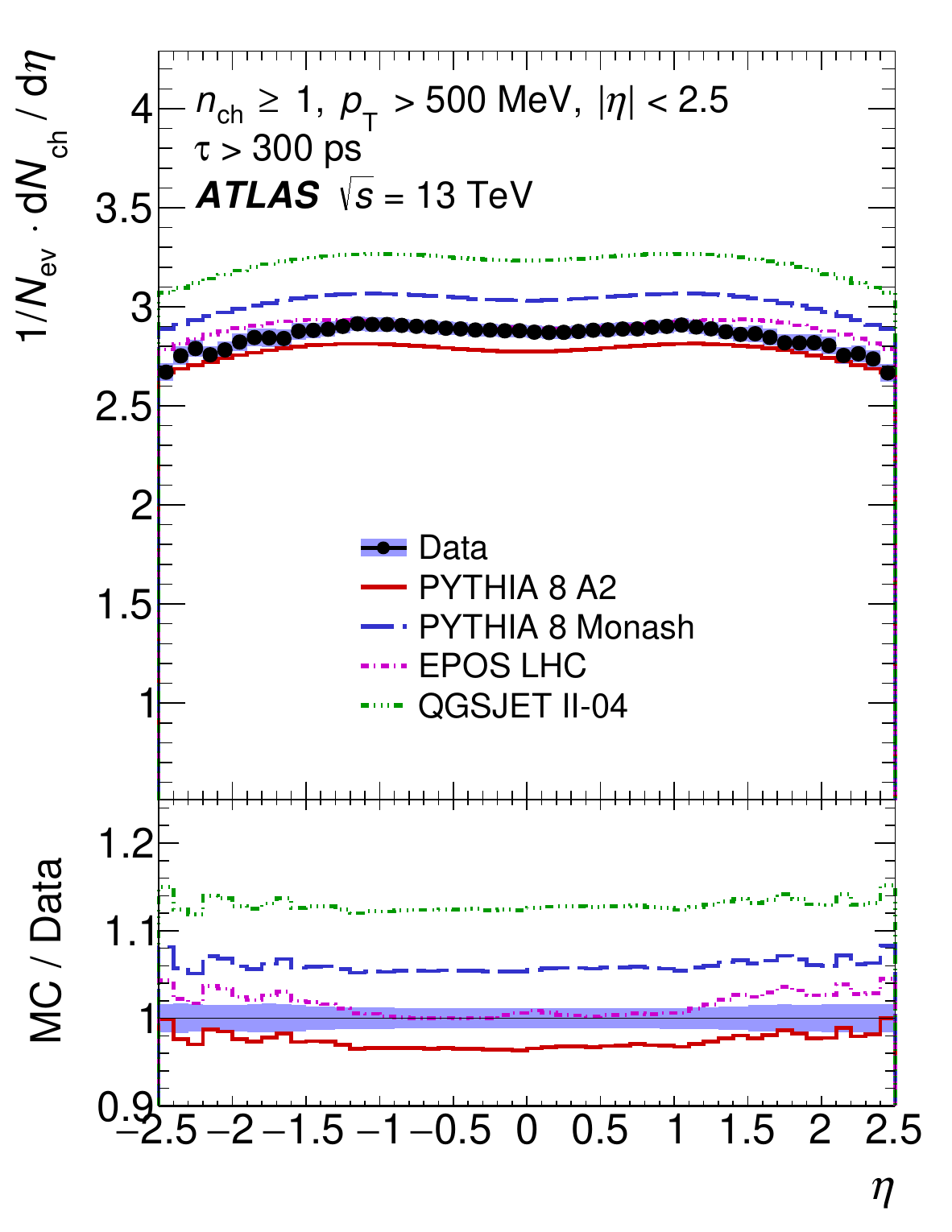}}
\subfloat[]{\includegraphics[width=0.5\textwidth]{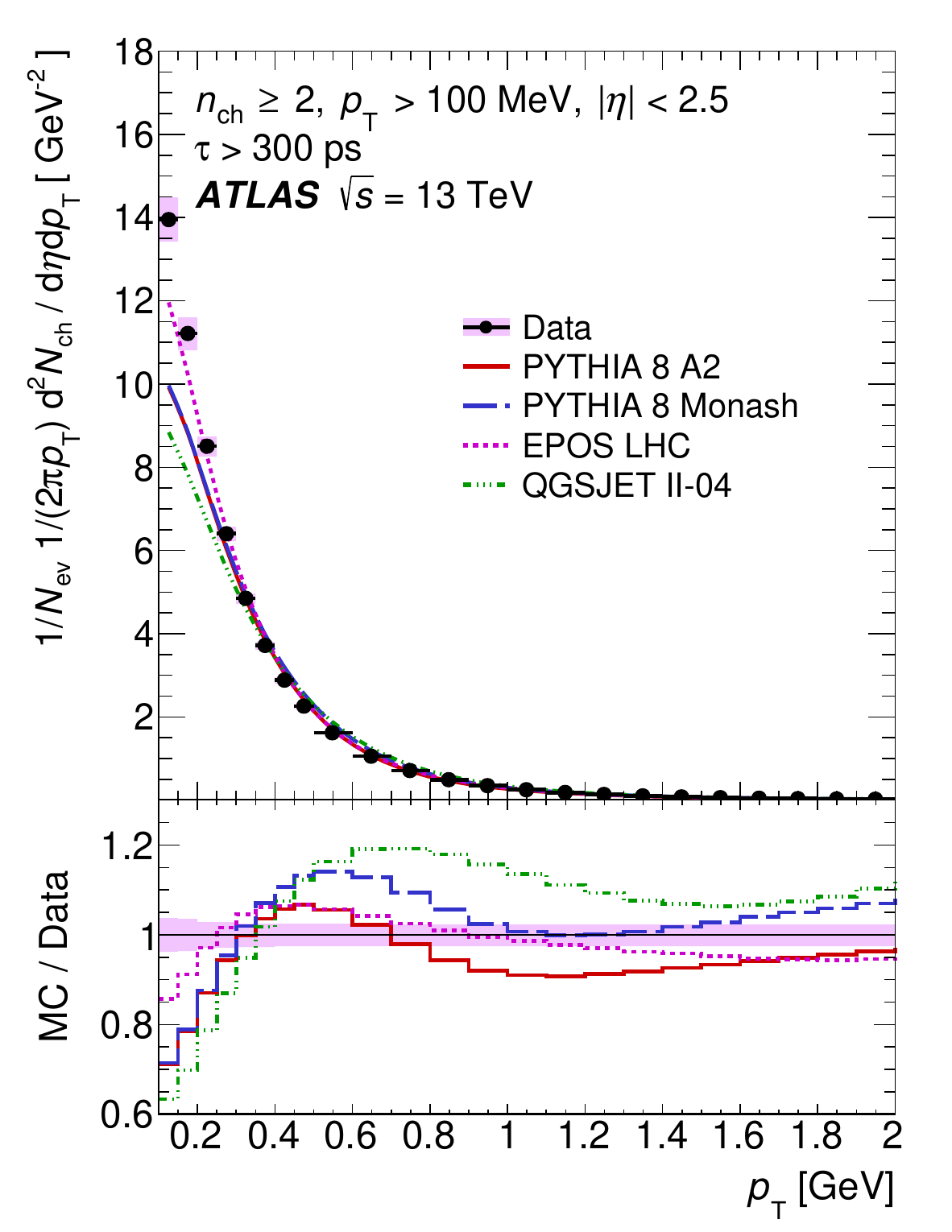}}
\end{center}
\caption{(a) Primary charged-particle multiplicities as a function of pseudorapidity in events with at least one primary charged particle with $\pt > 500$~\MeV and $|\eta| < 2.5$~\cite{STDM-2015-02}.
(b) Primary charged-particle multiplicities as a function of transverse momentum in events with at least two primary charged particles with $\pt > 100$~\MeV and $|\eta| < 2.5$~\cite{STDM-2015-17}.
The dots represent the data and the curves the different MC model predictions. The lower panels show the ratios of the predictions to the data.}
\label{fig:charged_part}
\end{figure}

The measured charged-particle multiplicities are shown in Figure~\ref{fig:charged_part}. The data are compared with predictions from various MC generators. The results highlight clear differences between MC models and the measured distributions.
Among the models considered, EPOS~\cite{Pierog:2013ria} reproduces the data the best, \textsc{Pythia}~8~\cite{Sjostrand:2007gs}  give reasonable descriptions of the data and \textsc{Qgsjet-ii}~\cite{Ostapchenko:2010vb} provides the worst description of the data.

\subsection{Underlying event studies}

A typical `hard' $pp$ collision studied at the LHC consists of a short-distance process and accompanying activity collectively termed the underlying event (UE).
Mechanisms that produce the UE include partons not participating in the hard-scattering process (beam remnants), radiation processes and additional hard and semi-hard scatters in the same $pp$ collision, termed multiple parton interactions (MPI).
Phenomenological models are required to describe these processes using several free parameters determined from experiment.

It is impossible to uniquely separate the UE from the hard scattering process on an event-by-event basis, but observables can be defined that are particularly sensitive to the properties of the UE\@. Typically, an object with high transverse momentum such as a $Z$ boson or the leading \pt{} charged-particle is identified. The UE activity is then characterised relative to the scale of the momentum transfer in the hard interaction and the azimuthal distribution of energy and particle flow.

\begin{figure}[!b]
\begin{center}
\subfloat[]{\includegraphics[width=0.25\textwidth]{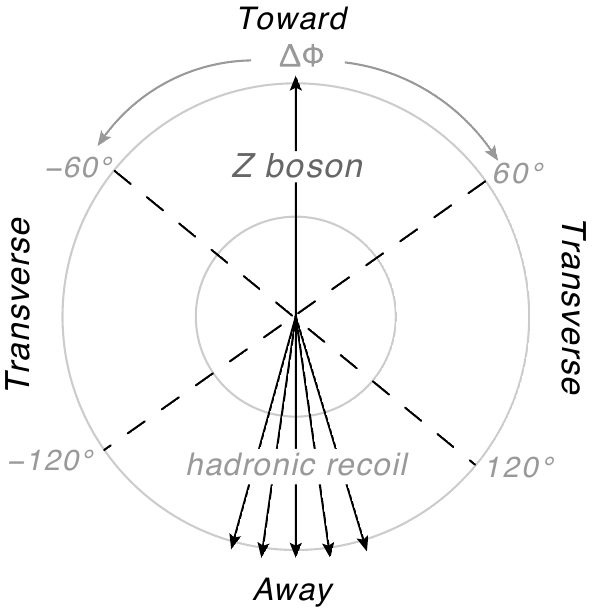}}
\hspace{0.5in}
\subfloat[]{\includegraphics[width=0.5\textwidth]{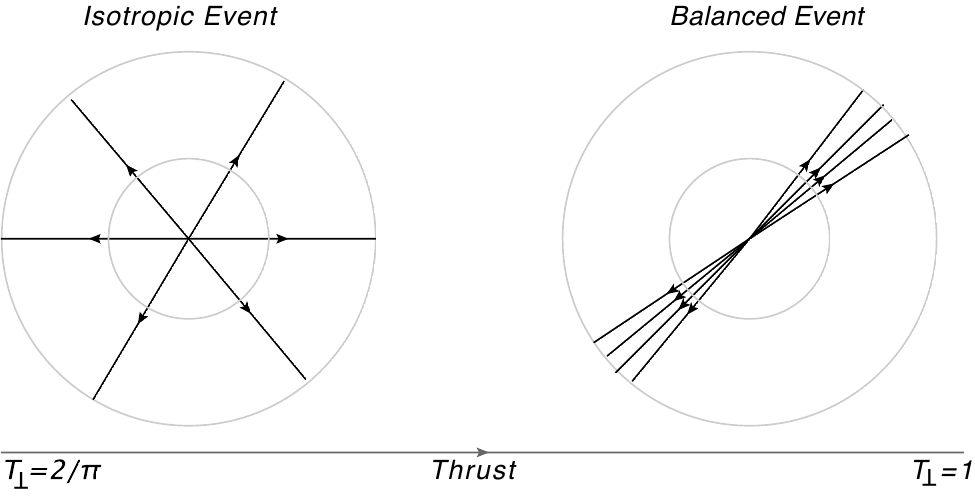}}
\end{center}
\caption{(a) Illustration of away, transverse, and towards regions in the transverse plane defined relative to the direction of a high transverse momentum object (a $Z$ boson, for example). (b) Illustration of an isotropic and a balanced event topology in the transverse plane with their corresponding values of thrust.}
\label{fig:ue_scheme}
\end{figure}

The ATLAS measurements of UE activity at $\sqrt{s} = 13$~\TeV exploit distributions constructed using charged particles with $|\eta| < 2.5$ and with $\pt > 500$~\MeV, in events with at least one such charged particle with transverse momentum above 1~\GeV~\cite{STDM-2016-07}, or in events containing two muons originating from the decay of a singly produced $Z$ boson~\cite{STDM-2017-28}.
These measurements use the established form of UE observables, in which the azimuthal plane of the event is segmented into several distinct regions with differing sensitivities to the UE (Figure~\ref{fig:ue_scheme}).
In particular, the two transverse regions, defined relative to the leading particle (either the $Z$ boson or the highest \pT track), are differentiated on an event-by-event basis by their scalar sum of charged-particle \pT. The one with the larger sum is labelled trans-max and the other trans-min. The trans-min region is most sensitive to the UE activity because it contains less activity from hard jets.
Several distributions are studied to understand the UE activity, including mean densities of charged-particle multiplicity and the mean scalar \pT sum of charged particles per unit $\eta$--$\phi$.

The topology of the tracks in the event can be further characterised by the transverse thrust
\begin{equation} %
T_{\perp} = \frac{ \sum_i | \vec{p}_{\textrm{T},i} \cdot \hat{n} | }{ \sum_i | \vec{p}_{\textrm{T},i} | }~,
\label{thrust}
\end{equation}
where the thrust axis $\hat{n}$ is the unit vector that maximizes $T_{\perp}$.
The transverse thrust has a
maximum value of one for a back-to-back dijet topology and a minimum value of $2/\pi$ for a circularly symmetric distribution of particles in the transverse plane, as illustrated in Figure~\ref{fig:ue_scheme}.

\begin{figure}[!t]
\begin{center}
\subfloat[]{\includegraphics[width=0.44\textwidth]{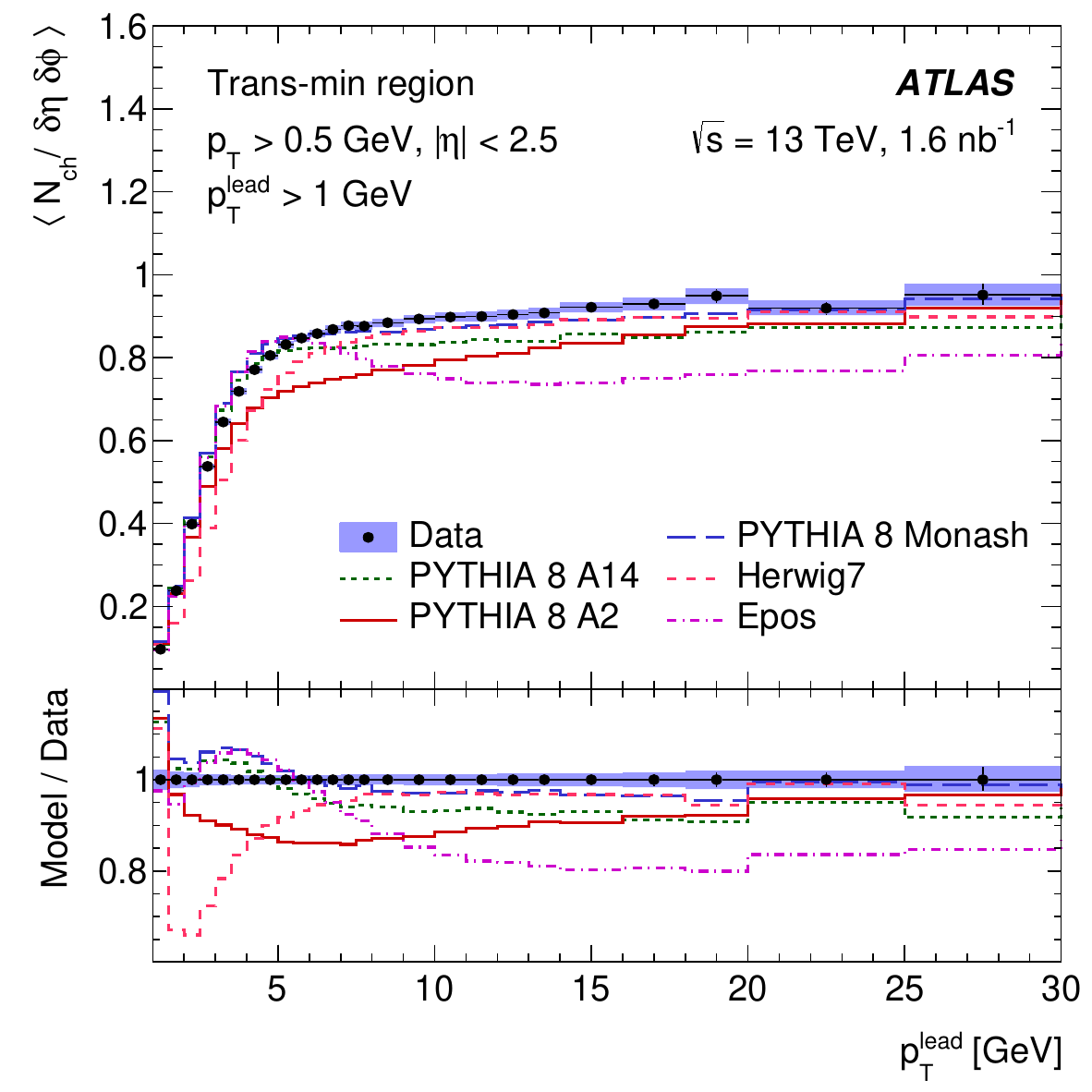}}
\hspace{0.5in}
\subfloat[]{\includegraphics[width=0.46\textwidth]{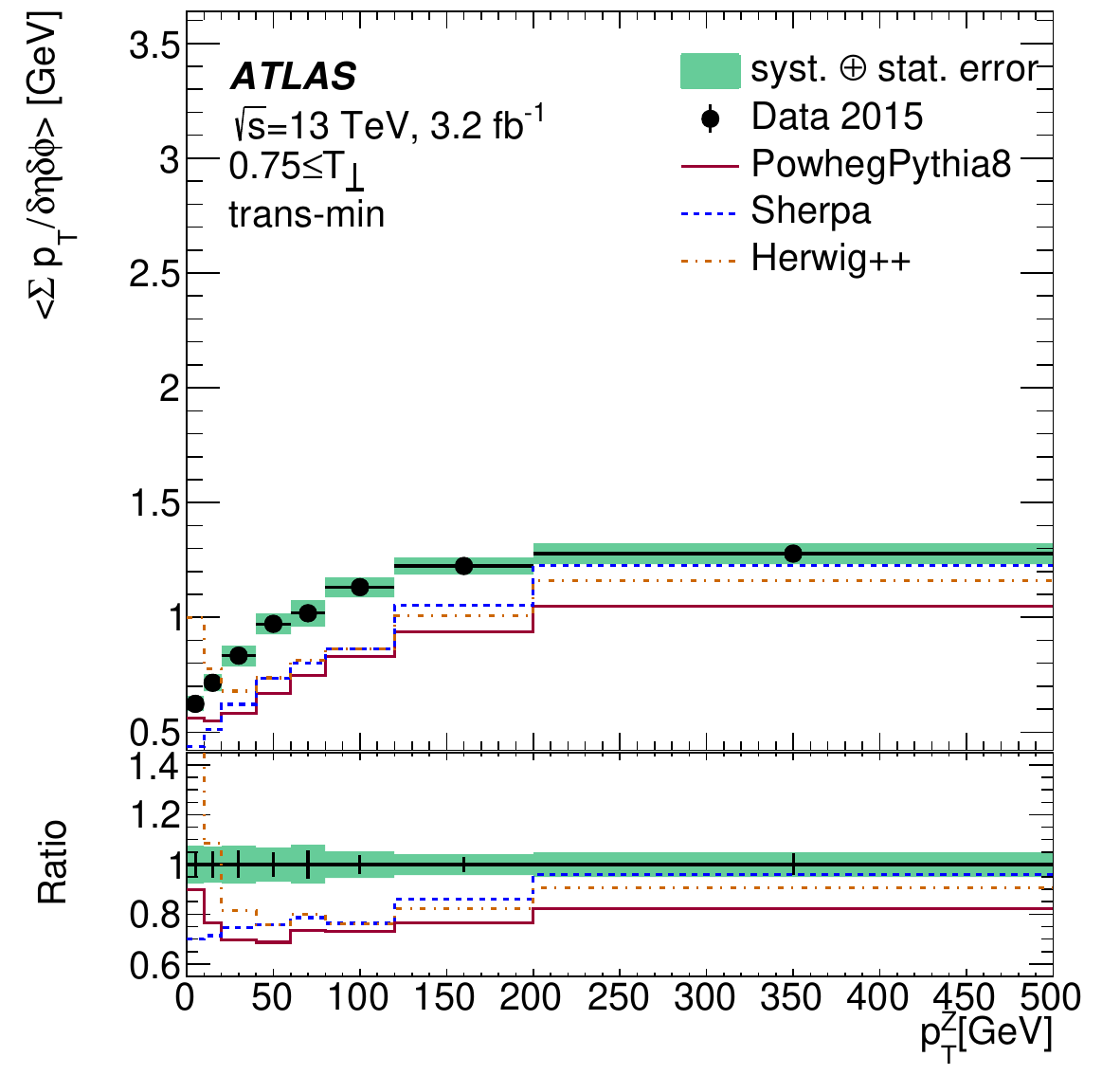}}
\end{center}
\caption{(a) Mean density of charged-particle multiplicity as a function of leading charged-particle \pT in the trans-min  azimuthal region~\cite{STDM-2016-07}. (b) Mean scalar \pT sum of charged particles as a function of the $Z$ boson \pT for $T_{\perp}\geq 0.75$ in the trans-min azimuthal region~\cite{STDM-2017-28}. The lower panels show the ratios of the predictions to the data.}
\label{fig:ue_results}
\end{figure}

Examples of measured UE distributions are shown in Figure~\ref{fig:ue_results}.
The prominent features are a turn-on effect, i.e., the rising activity as a function of the hard-scatter scale (here the $Z$ boson \pT or leading charged particle \pT), and a saturation of the activity at higher values of \pT.
Comparisons with predictions
from several commonly used MC generator configurations indicate that for most observables the models
show significant deviations from the data distributions regardless of the observable.
In particular,  events with higher values of $T_{\perp}$ show that the simulation of contributions other than MPI to the UE activity needs to be improved.

\section{Inclusive production of jets}
\label{sec:jets}
Precise measurements of processes with jets are crucial in understanding physics at hadron colliders. In QCD, jets are interpreted as resulting from the fragmentation of quarks and gluons produced in a short-distance hard scattering process.
Jet cross-sections provide valuable information about the strong coupling constant, $\alphas$, and the structure of the proton.
In addition, jet formation is a complex multi-scale problem, including important contributions from QCD effects that cannot be described by perturbation theory alone.
In the measurements described below, jets are identified using the anti-$k_t$ algorithm~\cite{Cacciari:2008gp,Cacciari:2011ma} with a radius parameter value of $R=0.4$, unless stated otherwise.

\subsection{Inclusive jet and dijet cross-section measurements}

\begin{figure}[h!]
\begin{center}
\subfloat[]{\includegraphics[width=0.48\textwidth]{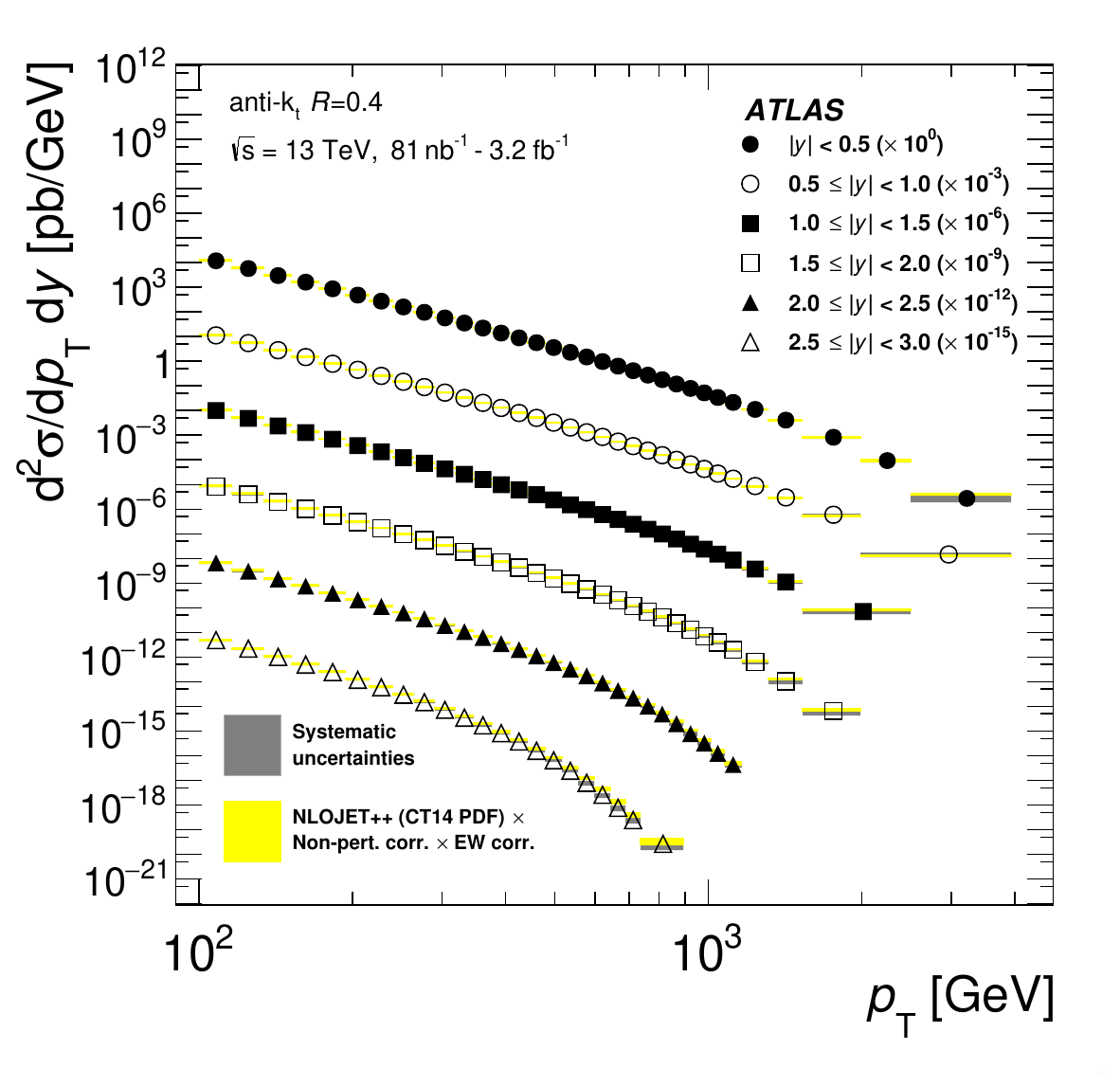}}
\subfloat[]{\includegraphics[width=0.48\textwidth]{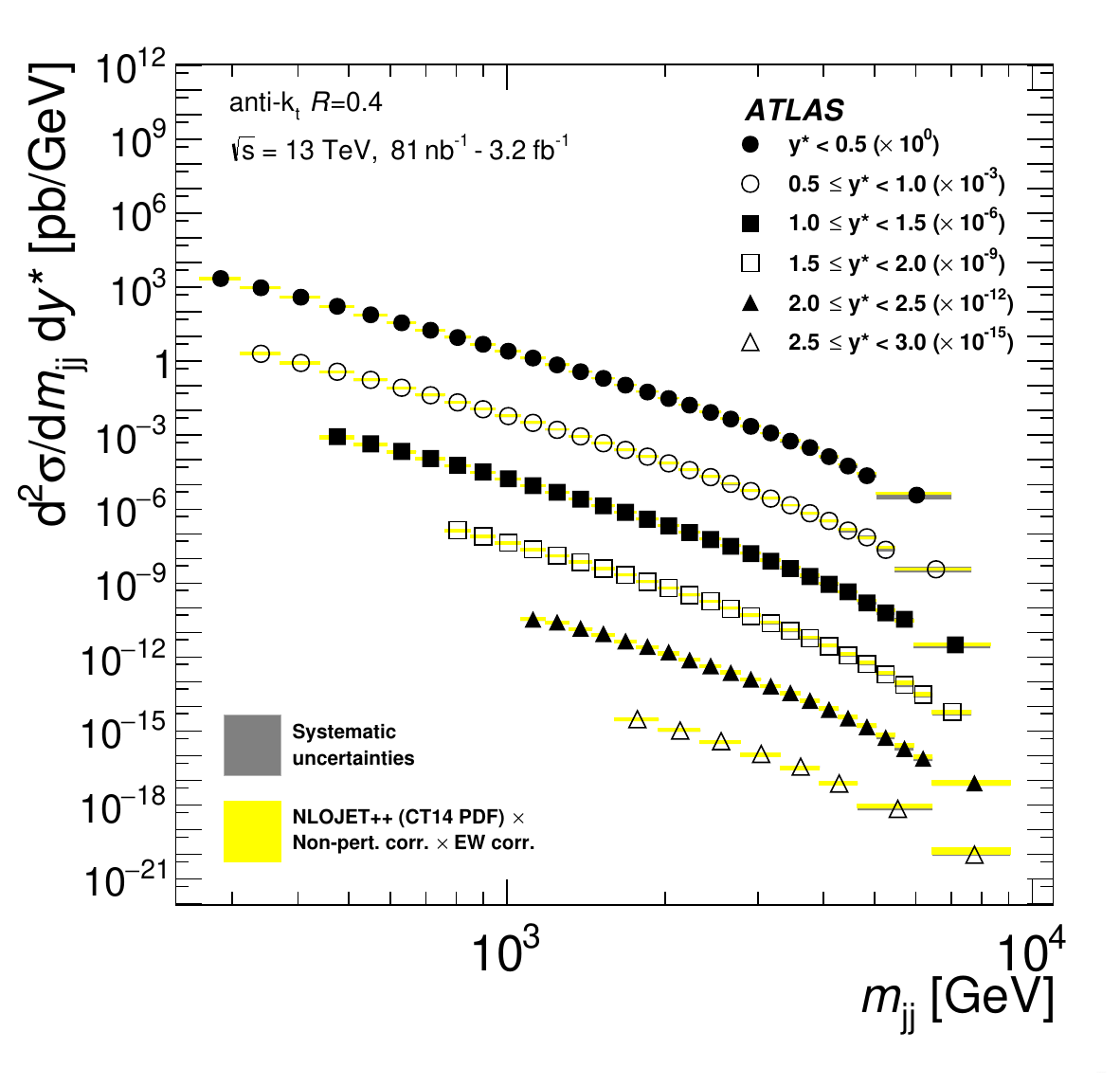}}\\
\subfloat[]{\includegraphics[width=0.75\textwidth]{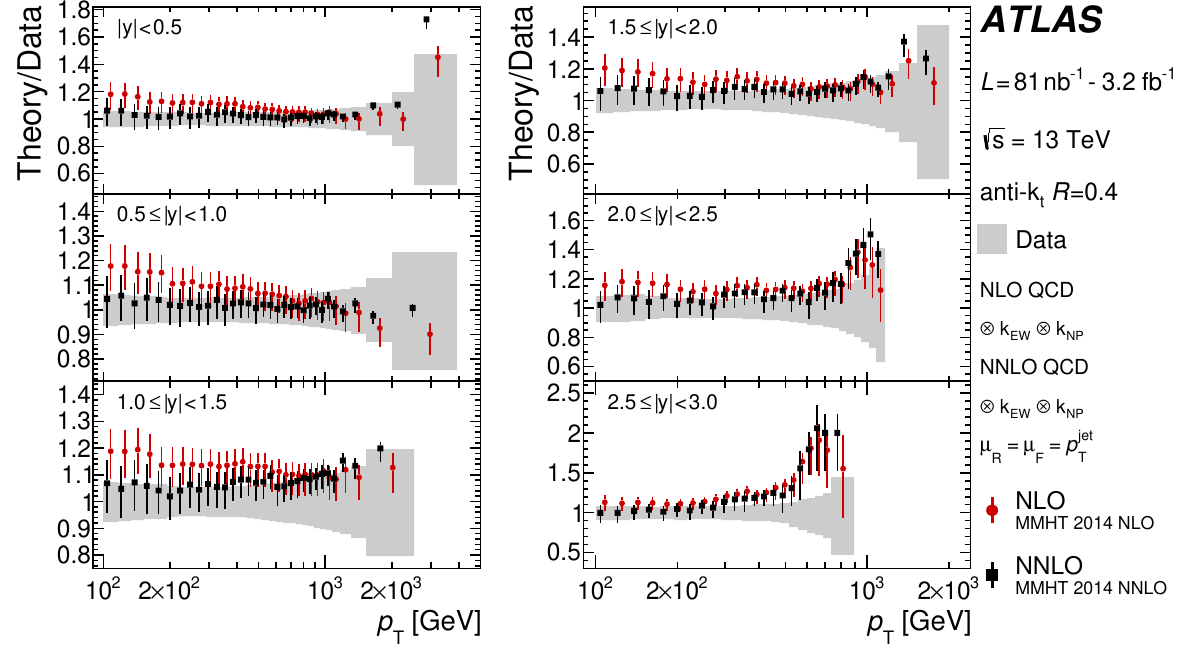}}
\end{center}
\caption{The measured (a) inclusive jet and (b) inclusive dijet cross-sections at $\sqrt{s} = 13$~\TeV, shown as a function of the jet transverse momentum or dijet invariant mass in several jet rapidity bins~\cite{STDM-2016-03}. (c) The ratios of NLO and NNLO pQCD predictions~\cite{Nagy:2003tz, Currie:2016bfm, Dittmaier:2012kx} to the measured inclusive jet cross-sections.
The theory uncertainties are shown by the lines and the shaded bands show the total data uncertainty including both the systematic and statistical uncertainties. }
\label{fig:jets}
\end{figure}

Inclusive jet and dijet cross-sections are measured in $pp$ collisions at $\sqrt{s} = 13$~\TeV in ATLAS~\cite{STDM-2016-03}. The measurements use a data sample with an integrated luminosity of 3.2 fb$^{-1}$ recorded in 2015.
The inclusive jet cross-sections are measured double-differentially as a function of the jet transverse momentum and rapidity.
The double-differential dijet production cross-sections are presented as a function of the dijet mass and the half absolute rapidity separation between the two leading jets.
Figure~\ref{fig:jets} shows the measured inclusive jet and dijet cross-sections and the corresponding ratios of the predictions to the data for the inclusive jet measurement.
Overall, fair agreement
between the measured cross-sections (that span several orders of magnitude) and the fixed-order NNLO pQCD calculations, corrected for non-perturbative and electroweak effects, is observed.
For example, in the case of jet cross-sections in individual jet rapidity bins independently, the $p$-values are in the percent range.
However, when
considering data points from all jet transverse momentum and rapidity regions in the inclusive jet
measurement, a significant tension between data and theory is observed.
Resolving this tension requires a good understanding of the correlations of the experimental and theoretical systematic uncertainties in jet \pT and rapidity.
A related jet measurement is performed by the CMS Collaboration~\cite{CMS-SMP-20-011}.

\subsection{Event shapes and azimuthal correlations in multijet events}
\label{sec:evshapes}

Event shapes are a class of observables that describe the dynamics of energy flow in multijet final states.
These observables are sensitive to different aspects of the theoretical description of these strong-interaction processes. They are defined to be infrared (soft and/or collinear) safe, which enables their calculation in pQCD\@.
They can therefore be used to precisely test pQCD calculations and additionally to extract the value of $\alphas$.
Hard, wide-angle radiation is studied by investigating the tails of the event-shape distributions. These configurations are sensitive to higher-order corrections to the dijet cross-section. Other regions of the event-shape distributions provide information about anisotropic, back-to-back configurations, which are sensitive to the details of the resummation of soft logarithms in the theoretical predictions.

\begin{figure}[!t]
\begin{center}
\includegraphics[width=1.0\textwidth]{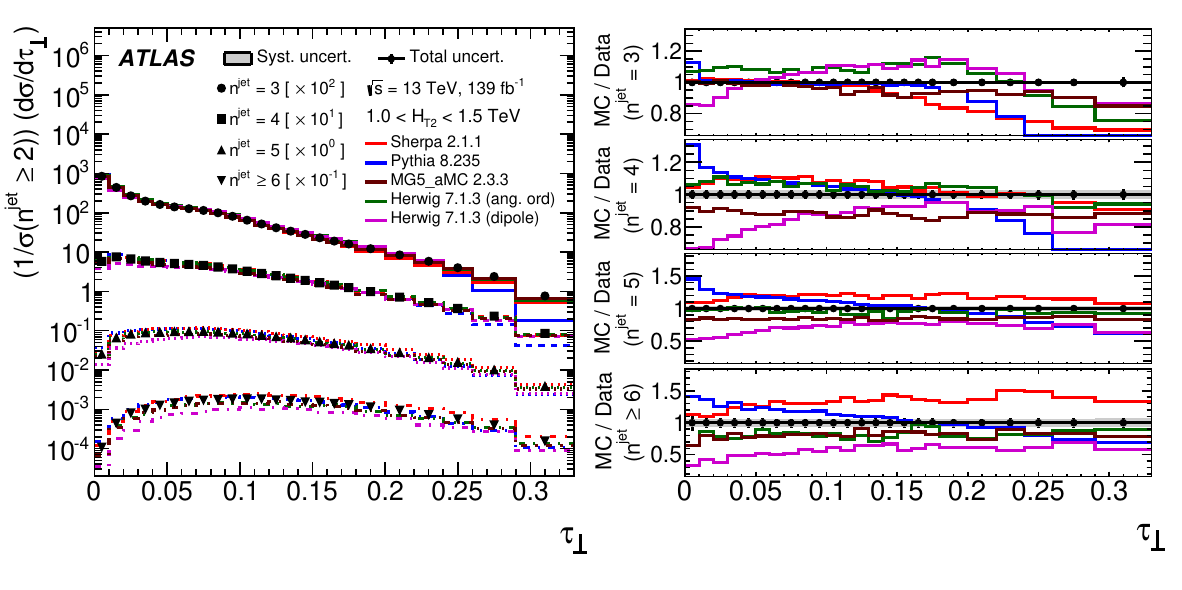}
\end{center}
\caption{Comparison between data~\cite{STDM-2019-02} and MC simulation~\cite{herwig2, Sjostrand:2014zea, Gleisberg:2008ta, Alwall_2014} as a function of the transverse thrust $ \tau_{\perp} = 1- T_{\perp}$ for different jet multiplicities. The panels on the right show the ratios between the MC and the data distributions. }
\label{fig:jet_shapes}
\end{figure}

Event-shape observables are measured in $pp$ collisions at the LHC by the ALICE, CMS and ATLAS Collaborations~\cite{ALICE:2012cor, CMS-SMP-12-022, CMS-EWK-11-021, STDM-2011-33, CMS-SMP-17-003}.
In the ATLAS study at $\sqrt{s} = 13$~\TeV, different event-shape variables are investigated to probe the properties of the multijet energy flow at the \TeV energy scale~\cite{STDM-2019-02}. The distributions of event-shape observables are normalised to the inclusive two-jet cross-section to reduce correlated experimental uncertainties.
Measurements are compared with fixed-order matrix elements matched to parton shower MC predictions.
An example of such an event-shape distribution, shown in Figure~\ref{fig:jet_shapes}, is the transverse thrust, $ \tau_{\perp} = 1- T_{\perp}$, where $T_{\perp}$ is defined according to Eq.~(\ref{thrust}).
Lower values of $ \tau_{\perp}$ indicate a back-to-back, `dijet-like' configuration, and higher values of $ \tau_{\perp}$ indicate a larger energy flow orthogonal to the thrust axis.
All the predictions qualitatively describe the main features of the data, but none of them gives a good description of all distributions within the experimental uncertainties. The discrepancies between data and all the MC samples investigated show that further refinement of the current MC predictions is needed to describe the data in some regions, particularly at high jet multiplicities. Moreover, these discrepancies show that these data provide a powerful testing ground for the understanding of the strong interaction at high energies.

A particularly interesting event-shape observable is the transverse energy--energy correlation (TEEC) function, defined as the transverse-energy-weighted distribution of the azimuthal differences between jet pairs in the final state, i.e.,
\[
\frac{1}{\sigma}\frac{\mathrm{d}\Sigma}{\mathrm{d}\cos\phi}
= \frac{1}{N} \sum_{n=1}^N \sum_{ij} \frac{E^n_{\mathrm{T}i}E^n_{\mathrm{T}j}}{\left(\sum_k E^n_{\mathrm{T}k}\right)^2}\delta(\cos \phi - \cos\varphi_{ij}),
\]
where the expression is valid for a sample of $N$ multijet events, labelled by the index $n$, and the indices $i$, $j$ and $k$ run over all jets in a given event. Here,
$\varphi_{ij}$ is the angle in the transverse plane between jet $i$ and jet $j$ and $\delta(x)$ is the Dirac delta function, which ensures $\phi = \varphi_{ij}$. The normalisation to the total dijet cross-section, $\sigma$, ensures that the integral of the TEEC function over $\cos\phi$ is unity.
The TEEC function is sensitive to gluon radiation and shows a clear dependence on the strong coupling constant.
The recent publication of the NNLO corrections to three-jet production in $pp$ collisions~\cite{mitov2022} provides an important improvement in the theoretical precision of predictions of these observables. In particular, the theoretical uncertainties due to the choice of the renormalisation and factorisation scales are significantly reduced as compared to NLO calculations.
This allows more precise tests of pQCD and an important reduction of the uncertainty in the determination of the $\alphas$.

\begin{figure}[!t]
\begin{center}
\includegraphics[width=0.7\textwidth]{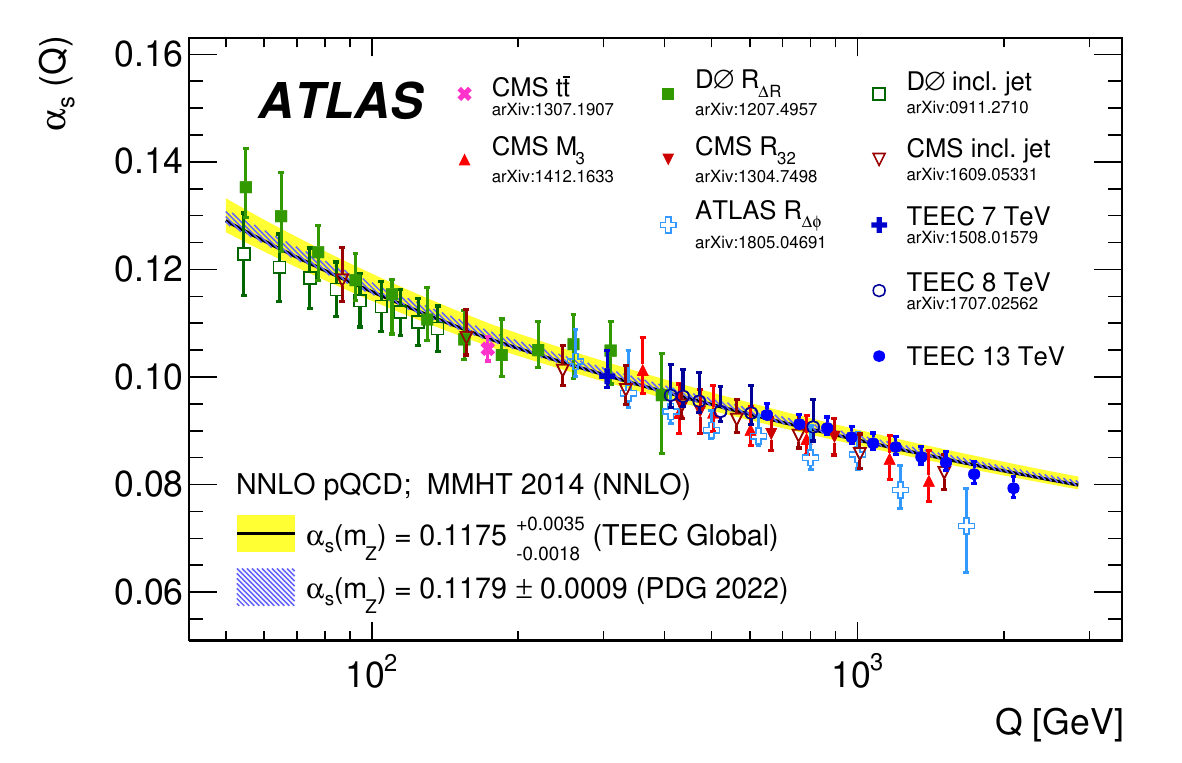}
\end{center}
\caption{Comparison of the values of $\alphas(Q)$ determined from fits to the TEEC functions with the QCD prediction using the world average as input (hatched band) and the value obtained from the global fit (solid band)~\cite{STDM-2018-51}. Results from previous analyses, both from ATLAS and from other experiments, are also included, showing an excellent agreement with the current measurements and with the world average.}
\label{fig:teec}
\end{figure}

The new ATLAS analysis of TEEC
performed at $\sqrt{s} = 13$~\TeV~\cite{STDM-2018-51} extends previous measurements~\cite{STDM-2014-03, STDM-2016-10} to higher energy scales $Q$ and improves the experimental precision.
High-energy multijet events are selected by requiring the scalar sum of \pT of the two leading jets, $H_{\textrm{T}2}$, to be above 1~\TeV, and the data are binned in this variable to study the scale dependence of these observables.
The agreement between data and NNLO pQCD predictions is good, thus providing a precision test of QCD at large $Q$.
A simultaneous fit to all TEEC distributions across different kinematic regions yields a value
$\alphas(m_Z) = 0.1175 \pm 0.0006~(\textrm{exp.})~^{+0.0034}_{-0.0017}~(\textrm{theo.})$.
Figure~\ref{fig:teec} presents $\alphas$ extracted from these fits differentially as a function of $Q$, showing a good agreement with the energy-scale dependence
of $\alphas$ predicted by the renormalisation group equation and with previous analyses.

A novel class of event shape observables was recently proposed to quantify the isotropy of collider events~\cite{Cesarotti:2020hwb}.
These observables, broadly called \emph{event isotropy}, measure how `far' a collider event is from a symmetric radiation pattern in terms of a Wasserstein distance metric~\cite{wasserstein1969markov}.
This distance is evaluated by solving optimal transport problems, using the `energy-mover's distance'~\cite{Komiske:2019fks}.
Event isotropies are shown to have increased sensitivity to isotropic multijet events when compared to other event shapes such as the transverse thrust.
They are capable of exposing a remote region of QCD phase space that is difficult to model and relevant to many searches for physics beyond the SM (BSM).

\begin{figure}[!t]
\begin{center}
\includegraphics[width=0.7\textwidth]{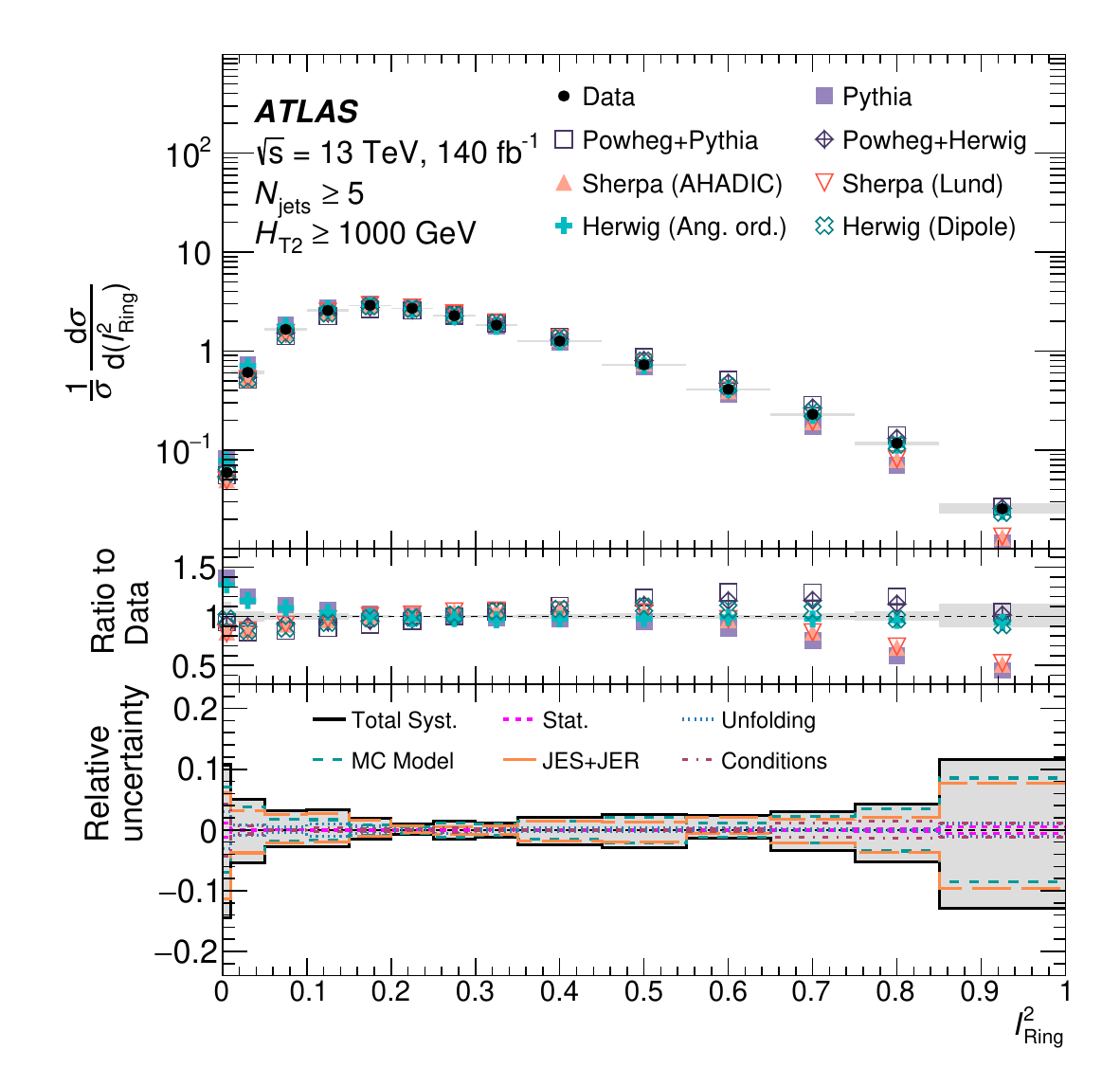}
\end{center}
\caption{The shape-normalised event-isotropy variable ($I^2_{\textrm{Ring}}$) cross-sections in data (closed circles)~\cite{STDM-2020-20}, compared with predictions from several MC generators~\cite{Sjostrand:2014zea, Bothmann:2019yzt, Chahal:2022rid, herwig2, Frixione_2007, Alioli_2010, Nason_2004}.
Events with $H_{\textrm{T}2}\geq 1$~\TeV and $N_{\textrm{jet}}\geq 5$ are presented.
The least isotropic dijet-like topology is near $I^2_{\textrm{Ring}}$ values of 0, and the most isotropic topology is near values of 1. The middle panel shows the ratio of the predictions to data while the bottom panel shows the relative uncertainty.
}
\label{fig:isotropy}
\end{figure}

ATLAS has measured cross-sections in multijet events at $\sqrt{s} = 13$~\TeV differentially relative to three event-isotropy observables in inclusive bins of jet multiplicity ($N_{\textrm{jet}}$) and $H_{\textrm{T}2}$~\cite{STDM-2020-20}.
The measured data are compared with the predictions of several state-of-the-art MC event
generators.
Figure~\ref{fig:isotropy} shows an example event isotropy variable measured by ATLAS in the region of $H_{\textrm{T}2}\geq 1$~\TeV and $N_{\textrm{jet}}\geq 5$.
Overall, agreement between the unfolded data and the simulated events tends to be best in balanced,
dijet-like arrangements and deteriorates in more isotropic configurations.

\subsection{Properties of jet formation and structure}

The study of the internal structure of jets has become a very active area of research at the LHC\@.
The large difference between the energy scale of the hard-scattered parton and the measured final-state hadrons creates a wide phase space for jet fragmentation processes.
To fully probe different regions of this phase space, a multitude of jet-substructure measurements is required.

\begin{figure}[t!]
\begin{center}
\subfloat[]{\includegraphics[width=0.5\textwidth]{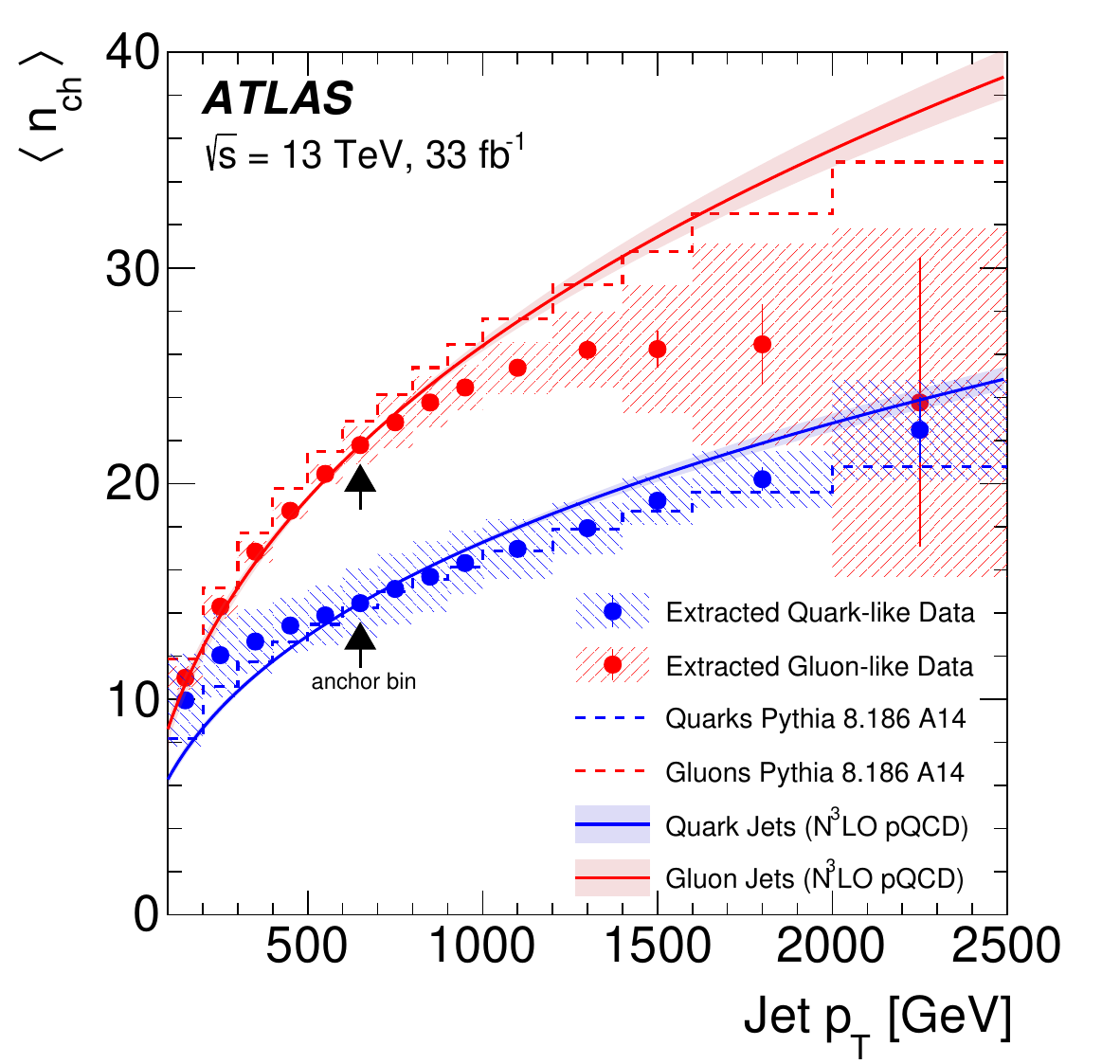}}
\subfloat[]{\includegraphics[width=0.45\textwidth]{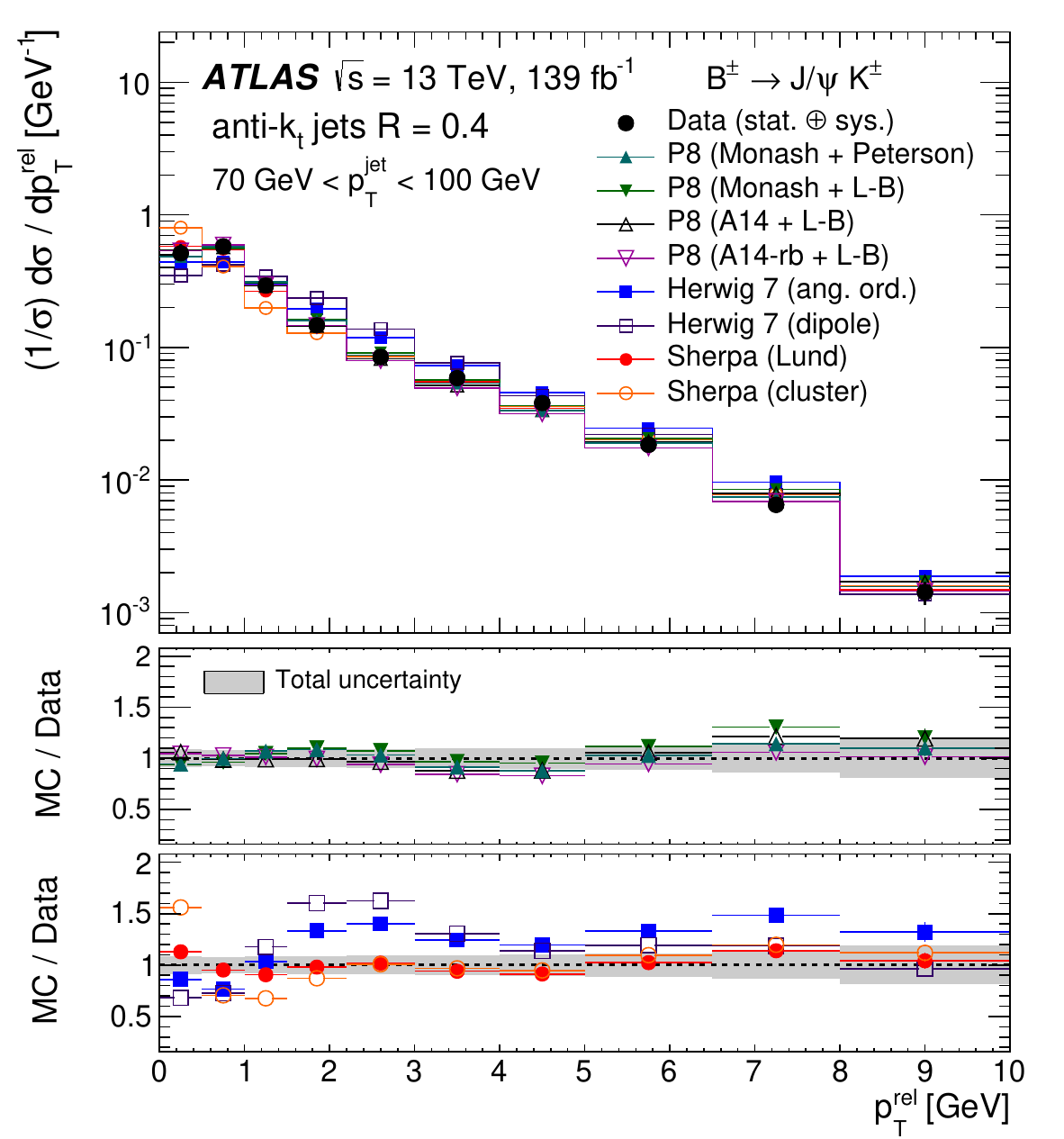}}
\end{center}
\caption{
(a) The dependence on jet transverse momentum of the mean charged-particle multiplicity for quark and gluon jets in data and in \textsc{Pythia}~8~\cite{Sjostrand:2007gs}, as well as from a calculation using pQCD~\cite{Capella:1999ms}. The calculation cannot predict the overall normalisation and therefore the prediction is normalised to the data in the sixth jet \pT bin, called the anchor bin and indicated by an arrow~\cite{STDM-2017-16}.
(b) Distribution of the
transverse momentum profile for $B$ mesons inside $b$-jets relative to the $b$-jet axis~\cite{STDM-2018-52},
together with different predictions from parton shower MC models~\cite{Sjostrand:2014zea, Bothmann:2019yzt, herwig2}. The lower panels show the ratios of the predictions to the data.}
\label{fig:jet_str}
\end{figure}

Basic properties of track-based jet fragmentation functions in $pp$ collisions at $\sqrt{s} = 13$~\TeV are measured by ATLAS~\cite{STDM-2017-16}.
Multiple jet properties, including the charged-particle multiplicity, the momentum fraction carried by charged particles, and angular properties of the radiation pattern inside jets are studied.
The forward and central jet spectra are considered separately to study distributions in quark- and gluon-induced jets, as presented in Figure~\ref{fig:jet_str}(a).
The simulations based on the \textsc{Pythia} fragmentation model provide a reasonable description of the quark-induced data across the jet \pt~range presented, but the gluon-induced jets have systematically fewer charged particles than the simulation.
In addition, measurement of the charged-particle multiplicity using model-independent jet labels (topic modelling)~\cite{Metodiev:2018ftz} provides a promising alternative to traditional extraction of quark- and gluon-induced jets using input from simulation.

In addition, ATLAS studies the fragmentation properties of jets containing $B$ mesons at $\sqrt{s} = 13$~\TeV~\cite{STDM-2018-52}.
The $B$ mesons are reconstructed using the decay of $B^{\pm}$ into $J/\psi K^{\pm}$, with the $J/\psi$ decaying into a pair of muons.
The measurement determines the longitudinal and transverse momentum profiles of the reconstructed $B$ mesons relative to the axes of the jets to which they are geometrically associated.
These distributions are measured in intervals of the jet transverse momentum, ranging from 50~\GeV to above 100~\GeV.
The results are compared with several MC predictions using different parton shower and hadronisation models. This is presented in Figure~\ref{fig:jet_str}(b).
Generally, the best description of the longitudinal profile is provided by the \textsc{Pythia}~8 and \textsc{Sherpa}~\cite{Gleisberg:2008ta} samples making use of the string hadronisation model, which provide similar descriptions for all values of the jet transverse momentum.

The observables sensitive to the fragmentation of $b$-quarks into $b$-hadrons are also measured in ATLAS from a sample of dileptonic top-quark pair (\ttbar) events~\cite{TOPQ-2017-19}.
The measurements provide a test of heavy-quark-fragmentation modelling at the LHC in a system where the top-quark decay products are colour-connected to the proton beam remnants.
The unfolded distributions (not shown) are compared with the predictions of several MC parton-shower generators and sets of tuned generator parameters (tunes).
The generators tuned to a combination of lepton- and hadron-collider measurements yield predictions that are found to agree with the observed data.

\begin{figure}[b!]
\begin{center}
\subfloat[]{\includegraphics[width=0.57\textwidth]{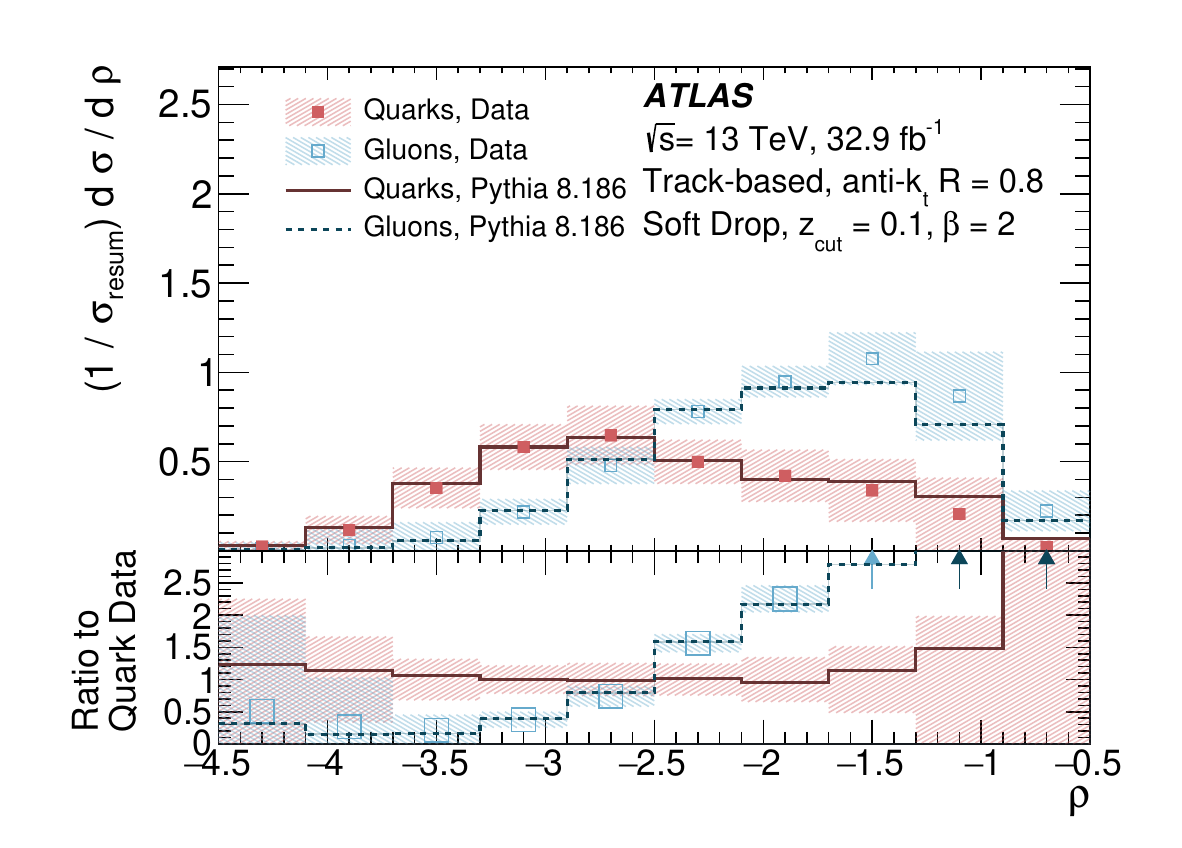}}
\subfloat[]{\includegraphics[width=0.42\textwidth]{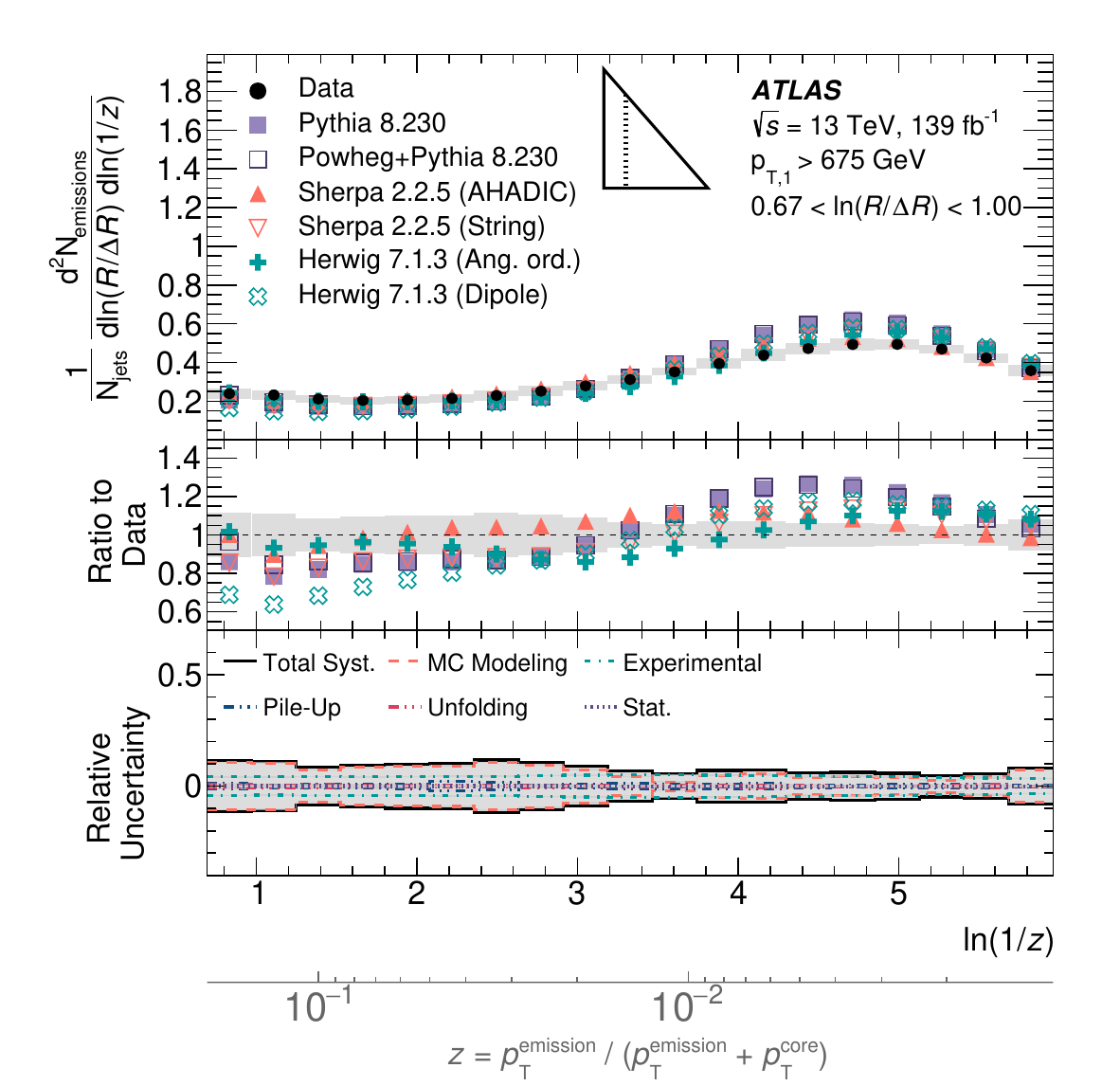}}
\end{center}
\caption{(a) Comparison of the quark and gluon unfolded relative jet mass distributions for the track-based soft-drop jet substructure measurement~\cite{STDM-2017-33}. The lower panel shows the ratio of the gluon data and predictions to those for quarks.
(b) Differential measurement of charged-particle activity inside jets in the Lund plane~\cite{STDM-2018-57}. Unfolded data are compared with particle-level simulation from several MC generators~\cite{Sjostrand:2007gs, Sjostrand:2014zea, Bothmann:2019yzt, Chahal:2022rid, herwig2, Frixione_2007, Alioli_2010, Nason_2004}. The uncertainty bands include all sources of systematic and statistical uncertainty. The middle panel shows the ratios of the predictions to data while the bottom panel shows the relative uncertainty.}
\label{fig:jet_groom}
\end{figure}

Grooming techniques systematically remove soft and wide-angle radiation, making the structure of the jet robust against contamination from pile-up, final-state radiation and the underlying event.
Jet substructure quantities are measured using jets groomed with the soft-drop grooming procedure~\cite{Larkoski:2014wba} in dijet events at $\sqrt{s} = 13$~\TeV with the ATLAS detector~\cite{STDM-2017-04, STDM-2017-33}.
Similar measurements in $pp$ collisions are performed by the CMS, STAR and ALICE Collaborations~\cite{CMS-SMP-16-010, STAR:2020ejj, ALICE:2022rdg}.
Jets are clustered using the anti-$k_t$ algorithm with radius parameter $R=0.8$.
Unfolded measurements of several substructure observables are provided for both the calorimeter-based observables and track-based observables.
For observables that are sensitive to the angular distribution of radiation within a jet, track-based observables are found to be more precise
than calorimeter-based observables, due to the better angular resolution of tracks.
The measurements are performed in different pseudorapidity regions, which are then used to extract quark and gluon jet shapes using the predicted quark and gluon fractions in each region.
An example jet substructure observable is the jet mass, defined as the norm of the four-momentum sum of constituents inside a jet.
The measurement of this observable, shown in Figure~\ref{fig:jet_groom}(a), is performed for a dimensionless version of the jet mass: the relative mass $\rho = \log{(m^2/\pT^2)}$.
Overall, all of the parton shower and analytical calculations provide a good description of the data in most regions of phase space.

Groomed large-radius jets ($R=1.0$) are also studied in ATLAS in events from inclusive multijet and \ttbar production~\cite{STDM-2017-34}.
Dedicated event selections are used to study jets produced by light quarks or gluons, and hadronically decaying top quarks and $W$ bosons.
The observables measured (not shown here) are sensitive to substructure, and therefore are typically used for tagging large-radius jets from boosted massive particles.
The data discriminate between the various MC models.
Overall, \textsc{Pythia}~8 for light-quark/gluon large-radius jet observables, and \textsc{Pythia}~8 matched to NLO QCD matrix element generators as well as \textsc{Sherpa} for top quark and $W$ boson large-radius jet observables, describe the data better than other models.
These measurements will be useful in improving the modelling of these substructure variables in MC generators. Since searches that utilise boosted topologies use these observables, or combinations of them, in tagging large-radius jets, a better modelling of them will help to increase the sensitivity of such searches.

In the soft gluon picture of jet formation, a quark or gluon radiates a haze of relatively low energy and statistically independent gluons.
As QCD is nearly scale-invariant, this emission pattern is approximately uniform in the two-dimensional space spanned by $\textrm{ln}(1/z)$ and $\textrm{ln}(1/\theta)$, where $z$ is the momentum fraction of the emitted gluon relative to the primary quark or gluon core and $\theta$ is the emission opening angle. This space is called the Lund jet plane~\cite{Andersson:1988gp}.
A measurement of the jet substructure based on the Lund jet plane is reported by ATLAS in Ref.~\cite{STDM-2018-57}.
The measurement is performed on an inclusive selection of dijet events, and their associated charged-particle tracks are used to construct the observables of interest. Several parton shower MC models are compared with the data, see Figure~\ref{fig:jet_groom}(b). No single model is found to be in agreement with the measured data
across the entire plane.

In a follow-up study ATLAS measures a differential cross-section of Lund subjet multiplicities in dijet events~\cite{STDM-2023-07}.
The Lund subjet multiplicity counts the number of subjets above a specified transverse momentum requirement in a jet’s angle-ordered clustering history.
The experimental precision achieved in the measurement allows tests of higher-order effects in QCD predictions. Most available predictions fail to accurately describe the measured data, particularly at large values of jet transverse momentum accessible at the LHC.

The unfolded ATLAS data on jet substructure provide a valuable input to help improve both perturbative and non-perturbative aspects of fragmentation modelling.
Including the present measurements in a future tune of the MC predictions can improve the description and reduce the theoretical uncertainties of many processes with jets in the final state.

\section{QCD studies based on  measurements with isolated photons}
\label{sec:photons}
Prompt photons with large transverse momenta constitute colourless probes of the hard interaction  with the highest reach in energy scale and provide another testing ground for pQCD in hadronic collisions. While not explored further in this section, these measurements have the potential to further constrain the parton distribution functions in the proton, particularly the gluon density, within a global QCD fit.
Prompt photons are defined as those that are not secondaries from hadron decays.
Prompt-photon production via hadron collisions is understood to proceed via two processes: the photon may arise directly from the hard interaction (direct process) or the photon may be emitted in the fragmentation of a high transverse momentum parton (fragmentation process).
Due to the abundance of photons from neutral-hadron decays and the contribution from the fragmentation process, prompt-photon production in hadron collisions is studied by requiring the photons to be isolated.
An isolation requirement is also essential in theoretical calculations to avoid divergencies in the matrix elements when the photon is collinear with a parton. This is achieved by using the method based on the Frixione criterion~\cite{Frixione:1998jh}.

\begin{figure}[!t]
\begin{center}
\subfloat[]{\includegraphics[width=0.54\textwidth]{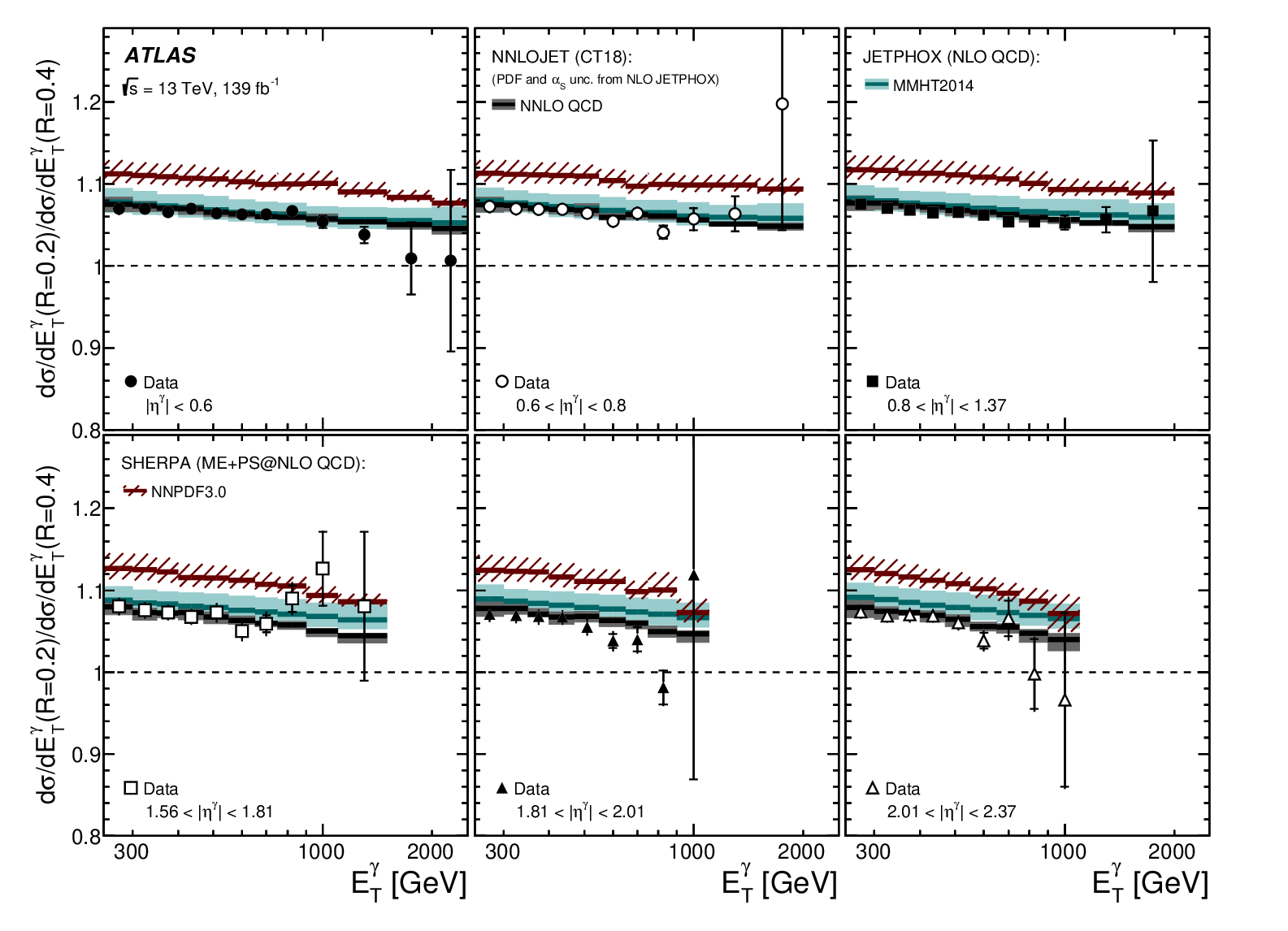}}
\subfloat[]{\includegraphics[width=0.45\textwidth]{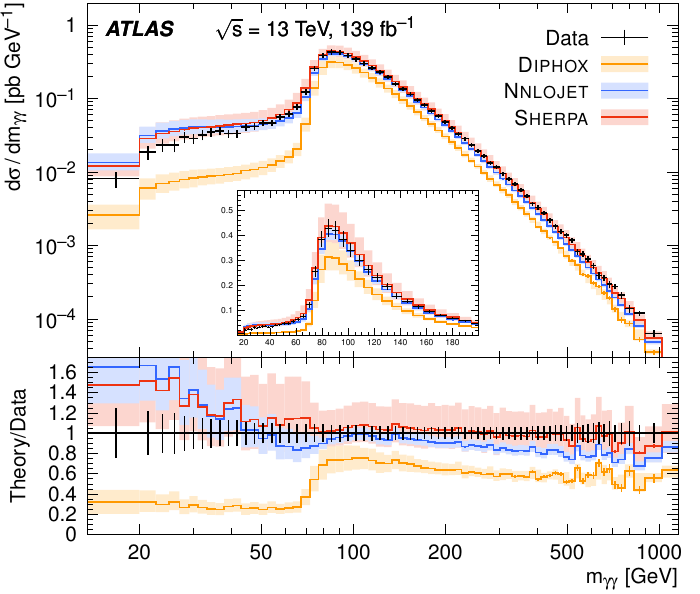}}
\end{center}
\caption{(a) Measured ratios of the differential cross-sections for inclusive isolated-photon production for isolation-cone radii of $R=0.2$ and $R=0.4$ at $\sqrt{s} = 13$~\TeV as functions of the photon transverse energy in different regions of photon pseudorapidity~\cite{STDM-2018-39}.
(b) Differential cross-sections for prompt photon pair production at $\sqrt{s} = 13$~\TeV  measured as  a function of diphoton invariant mass, $m_{\gamma\gamma}$~\cite{STDM-2017-30}.
The measurements are compared with various theoretical predictions~\cite{Catani:2002ny, Aurenche:2006vj, Chen:2022gpk, Bothmann:2019yzt, Hoeche:2009xc, Binoth:1999qq, Gehrmann:2018szu}.
The shape of the $m_{\gamma\gamma}$ distribution in (b) is governed by the transverse-momentum requirements placed on the individual photons, with the low-mass region being suppressed and only populated through $\gamma\gamma$+multi-jet configurations. Such configurations are not modelled well at NLO accuracy (DIPHOX curve).
}
\label{fig:photons}
\end{figure}

Differential cross-sections for inclusive isolated-photon and photon pair production in $pp$ collisions at $\sqrt{s} = 13$~\TeV are measured by ATLAS~\cite{STDM-2017-29, STDM-2018-39, STDM-2017-30}.
For inclusive photon production, the cross-sections are measured as functions of the photon transverse energy in different regions of photon pseudorapidity.
In addition, the dependence of the inclusive-photon production on the photon isolation is investigated by measuring the fiducial cross-sections as functions of the isolation-cone radius ($R$) and the ratios of the differential cross-sections with different radii in different regions of photon pseudorapidity~\cite{STDM-2018-39}. Measuring ratios provides a stringent test of pQCD with reduced experimental and theoretical uncertainties.
Photon pair production allows uniquely precise studies in events with two vector bosons. Differential cross-sections are measured as functions of several observables of the diphoton system, including the transverse momenta of the leading and sub-leading photon, the invariant mass and transverse momentum of the diphoton system.
For all of these single-photon and photon-pair measurements, good agreement is generally found with the predictions at the highest theoretical precision, as presented in Figure~\ref{fig:photons}.
The improvement observed when taking into account higher-order terms beyond NLO is impressive
and (only) fixed-order NNLO calculation, as implemented by \textsc{Nnlojet}~\cite{Chen:2019zmr}, give a satisfactory description of both inclusive photon and diphoton data in pQCD.

\begin{figure}[t!]
\begin{center}
\includegraphics[width=0.53\textwidth]{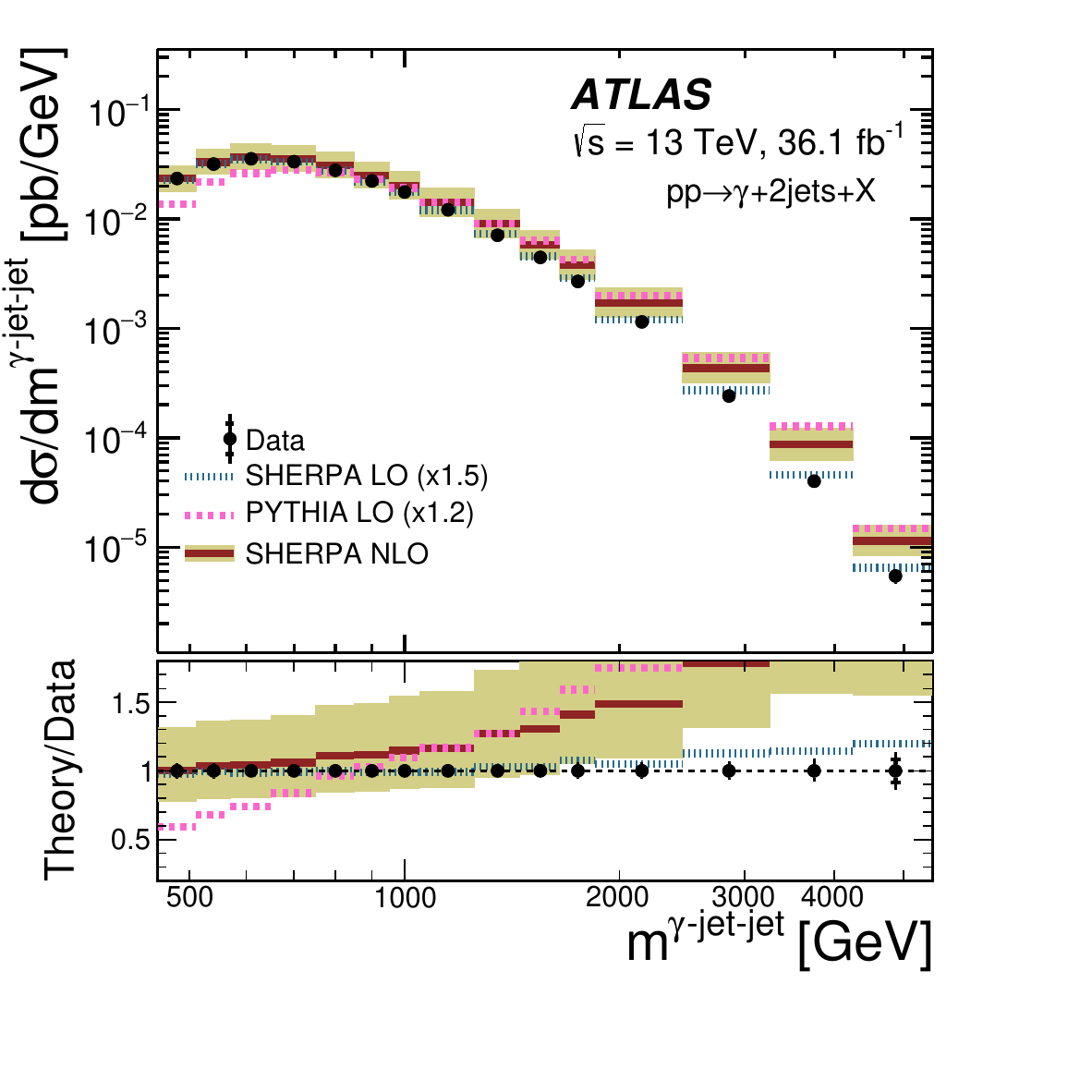}
\end{center}
\vspace*{-10mm}
\caption{Measured cross-sections for isolated-photon plus two-jet production (dots) as a function of the invariant mass of the photon and jets~\cite{STDM-2017-32}. Various theory predictions~\cite{Sjostrand:2007gs,Gleisberg:2008ta,Bothmann:2019yzt, Hoeche:2009xc} are also shown (horizontal lines). The lower panel shows the ratios of the predictions to the data.}
\label{fig:photon_jets}
\end{figure}

The production of prompt photons can be further studied using the jet dynamics in events with at least one hard jet, e.g., via the measurements of angular correlations between the photon and jets.
Measurements of the cross-sections for the production of an isolated photon in association with one or two jets at $\sqrt{s} = 13$~\TeV are provided by ATLAS~\cite{STDM-2017-01, STDM-2017-32}.
Cross-sections are measured as functions of a variety of observables, including angular correlations and invariant masses of the objects in the final state.
Measurements are also performed in phase-space regions enriched in each of the two underlying physical mechanisms, namely direct and fragmentation processes.
The tree-level plus parton-shower predictions (normalised to the integrated measured cross section) and the NLO QCD predictions are compared with the measurements. The multi-leg NLO QCD plus parton-shower calculations of predictions from \textsc{Sherpa} describe the data adequately in shape and normalisation except for regions of phase space such as those with high values of the invariant mass of the photon and jets (see Figure~\ref{fig:photon_jets}), where the predictions overestimate the data.

\section{Strong and electroweak production of single gauge bosons}
\label{sec:singlebos}
Measurements of single gauge boson production provide an excellent probe of pQCD and of the proton structure.
In association with jets, they become a probe of higher-order QCD corrections. Measurements of jet flavour activity provide
insights into gluon splitting and into the proton structure functions (PDF) of heavier quarks.
The production of gauge bosons with jets also constitutes one of the most important backgrounds for Higgs boson measurements and for various BSM searches, and is hence considered a very important input for the tuning of MC simulations.
Single gauge boson prodution can also be used to explore EW physics and to precisely measure fundamental SM parameters  (see Section~\ref{sec:params}).

With the increased centre-of-mass energy in Run~2, the LHC experiments can probe more energetic phase spaces. New reconstruction and analysis techniques allow for more precise measurements. This goes hand-in-hand with improvements in the theory sector, both in fixed-order calculations and in multi-leg ME+PS generators~\cite{PMGR-2021-01}.

The fiducial phase space for these  analyses typically requires leptons, usually electrons or muons, $\ell$, with $\etal < 2.5$  and
minimum \ptl, in the range of $25$--$30$~\GeV. In the \Z\  case,~\footnote{In the following, \Z\ refers implicitly to neutral current \Zgs exchange including interference effects.}
a  window on the dilepton mass \mll\ of $\pm 20$--$25$~\GeV  is selected around the $Z$ mass,
whereas a typical \W\ selection requires $\met > 25$--$30$~\GeV and a minimum transverse mass \mtw\ of $50$--$60$~\GeV.
Systematic uncertainties in inclusive $W$ and $Z$ distributions are typically dominated by electron and muon reconstruction and calibration,
whereas the systematic uncertainties in distributions of jets or hadrons produced in association with a gauge boson are typically
dominated by jet calibration and the identification efficiency for heavy-flavour hadrons or jets.

\subsection{Inclusive \W\ and \Z\ production in early Run~2 data}
A small amount of the first Run~2 data taken in 2015, 81~\ipb, was used to measure fiducial cross-sections for \Wp, \Wm\ and \Z\ production at the new
centre-of-mass energy~\cite{STDM-2015-03}.
\W(\Z) fiducial cross-sections were measured with a systematic precision of $2(1)\%$
and a luminosity uncertainty of $2\%$, as shown in Figure~\ref{fig:wz:xsec}. Their ratios are determined with a precision of just under $1\%$ and $2\%$ for $\sigma_{\Wp}/\sigma_{\Wm}$ and $\sigma_{\Wpm}/\sigma_{\Z}$ respectively.
The measured cross-sections agree in general with predictions of NNLO accuracy in pQCD~\cite{Camarda_2020,Camarda_2021,Camarda_2022,Camarda_2023,Bondarenko_2023} using NLO EW corrections~\cite{Dittmaier_2002,Baur_2004,Calame_2006,Arbuzov_2008,Baur_2002,Dittmaier_2014,Baur_1998,Baur_2002,Dittmaier_2010,Arbuzov_2006} and various NNLO PDF sets. The $\sigma_{\Wp}/\sigma_{\Wm}$ ratio allows the best distinction between the PDF sets.
The systematic precision of the 13~\TeV \Z\ cross-section measurement has been slightly improved using a larger data sample of 36.1~\ifb~\cite{STDM-2018-14} (see below). The improved systematic precision of $0.5\%$ in the $W$ channel together with a reduced uncertainty of $1\%$ in the integrated luminosity, have allowed even more precise cross-section measurements using a low pile-up data sample, corresponding to an integrated luminosity of 338~\ipb~\cite{STDM-2018-17}.  Figure~\ref{fig:wz:xsec}(a), also shows a recent measurement of the \W\ and \Z\ cross-sections performed with 25~\ipb\ of pp collision data taken at an energy of 5.02~\TeV~\cite{HION-2018-02}.

\begin{figure}[t!]
\begin{center}
\subfloat[]{\includegraphics[width=0.49\textwidth]{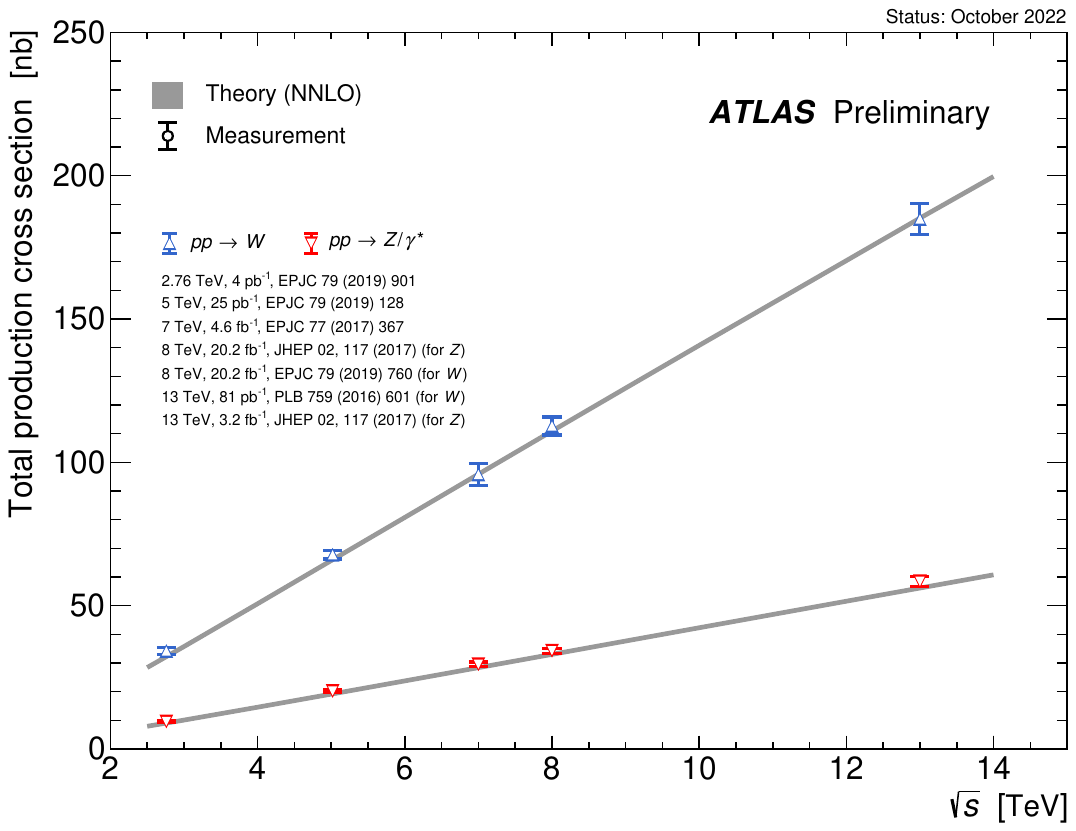}}
\subfloat[]{\includegraphics[width=0.49\textwidth]{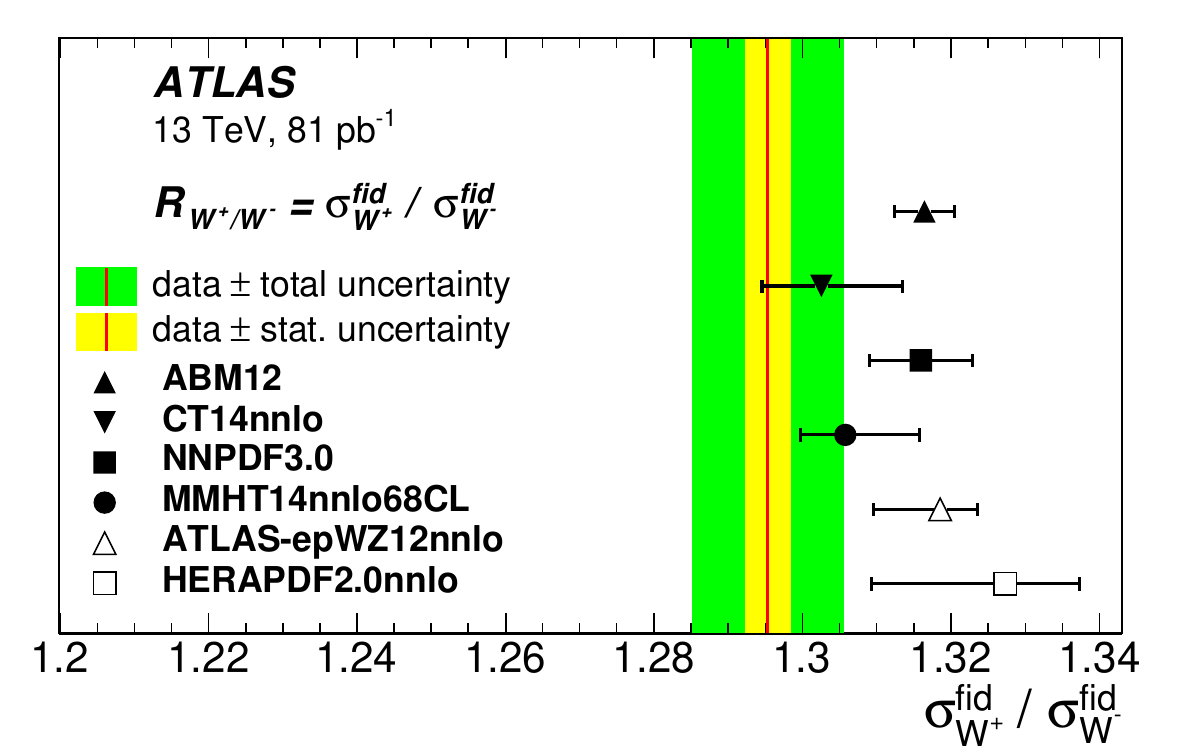}}
\end{center}
\caption{(a) Measured inclusive \W\ and \Z\ cross-sections as a function of the $pp$ centre-of-mass energy~\cite{ATL-PHYS-PUB-2023-039} and (b) $\sigma_{\Wp}/\sigma_{\Wm}$  at 13~\TeV compared with NNLO predictions~\cite{Camarda_2020} using various PDF sets~\cite{STDM-2015-03}.} \label{fig:wz:xsec}
\end{figure}

\subsection{\W\ and \Z\ transverse momentum and \phistar}
\label{sec:ptv}
The \Z\ transverse momentum, \ptz\ is an excellent probe of initial-state quark and gluon emission and of intrinsic parton transverse momentum. Low-\ptz\ ranges are typically modelled via resummed approaches whereas high-\ptz\ domains are described by perturbative QCD\@. A partial data sample of 36.1~\ifb\ is used to perform a measurement of \ptz\ and the alternate variable \phistar, calculated from angular variables~\cite{Banfi_2011}, normalised to the total fiducial cross-section.  A precision of $0.2\%$ is reached for low values of \ptz. A prediction by \pythiaeight~\cite{Sjostrand:2014zea} at LO in QCD\@, supplemented by a parton shower, and  NLO descriptions by \powpyeight~\cite{Nason_2004,Frixione_2007,Alioli_2010,Alioli:2008gx}, both tuned on ATLAS 7~\TeV data (AZ/AZNLO tune)~\cite{STDM-2012-23}, provide a good description in the low and medium-energy range, see Figure~\ref{fig:wz:zpt}(a). The high-\ptz\ range is well described by a fixed-order NNLO calculation by NNLOjet~\cite{NNLOjet2016}. The best prediction is provided by the  fixed-order \radish program at NNLO+N$^3$LL~\cite{Bizon2018,Bizon2019}, which agrees with the data over the full \ptz and \phistar\ spectra, except for a small region at very low \ptz\ that is sensitive to non-perturbative effects.

A low pile-up data sample corresponding  to 338~\ipb is used to derive precise cross-sections as a function of \ptz\ and \ptw\ in the regime $\pt < 100~\GeV$~\cite{STDM-2018-17}. The data is described reasonably well by $W$ and $Z$ predictions at NNLO+NNLL in pQCD (see Figure~\ref{fig:wz:zpt}(b) for \ptw). The two generators tuned to  7~\TeV ATLAS data describe reasonably well the low-\ptw\ regime but fail to describe data with $\ptw > 40~\GeV$.

\begin{figure}[t!]
\begin{center}
\subfloat[]{\includegraphics[width=0.53\textwidth]{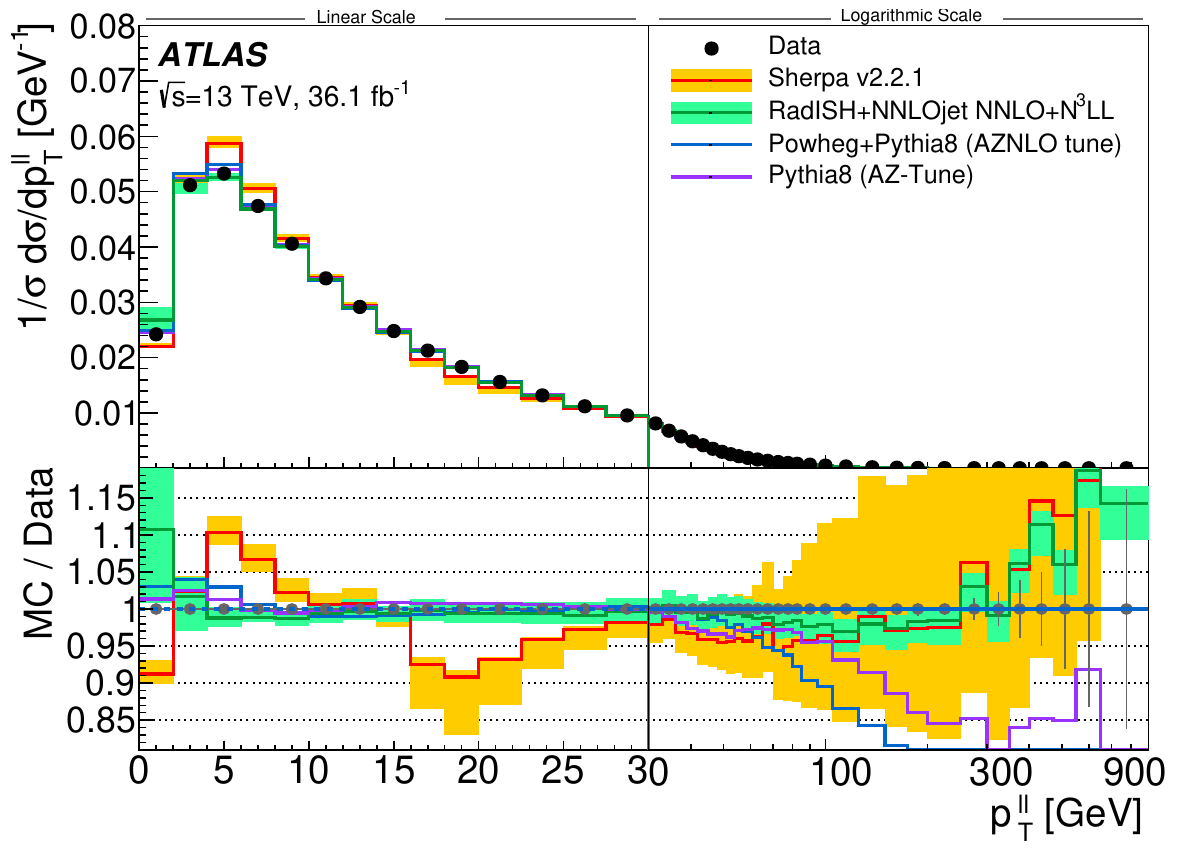}}
\subfloat[]{\includegraphics[width=0.45\textwidth]{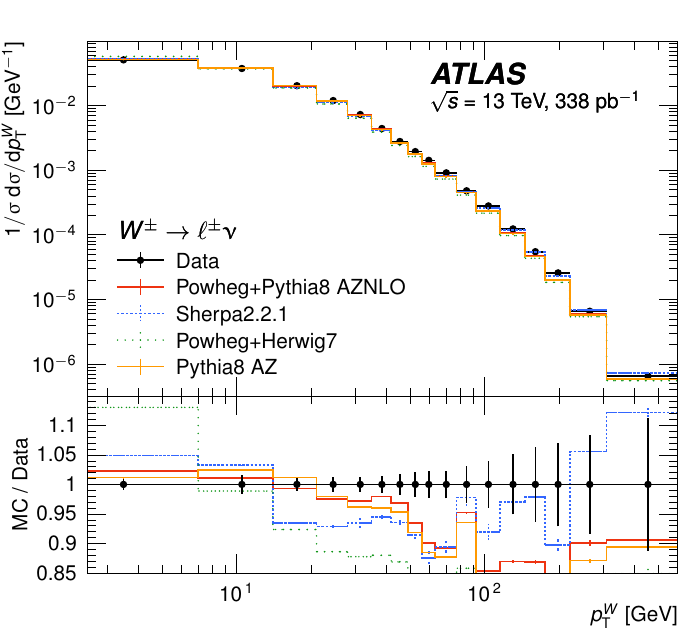}}
\end{center}
\caption{(a) Unfolded normalised distributions of \ptz~\cite{STDM-2018-14} and (b) \ptw~\cite{STDM-2018-17}, compared with various predictions. The lower panels show the ratios of the predictions~\cite{Sjostrand:2014zea,Nason_2004,herwig2,Bothmann:2019yzt,Bizon2019} to the data.}
\label{fig:wz:zpt}
\end{figure}

\subsection{Precise 2D $Z$ cross-section measurement in full phase space}
\label{sec:DY2D}

The 5-dimensional differential $Z$ (or $W$) cross-section $\frac{\mathrm{d}\sigma}{\mathrm{d}\pt \mathrm{d} y \mathrm{d} m \mathrm{d} \cos\theta \mathrm{d}\phi}$ with lepton angles $\theta$ and $\phi$ in the Collins--Soper frame~\cite{Collins:1977iv}  can be described as the product of an unpolarized cross-section $\frac{\mathrm{d}\sigma^{U+L}}{\mathrm{d}\pt \mathrm{d} y \mathrm{d} m}$ with the sum of spherical harmonic polynomials multiplied by eight angular coefficients~\cite{Mirkes:1992hu}.
The Run~1 $\sqrt{s} = 8$~\TeV\ data sample with an integrated luminosity of 20.2~\ifb\ was used previously to extract the angular coefficients~\cite{STDM-2014-10} as a function of \ptz\ and \yll. A novel measurement using the same data sample~\cite{STDM-2018-05}  now also extracts the unpolarized cross-section as a function of \ptz\ and \yll\ in a complex fit with templates corresponding to the spherical polynomials. The measurement is corrected for lepton acceptance effects, enabling more precise theoretical interpretations than a classic fiducial measurement. The differential cross-sections are determined at percent accuracy level. The uncertainties are  statistically dominated, followed by the one order smaller lepton identification and calibration uncertainties. The much smaller PDF uncertainties constitute the only non-negligible theory uncertainty, as the QCD uncertainties are negligible per design. QED/EWK effects break the factorisation assumption underlying the above expression of the differential cross-section but the uncertainty on the QED FSR radiation is contained in the lepton uncertainties and the contributions of other higher-order QED/EWK corrections, such as initial-finalstate interference diagrams, are expected to be negligible at the $Z$ pole~\cite{STDM-2014-10,Jadach_2019}.
The measurements are consistent with state-of-the-art QCD perturbative predictions based on $q_{\textrm T}$-resummation at approximate N$^4$LL accuracy matched to fixed-order $\mathcal{O}(\alphas^3)$ calculations at high \ptz~\cite{Camarda_2020,Camarda_2021,Camarda_2023}  (see Figure~\ref{fig:z2d}).

\begin{figure}[t!]
\begin{center}
\subfloat[]{\includegraphics[width=0.49\textwidth]{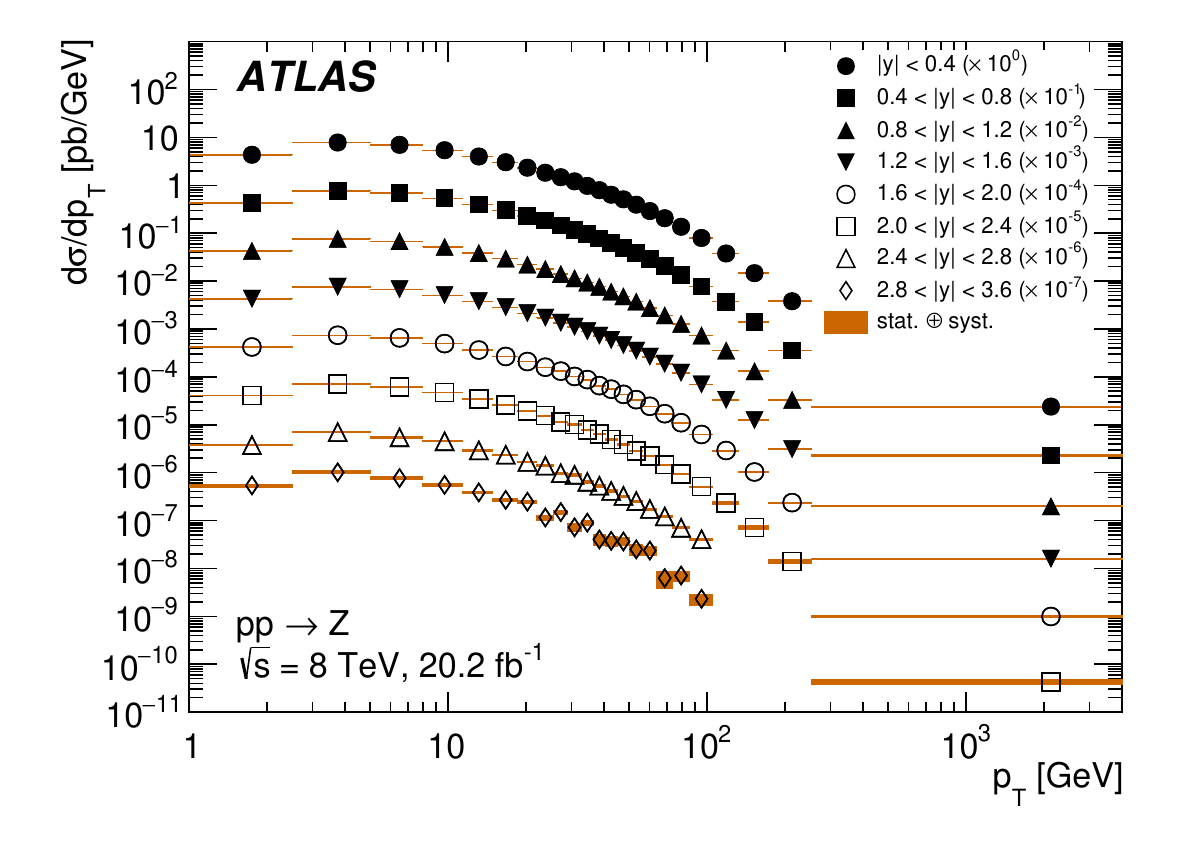}}
\subfloat[]{\includegraphics[width=0.49\textwidth]{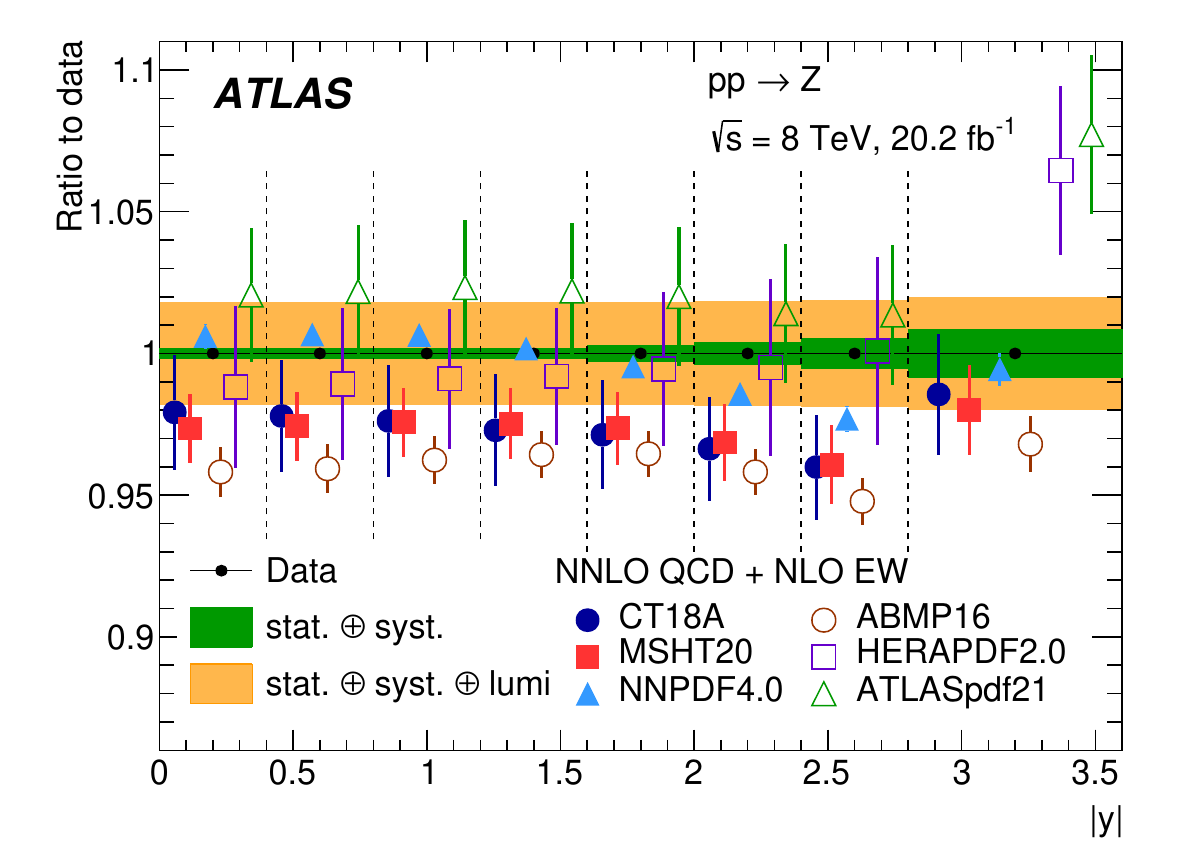}}
\end{center}
\caption{(a) Measured absolute differential cross-sections as a function of \ptz\ for each |$y$| bin and (b) ratio comparisons between the differential cross measurements as a function of |$y$| and NNLO QCD predictions obtained from \textsc{DYTurbo}~\cite{Camarda_2020,Camarda_2021,Camarda_2023} using different NNLO PDF sets~\cite{STDM-2018-05}.}\label{fig:z2d}
\end{figure}

\subsection{\Z\ bosons in association with highly energetic jets}
The measurement of \Z\ boson production in association with high-energy jets, provides a powerful probe of perturbative QCD and its interplay with higher-order EW processes~\cite{Rubin10, Prestel16, Boughezal15, Sjoestrand14,Krauss_2014,Kallweit_2016}, especially the collinear emission of a $Z$ boson from a dijet configuration and combined higher-order QCD and EWK correction in back-to-back Z+jets constellations~\cite{Kallweit_2016}, with a clean experimental signature from the leptonic \Z\ decay. While a Run~2 measurement of \Zjets\ cross-sections with a partial  data sample of 3.2~\ifb\ provided an early  probe of pQCD predictions for the new centre-of-mass energy~\cite{STDM-2016-01}, the full Run~2 data sample allows much higher energies to be probed~\cite{STDM-2018-49}. For very high-\pt\ jets, a collinear enhancement is expected in the angle between the \Z\ boson and the closest jet. The measurement focuses on the study of two topologies in events with a leading jet with $\pt > 500$~\GeV: events where the jet and the $Z$ boson are back-to-back, and those where they are collinear. Distinct patterns in jet multiplicities, momentum ratios,  and angular distributions are observed. The systematic uncertainty of typically $5\%$ is dominated by the jet calibration uncertainty and statistical uncertainties in differential distributions are of similar size.
Figure~\ref{fig:wz:zjets}(a) shows the transverse momentum of the leading jet. A good modelling by NLO multi-leg generators~\cite{Bothmann:2019yzt,Frederix:2012ps,PMGR-2021-01} is observed.
Fixed-order NNLO predictions~\cite{nnlojet1,nnlojet2} agree with the data at a high level of precision. The slight overestimate at very high jet transverse momenta could be due to missing NLO EW corrections. Figure~\ref{fig:wz:zjets}(b) shows the minimum angle between the \Z\ and the closest jet with $\pt > 100~\GeV$, a quantity also probed with \W\ events in Run~1 data~\cite{STDM-2015-16}. It shows the clear collinear enhancement of events with a low-energy \Z\ boson ($\Delta R(Z,j) <1.4$) and the back-to-back region $\Delta R(Z,j) \sim \pi$ where a high-momentum \Z\ boson recoils against the high-\pt\ jet.

\begin{figure}[h!]
\begin{center}
\subfloat[]{\includegraphics[width=0.49\textwidth]{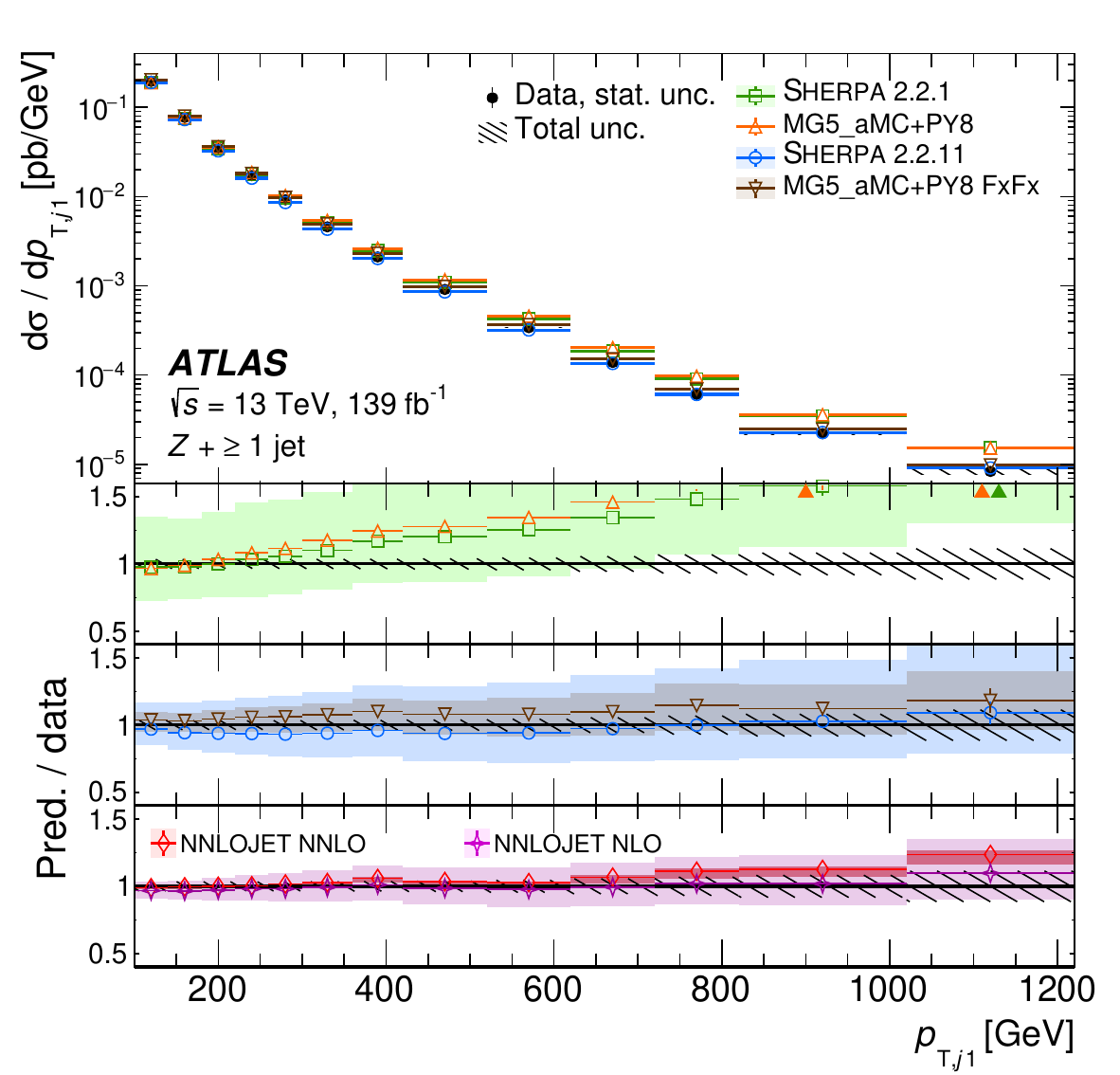}}
\subfloat[]{\includegraphics[width=0.49\textwidth]{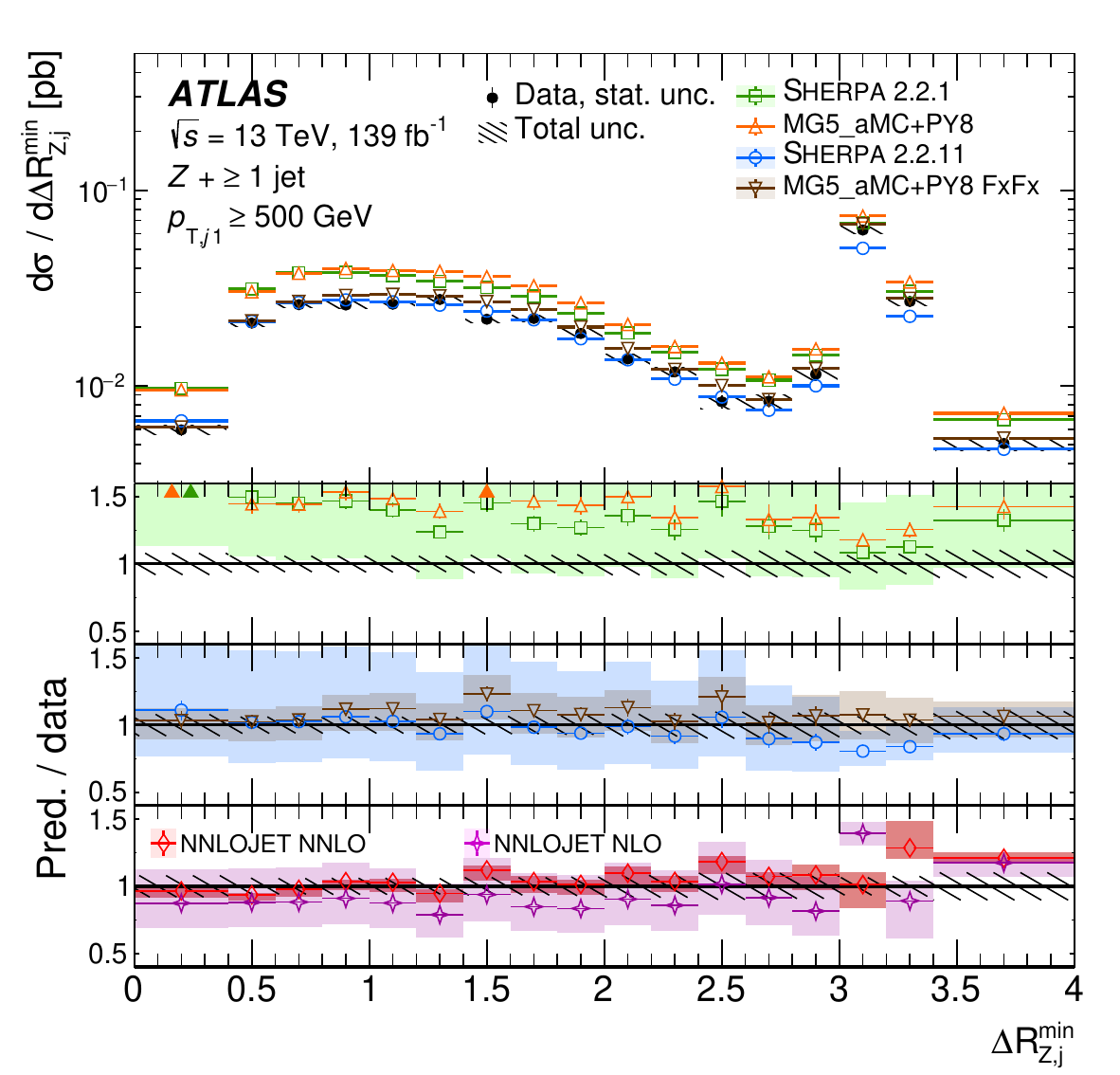}}
\end{center}
\caption{\Zjets\ cross-section as a function of (a) the \pt\ of the leading jet and (b) the angle between the \Z\ and the closest jet for events with $\ptlj > 500~\GeV$~\cite{STDM-2018-49}. The lower panels show the ratios of the predictions~\cite{Bothmann:2019yzt,Alwall:2014hca,PMGR-2021-01,Frederix:2012ps,nnlojet1,nnlojet2} to the data.}\label{fig:wz:zjets}
\end{figure}

\subsection{\Z\ bosons in association with $b$-jets}
Measurements  of \Z\ bosons produced in association with $b$- and $c$-jets are interesting not the least because
theoretical calculations are confronted with an choice of flavour and mass schemes, which converge as more higher orders are included~\cite{Fabres09,Campbell04,Maltoni12}. They also provide an important test of $b$ and $c$ quark PDFs.
Inclusive and differential cross-sections are measured for various observables for events with at least one $b$-jet, at least two $b$-jets (see Figure~\ref{fig:wz:zhf}(a))  and at least one $c$-jet (see Figure~\ref{fig:wz:zhf}(b))~\cite{STDM-2018-43} with precisions of $6\%$, $9\%$ and $13\%$ respectively. The extraction of the  \Zjets\ backgrounds with different parton flavours is performed via a fit to the flavour-tagging discriminant in each bin of the observable.
The  observables are compared with a variety of predictions of different orders in QCD, different flavour schemes and different PDFs.
The best overall description is provided by 5-flavour scheme (5FS) multi-leg generators~\cite{Frederix:2012ps,Bothmann:2019yzt} and
5FS NNLO predictions with the `flavour-dressing' approach~\cite{Gauld:2022lem}. Calculations in the 4-flavour scheme are found to be not
suitable for selections with at least one $b$-jet, but can produce acceptable estimates for final states with at least two $b$-jets. None of the predictions describes the full range of the \mbb\ distribution and all generators underestimate the $\Z+c$-jet  cross-sections. PDFs with different intrinsic-charm content~\cite{Brodsky:1980pb} are compared with PDF-sensitive $\Z+c$-jet distributions but no significant difference between the various PDFs is found.

In very high-energy events with at least two $b$-jets, a topology that can constitute a major background in the search for massive BSM particles, the two $b$-jets may not be resolved into two separate jets. Instead, they may be reconstructed as a single large-radius jet. Reference~\cite{STDM-2017-37} shows results derived on a partial data sample of 36~\ifb, where `trimmed' anti-$k_t$ jets with $\pt > 200$~\GeV with a radius parameter $R=1.0$~\cite{JETM-2018-02, JETM-2018-06} are required to have two $b$-tagged sub-jets with $R = 0.2$. The uncertainties in the measurements are about $40\%$ over large parts of the phase space and statistically dominated. Predictions using the 5FNS scheme are found to model the data best.

\begin{figure}[h!]
\begin{center}
\subfloat[]{\includegraphics[width=0.49\textwidth]{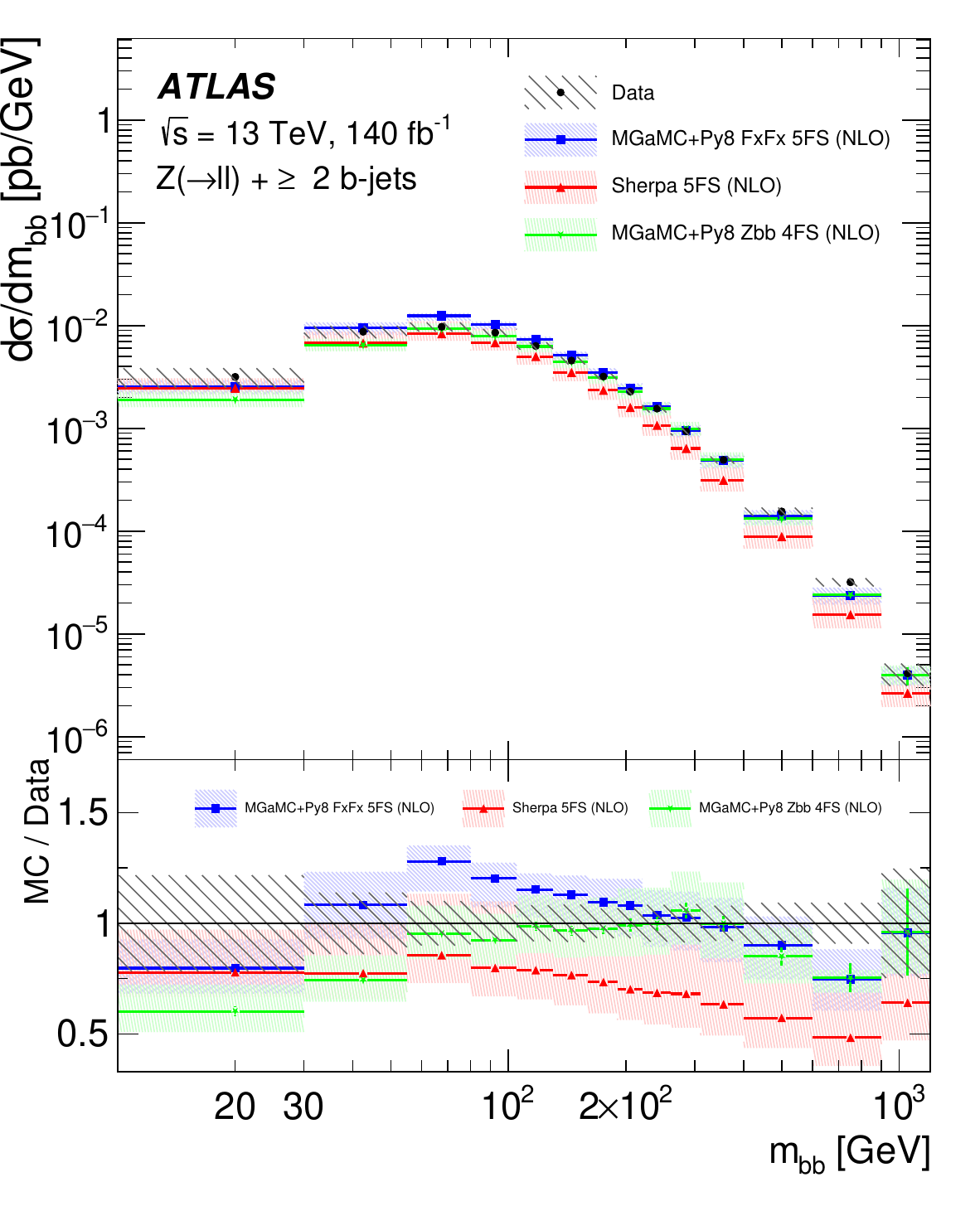}}
\subfloat[]{\includegraphics[width=0.49\textwidth]{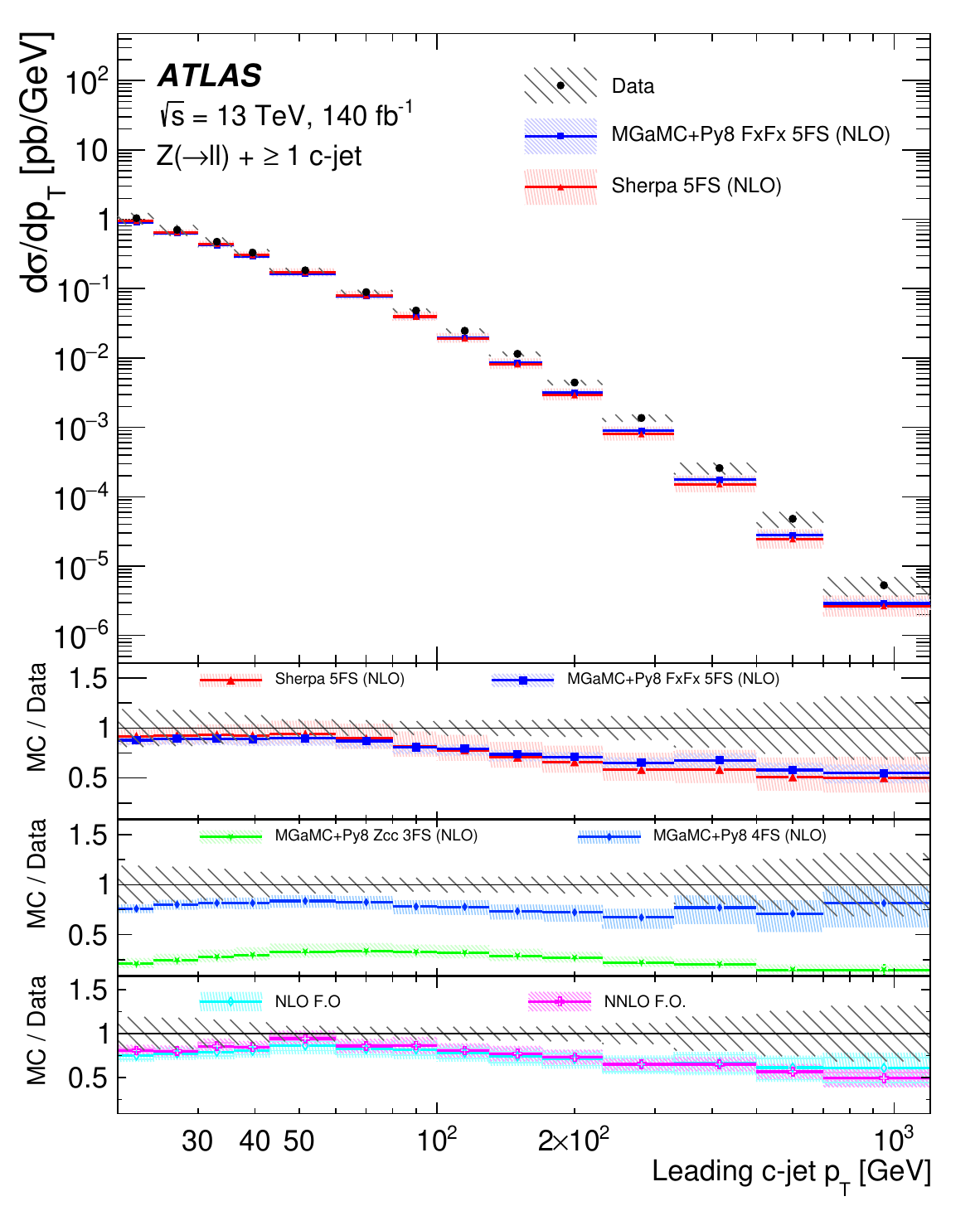}}
\end{center}
\caption{Cross-section as a function of (a) \mbb\ in events with at least two $b$-jets  and  (b) the transverse momentum of the leading $c$-jet~\cite{STDM-2018-43}. The lower panels show the ratios of the predictions~\cite{Frederix:2012ps,Bothmann:2019yzt,Gauld:2022lem} to the data.} \label{fig:wz:zhf}
\end{figure}

\subsection{$W$ boson in association with a $D$ meson}

The production of $W + c$  is an excellent probe of the comparatively less constrained strange quark PDF of the proton~\cite{Ubiali20}.
In the analysis of Ref.~\cite{STDM-2019-22}, this process is identified by explicit reconstruction of a $D^\pm$ or a $D^{*\pm}$ meson from the tracks of their charged decay products in a fiducial phase space of $\pt(D^{(*)}) > 8$~\GeV and $|\eta|(D^{(*)}) < 2.2$, in association with a leptonically decaying $W$ boson.
For the targeted signal, the $W$ and $D$ meson have opposite-sign charge (OS). On the other hand, most backgrounds including $W+g(c\bar{c})$ production, have no preferred charge relation. Therefore, the signal is extracted as the difference between OS and same-sign (SS) distributions.
Inclusive and differential cross-sections as a function of $\pt(D)$ and $\eta(\ell)$ are measured via profile-likelihood (pLLH) fits of folded theory to the OS and SS $D^{(*)}$ mass distributions. In addition the $W$ charge ratios are computed. The  percentage-level uncertainties, dominated by secondary-vertex reconstruction and signal modelling, are at the level of the PDF uncertainties.
Figure~\ref{fig:wz:wcharm} compares the pseudorapidity of the $D^{(*)}$ meson and the charge ratio with \mgamc~2.9.3~\cite{Alwall:2014hca} predictions using different PDF sets. The measurements show a broader distribution than the nominal predictions but are consistent with the predictions when PDF uncertainties are included. A key result is the $W^+/W^-$ charge ratio that is sensitive to differences between the strange- and anti-strange quark PDFs. Here the results are found to be compatible with PDF fits that constrain the strange-quark sea to be symmetric.

\begin{figure}[h!]
\begin{center}
\subfloat[]{\includegraphics[width=0.45\textwidth]{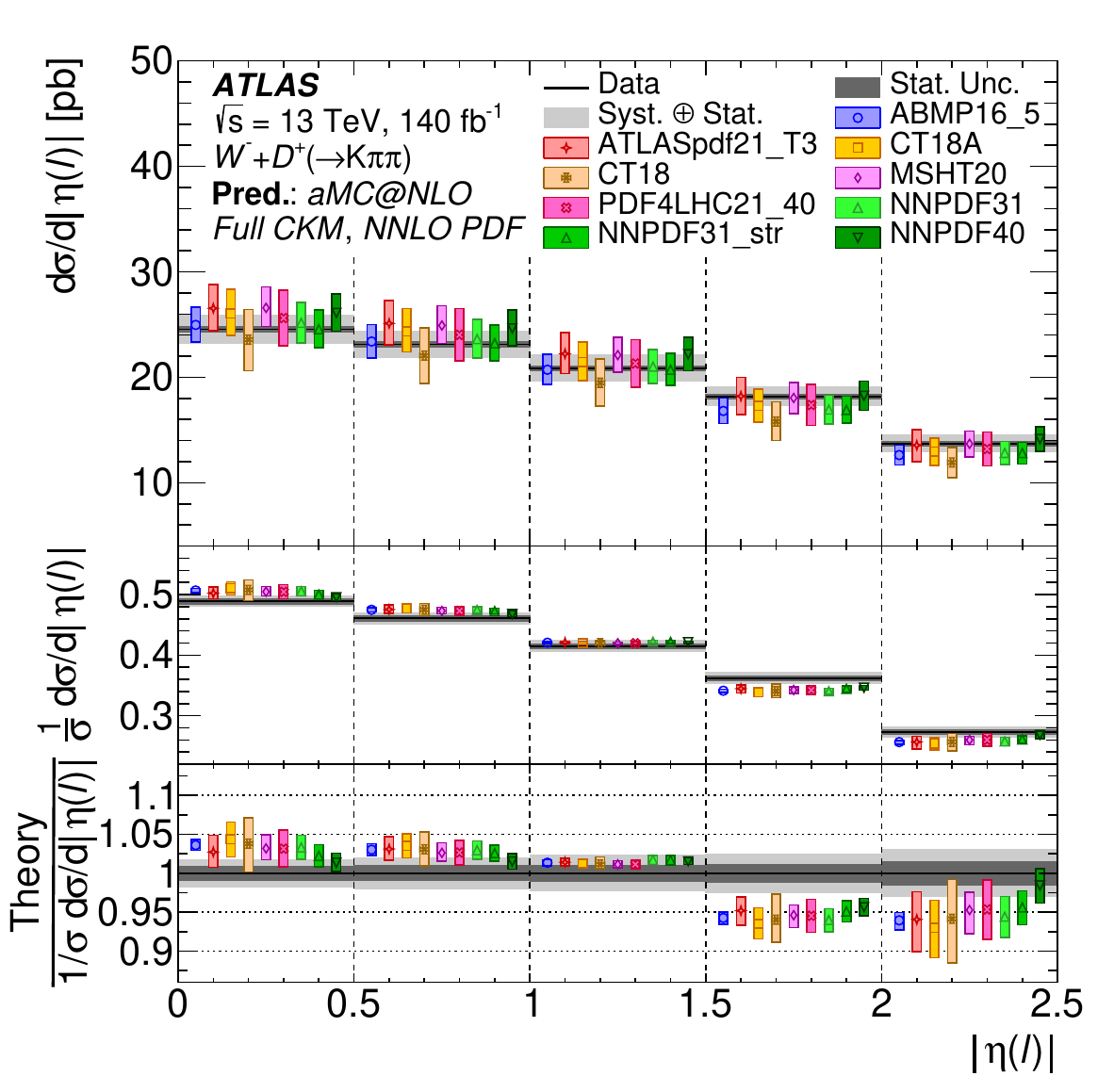}}
\subfloat[]{\includegraphics[width=0.53\textwidth]{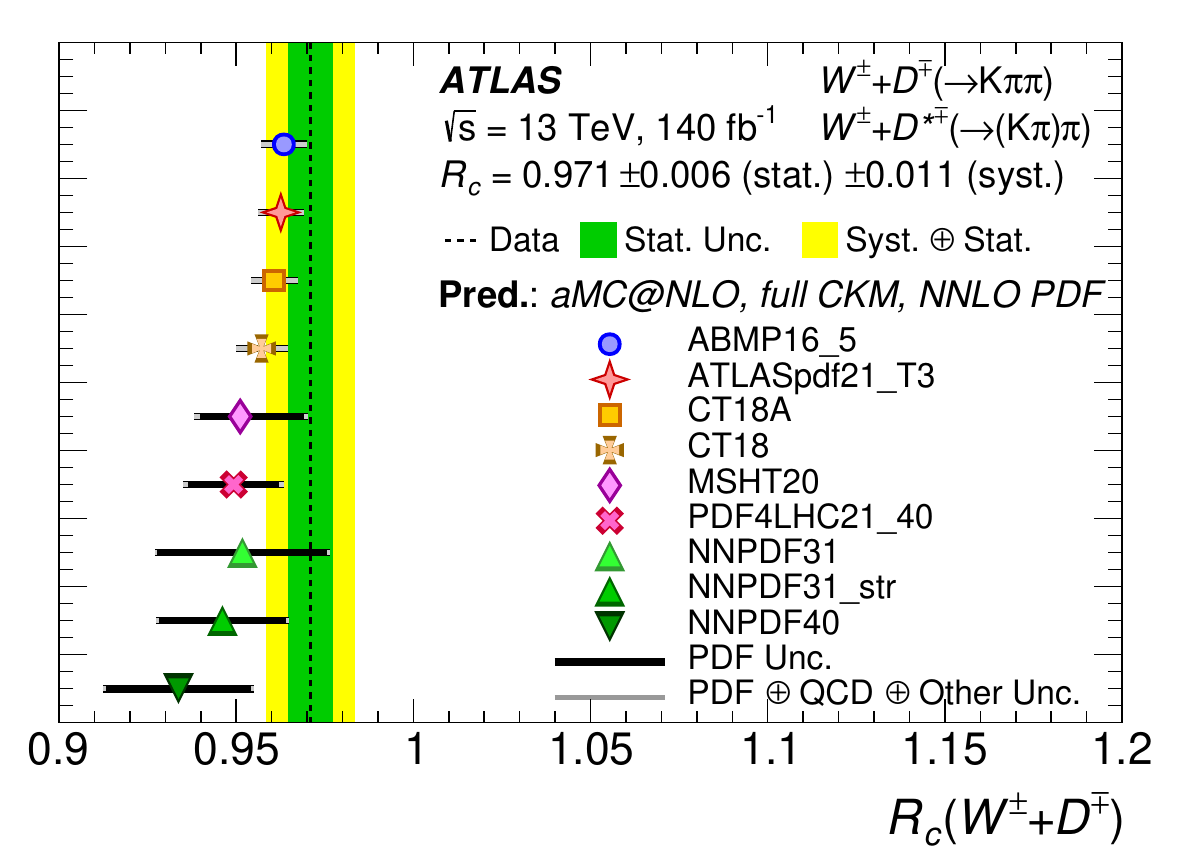}}
\end{center}
\caption{Measurements of (a) $\eta$ of the $D^{+}$ meson and (b) the $W$ charge ratio, compared with \mgamc\ predictions~\cite{Alwall:2014hca} using different PDF sets~\cite{STDM-2019-22}.}\label{fig:wz:wcharm}
\end{figure}

\subsection{Determination of PDFs from diverse ATLAS measurements}

ATLAS has presented the first comprehensive and comparative NNLO perturbative QCD analysis of a number of
data samples with sensitivity to parton distributions~\cite{STDM-2020-32}.
The data sets used are:
inclusive $W$ and $Z$ cross sections~\cite{STDM-2012-20} and inclusive jets~\cite{STDM-2013-11} at $\sqrt{s} = 7$~TeV,
inclusive $Z$~\cite{STDM-2016-04}, inclusive $W$~\cite{STDM-2017-13}, $W$+jets~\cite{STDM-2016-14}, $Z$+jets~\cite{STDM-2016-11},
top-pair production~\cite{TOPQ-2015-06,TOPQ-2015-07}, inclusive isolated photons~\cite{STDM-2017-12} and inclusive jets~\cite{STDM-2015-01} at $\sqrt{s} = 8$~TeV,
and top-pair production~\cite{TOPQ-2018-15},  and inclusive jets~\cite{STDM-2016-03} at $\sqrt{s} = 13$~TeV, in addition to HERA data~\cite{H1:2015ubc}.
Correlations between the systematic uncertainties of the different analyses are preserved.
The novel {\textit ATLASpdf21} PDF set is  extracted via the xFitter framework~\cite{Alekhin:2014irh} using predictions at NNLO in pQCD\@. The impact of the various data samples and their correlation is studied. The addition of the ATLAS data to the HERA data brings this PDF much closer to the global PDFs, as shown in Figure~\ref{fig:pdf}. The strange-quark PDF at low values of $x \lesssim 0.01$ is found to be less suppressed than assumed in PDFs from before the LHC and found to be more in line with modern PDFs at higher $x \gtrsim 0.1$, as shown in Figure~\ref{fig:pdf}(b).

\begin{figure}[!t]
\begin{center}
\subfloat[]{\includegraphics[width=0.49\textwidth]{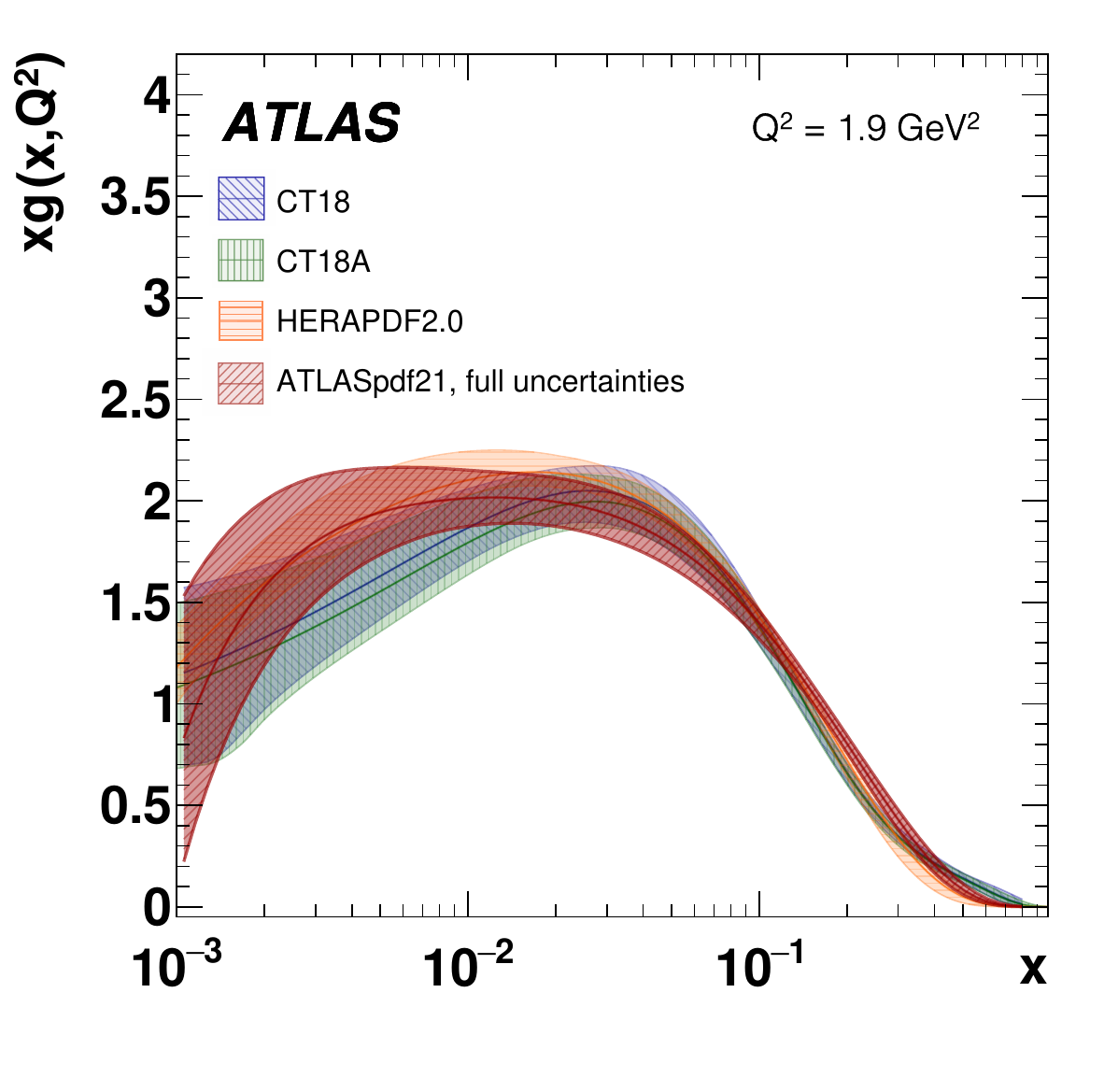}}
\subfloat[]{\includegraphics[width=0.49\textwidth]{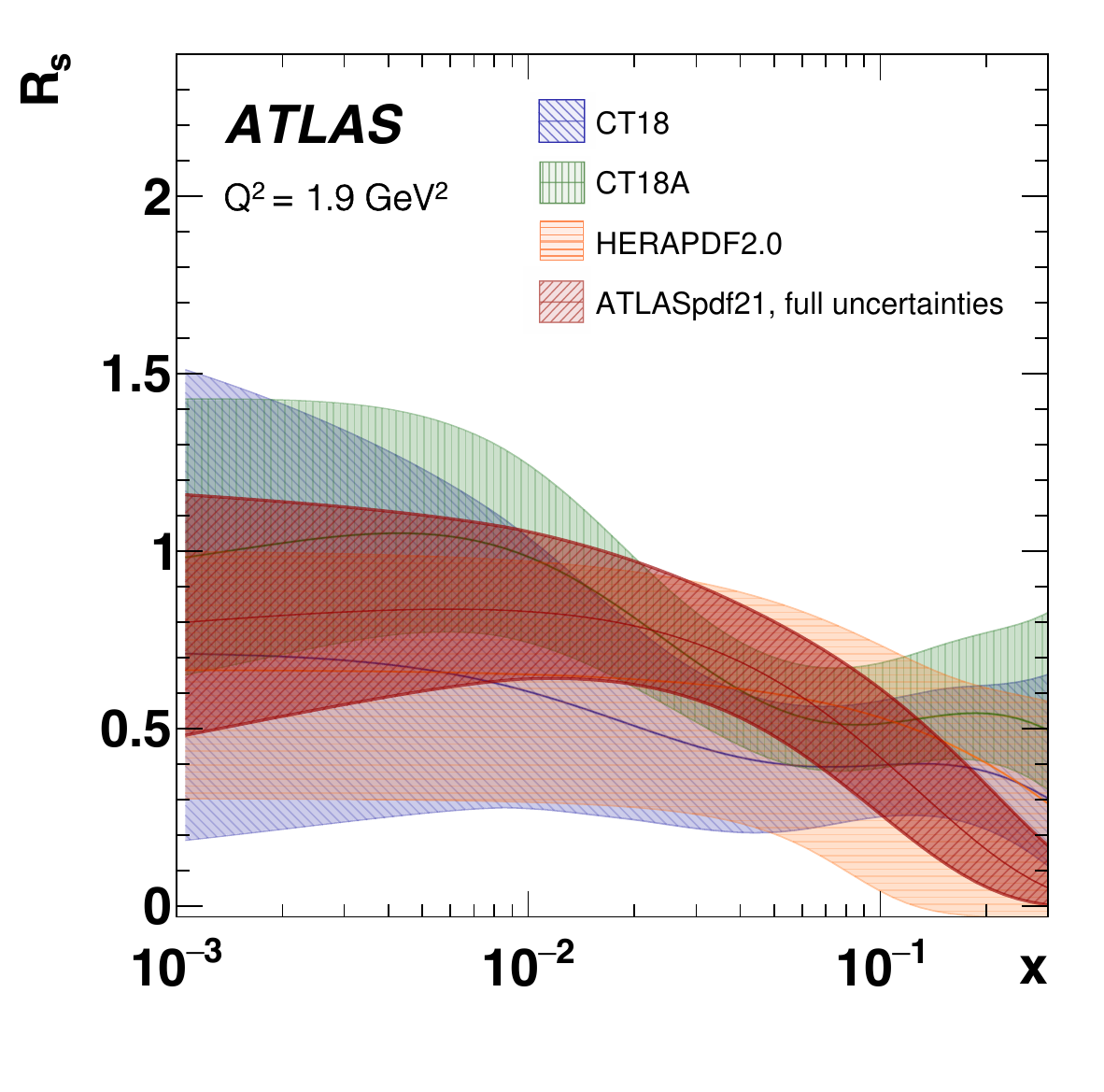}}
\end{center}
\caption{(a) The gluon density $xg$  and (b) the strangeness-suppression $R_s = x(s+\bar{s})/x(\bar{u}+\bar{d})$  distributions at a low scale of $Q^2 = 1.9$~\GeV$^2$ of the ATLASpdf21 fit~\cite{STDM-2020-32} compared with other PDF sets~\cite{Hou_2021,H1:2015ubc}.}\label{fig:pdf}
\end{figure}

\subsection{Electroweak production of dijets in association with a \Z\ boson}
\label{sec:VBF}

While the production of weak bosons in association with jets proceeds largely through the strong interaction, it is possible to access the purely EW production of weak bosons with a dijet system. The EW production of a single weak boson is defined by the $t$-channel exchange of such a boson and is very sensitive to the VBF production mechanism~\cite{STDM-2017-27}. The SM triple-gauge coupling (TGC) involved could be enhanced or altered in BSM scenarios. Measurements of this process hence provide a fundamental test of the EW sector of the SM, similar  to the diboson processes discussed in Section~\ref{sec:EWK_measurements}. The largest challenge of the measurement is the large background from strong \Ztwojets\ production. To enrich the EW production, the $Z$~boson is selected as centred between two \textit{tagging jets} with a high invariant mass $m_{jj}$ and a large rapidity gap between the tag jets without central jet activity. Inclusive and differential cross-sections of four characteristic observables are extracted for the EW  \Zjj\ process (see Figure~\ref{fig:zjj:diff}) and, with a relaxed \mjj\ selection, for the strong \Zjj\ process~\cite{STDM-2017-27}. The EW \Zjj\ results, with an inclusive precision of $6.5\%$, agree well with predictions from \herwigvbfnlo~\cite{herwig1,herwig2,vbfnlo}, while the strong \Zjj\ production is most precisely modelled by \mgfpy~\cite{Alwall:2014hca}. The results are also used to constrain Wilson coefficients of dimension-6 effective field theory (EFT) operators~\cite{Brivio:2017vri} (see Section~\ref{sec:smeft}). Overall, the constraints are weaker than the ones derived with \WW\ and \WZ\ selections (see Section~\ref{sec:EWK_measurements}) but become stronger if only the SM-EFT interference terms are considered, which have  a linear effect on the cross-section. The analysis shows a unique sensitivity to the interference between the SM and CP-odd EFT amplitudes.

\begin{figure}[h!]
\begin{center}
\subfloat[]{\includegraphics[width=0.52\textwidth]{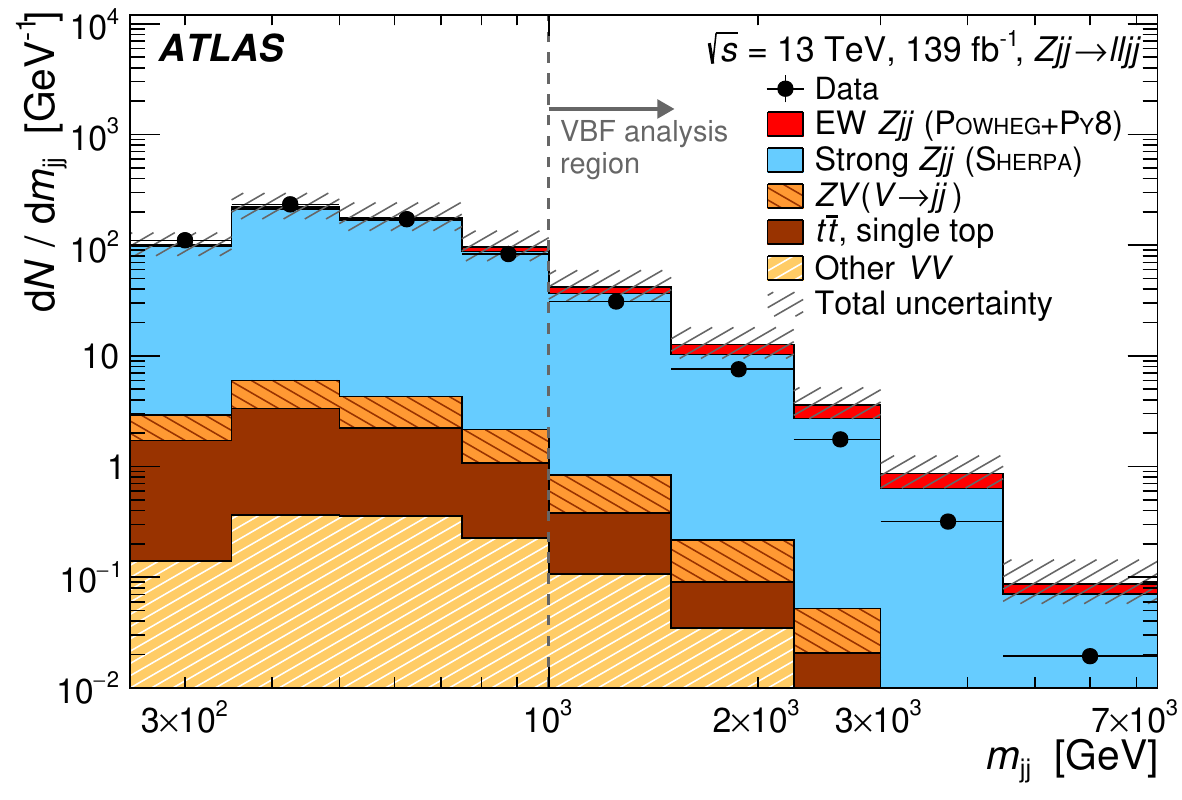}}
\subfloat[]{\includegraphics[width=0.46\textwidth]{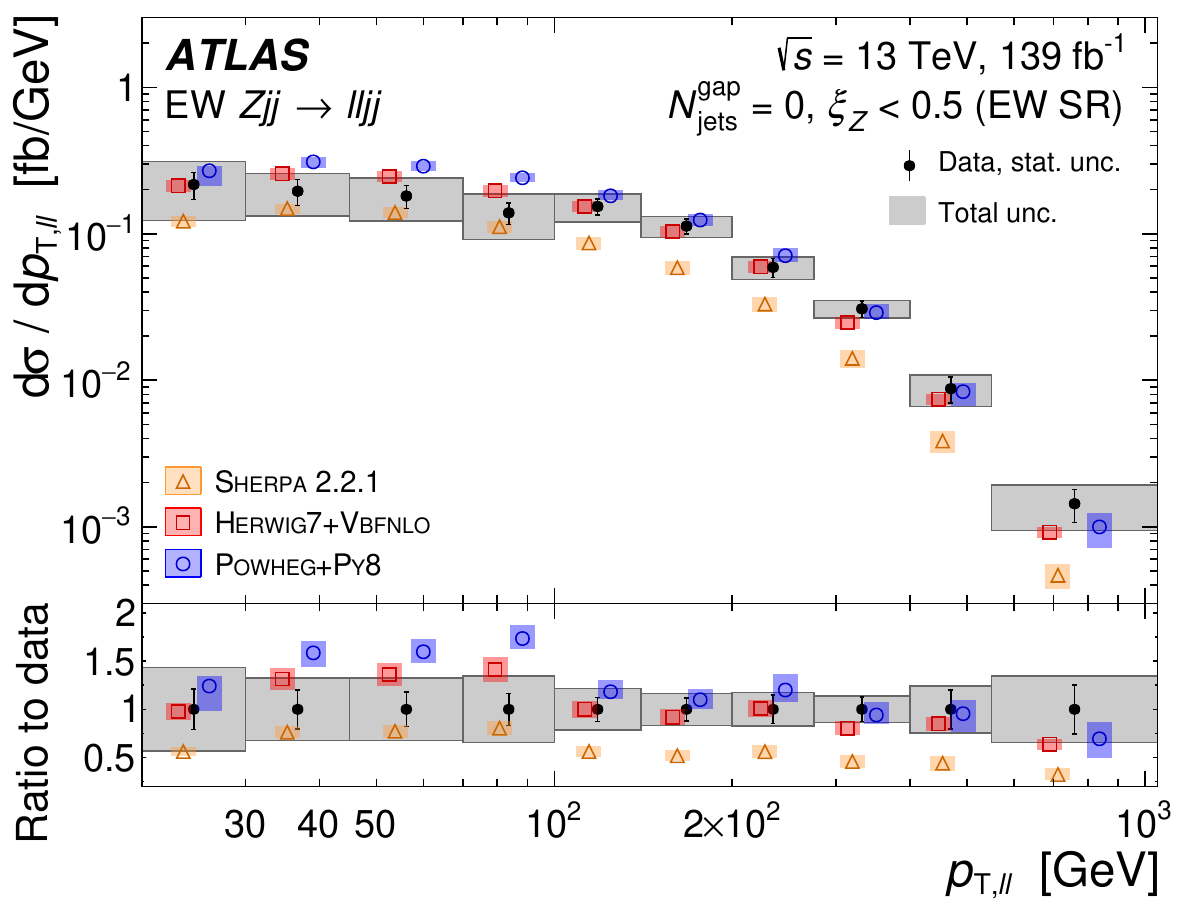}}
\end{center}
\caption{(a) Event yields as a function \mjj\ and  (b) measured cross-sections for EW $Zjj$ production as a function of \ptz~\cite{STDM-2017-27}. The lower panel shows the ratios of the predictions~\cite{Bothmann:2019yzt,Alioli_2010,herwig2,vbfnlo} to the data.}\label{fig:zjj:diff}
\end{figure}

\section{Strong and electroweak production of two  gauge bosons}
\label{sec:EWK_measurements}

Measurements of diboson production provide an excellent probe of pQCD and the gauge structure of the SM, with sensitivity to $ZWW$ and $\gamma WW$ TGCs. The high-energy tails of differential distributions are sensitive to new physics contributions, often parameterised by anomalous TGCs or in a model-independent way using the EFT framework (see Section~\ref{sec:smeft}). Polarisation measurements of the massive EW bosons further probe the SM gauge structure and details of the EW symmetry breaking (EWSB) mechanism. The increased centre-of-mass energy and large integrated luminosity of the 13~\TeV data sample allows the first observation of the EW production of two gauge bosons, which includes VBS processes with quartic gauge couplings (QGC) and $s$- and $t$-channel exchanges of a gauge or Higgs boson that regularise the amplitudes~\cite{Chanowitz88}. These processes provide a further probe of the EW theory and allow  model-independent searches for new physics via the EFT framework.

While the strong production of two gauge bosons had already been observed at lower energies, the increased centre-of-mass energy allows more sophisticated analysis techniques to be applied, to explore higher-energy phase spaces and to probe additional physics aspects.
The sensitivity to BSM physics is improved and combined EFT constraints are derived
based on published differential cross-sections (see Section~\ref{sec:smeft}).
The measurement of strong production of diboson events  in association with jets allows better control of the major backgrounds
for the observation of the EW diboson production in 13~\TeV data (see section ~\ref{sec:vbs}).
The experimental progress is accompanied by the theoretical advancements in both fixed-order calculations~\cite{Grazzini18,Dittmaier_2023,vbfnlo} and MC generators~\cite{Bothmann:2019yzt,Monni_2020,Alwall:2014hca}.
All diboson measurements in this review use leptonic \W\ and \Z\ decay modes with selections similar to those used in
Section~\ref{sec:singlebos}.

Figure~\ref{fig:diboson} shows an overview of the ATLAS diboson cross-section measurements. The figure demonstrates the significant step in precision with the higher centre-of-mass energy and the large data sample.

\begin{figure}[h!]
\begin{center}
\includegraphics[width=0.98\textwidth]{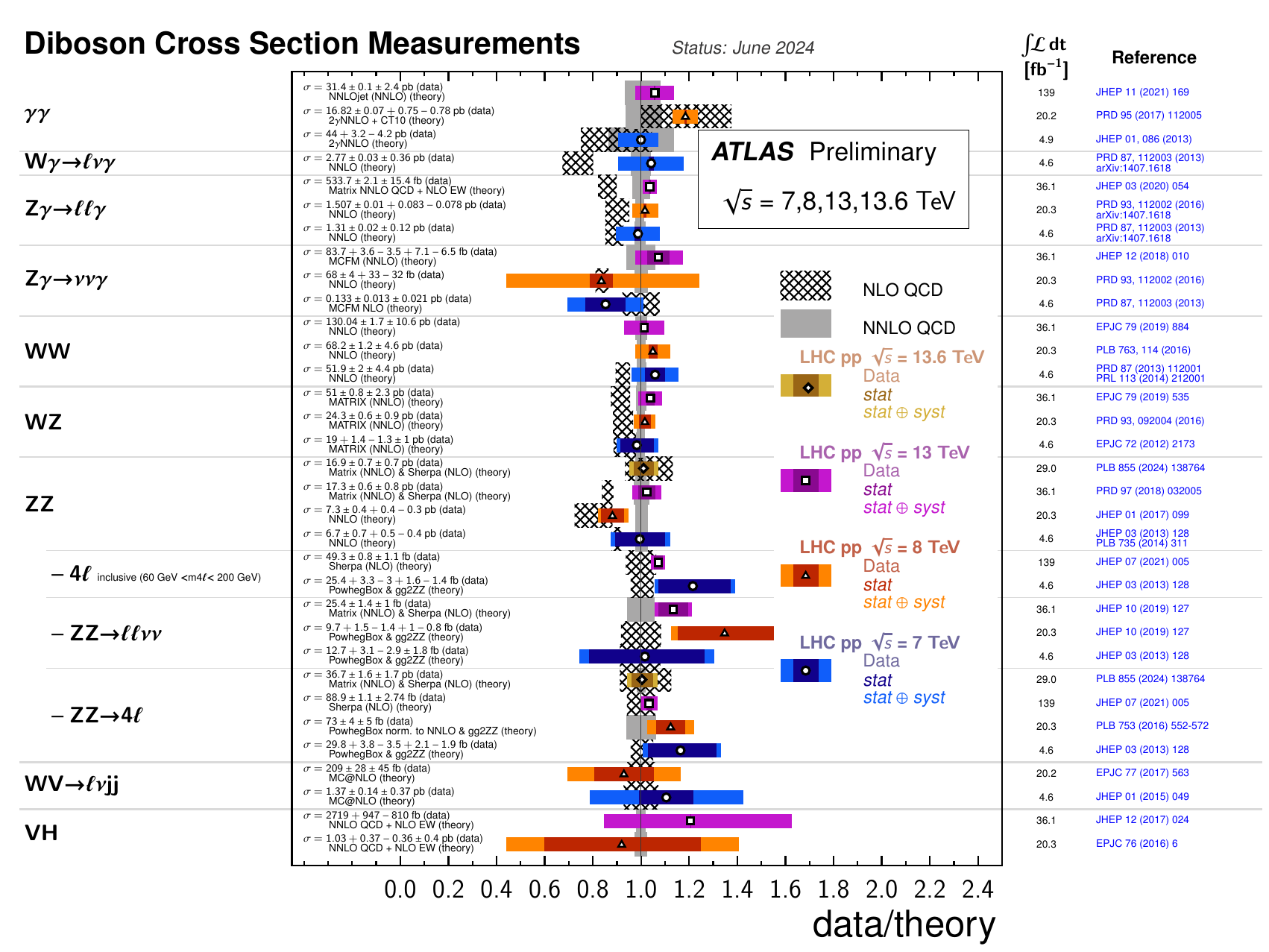}
\end{center}
\caption{Overview of ATLAS diboson cross-section measurements. The results discussed in this review are shown with a square marker~\cite{ATL-PHYS-PUB-2023-039}.}
\label{fig:diboson}
\end{figure}

\subsection{\WpmZ\ production and observation of joint-polarisation states}
\label{sec:EWK1}
Inclusive and differential  \WpmZ\ production cross-sections are measured in leptonic decays in a partial Run~2 data sample of 36~\ifb\ as reported in Ref.~\cite{STDM-2018-03}. Inclusive cross-sections are measured with a precision of $7\%$ and agree with predictions from the MATRIX framework at NNLO in QCD~\cite{Grazzini16,Grazzini15a}. Differential cross-sections are fairly well described by the theory predictions, except for high jet multiplicities. The MATRIX calculations show the best agreement with the data. In addition, the longitudinal polarisation fractions of the \W\ and \Z\ bosons are measured based on the angles between the gauge bosons and their  decay products and they are found to be in agreement with the SM predictions. The transverse \WZ\ mass, $m_{\textrm T}^{WZ}$ (see Figure~\ref{fig:wz:diff}(a)) is used to extract strong constraints on dimension-6 EFT parameters.

\begin{figure}[h!]
\begin{center}
\subfloat[]{\includegraphics[width=0.45\textwidth]{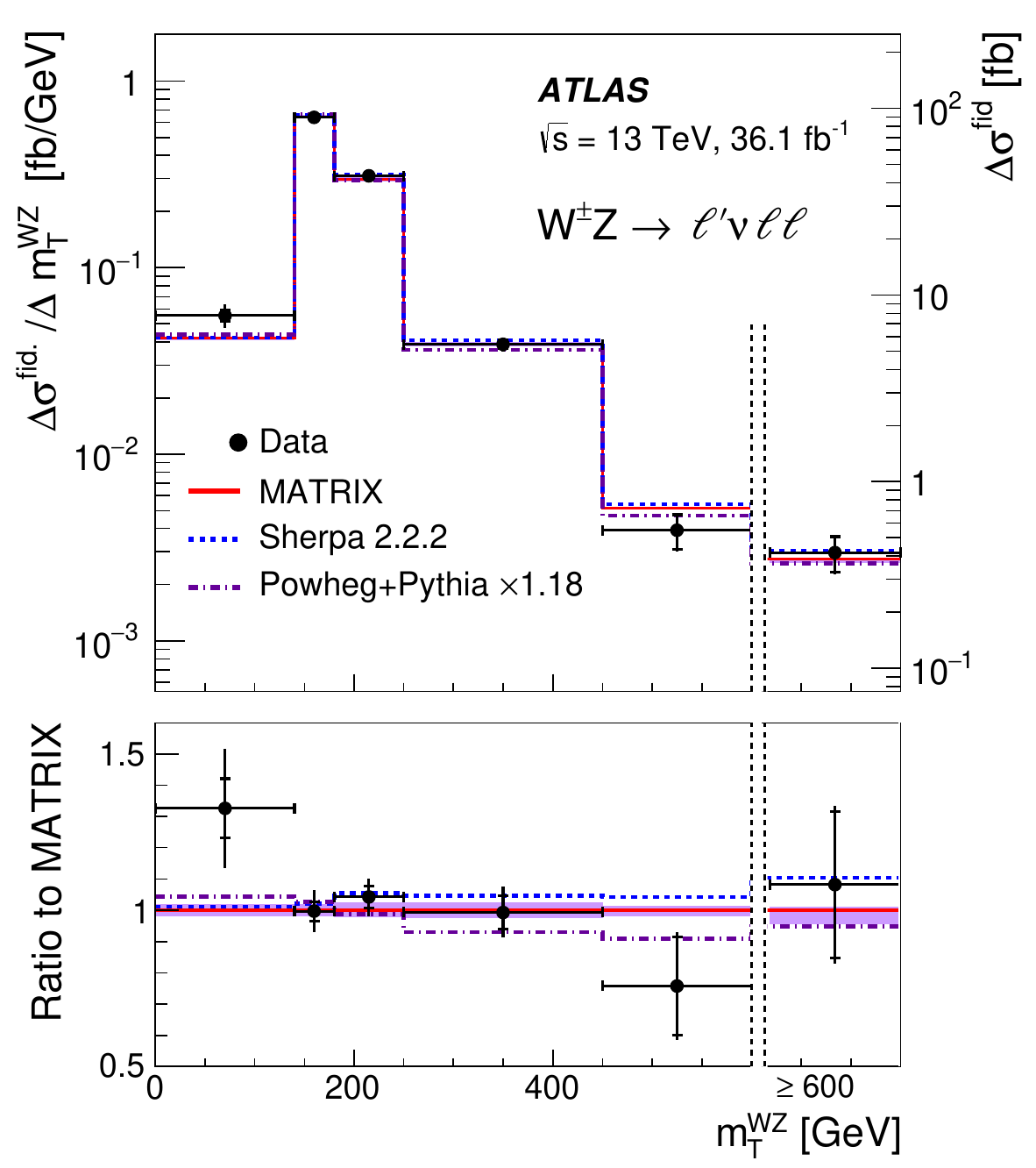}}
\subfloat[]{\includegraphics[width=0.53\textwidth]{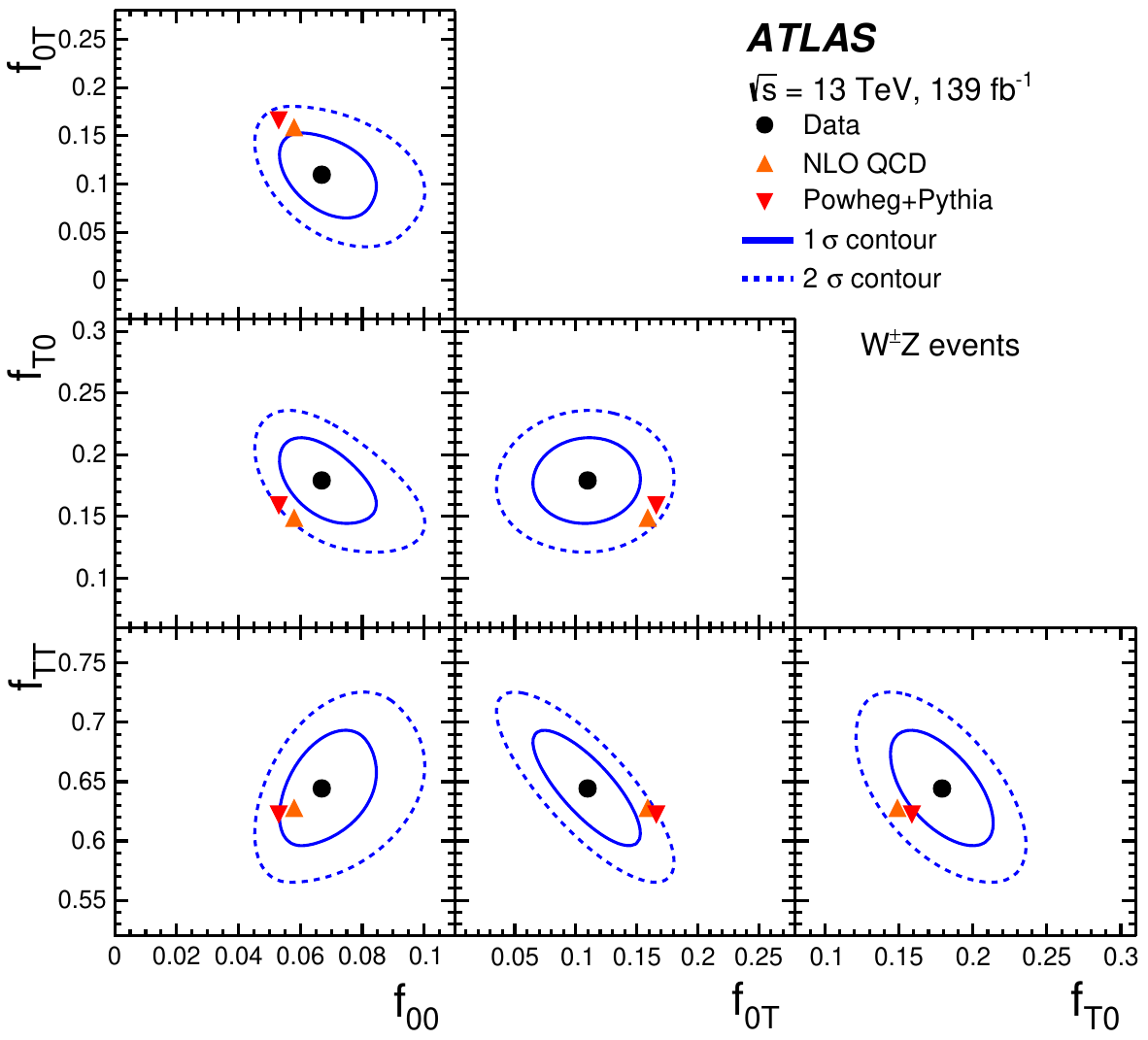}}
\end{center}
\caption{(a) Fiducial $WZ\to \ell\ell\ell\nu$ cross-section as a function of $m_{\textrm T}^{WZ}$~\cite{STDM-2018-03}. The lower panel shows the ratio of the data and \powpyeight~\cite{Nason11,Nason14} and \Sherpa~2.2.2~\cite{Bothmann:2019yzt} predictions to the MATRIX prediction~\cite{Grazzini16,Grazzini15a}.(b) Measured joint helicity fractions of the \W\ and \Z\ bosons~\cite{STDM-2022-01} compared with NLO QCD fixed-order~\cite{Denner20} and MC predictions~\cite{Nason11,Nason14}. The components $f_{00}$, $f_{TT}$, $f_{0T}$ and $f_{T0}$ indicate combinations of longitudinal (0) and transverse (T) polarisation.}\label{fig:wz:diff}
\end{figure}

The full Run~2 data sample is used to measure the joint longitudinal/transverse polarisation states of \W\ and \Z\ bosons in \WpmZ\  production~\cite{STDM-2022-01},
which are  sensitive to both the EW gauge symmetry structure and the particular way
it is spontaneously broken~\cite{Azatov17,Panico18}. No distinction is made between the two transverse helicity states.
To obtain the complete kinematics, the neutrino $p_z$ component is reconstructed using an NN regression. A deep NN (DNN) classifier is trained to separate the four joint helicity states.
The measured joint helicity fractions (see Figure~\ref{fig:wz:diff}(b)) are in agreement with SM NLO QCD fixed order~\cite{Denner20} and \powpyeight~\cite{Nason11,Nason14} predictions.
Individual helicity fractions of the \W\ and \Z\ bosons are also measured and found to be consistent with joint helicity fractions within the
expected amount of correlation. All helicity fractions are also measured separately in $W^+Z$ and $W^-Z$ events.
Inclusive and differential cross-sections for several kinematic observables sensitive to polarisation are measured and agree best with the
\powpyeight prediction normalised to the NNLO QCD prediction by MATRIX~\cite{Grazzini15a}.

Reference~\cite{STDM-2020-01} reports a further probe of the gauge structure of the SM, by selecting  \WpmZ\ events in kinematic domains with large $Z$ but small $WZ$ transverse momentum where the fraction of events with two longitudinally polarised gauge bosons is enhanced. The selection is used to study the energy dependence of diboson polarisation  and the suppression of events with two transverse-polarised gauge bosons  for small rapidity differences between the two gauge bosons~\cite{FRIXIONE19923,Baur_1994}. The results are found to agree with the SM predictions.

\subsection{\WpWm\  production}
\label{sec:EWK2}
A measurement of \WpWm\  production cross-sections~\cite{STDM-2017-24} is performed in the $e^\pm\mu^\mp$ final state, based on a partial data sample of 36~\ifb.  The number of events due to top-quark pair production, the largest background, is reduced by rejecting events containing jets with a transverse momentum exceeding 35~\GeV. The inclusive fiducial cross-section, six differential distributions and the cross-section as a function of the jet-veto transverse momentum threshold are measured and compared with several theoretical predictions. Constraints on anomalous EW gauge boson self-interactions are derived, using the transverse momentum of the leading lepton (see Figure~\ref{fig:ww:diff}(a)) in a dimension-6 EFT framework.

A complementary measurement~\cite{STDM-2018-34} is targeting \WpWm\ production
in associations with jets with a transverse momentum of at least 30~\GeV. Two additional measurements use a subselection with high-transverse-momentum jets of $\pt > 200$~\GeV.
The background from top-quark pair production is considerably reduced by rejecting events containing jets with $b$-hadron decays. The fiducial \WpWm\  cross-section is determined with an uncertainty of $10\%$ in a maximum-likelihood fit.  Differential cross-sections (see Figure~\ref{fig:ww:diff}(b)) are measured as a function of twelve observables that comprehensively describe the kinematics of \WpWm\  events. Excellent agreement is observed with state-of-the-art MC generators~\cite{Sjostrand:2014zea,Frederix:2012ps,Bothmann:2019yzt,Hamilton_2016}, where the $gg$-initiated component is modelled by \sherpa~2.2.2 and with NNLO(QCD)+NLO(EW) fixed-order calculations by MATRIX~\cite{Grazzini18,Grazzini_2020b,Kallweit_2015}. Improved limits on the EFT Wilson coefficient $c_W$  are obtained compared to earlier inclusive measurements~\cite{STDM-2017-24} if quadratic terms are neglected,
but they are still weaker than those obtained from \Zjj\ events~\cite{STDM-2017-27}.

\begin{figure}[h!]
\begin{center}
\subfloat[]{\includegraphics[width=0.45\textwidth]{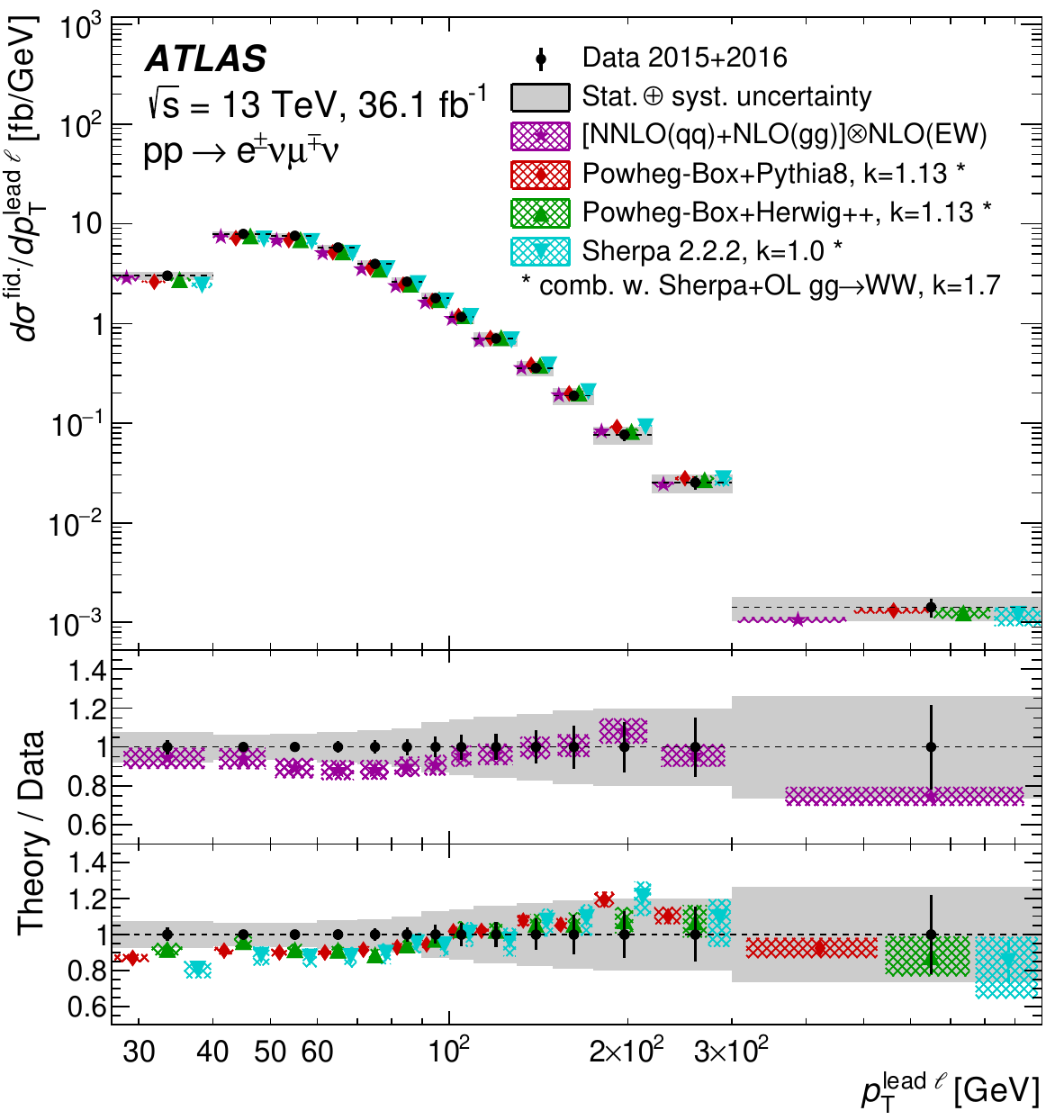}}
\subfloat[]{\includegraphics[width=0.53\textwidth]{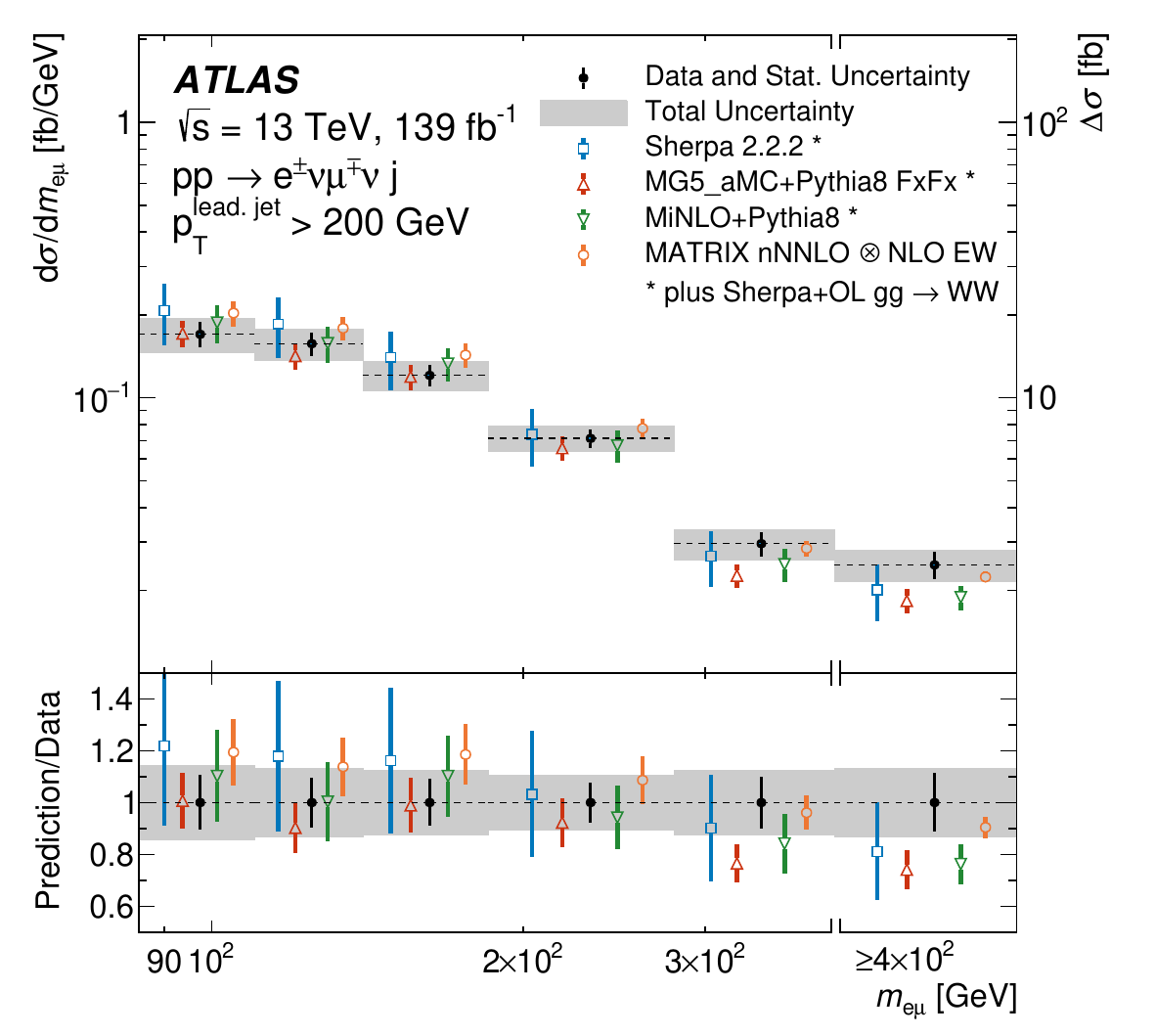}}
\end{center}
\caption{(a) Fiducial $\WpWm \to e\mu$ cross-section as a function of the transverse momentum of the leading lepton~\cite{STDM-2017-24}  and  (b) differential cross-section as a function of $m_{e\mu}$ in the high-\pt\ (jet) phase space~\cite{STDM-2018-34}. The lower panels show the ratios of the predictions~\cite{herwig1,Sjostrand:2014zea,Frederix:2012ps,Bothmann:2019yzt,Hamilton_2016,Grazzini18,Grazzini_2020b,Kallweit_2015} to the data.}
\label{fig:ww:diff}
\end{figure}

\subsection{Measurement of the \ZZ\ cross-sections}
\label{sec:zz}
The comparably rare production of two on-shell \Z\ bosons decaying leptonically is dominantly due to $t$-channel $q\bar{q}$-initiated processes and a 10\%--20\% $gg$-initiated component~\cite{STDM-2018-30}. While TGCs between neutral bosons do not exist in the SM, they may be introduced as anomalous TGCs via BSM processes. Moreover, the process constitutes an important background to $H\to ZZ$ and, in association with jets, to the EW $ZZjj$ production.
Inclusive and differential cross-sections are measured in a partial 13~\TeV data sample in final states with electrons or muons ($4\ell$)~\cite{STDM-2016-15} and in a \met-based selection targeting final states with two electrons or muons and two neutrinos ($\ell\ell\nu\nu$)~\cite{STDM-2017-03}, the latter requiring $\met> 110$~\GeV. While the $\ell\ell\nu\nu$ analysis has a reduced phase space compared to the $4\ell$ final state, it profits from the higher branching ratio of $Z\to\nu\nu$ and simple reconstruction of possibly nearby charged leptons at high momentum. The inclusive cross-sections are  measured with a $5\%$ ($7\%)$ total precision in the $4\ell$ ($\ell\ell\nu\nu$) final states, with similar statistical and systematic contributions, and are in agreement with NNLO predictions by MATRIX~\cite{Grazzini15a}.
The differential cross-sections measured for $4\ell$ and $\ell\ell\nu\nu$ final states show a reasonable
agreement with the MC generators \sherpa~2.2~\cite{Gleisberg:2008ta,Bothmann:2019yzt} and \powpyeight~\cite{Nason11,Nason14,Sjostrand:2007gs} and with the NNLO fixed-order predictions by MATRIX. Whereas for $4\ell$  the $gg$-initiated process is modelled by \sherpa~2.1 for both MC generators,for $\ell\ell\nu\nu$, $\textsc gg2vv$~3.1.6~\cite{Kauer_2012,Kauer_2013,Sjostrand:2007gs} is used for the \powpyeight prediction. The transverse momentum of the leading  $Z$ boson for $4\ell$ and of the
$Z$ boson decaying into charged leptons $p_{\textrm T}^{\ell\ell}$ in $\ell\ell\nu\nu$ (see Figure~\ref{fig:zz:diff})
are used to extract constraints on EFT parameters, including four dimension-8 operators describing aTGC interactions of neutral gauge bosons.
Constraints from the $\ell\ell\nu\nu$  final state are found to be more stringent than the ones from the $4\ell$  final state
due to the higher-energy reach of the former.

\begin{figure}[h!]
\begin{center}
\subfloat[]{\includegraphics[width=0.435\textwidth]{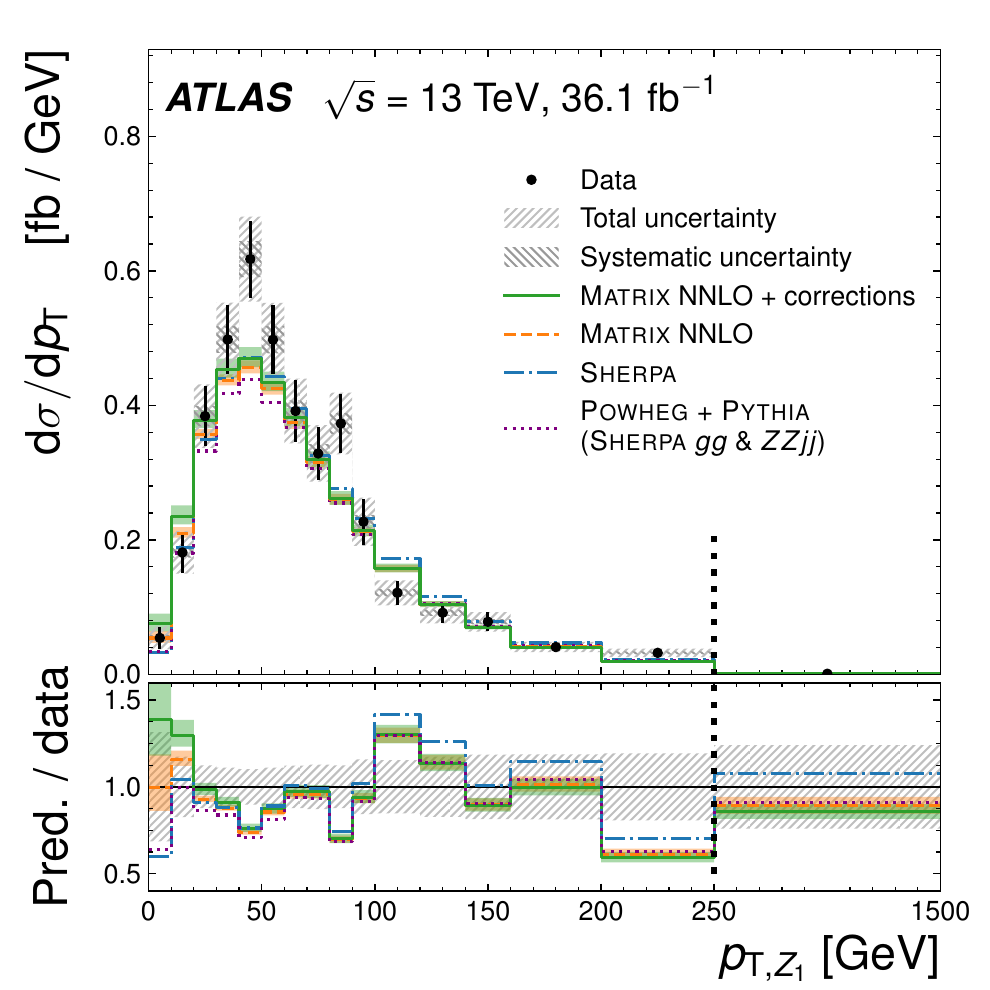}}
\subfloat[]{\includegraphics[width=0.545\textwidth]{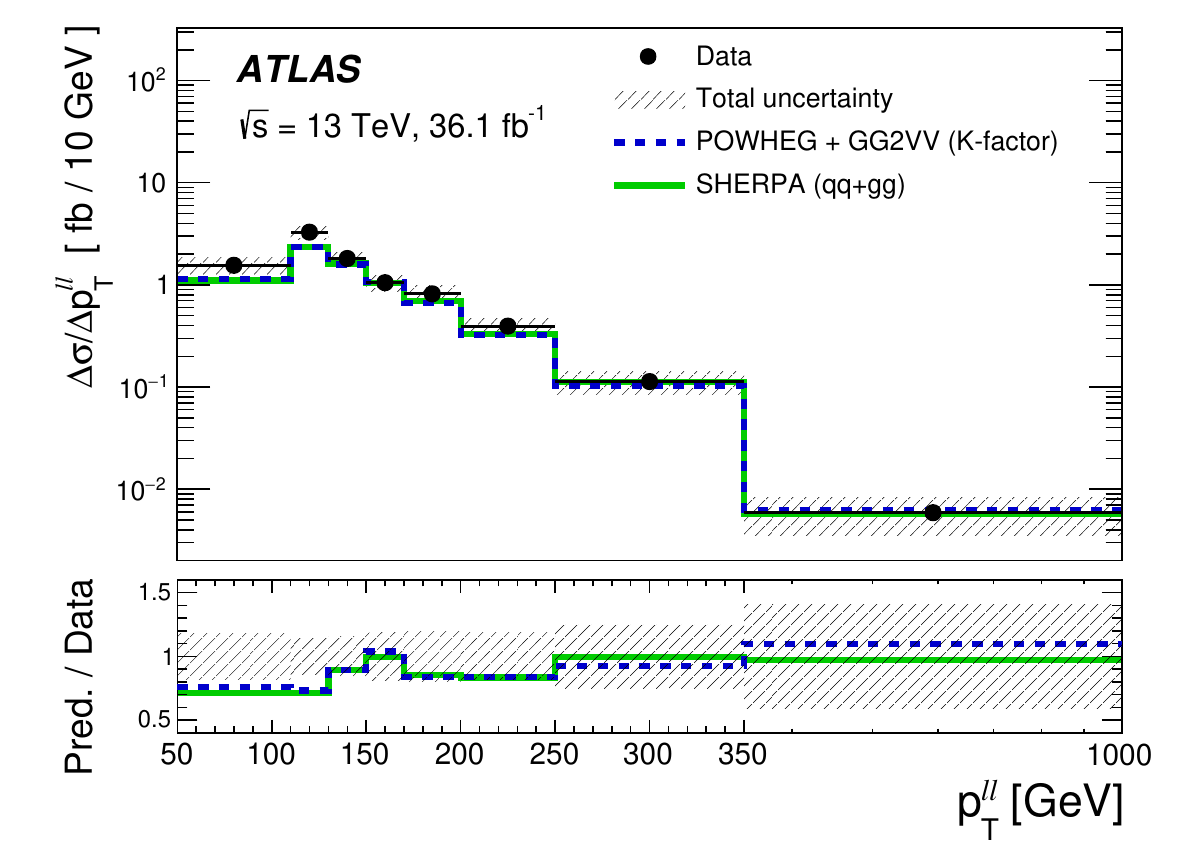}}
\end{center}
\caption{Differential cross-sections as a function  of (a) $p_{\mathrm{T},Z1}$ in $\ell\ell\ell\ell$ final states~\cite{STDM-2016-15} and  (b) $\pt^{\ell\ell}$ in $\ell\ell\nu\nu$ final states~\cite{STDM-2017-03}. The lower panels show the ratios of the predictions~\cite{Bothmann:2019yzt,Nason14,Kauer_2013,Sjostrand:2007gs,Grazzini15a} to the data.}\label{fig:zz:diff}
\end{figure}

A follow-up study on the full Run~2 data sample~\cite{STDM-2021-05} establishes a $4.3~\sigma$ evidence for the pair production of jointly
longitudinally polarised $Z$ bosons, using a pLLH fit to the output of a boosted decision tree (BDT) trained on angular variables in the $ZZ$ system. %
Moreover, the differential $ZZ$ cross-section is measured as a function of a
CP-sensitive \textit{Optimal Observable} $OT_{yz,1}T_{yz,3}$ based on CP-sensitive polar and azimuthal angles of both $Z$ boson systems
(see Figure~\ref{fig:zz:pol}(a)).
The measured cross-section is used to constrain the CP-odd neutral TGCs $f_Z^4$ and $f_Y^4$. No significant deviation from the SM is observed.

\begin{figure}[h!]
\begin{center}
\subfloat[]{\includegraphics[width=0.52\textwidth]{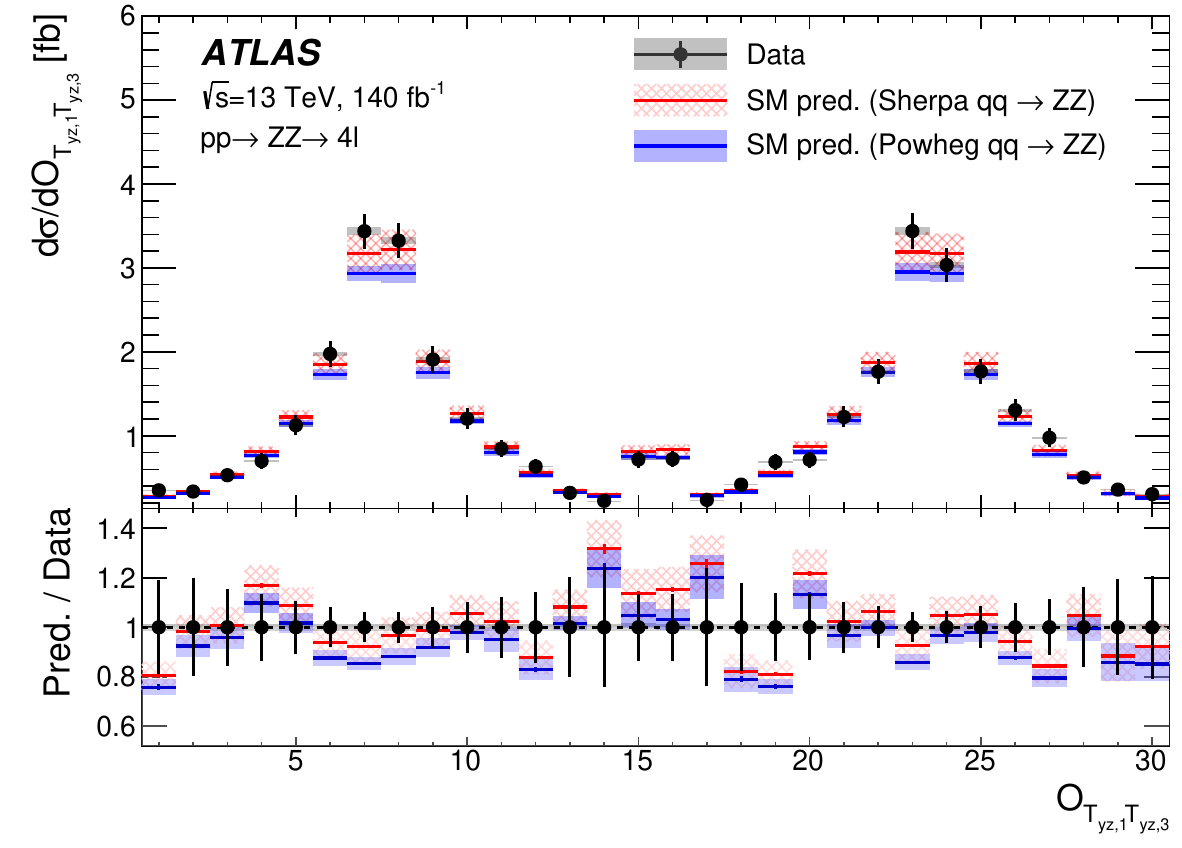}}
\subfloat[]{\includegraphics[width=0.46\textwidth]{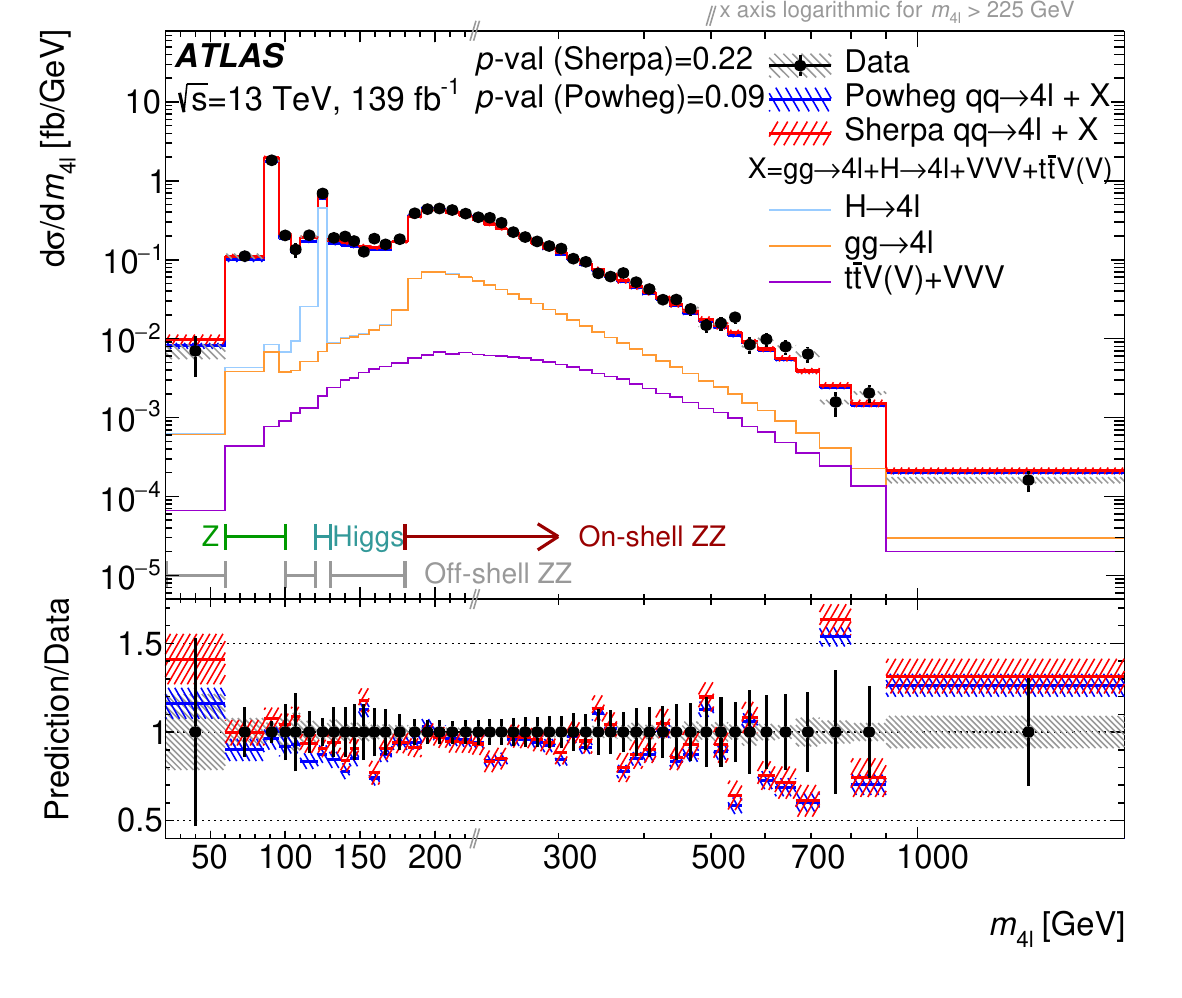}}
\end{center}
\caption{(a) Differential $ZZ \to 4\ell$ cross-section as a function of the Optimal Observable $OT_{yz,1}T_{yz,3}$~\cite{STDM-2021-05} and  (b) $4\ell$  cross-section as a function of $m_{4\ell}$~\cite{STDM-2018-30}. The lower panels show the ratios of the predictions~\cite{Bothmann:2019yzt,Nason14} to the data.} \label{fig:zz:pol}
\end{figure}

\subsection{Measurements of the 4-lepton cross-section and polarisation}
\label{sec:EWK4}
The on-shell \ZZ\ measurement in Section~\ref{sec:zz} in final states with four charged leptons can be extended to the
whole four-lepton invariant mass range $m_{4\ell}$.
The full Run~2 data sample is used to measure inclusive cross-sections in four kinematic regions, $Z\to 4\ell$, $H\to 4\ell$,
off-shell \ZZ\ and on-shell \ZZ, with a precision of 3\%--7\%~\cite{STDM-2018-30}.
In addition, differential cross-sections are obtained for six observables separately in the four $m_{4\ell}$ regions (see Figure~\ref{fig:zz:pol}(b)).
They are found to be reasonably modelled by \sherpa~2.2.2~\cite{Cascioli_2014} and \powpyeight~\cite{Nason11}, with the $gg$-initiated component modelled by \sherpa~2.2.2, and are used to derive constraints on  22 EFT parameters, both excluding and including the quadratic EFT contributions.

\subsection{Measurements of the \Zg\ cross-sections, inclusively and in association with jets}
\label{sec:EWK5}
Similarly to the \ZZ\ case, associated \Zg\ production has no TGC terms in the SM, however
BSM effects could contribute via anomalous TGCs. The full Run~2 data sample is used to select final states with two electrons or muons and one prompt isolated photon with $\pt > 30$~\GeV, with a kinematic selection to reduce photons originating from the \Z\ decay~\cite{STDM-2018-04}.
The fiducial cross-section is measured with a precision of $3\%$, making this the most precisely measured diboson final state. Results are found to be consistent with NNLO QCD predictions~\cite{Grazzini15b, Campbell17} from MATRIX~\cite{Grazzini18}. Differential cross-sections for six observables are measured and are in agreement with NLO multi-leg generator predictions from \sherpa~2.2.8~\cite{Krause:2017nxq,Bothmann:2019yzt} and  \mgamc~2.2.3~\cite{Alwall:2014hca} and with MATRIX predictions at NNLO QCD~\cite{Grazzini15b}, except for some phase space regions at low $m(\ell\ell\gamma)$ and low azimuthal distance $\Delta\phi(\ell\ell,\gamma)$ between $Z$ and $\gamma$ that are underpredicted by MATRIX\@. The predictions use the Frixione smooth-cone photon isolation criterion~\cite{Frixione:1998jh} (see also section~\ref{sec:photons}).
A further analysis~\cite{STDM-2020-14} measures thirteen 1D and five 2D differential cross-sections for \Zg+jets events with jet $\pt > 30$~\GeV (50~\GeV) for $\eta < 2.5\,(> 2.5)$ with a precision of 4\%--10\% (see Figure~\ref{fig:zgam:diff}(a)). The jet activity is well described by \sherpa~2.2.11 at NLO~\cite{Krause:2017nxq,Bothmann:2019yzt}, \mgamc~2.2.3~\cite{Alwall:2014hca}  and MATRIX~\cite{Grazzini15b} calculations, whereas LO \sherpa~2.2.4~\cite{Bothmann:2019yzt} and $\textsc MiNNLOps$~\cite{Monni_2020} yield a worse description

A measurement based on a partial data sample of 36~\ifb\ is performed in final states with an isolated prompt  photon and \met\ to target
$Z(\to\nu\nu)\gamma$ production~\cite{STDM-2017-18}, requiring $E_{\textrm T}(\gamma) > 150$~\GeV and $\met > 150$~\GeV  to
exceed  the photon trigger threshold and to reduce the backgrounds. In this high-\pt\ phase space, integrated and differential cross-sections are measured, for a selection inclusive in jets and a selection that vetoes jets.
Figure~\ref{fig:zgam:diff}(b) shows as an example the $E_{T}(\gamma)$ distribution in the exclusive $N_\mathrm{jets}=0$ selection that is used to extract constraints on EFT parameters related to neutral TGCs more stringent than those derived with \ZZ\ on the same data sample~\cite{STDM-2017-03}.
The unfolded cross-sections agree with NLO \sherpa~2.2.2~\cite{Krause:2017nxq,Bothmann:2019yzt} and \mgamc~2.2.3~\cite{Alwall:2014hca} simulations and fixed-order predictions at NNLO QCD~\cite{Campbell11}.

\begin{figure}[h!]
\begin{center}
\subfloat[]{\includegraphics[width=0.57\textwidth]{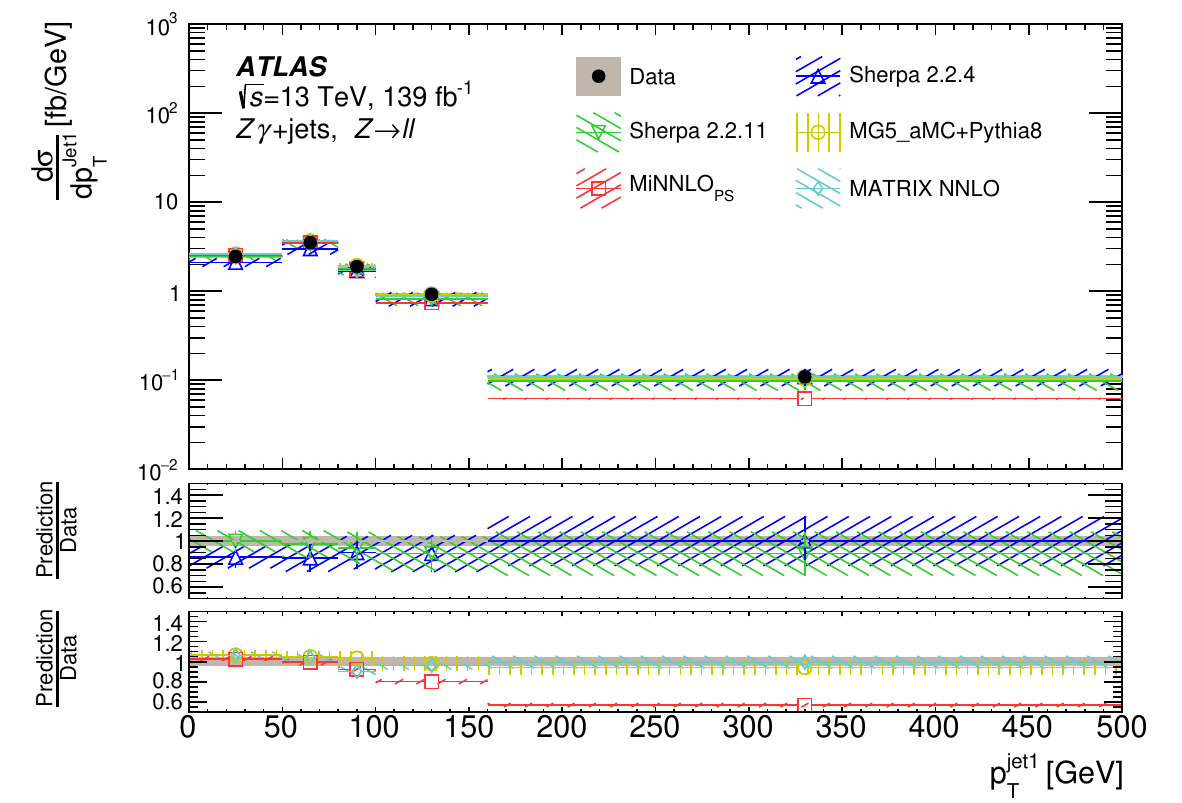}}
\subfloat[]{\includegraphics[width=0.41\textwidth]{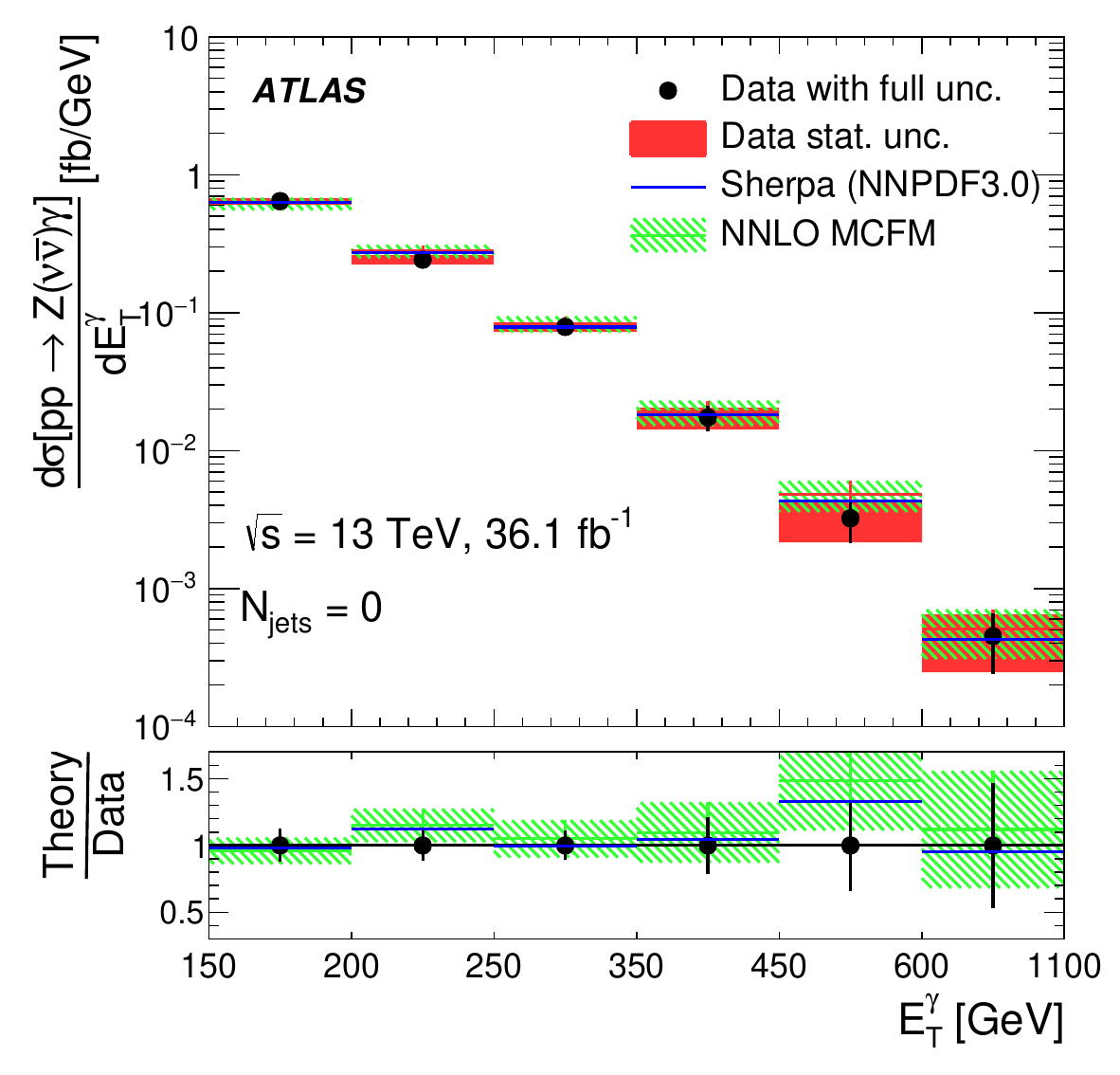}}
\end{center}
\caption{Differential $Z\gamma$ cross-sections as a function of (a) the transverse momentum of the leading hadronic jet~\cite{STDM-2018-04} and  (b) $E_{T}(\gamma)$~\cite{STDM-2017-18}.  The lower panels show the ratios of the predictions~\cite{Krause:2017nxq,Bothmann:2019yzt,Krause:2017nxq,Alwall:2014hca,Monni_2020,Grazzini15b,Campbell11} to the data.}\label{fig:zgam:diff}
\end{figure}

\subsection{Combined SMEFT analysis}
\label{sec:smeft}

Results from the analyses of EW \Zjj~\cite{STDM-2017-27}, \WpWm~\cite{STDM-2017-24}, \WpmZ~\cite{STDM-2018-03} and  $Z\to4\ell$~\cite{STDM-2018-30}, as reviewed in sections~\ref{sec:VBF}, \ref{sec:EWK2}, \ref{sec:EWK1} and~\ref{sec:zz}, respectively, are combined in a simultaneous maximum-likelihood fit to 15 EFT parameters~\cite{ATL-PHYS-PUB-2021-022} within a SMEFT framework~\cite{Brivio:2017vri}, using the EFT expansion restricted to the leading dimension~6 and dimension~8 terms:
\begin{eqnarray}{\cal L}_\text{SMEFT} = {\cal L}_\text{SM}^{(4)} + \sum\limits_i \frac{c_i^{(6)}}{\Lambda^2}O_i^{(6)} + \sum\limits_j \frac{c_j^{(8)}}{\Lambda^4}O_j^{(8)} \,,\label{eq:smeft}
\end{eqnarray}
where $c_i$ are the dimensionless Wilson coefficients and $O_i^d$ the gauge-invariant combinations of SM fields with an energy dimension $d$.
All measurements agree with the SM expectation at the level of about two standard deviations or better.
Assuming a mass scale $\Lambda = 1$~\TeV, the coefficients $c^{(3)}_{Hq}$  and $c_W$ and five additional linear combinations of coefficients are constrained to be smaller than one (see Figure~\ref{fig:smeft}).
This combination constitutes an additional step towards an ATLAS global SMEFT interpretation.

\begin{figure}[h!]
\begin{center}
\includegraphics[width=0.80\textwidth]{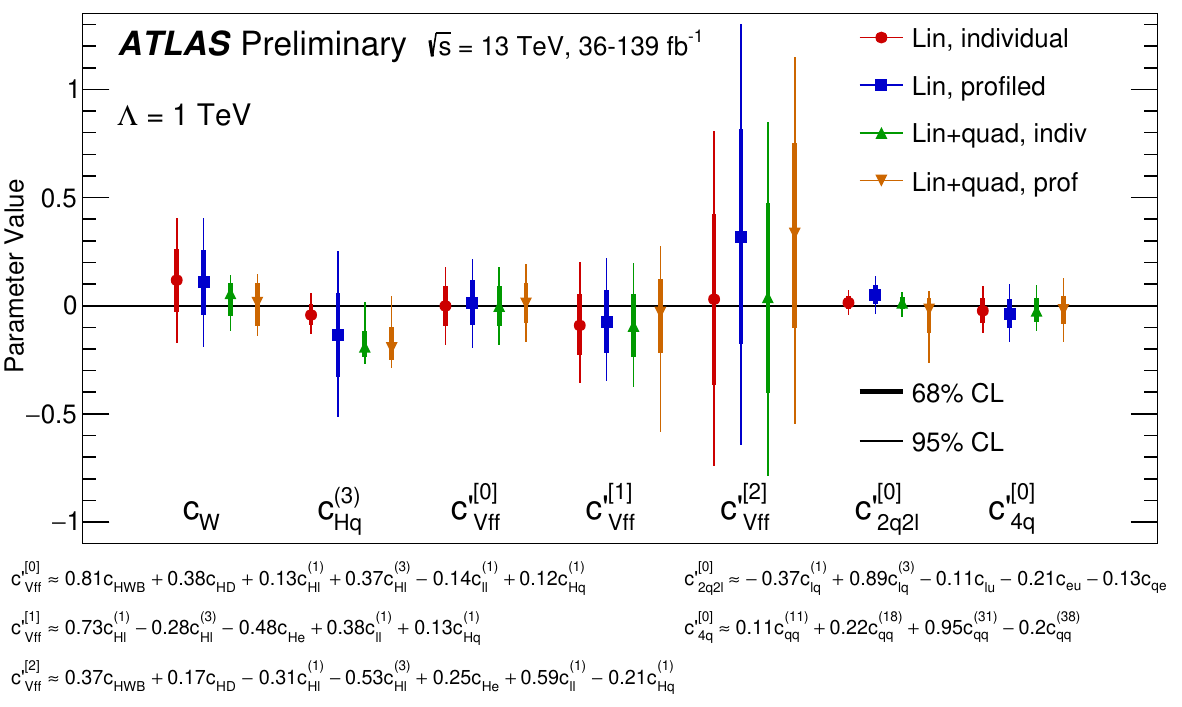}\\
\includegraphics[width=0.80\textwidth]{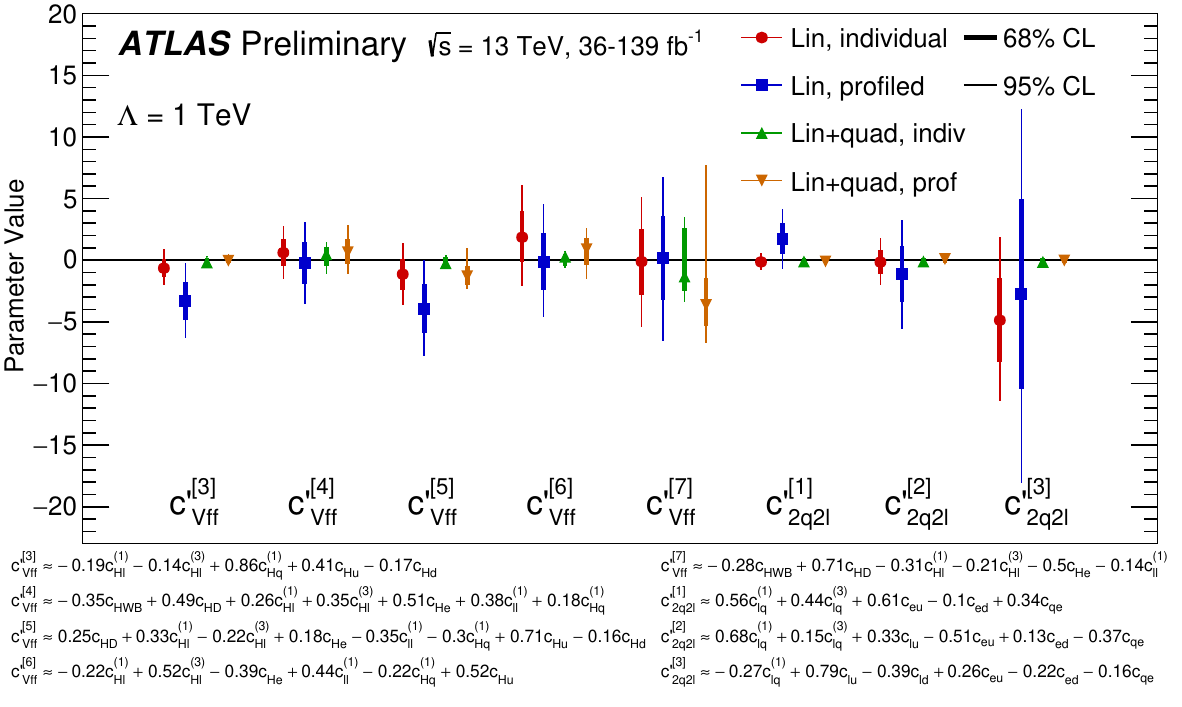}
\end{center}
\caption{Confidence intervals for the 15 parameters included in the combined SMEFT fit. Results are quoted both for fits linear in the parameters ({\textit lin}) and for fits that also take into account quadratic contributions({\textit lin+Quad}), for fits of individual parameters, while fixing other parameters to zero ({\textit indiv}) and for the combined fit, in which the remaining 14 parameters are profiled ({\textit prof}).~\cite{ATL-PHYS-PUB-2021-022}.}\label{fig:smeft}
\end{figure}

Confidence intervals obtained for individual
parameters, while fixing other parameters to zero, are furthermore compared with the results from the combined fit,
in which the remaining 14 parameters are profiled.

\subsection{Observation of electroweak production of two gauge bosons}
\label{sec:vbs}

The EW production of a diboson system in association with a
dijet system, EW $VVjj$, is related to the EW production of
single gauge bosons  discussed in Section~\ref{sec:VBF}.
Through its VBS component, it is sensitive to
QGC and details of the gauge structure with $s$- and $t$-channel
exchanges of gauge and Higgs bosons.
Figure~\ref{fig:vbs:gen} shows example Feynman diagrams for EW VBS, EW non-VBS and QCD
$VVjj$ production in the $\Wpm\Wpm jj$  channel. For other $VVjj$ processes, additional gg-initiated diagrams contribute
which are not accessible for $\Wpm\Wpm jj$.

Similarly to the diboson measurements, the analyses typically focus on
the leptonic decays of the outgoing heavy bosons (into $e$, $\mu$, or
$\nu$) or detect isolated photons. The EW production is enriched by
requiring the presence of two tagging jets with large invariant mass
$m_{jj}$ and large rapidity gap, which are not identified as
$b$-jets. The gauge boson decay products are typically expected to
be centred between the two tagging jets.
Theory calculations have become available at NLO QCD +EW and feature significant EWK corrections of
-12\% or larger~\cite{Jaeger09,Biedermann_2017,Dittmaier_2023,Denner_2019,Denner_2020,Denner_2022}.

Advanced machine-learning and fitting techniques are employed to overcome the  major challenge of separating the signal from its main background, the strong production of two gauge bosons in association with jets (see Sections~\ref{sec:EWK1}--\ref{sec:EWK5} and Figure~\ref{fig:vbs:gen}). The predictions for these backgrounds are typically not sufficiently accurate in the VBS phase space and need to be adjusted in a data-driven way. The challenges are typically addressed by designing a strong-$VVjj$ control region (CR)  and, if applicable, an additional background CR\@. The EW $VVjj$ signal is then extracted from a combined fit to the signal (SR) and control region  of the $m_{jj}$ distribution or from a multivariant discriminant trained to separate the EW $VVjj$ component.

\begin{figure}[ht!]
\begin{center}
\subfloat[]{\includegraphics[width=0.32\textwidth]{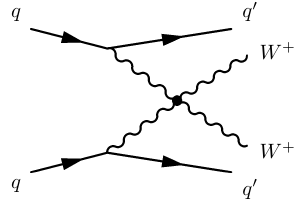}}
\subfloat[]{\includegraphics[width=0.32\textwidth]{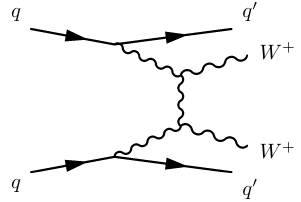}}
\subfloat[]{\includegraphics[width=0.32\textwidth]{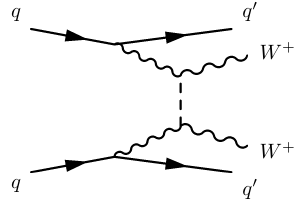}}\\
\subfloat[]{\includegraphics[width=0.32\textwidth]{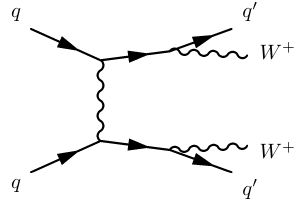}}
\subfloat[]{\includegraphics[width=0.32\textwidth]{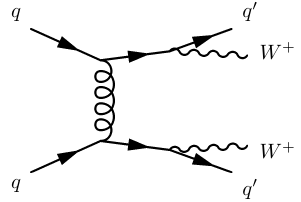}}\\
\end{center}
\caption{Example Feynman diagrams for EW $\Wpm\Wpm jj$  production with VBS via (a) a quartic gauge boson vertex, (b) a $t$-channel exchange of a gauge boson or (c) a Higgs boson, (d) a non-VBS  process  and (e) a Feynman diagram for QCD $VVjj$ production with strong interaction vertices~\cite{STDM-2013-06}.}
\label{fig:vbs:gen}
\end{figure}

The golden channel is the EW production of same-charge
$\WpmWpm jj$, as the strong background is significantly reduced
compared to all other diboson combinations. After first evidence in
the 8~\TeV data sample~\cite{STDM-2013-06}, the higher centre-of-mass
energy in Run~2 enabled the observation of this process in partial
CMS~\cite{CMS-SMP-17-004} and ATLAS~\cite{STDM-2017-06} data
samples. Moreover, ATLAS has used the full Run~2 data sample to publish more
precise inclusive and differential $\WpmWpm jj$ cross
sections~\cite{STDM-2018-32}.  The EW $\WpmWpm jj$ signal is
extracted via a fit to the \mjj\ distribution (see
Figure~\ref{fig:vbs:obs1}(a)) with a 10$\%$ precision using the full
Run~2 data.
Cross-sections are in agreement with LO \mgamc~2.6.7+\herwigseven~\cite{Alwall:2014hca,herwig1,herwig2}, LO \mgamc~2.6.7+\pythiaeight~\cite{Sjostrand:2014zea}, LO \sherpa~2.2.11~\cite{Bothmann:2019yzt} and  \powpyeight~\cite{J_ger_2011}, using the VBS
approximation~\cite{Jaeger09}. Differential cross-sections are
extracted by fits to \mjj(\mll) in each bin of the variable of interest
(see Figure~\ref{fig:vbs:obs1}(b)). Moreover, the \mll\ distribution
is used to constrain eight dimension-8 EFT operators and the
transverse-mass distribution is used to derive limits on
doubly charged Higgs boson production~\cite{Georgi:1985nv}.

\begin{figure}[ht!]
\begin{center}
\subfloat[]{\includegraphics[width=0.56\textwidth]{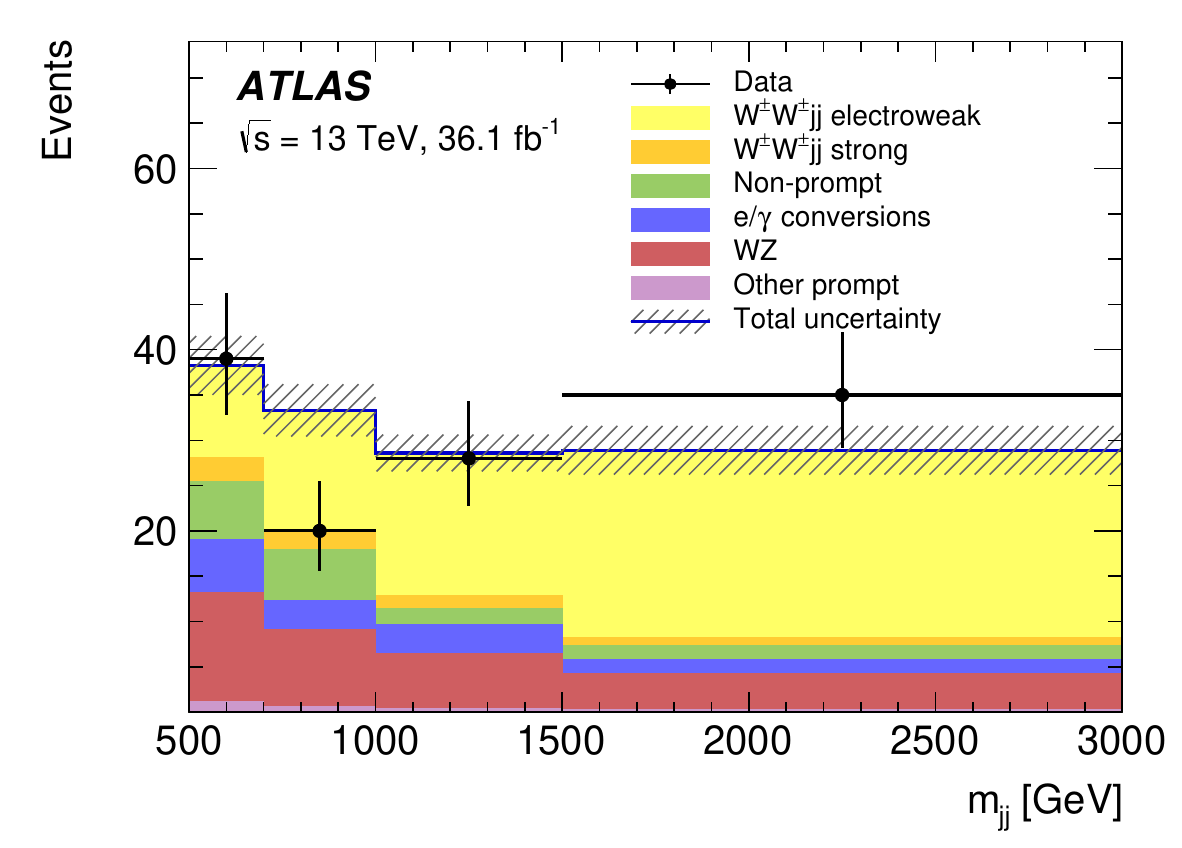}}
\subfloat[]{\includegraphics[width=0.43\textwidth]{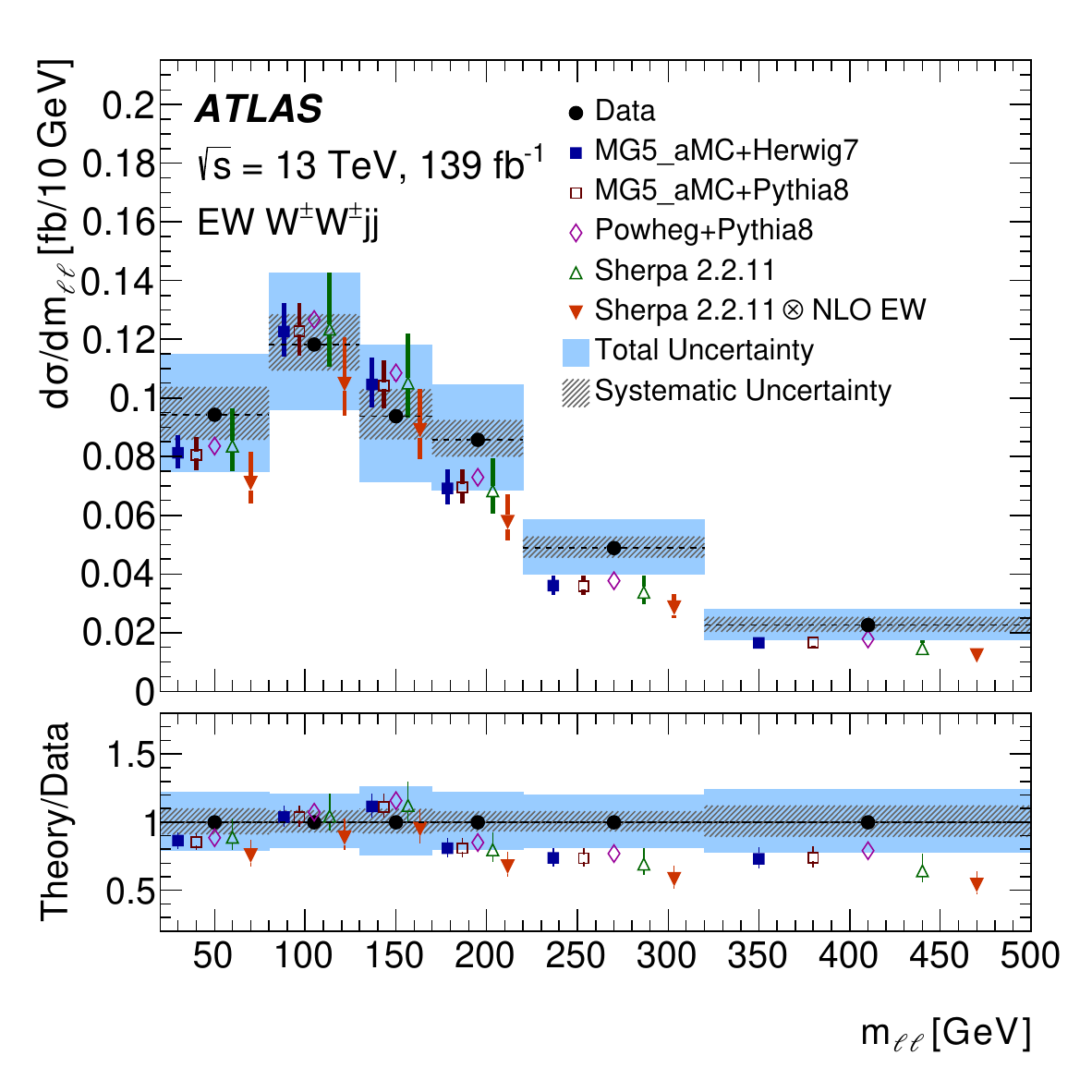}}
\end{center}
\caption{(a) Post-fit yields in the EW $\WpmWpm jj$ signal region as a function of \mjj~\cite{STDM-2017-06}  and (b) EW $\WpmWpm jj$ cross-section as a function of \mll~\cite{STDM-2018-32}. The lower panel in shows the ratios of the predictions~\cite{Alwall:2014hca,Sjostrand:2014zea,herwig1,Bothmann:2019yzt,J_ger_2011,Biedermann_2017} to the data.}
\label{fig:vbs:obs1}
\end{figure}

The more challenging EW production of two oppositely charged $W$ bosons, $\Wp\Wm jj$ is also observed in the full Run~2 data sample~~\cite{STDM-2022-06}
using a pLLH fit to an NN that discriminates between EW and QCD $\Wp\Wm jj$ production.
The inclusive cross-section is  measured with a statistically-dominated precision of 18.5~$\%$ and is in agreement with SM predictions derived with \powhegbox~V2~\cite{Nason_2004,Alioli_2010,Frixione_2007}.
Similar matrix elements to EW $WWjj$ production are probed in the photon-induced $WW$
process~\cite{STDM-2017-21} that was also observed and which is discussed in Section~\ref{ggWW}.

The large size of the ATLAS full Run~2 data sample has also allowed for the first time the observation of EW $VVjj$ production modes with one or two neutral gauge bosons in the final state:
$WZjj$~\cite{STDM-2017-23,WZjjRun2};$ZZjj$~\cite{STDM-2017-19}, which was followed by a measurement of a
region with enhanced EW $\ell\ell\ell\ell jj$
production~\cite{STDM-2020-02};
$W\gamma jj$ with leptonic $W$ boson decays~\cite{STDM-2018-31};
$Z\gamma jj$ using the invisible decay
$Z\to\nu\bar{\nu}$~\cite{EXOT-2021-17} with additional measurements in
the complementary large-$\pt^\gamma$ component~\cite{STDM-2018-59} and in
$Z\to \ell\ell$ decays~\cite{STDM-2018-36}. These channels are discussed in more detail in the following.

A first observation of the comparatively rare EW $WZjj$  process has been derived from a partial Run~2 data sample,
based on a BDT in a large-\mjj\ SR (see Figure~\ref{fig:vbs:obs3}(a))
with a statistically dominated 25\%--30\% total uncertainty~\cite{STDM-2017-23}.
Integrated and differential cross sections for the EW $WZjj$  process are derived as well on the full Run-2 data set at
an improved precision of $19\%$~\cite{WZjjRun2} and are found in agreement with LO SM predictions by \mgfpluspy~\cite{Alwall:2014hca}
and \Sherpa~2.2.12~\cite{Bothmann:2019yzt}.
The 2D distribution of the BDT score and $m_T^{WZ}$ is used to constrain dim~8 EFT operators.

The more abundant but also more challenging EW $W\gamma jj$ signal~\cite{STDM-2018-31} is extracted via a fit to a NN discriminant.
The fiducial cross section is measured with a comparable precission of 19$\%$ and differential cross sections are also  measured.
As with EW $WZjj$, the  measurements are found in agreement with LO SM predictions by \mgfpluspy~\cite{Alwall:2014hca}  and \Sherpa~2.2.12~\cite{Bothmann:2019yzt}.
The results are  used to derive constraints on dim~8 EFT operators, including the  first LHC constraints on the
coefficients $f_{T3}$ and $f_{T4}$ of dim-8 tensor-type operators.

EW $VVjj$ production with purely neutral gauge bosons in the final state is of interest as in the SM it cannot evolve via purely neutral TGCs or purely neutral QGCs. The comparatively small EW $ZZjj$ cross-sections are measured with a precision of 11\% (28\%) in the $\ell\ell\ell\ell jj$ ($\ell\ell\nu\nu jj$) channel  using a pLLH fit performed on the output of  BDTs in high-\mjj\ SRs and additional CRs (see Figure~\ref{fig:vbs:obs3}(b))~\cite{STDM-2017-19}.
They  are in agreement with \powhegbox~V2, reweighted in $m_{jj}$ based on \mgamc. Joint QCD+EW differential $\ell\ell\ell\ell jj$ cross-sections are extracted in a fiducial region with enhanced EW $ZZjj$ production and  compared  with QCD predictions at NLO from \sherpa~2.2.2~\cite{Bothmann:2019yzt} and  \mgamc\ combined with LO EW predictions by \MGNLO+\PYTHIA[8]. In addition, \mjj\ and \mll\ distributions are used to extract limits on dimension~8 EFT parameters~\cite{STDM-2020-02}.

\begin{figure}[t!]
\begin{center}
\subfloat[]{\includegraphics[width=0.29\textwidth]{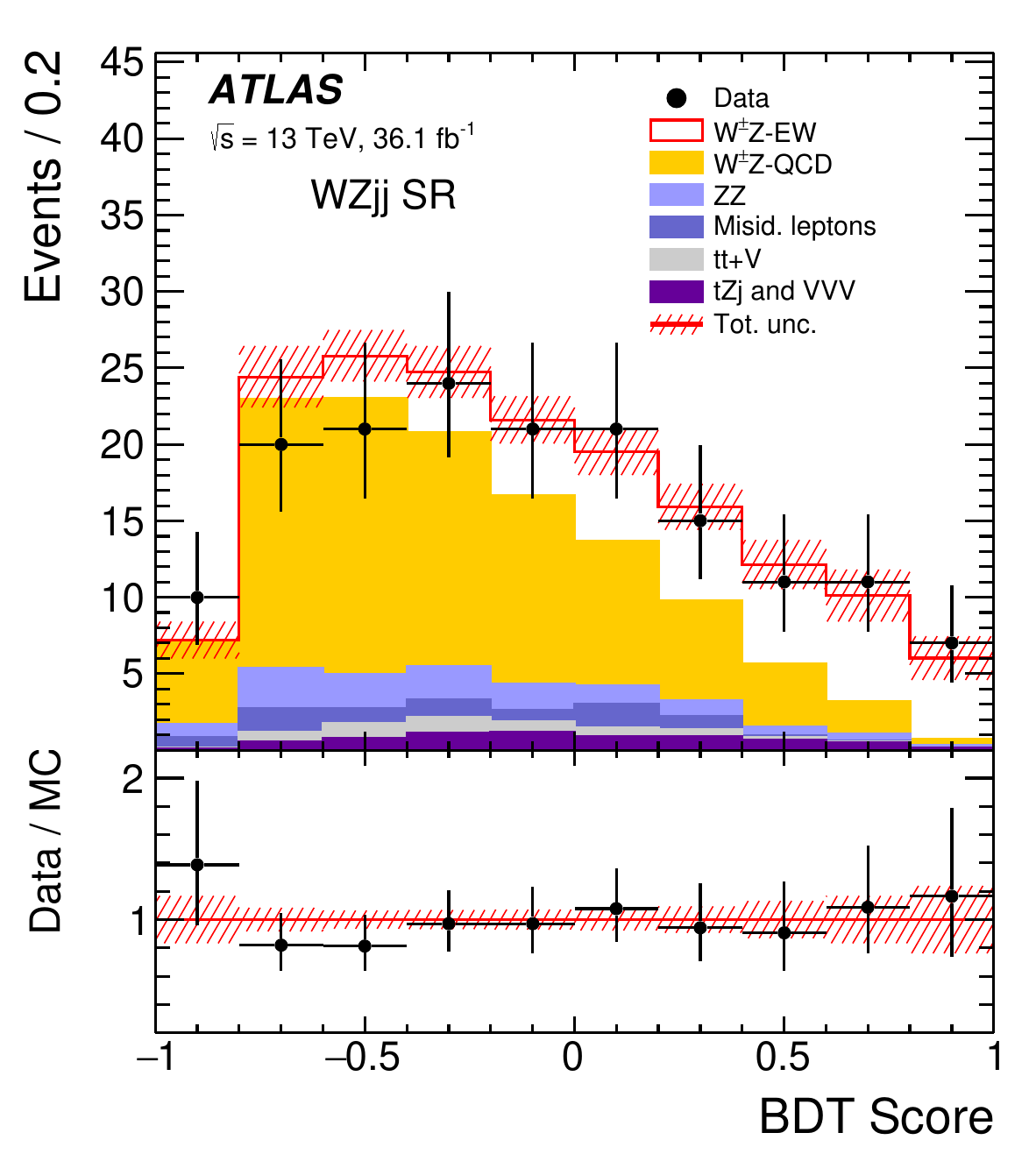}}
\subfloat[]{\includegraphics[width=0.35\textwidth]{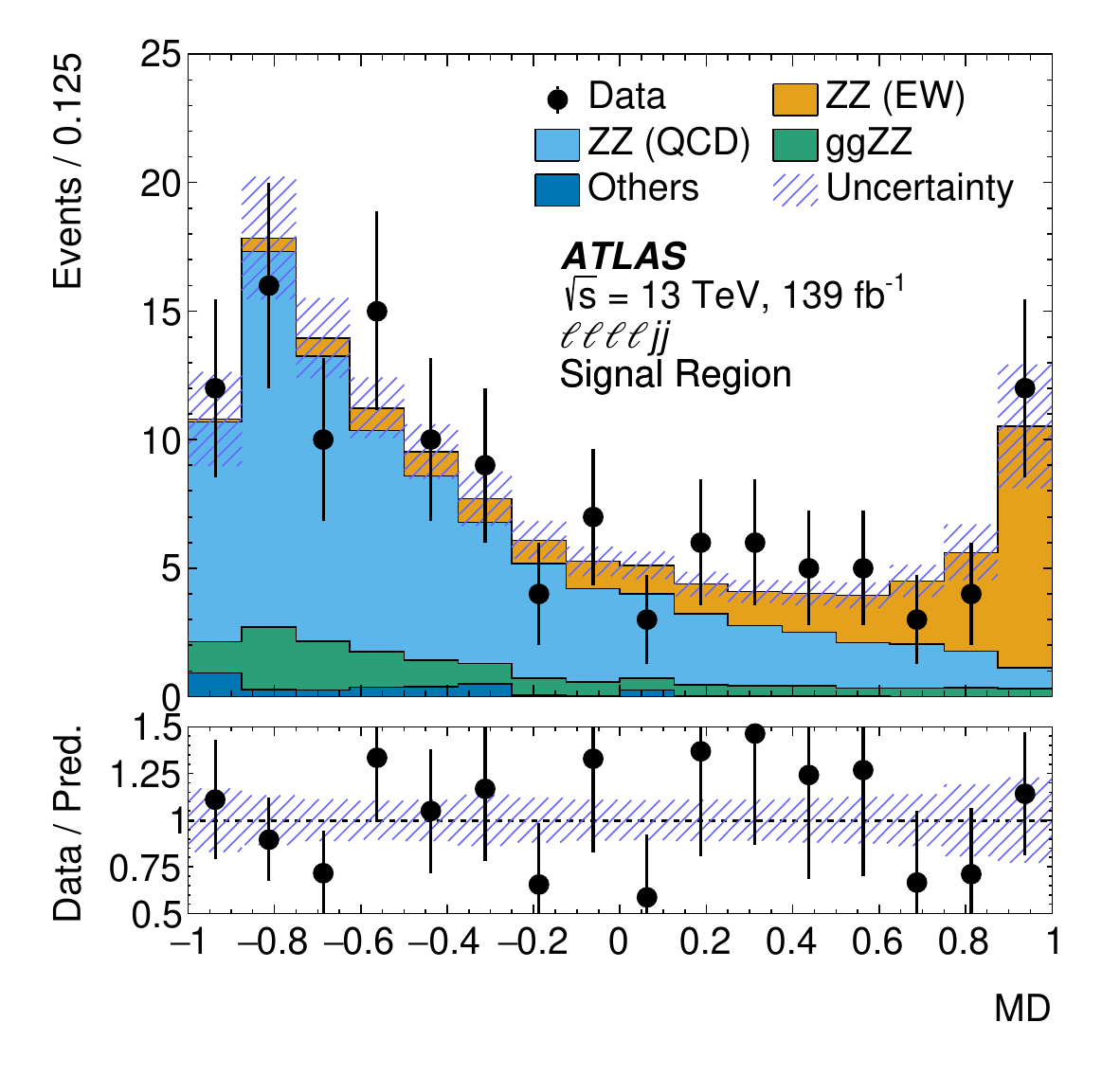}}
\subfloat[]{\includegraphics[width=0.34\textwidth]{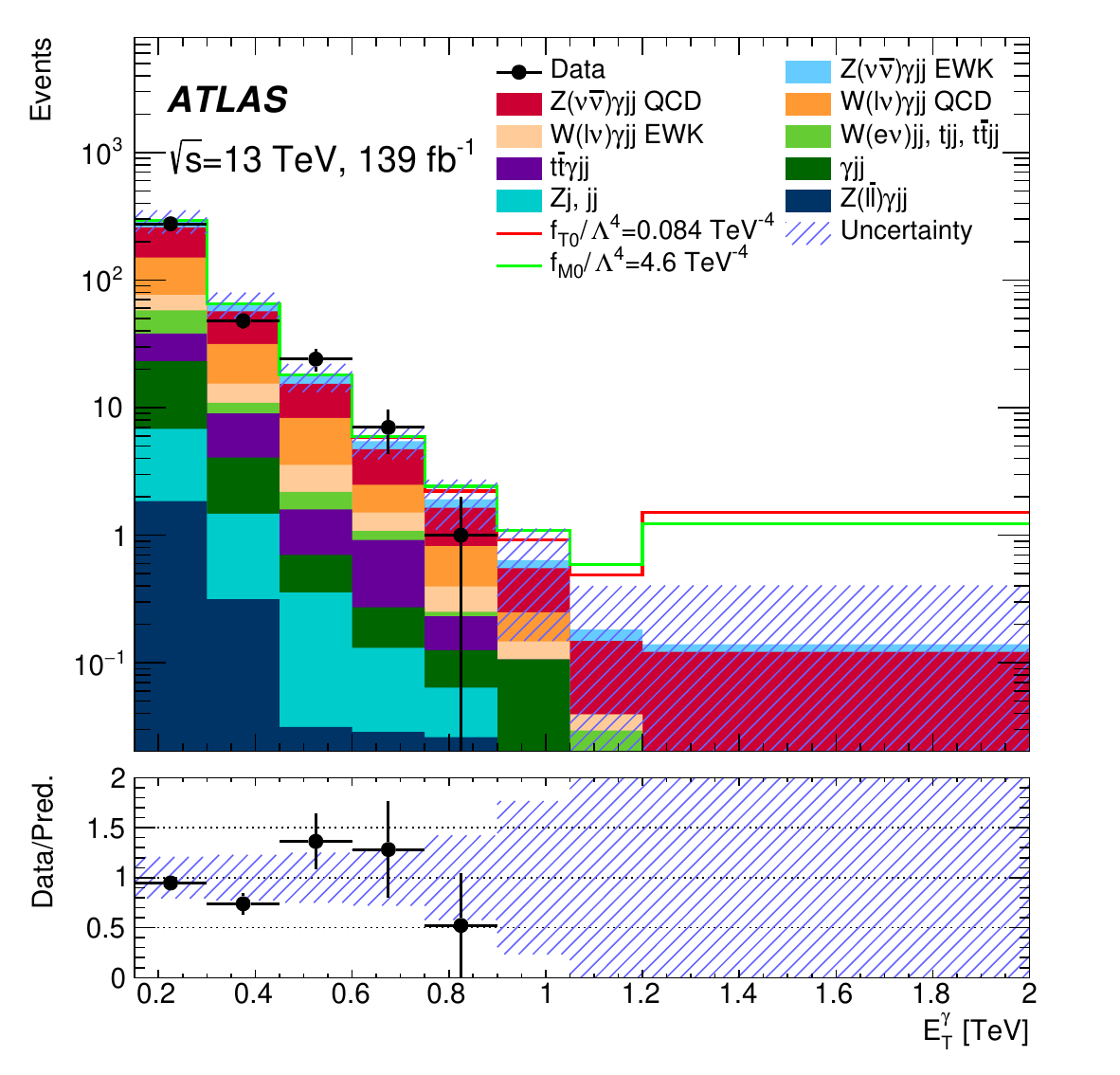}}
\end{center}
\caption{ (a) Post-fit distributions of the BDT score in the $WZjj$ signal region~\cite{STDM-2017-23}, (b) post-fit event yields as a function of the BDT score in the EW $ZZjj$  $\ell\ell\ell\ell jj$ SR~\cite{STDM-2017-19} and (c) event yield as a function of $E_{\textrm T}^\gamma$ for an EW $Z(\to \nu\bar{\nu})\gamma jj$ selection~\cite{STDM-2018-59}.  The lower panels show the ratios of the data to the predictions~\cite{Gleisberg:2008ta, Jaeger_2014,Alwall:2014hca,vbfnlo}.}\label{fig:vbs:obs3}
\end{figure}

The larger EW $Z(\to\ell\ell)\gamma jj$ cross-sections are extracted from a pLLH fit to \mjj\ in an EW $Z\gamma jj$-enriched SR and in a CR, with a statistically limited precision of 14$\%$~\cite{STDM-2018-36}.  The cross-section is found to be in agreement with the LO predictions of \mgamc~2.6.5. Differential cross-sections are derived for the SR enriched in EW $Z(\to\ell\ell)\gamma jj$ and for a more extended fiducial region with a relaxed cut on \mjj\ and are found to be consistent with predictions of \mgamc~2.6.5~\cite{Alwall:2014hca} (EW $Z\gamma jj$) + \Sherpa~2.2.11~\cite{Bothmann:2019yzt} (QCD $Z\gamma jj$).
The EW $Z\gamma jj$ process with invisible $Z$ decays is even more frequent but more challenging to separate from background processes. Final states with low $E_{\textrm T}^\gamma$ are triggered via the presence of large $\met$.  The EW $Z(\to \nu\bar\nu)\gamma jj$ cross-sections are extracted  via a combined fit to \mjj\ in several CRs and in a high-\mjj\ SR~\cite{EXOT-2021-17}. Final states with high-$E_{\textrm T}^\gamma$ can be  triggered by the photon instead.  In this case, the EW $Z(\to \nu\bar{\nu})\gamma jj$ component is extracted  in a fit to a BDT score~\cite{STDM-2018-59}. The combined EW $Z(\to \nu\bar{\nu})\gamma jj$ cross-section is measured with a precision of 22$\%$ and is compatible with predictions from \mgamc~2.6.5~\cite{Alwall:2014hca} with NLO corrections from VBFNLO~\cite{vbfnlo}. The $E_{\textrm T}^\gamma$ distribution (see Figure~\ref{fig:vbs:obs3}(c)) is used to constrain dimension~8 EFT operators.

Figure~\ref{fig:vbfvbs} shows an overview of EW measurements relevant for SM or BSM triple and quartic gauge couplings: EW production of single gauge bosons and gauge boson pairs  and triboson measurements. The figure demonstrates the significant step in precision with the higher centre-of-mass energy and the large data sample and the overall good agreement with the SM predictions.

\begin{figure}[b!]
\begin{center}
\includegraphics[width=0.98\textwidth]{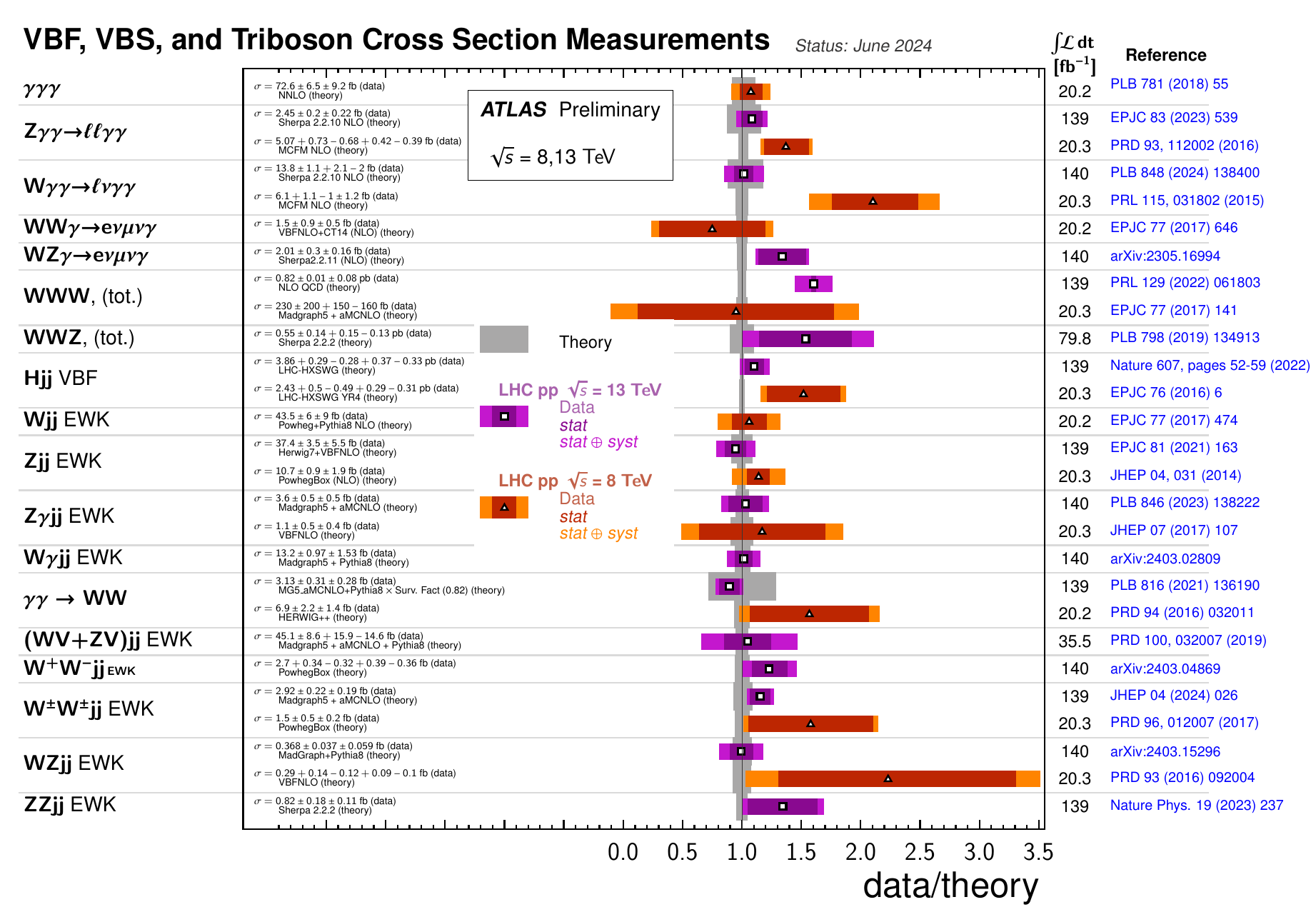}
\end{center}
\caption{Overview of ATLAS measurements of EW production of single gauge bosons and gauge boson pairs and of triboson production~\cite{ATL-PHYS-PUB-2023-039}. The results discussed in this review are shown with a square marker~.}
\label{fig:vbfvbs}
\end{figure}

\section{Production of three gauge bosons}
\label{sec:vvv}

The production of three gauge bosons is a sensitive probe of the SM gauge structure and among the rarest processes measured at the LHC~\cite{Belanger92}. The increased centre-of-mass energy and large integrated luminosity of the 13~\TeV data sample allowed the first observation of the production of three heavy gauge bosons.
Figure~\ref{fig:vbfvbs} shows an overview of all ATLAS triboson measurements.
Figure~\ref{fig:tribos:gen} shows examples of LO triboson production in the SM.

\begin{figure}[h!]
\begin{center}
\subfloat[]{\includegraphics[width=0.32\textwidth]{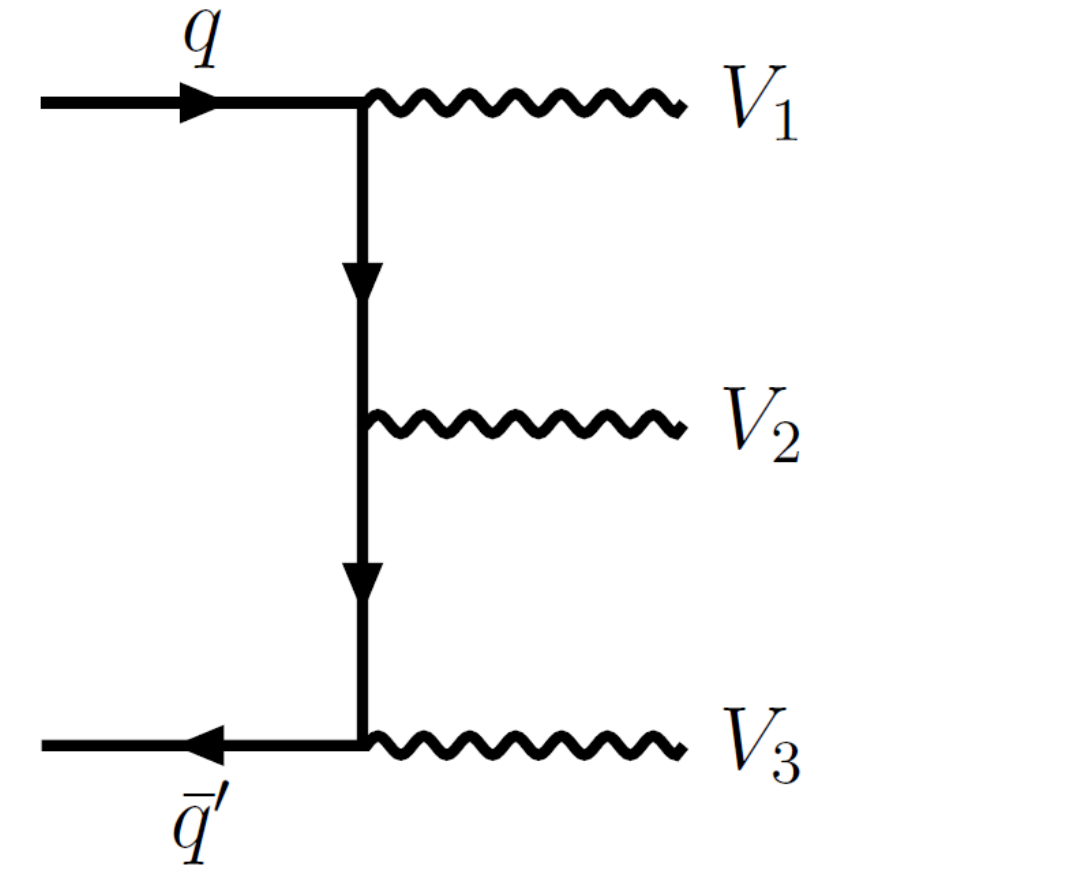}}
\subfloat[]{\includegraphics[width=0.32\textwidth]{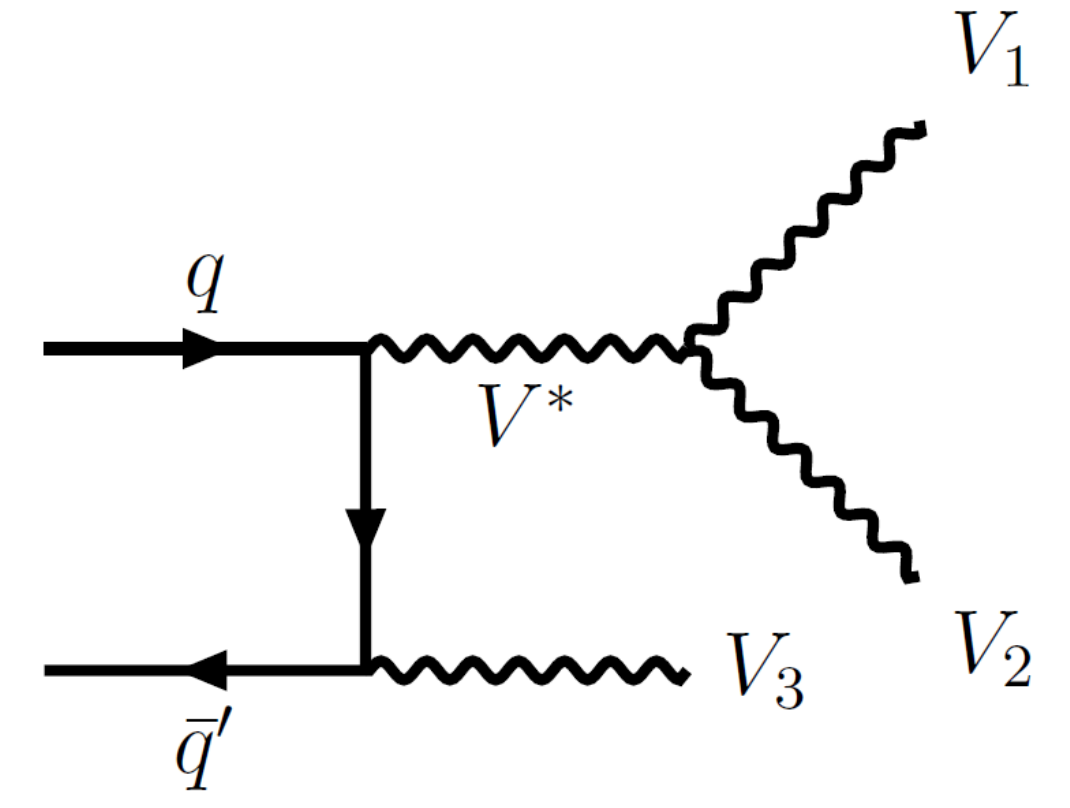}}\\
\subfloat[]{\includegraphics[width=0.32\textwidth]{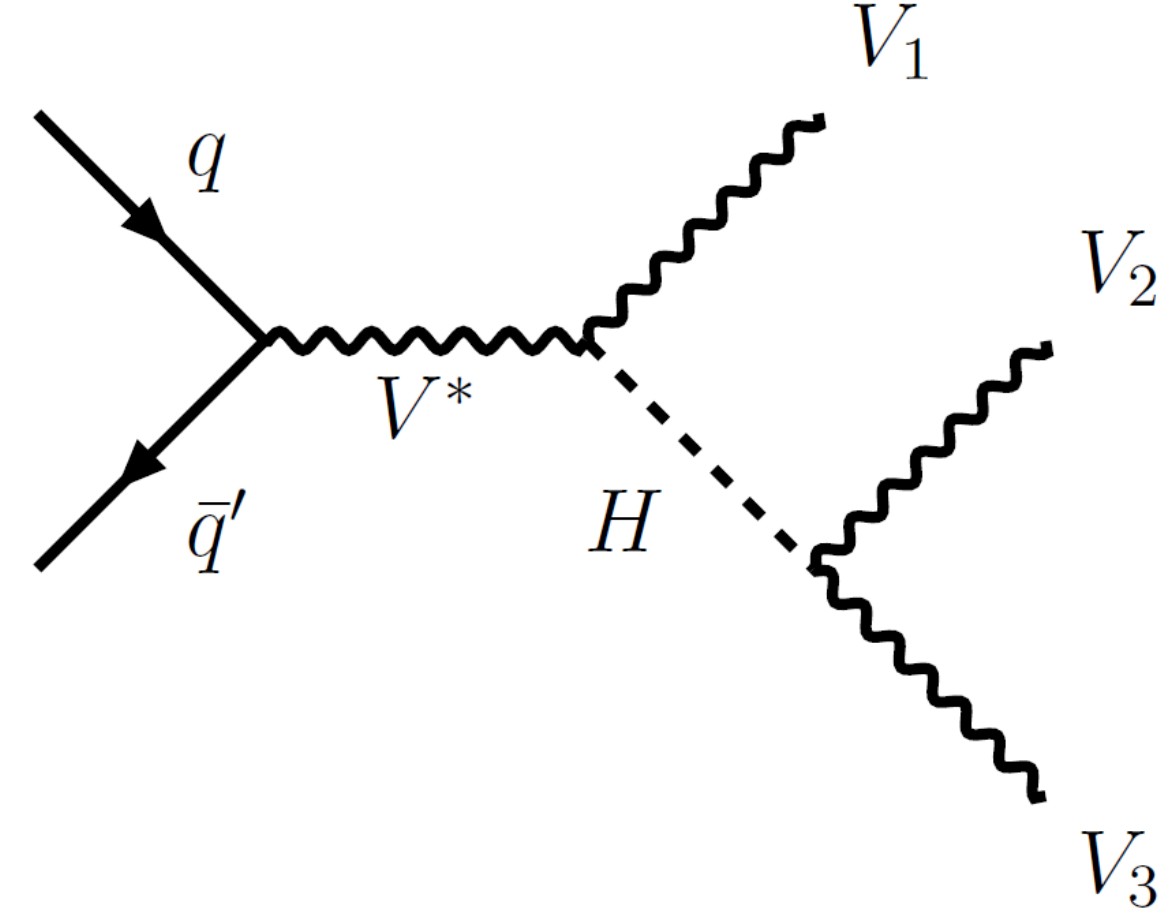}}
\subfloat[]{\includegraphics[width=0.32\textwidth]{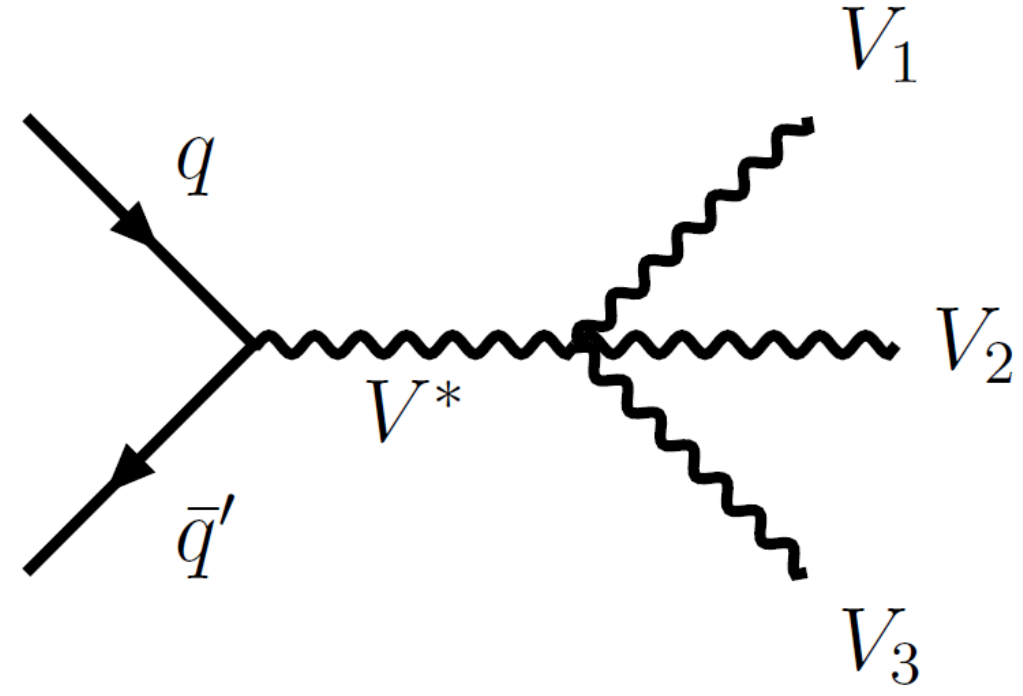}}\\
\end{center}
\caption{Example Feynman diagrams for the production of three massive vector bosons, including (a) $t$-channel production,
(b) and (c) diagrams sensitive to triple gauge couplings and (d) diagrams sensitive to quartic gauge couplings~\cite{STDM-2017-22}.}
\label{fig:tribos:gen}
\end{figure}

Triboson production involving one or more \W\ bosons can proceed through $t$-channel processes, but also diagrams involving TGCs or QGCs.
Figure~\ref{fig:tribos:gen} shows example Feynman diagrams for triboson production.
For neutral gauge bosons, only $t$-channel processes contribute in the SM\@. As discussed before, diagrams with TGCs or QGCs are interesting as they are susceptible to enhancements from BSM physics, leading to anomalous couplings.
With the Run~1 LHC data only the production of the combinations $\gamma\gamma\gamma$~\cite{STDM-2016-06} and \Zgg~\cite{STDM-2014-01} was observed. The higher centre-of-mass energies and the large Run~2 data sample allowed the observation of three additional processes: \Wgg~\cite{STDM-2018-33}, \WWW~\cite{STDM-2019-09} and \WZg~\cite{STDM-2019-17}. The \Zgg\ production was measured for the first time in a phase space dominated by the initial-state radiation contribution~\cite{STDM-2021-09}.

The first observation of \WWW\ production~\cite{STDM-2019-09}  is based on final states with two same-charge leptons and at least two jets ($\ell\nu\ell\nu jj$) and with three leptons ($\ell\nu\ell\nu\ell\nu$), excluding opposite-sign same-flavour pairs. The signal is extracted via a fit to multivariate classifiers in four signal regions and to $m_{\ell\ell\ell}$ in three $WZ$ CRs (see Figure~\ref{fig:vvv:res}(a)).  The measured cross-section is $2.6\sigma$ above the SM prediction, calculated at NLO in QCD and at LO EW accuracy~\cite{Bothmann:2019yzt,Hoeche:2014rya,Luisoni:2013cuh}.

The \WZg\ signal~\cite{STDM-2019-17} is selected via a trilepton+$\gamma$ final state, with one lepton pair consistent with coming from a \Z\ decay. The signal is extracted via a combined fit to the SR and of $ZZ\gamma$ and $ZZ(e\to\gamma)$ CRs. The resulting cross-section is consistent with the SM prediction from \sherpa~2.2.11~\cite{Bothmann:2019yzt} within $1.5\sigma$ (see Figure~\ref{fig:vvv:res}(b)).

The \Wgg~\cite{STDM-2018-33} signal is selected via $e\nu\gamma\gamma$ and $\mu\nu\gamma\gamma$ final states. The signal is extracted via a combined fit to the SR and a top-quark  ($tt\gamma$, $tW\gamma$, $tq\gamma$) CR\@. The extracted cross-section is in excellent agreement with the prediction from \sherpa~2.2.10~\cite{Bothmann:2019yzt} (see Figure~\ref{fig:vvv:res}(c)).

The \Zgg~\cite{STDM-2021-09} signal is selected in final states with two isolated photons and two electrons/muons.
The final-state radiation contribution is suppressed by requirements on 2-body and 3-body subsystem masses. The integrated cross-section is measured with a precision of 12\% and is in agreement with the SM predictions from \sherpa~2.2.10~\cite{Bothmann:2019yzt} and \mgamc~2.7.3~\cite{Alwall:2014hca}. Differential cross-sections are measured in six variables and found to be in agreement with the predictions. The distribution of $\pt^{ll}$ is used to extract constraints on eight dimension~8 EFT operators.

\begin{figure}[t!]
\begin{center}
\subfloat[]{\includegraphics[width=0.29\textwidth]{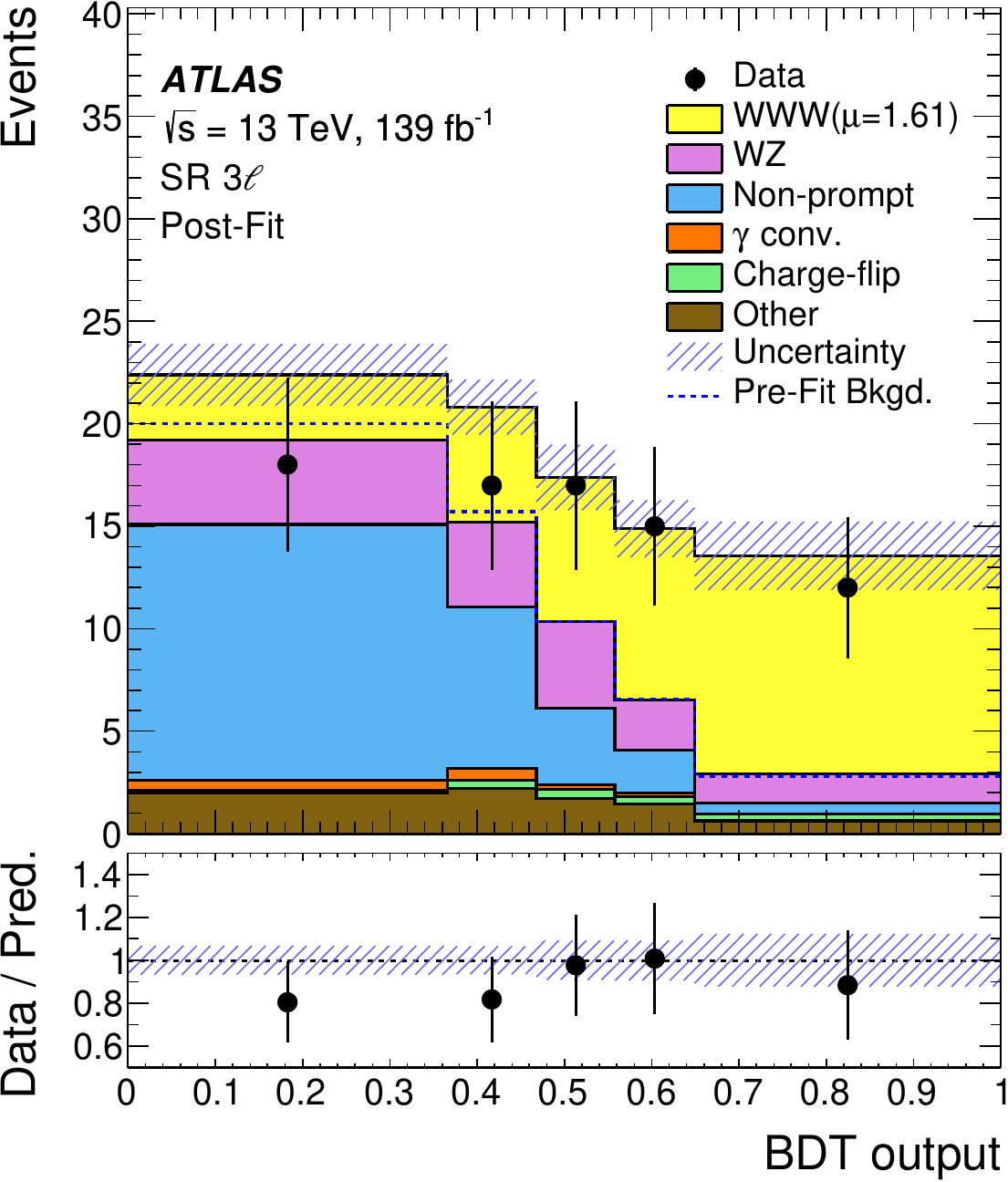}}
\hspace{0.3cm}
\subfloat[]{\includegraphics[width=0.31\textwidth]{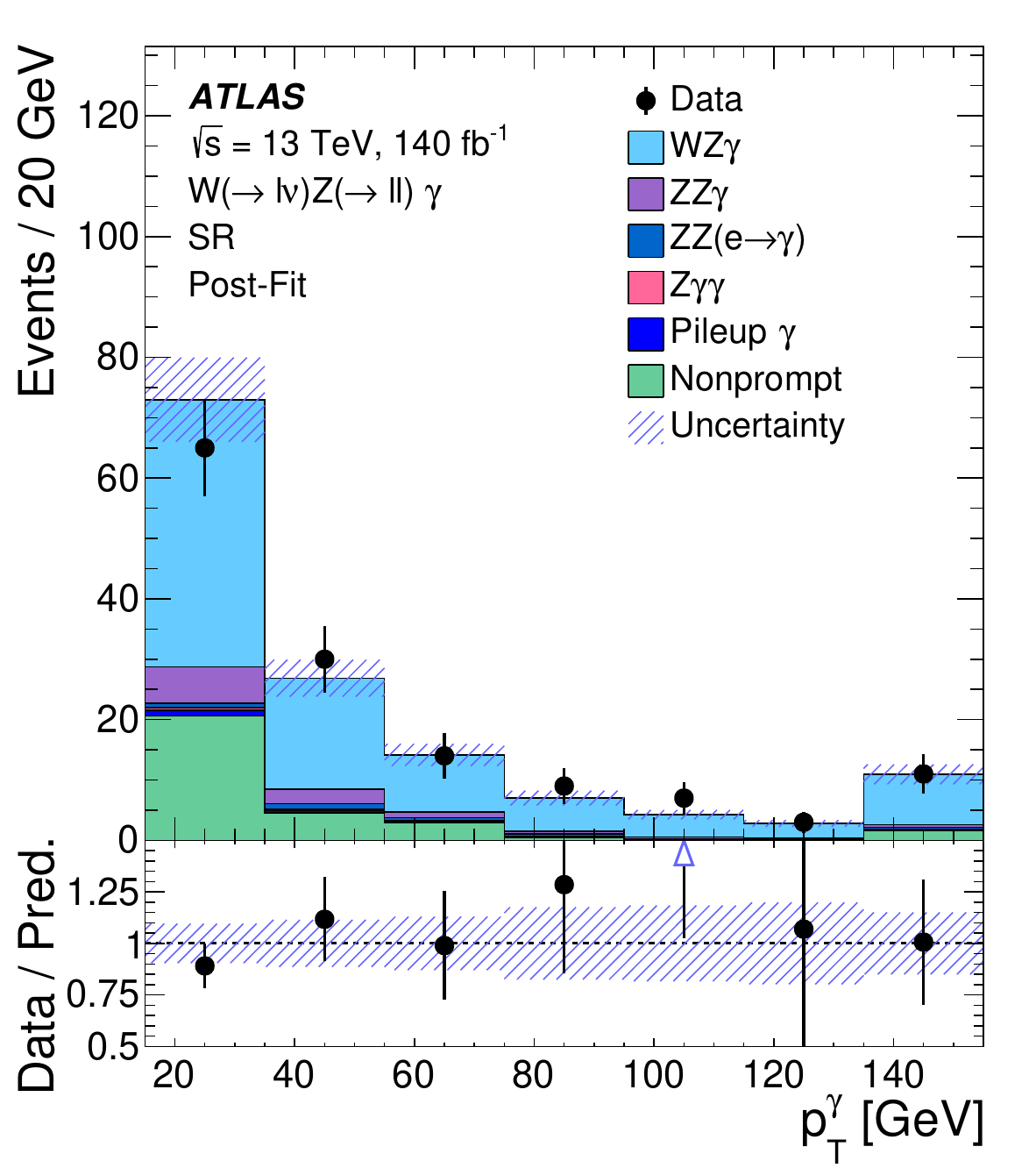}}
\subfloat[]{\includegraphics[width=0.38\textwidth]{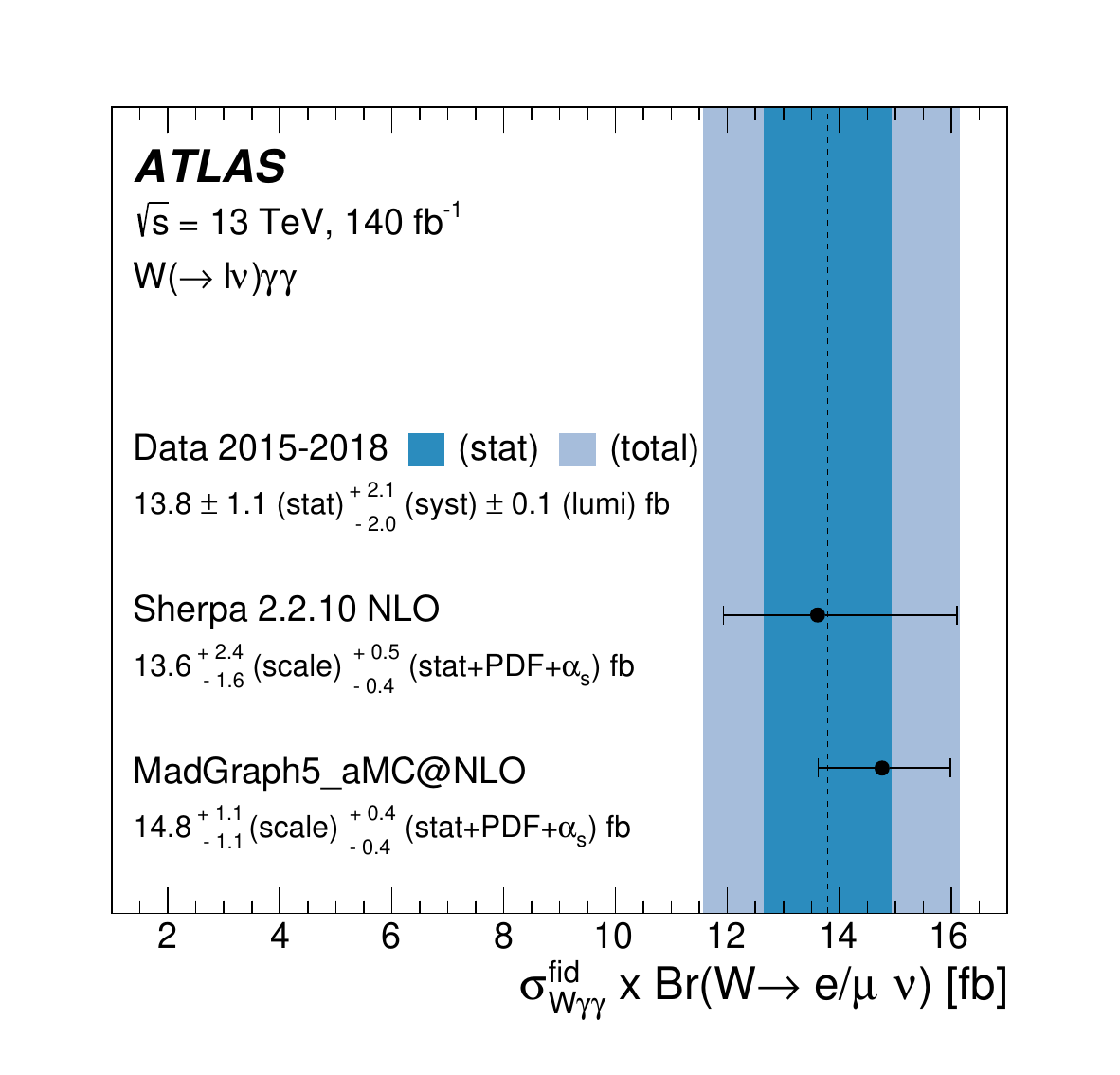}}
\end{center}
\caption{(a) Post-fit \WWW\ BDT score in the 3-lepton chanel~\cite{STDM-2019-09}, (b) distribution of photon \Et\ in the $WZ\gamma$ SR~\cite{STDM-2019-17} and (c) the measured fiducial $W(\to e\nu/\mu\nu)\gamma\gamma$ integrated cross-section compared with theory predictions~\cite{STDM-2018-33}.  The lower panels show the ratios of the data to the predictions~\cite{Bothmann:2019yzt,Alwall:2014hca}.}
\label{fig:vvv:res}
\end{figure}

\FloatBarrier



\section{Photon--photon interactions}
\label{sec:gammagamma}

Beams of protons and ions accelerated to \TeV energies at the LHC provide an opportunity to study not only the strong interactions between hadrons, but also processes involving photons in the initial state.
This is due to the presence of intense EM fields associated with the colliding hadrons.
The EM interactions are dominant at large impact parameters, $b > 2R$, where $R$ is a typical radius of the charge distribution. Therefore such collisions are also referred to as ultraperipheral collisions (UPC)~\cite{Baltz:2007kq, Klein:2020fmr}.

The EM fields associated with the ultrarelativistic hadrons can be treated as fluxes of quasi-real photons according to the equivalent photon approximation formalism~\cite{Baltz:2007kq}.
Since each photon flux scales as $Z^2$, where $Z$ is the atomic number, the two-photon luminosities are significantly enhanced for heavy ion beams, up to $Z^4 = 4.5\cdot 10^7$ in the case of Pb+Pb collisions.
The photon energy spectra follow a power-law behaviour ($E^{-1}$) up to energies of the order of $E \approx \gamma/R$ (where $\gamma$ is the relativistic Lorentz factor of the proton or ion),
beyond which the photon flux is exponentially suppressed.
Hence, the initial photon spectrum is harder for smaller charges, which favours proton over Pb beams in the production of final states with large invariant masses, such as $W$ boson pairs.

\subsection{Production of lepton pairs}

Among the possible set of photon-induced reactions, the exclusive production of lepton pairs from photon--photon collisions, i.e., $\gamma\gamma\to\ell\ell~(\ell=e,~\mu)$, is the most elementary process.
It is a particularly effective tool to study the photon flux and production cross-sections, and to investigate the effects of nuclear break-up in UPC heavy-ion collisions, or the modelling of strong-force interactions between scattered protons, which suppress cross-sections by factors known as soft-survival probabilities~\cite{Harland-Lang:2015cta}.

A measurement of the cross-sections for exclusive dimuon production, $pp \to p(\gamma\gamma\to\mu\mu)p$, at
$\sqrt{s} = 13$~\TeV is performed, using a partial Run 2 data sample corresponding
to an integrated luminosity of 3.2~fb$^{-1}$~\cite{STDM-2016-13}.
To select exclusive $\gamma\gamma\to\mu\mu$  candidates, a veto on additional charged-particle track
activity is applied.
The fiducial cross-section in the dimuon invariant mass range between 12~\GeV and 70~\GeV and differential cross-sections as a function of the dimuon invariant mass, are measured.

The observation of forward proton scattering in association with muon or electron pairs
produced via photon--photon fusion, $pp \to p(\gamma\gamma\to\ell\ell)p^{(*)}$, is also performed by ATLAS~\cite{STDM-2018-16}, in a similar way to the CMS and TOTEM analyses~\cite{CMS-PPS-17-001}.
Proton--proton collision data recorded at $\sqrt{s} = 13$~\TeV are analysed,
corresponding to an integrated luminosity of 15~fb$^{-1}$.
The $p^{(*)}$ indicates that the other final-state proton either stays intact (but is undetected) or fragments to a low mass hadronic system after emitting a photon.
One of the scattered protons is detected by the AFP~\cite{ATL-PHYS-PUB-2017-012}
while the leptons are reconstructed by the central ATLAS detector, as shown in Figure~\ref{fig:paper_fig_1}.
This figure demonstrates that the proton energy loss $\xi$ measured in the AFP spectrometer is compatible with the proton energy loss calculated based on lepton kinematics.
Both ATLAS $pp \to p(\gamma\gamma\to\ell\ell)p^{(*)}$ measurements at $\sqrt{s} = 13$~\TeV are compared with theoretical predictions that include corrections for soft-survival effects~\cite{Harland-Lang:2015cta,Dyndal:2014yea}. These predictions are in reasonable agreement with the measured cross-sections~\cite{STDM-2016-13, STDM-2018-16}.

\begin{figure}[!t]
\centering
\includegraphics[width=0.68\textwidth]{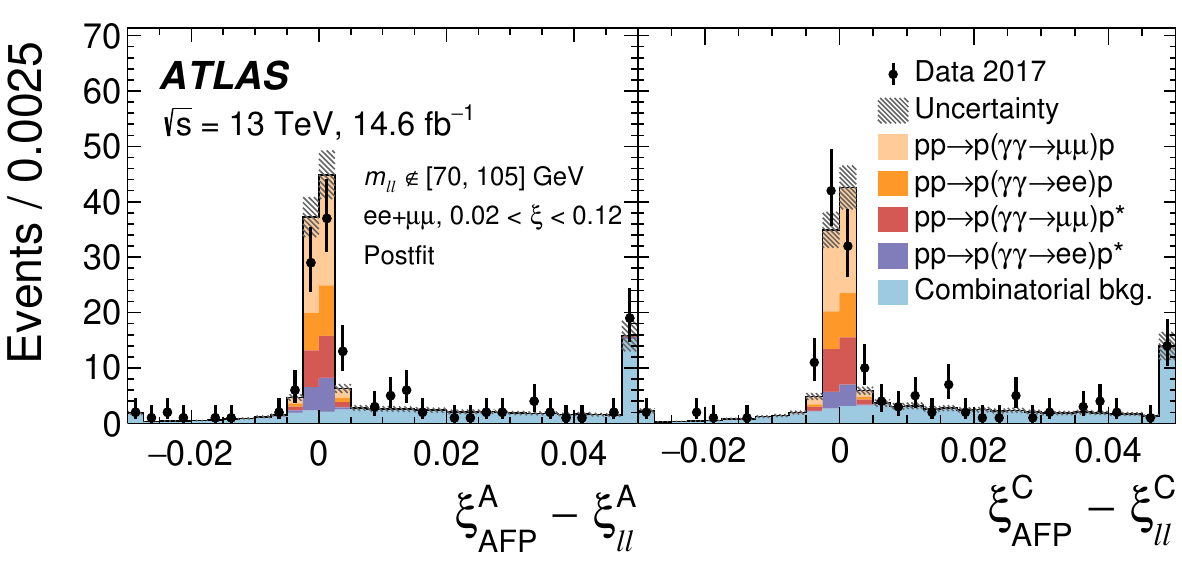}%
\caption{
\label{fig:paper_fig_1}
Distributions of the difference of proton energy loss  for the process $pp \to p(\gamma\gamma\to\ell\ell)p^{(*)}$ measured in the AFP spectrometer ($\xi_{\textrm{AFP}}$) and the expected proton energy loss based on lepton kinematics ($\xi_{\ell\ell}$)
for the two detector sides (labeled as A and C)~\cite{STDM-2018-16}.
The simulated predictions are normalised to data to illustrate the expected signal composition.
The right-most bin in each histogram contains the overflow entries.
}
\end{figure}

Exclusive dilepton production, $\textrm{Pb+Pb} \to \textrm{Pb}^{(*)}(\gamma\gamma\to \ell\ell)\textrm{Pb}^{(*)}$, via both electron-pair and muon-pair final states, is also measured by ATLAS, by utilising up to 1.7~nb$^{-1}$ of Pb+Pb data recorded at $\sqrt{s_{NN}}~=~5.02$~\TeV~\cite{HION-2021-16, HION-2016-02}.
The events are categorised relative to the presence of forward neutrons emitted as a result of Pb ion excitation (Pb$^{*}$) due to multiple Coulomb interactions accompanying the dilepton production process.
Such neutrons are detected via the zero-degree calorimeters~\cite{Jenni:1009649}.
Differential cross-sections in a fiducial acceptance  are presented as a function of several dilepton kinematic variables, and compared with theory calculations~\cite{Klein:2016yzr,Harland-Lang:2018iur}.
In particular, the muon kinematics can be used to estimate the initial photon energies, $k_1$ and $k_2$: $k_{1,2} = (1/2)m_{\mu\mu} \exp(\pm y_{\mu\mu})$, where $m_{\mu\mu}$ is the dimuon invariant mass and $y_{\mu\mu}$ is the dimuon rapidity.
Since the two photons are emitted independently, each event can be characterised by the maximum and minimum photon energies $k_{\textrm{max}}$ and $k_{\textrm{min}}$, where $k_{\textrm{max}}$ is the larger of the two photon energies.
Generally, as shown  in Figure~\ref{fig:dileptons}(a), good agreement is found but some systematic differences are seen, which may be explained by deficiencies in the modelling of the incoming photon flux.

\begin{figure}[t!]
\begin{center}
\subfloat[]{\includegraphics[width=0.43\textwidth]{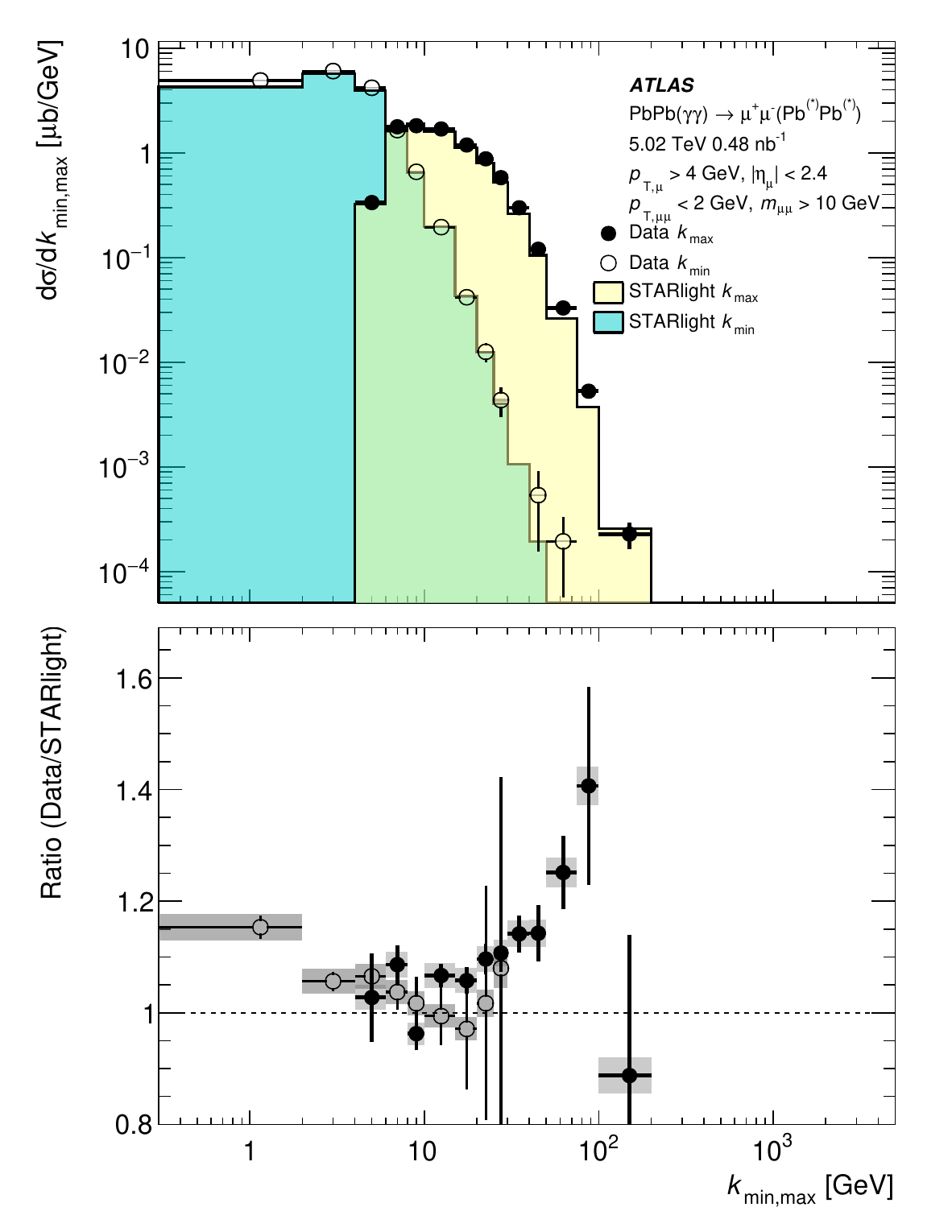}}
\subfloat[]{\includegraphics[width=0.55\textwidth]{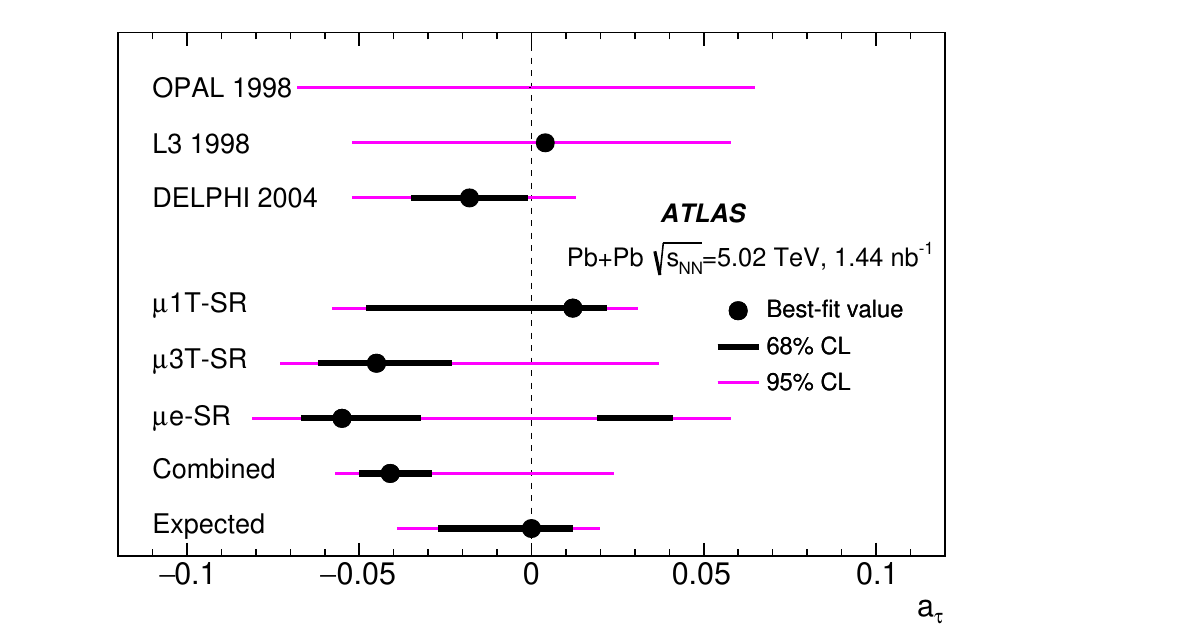}}
\end{center}
\caption{(a) Differential cross-sections for exclusive dimuon production in Pb+Pb UPC as a function of the maximum photon energy ($k_{\textrm{max}}$) and minimum photon energy ($k_{\textrm{min}}$)~\cite{HION-2016-02}. The lower panel shows the ratio of the data to the predictions.
(b) ATLAS measurements of $\tau$-lepton anomalous magnetic moment ($a_{\tau}$) from fits to individual signal regions, and from the combined fit~\cite{STDM-2019-19}.
These are compared with existing measurements from various experiments at LEP.
}
\label{fig:dileptons}
\end{figure}

The production of $\tau$-lepton pairs in Pb+Pb UPC provides a highly interesting opportunity to study the EM properties of the $\tau$-lepton.
The $\gamma\gamma\to\tau\tau$ channel is challenging due to hadronic backgrounds and neutrinos in $\tau$-lepton decays diluting visible final-state kinematics.
The ATLAS and CMS Collaborations report the observation of the $\gamma\gamma\to\tau\tau$ process in Pb+Pb UPC~\cite{STDM-2019-19, CMS-HIN-21-009}, where semileptonic $\tau\tau$ decays into a muon and charged-particle track(s) are exploited. The measurements are found to be compatible with SM predictions, with a signal strength of $\mu_{\tau\tau} = 1.03^{+0.06}_{-0.05}$ measured by ATLAS\@.
The measured signal event properties are used to set constraints on the $\tau$-lepton anomalous magnetic moment, $a_{\tau}$, via parameterization of the $\tau\tau\gamma$ coupling in LO QED calculations by $F_1(q^2)\gamma^{\mu} + F_2(q^2)\frac{i}{2m_{\tau}}\sigma^{\mu\nu}q_{\nu}$, where $q_{\nu}$ is the photon four-momentum, $\sigma^{\mu\nu} = \text{i}[\gamma^\mu, \gamma^\nu]/2$ the spin tensor, and the form factors satisfy $F_1(q^2\to 0)=1$ and $F_2(q^2\to 0)=a_{\tau}$.
The precision of the ATLAS measurement, corresponding to $-0.057 < a_{\tau} < 0.024$ at 95\% confidence level (CL), is similar to the most precise single-experiment measurement by the DELPHI Collaboration at LEP~\cite{DELPHI:2003nah} (see Figure~\ref{fig:dileptons}(b)).
The ATLAS result represents the first use of hadron-collider data to test the EM properties of the $\tau$-lepton.

\subsection{Light-by-light scattering}

Light-by-light (LbyL) scattering, $\gamma\gamma\to\gamma\gamma$, is a rare process in the SM that proceeds at lowest order in quantum electrodynamics (QED) via virtual one-loop box diagrams involving charged fermions (leptons and quarks) and $W$ bosons.
LbyL scattering via an electron loop can be precisely, albeit indirectly and in a different phase-space region, tested in measurements of the anomalous magnetic moment of the electron and muon~\cite{Hanneke:2008tm, Muong:2021ojo}.
The $\gamma\gamma\rightarrow \gamma\gamma$ reaction can also be studied in photon scattering in the Coulomb field of a nucleus (Delbr{\"u}ck scattering)~\cite{Wilson:1953zz} and in the photon splitting process~\cite{Akhmadaliev:2001ik}.

An alternative way by which LbyL interactions can be studied is by using Pb+Pb UPC events at the LHC~\cite{Enterria:2013yra,Klusek-Gawenda:2016euz}.
In such a case, the final-state signature of interest is the exclusive production of two photons,
$\textrm{Pb+Pb} \to \textrm{Pb}^{(*)}(\gamma\gamma\to \gamma\gamma)\textrm{Pb}^{(*)}$, where the diphoton final state is measured in the detector surrounding the Pb+Pb interaction region, and the incoming Pb ions survive the EM interaction, with a possible EM excitation.
Hence, one expects that two low-energy photons be detected with no further activity in the central detector.
In particular, no reconstructed charged-particle tracks originating from the Pb+Pb interaction point are expected, as demonstrated in Figure~\ref{fig:lblED}.
The exclusive diphoton final state can also be produced via the strong interaction through a quark loop in the exchange of two gluons in a color-singlet state~\cite{Khoze:2004ak}. This central exclusive production (CEP) process, $gg \to\gamma\gamma$, is treated as a background in the studies described below and is determined using a dedicated control region in the data.

\begin{figure}[!t]
\centering

\includegraphics[width=0.8\textwidth]{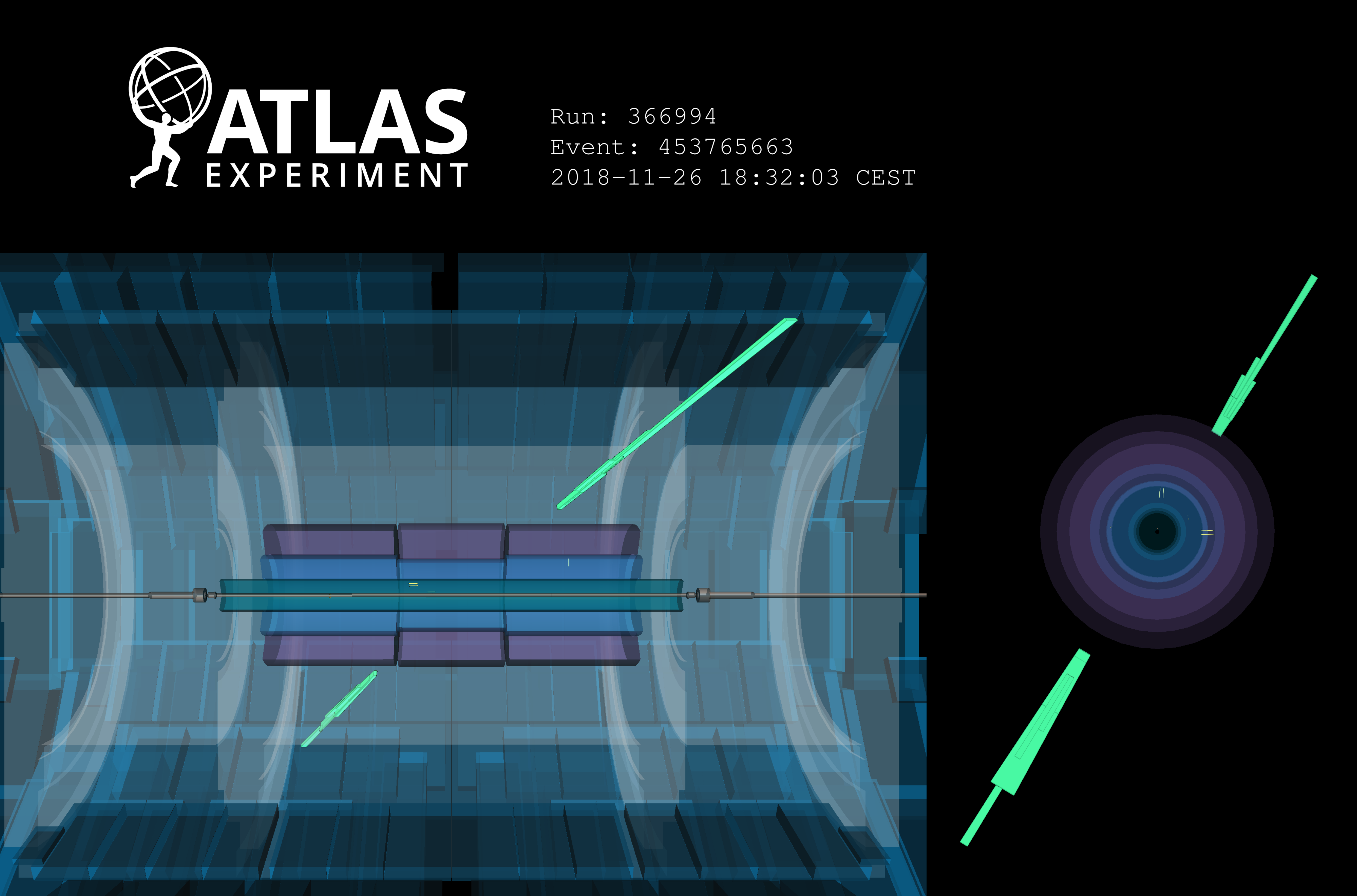}
\caption{
\label{fig:lblED} Event display for an exclusive $\gamma\gamma\to\gamma\gamma$ candidate recorded in Pb+Pb Run~2 data by ATLAS\@.
Two back-to-back photons with an invariant mass of 29~\GeV and no additional activity in the detector are shown.}
\end{figure}

The first direct evidence of the LbyL process in Pb+Pb UPC at the LHC was established by the  ATLAS~\cite{HION-2016-05} and CMS~\cite{CMS-FSQ-16-012} Collaborations.
The evidence was obtained from Pb+Pb data recorded in 2015 at a centre-of-mass energy of $\sqrt{s_{NN}}~=~5.02$~\TeV with integrated luminosities of 0.5~$\textrm{nb}^{-1}$~(ATLAS) and  0.4~$\textrm{nb}^{-1}$~(CMS).
Exploiting a data sample of Pb+Pb collisions collected in 2018 at the same centre-of-mass energy with an integrated luminosity of 1.7~$\textrm{nb}^{-1}$, the ATLAS Collaboration observed LbyL scattering with a significance of  $8.2$ standard deviations~\cite{HION-2018-19}.

In the combined 2015 and 2018 Pb+Pb data analysis~\cite{HION-2019-08}, ATLAS studied the LbyL scattering with improved precision and more detail. In addition to the fiducial cross-section, ATLAS measures the differential cross-sections as a function of several properties of the final-state photons (see Figure~\ref{fig:lbyl}).
All measured cross-sections are consistent within two standard deviations with the SM theory (LO QED) predictions for LbyL scattering~\cite{Harland-Lang:2018iur}.
The inclusion of NLO QED and NLO QCD corrections~\cite{AH:2023kor} reduces, but does not eliminate, the small tension with theoretical predictions.
The result explores a broader range of diphoton masses, increasing the expected signal yield by about 50\% in comparison to the previous ATLAS measurements.

\begin{figure}[t!]
\begin{center}
\subfloat[]{\includegraphics[width=0.45\textwidth]{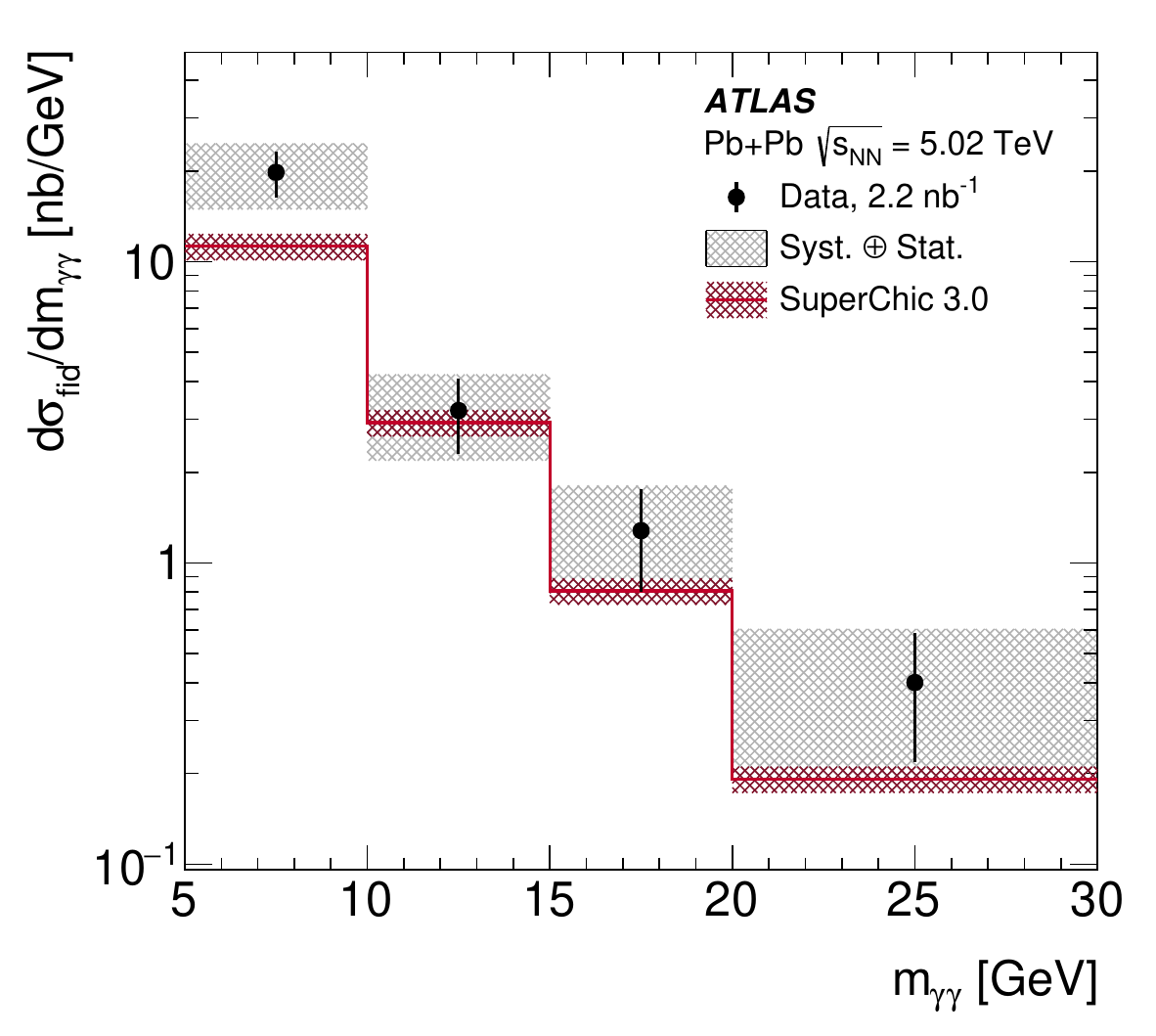}}
\subfloat[]{\includegraphics[width=0.45\textwidth]{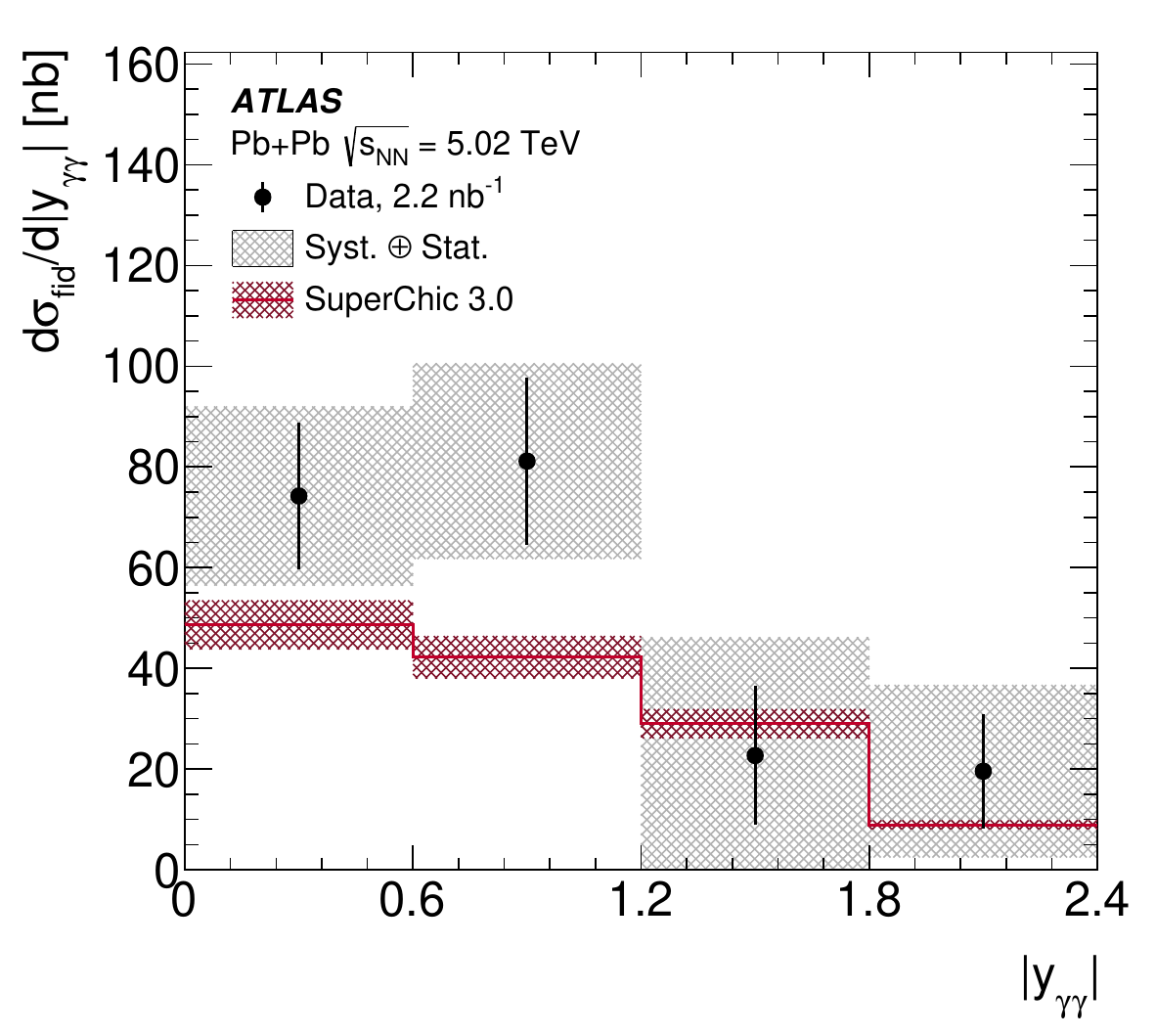}}
\end{center}
\caption{Measured differential fiducial cross-sections of $\gamma\gamma\to\gamma\gamma$ production in Pb+Pb collisions at $\sqrt{s_{NN}}~=~5.02$~\TeV for (a) diphoton invariant mass and (b) diphoton absolute rapidity~\cite{HION-2019-08}. The measured cross-section values are shown as points with error bars giving the statistical uncertainty and the bands indicating the size of the total uncertainty. The results are compared with the prediction from the \textsc{SuperChic}~3 MC generator~\cite{Harland-Lang:2018iur} (solid line) with bands denoting the theoretical uncertainty. }
\label{fig:lbyl}
\end{figure}

The measurement of LbyL scattering is sensitive to BSM processes, such as `axion-like' particles. These are hypothetical pseudoscalar particles with typically weak interactions with SM particles.
The diphoton invariant mass distribution reported by ATLAS is used to set limits on the production of axion-like particles~\cite{HION-2019-08}. This result provides the most stringent limits to date on axion-like particle production for masses in the range of 6--100~\GeV.

\subsection{Exclusive $W$ boson pair production}
\label{ggWW}

The study of $W$ boson pair production from the interaction of incoming photons ($\gamma\gamma\to WW$) offers a unique window to a wide range of physical phenomena. In the SM, the $\gamma\gamma\to WW$ process proceeds through trilinear and quartic gauge-boson interactions. This process is unique in that, at leading order, it only involves diagrams with self-couplings of the electroweak gauge bosons, as shown in Figure~\ref{fig:yyWW-feynman-diagrams}.

\begin{figure}[!t]
\centering
\subfloat[]{\includegraphics[width=0.25\textwidth]{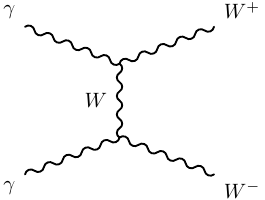}}\,\;\;\;\;\hspace{2cm}
\subfloat[]{\includegraphics[width=0.25\textwidth]{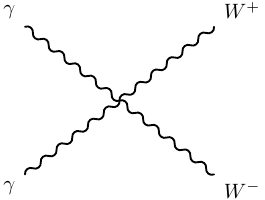}}
\caption{
\label{fig:yyWW-feynman-diagrams} The leading-order Feynman diagrams contributing to the $\gamma\gamma\to WW$ process are (a) the $t$-channel diagram proceeding via the exchange of a $W$ boson between two $\gamma WW$ vertices and (b) a diagram with a quartic $\gamma\gamma WW$ coupling). In addition, a $u$-channel diagram exists (not shown), which also proceeds via two $\gamma WW$ vertices.}
\end{figure}

ATLAS has studied the $pp \to p^{(*)}(\gamma\gamma\to WW)p^{(*)}$ reaction at $\sqrt{s}=13$~\TeV using full Run 2 data sample~\cite{STDM-2017-21}.
Previously, the ATLAS and CMS Collaborations found evidence for $\gamma\gamma\to WW$ production with the Run 1 data, ATLAS by using 8~\TeV\ $pp$ collisions~\cite{STDM-2015-10} and CMS by combining their 7~\TeV\ and 8~\TeV\ $pp$ collision data~\cite{CMS-FSQ-12-010,CMS-FSQ-13-008}.

Events with leptonic $W$ boson decays into $e\nu\mu\nu$ final states
are selected by requiring that no tracks except those of the two charged leptons are associated
with the production vertex, following the strategy developed in the previous $pp \to p(\gamma\gamma\to\ell\ell)p$ measurements~\cite{STDM-2016-13}.
The modelling of the hadronic activity in quark- and gluon-induced background processes, and uncorrelated activity from additional $pp$ interactions, is constrained using same-flavour $Z \to \ell\ell$ events in data, reducing the associated uncertainties by a significant amount.
The background-only
hypothesis is rejected with a significance of 8.4 standard
deviations whereas well above 5 standard
deviations
was expected.
The signal strength and the cross-section for
the sum of elastic and dissociative production mechanisms are
measured.
The cross-section for the $\gamma\gamma \to WW$ process
is measured in a fiducial volume close to the acceptance of the detector.
The measured cross-section is found to be in agreement with the SM prediction and may serve as input into future EFT interpretations.

The measurements of rare EW processes in two-photon interactions ($\gamma\gamma \to \gamma\gamma$, $\gamma\gamma \to WW$) are statistically limited, hence opening the possibility for substantial improvements with the future LHC runs.



\section{Measurements of fundamental parameters of the SM}
\label{sec:params}

With the discovery of the Higgs boson~\cite{HIGG-2012-27,CMS-HIG-12-028} and the measurement of its mass, the EW sector of the SM is overconstrained~\cite{haller2022status}, such that precise measurements of fundamental parameters can serve as a probe of the SM, and a means to search for new physics in a model-independent way. In the QCD sector, the SM can precisely predict the energy dependence of the strong coupling but relies on experimental input to determine its value at a reference scale~\cite{STDM-2023-01}.
During LHC Run~2, ATLAS  performed a range of precice measurements of fundamental parameters of the SM, not only on $\sqrt{s} = 13$~\TeV data but also on the  $\sqrt{s} = 7$~\TeV and  $\sqrt{s} = 8$~\TeV data samples. The latter  profited from the more precise predictions, more recent PDF sets and advanced statistical methods, available during Run~2, while at the same time benefitting from lower pile-up and lower trigger thresholds in  the Run~1 data samples.

\subsection{Reanalysis of the W mass measurement}
The mass of the $W$ boson, $m_W$, is one of the fundamental parameters of the EW sector of the SM and affects the Higgs boson and top-quark masses $m_H$ and $m_t$ via radiative corrections~\cite{Awramik_2004,Sirlin_1980,Ciuchini_2013,ParticleDataGroup:2024cfk}. Figure~\ref{fig:mWfit} demonstrates this interdependence by comparing the  direct ATLAS measurements of $m_W$~\cite{STDM-2019-24} and the LHC combination  of $m_t$~\cite{mtop_2024} with the indirect predictions from the ATLAS $m_H$  measurement~\cite{ATLAS_mH_2023} and from the EW fit~\cite{Haller_2018}.

\begin{figure}[b!]
\begin{center}
\includegraphics[width=0.60\textwidth]{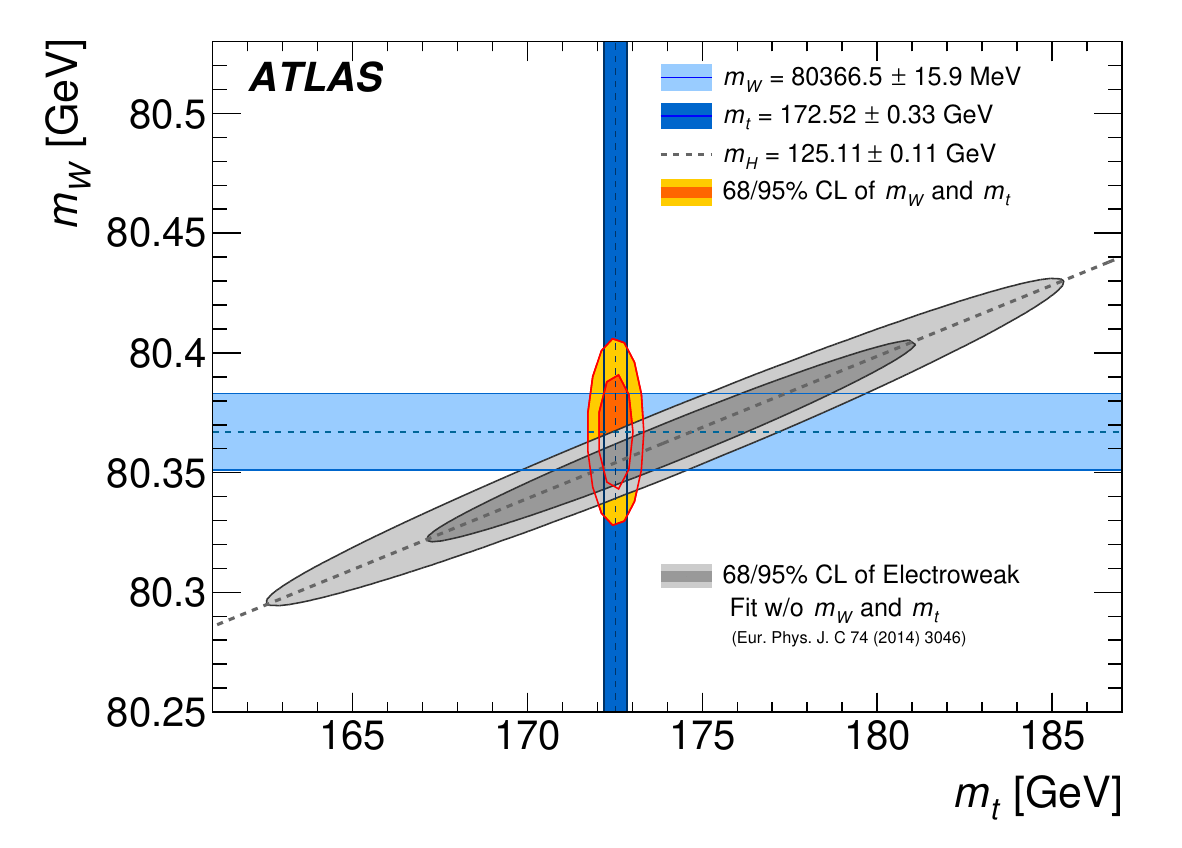}
\end{center}
\caption{The 68$\%$ and 95$\%$ confidence level contours of the $m_W$ and $m_t$ indirect determinations from the global electroweak fit~\cite{Haller_2018}, compared to the 68$\%$ and 95$\%$ confidence-level contours of the present ATLAS measurement of $m_W$~\cite{STDM-2019-24} and the LHC measurement of $m_t$~\cite{mtop_2024} and to the ATLAS measurement of $m_H$~\cite{ATLAS_mH_2023}.} \label{fig:mWfit}
\end{figure}

The first $m_W$ measurement at the LHC was performed by ATLAS~\cite{STDM-2014-18} on the Run~1 $\sqrt{s} = 7$~\TeV data, which has now been reanalyzed~\cite{STDM-2019-24} in the context of a significant tension with the precise measurement from the CDF Collaboration~\cite{CDF:2022hxs}.  The $W$ boson mass is extracted from template fits to the \ptl\ and \mtw\ distributions such that a good modelling of the charged Drell-Yan process, including QCD and EW corrections, is crucial for this measurement~\cite{Carloni_Calame_2017,Dittmaier_2002}. The baseline simulation by \powpyeight~\cite{Nason_2004,Frixione_2007,Alioli_2010}, with the AZNLO tune~\cite{STDM-2012-23} which effectively extrapolates from the precisely measured \ptz\ to the \ptw\ distribution has been independently validated in the relevant low-\ptw\ region in recent high-precision measurements on low-pileup data, as described in Section~\ref{sec:ptv}.
As in ~\cite{STDM-2014-18}, the final combination is dominated by the more precise \ptl\ measurement. While the original analysis used sequential fits with templates altered according to the systematic uncertainties, the reanalysis uses a simultaneous fit with a detailed model of statistical and systematic uncertainties and their correlations and a more advanced proton PDF as a baseline. This results in a shift of the central value within the uncertainty of the first publication and a reduction of the total uncertainty by 3~MeV. The updated $m_W$ measurement is:  $m_W =$ 80\,366.5 $\pm$ 9.8(stat.) $\pm$ 12.5 (syst.)~\MeV $=$  80\,366.5 $\pm$ 15.9~\MeV. The systematic uncertainty is dominated by PDF uncertainties, missing higher-order EW corrections and by electron and muon calibration uncertainties.
Figure~\ref{fig:mW}(a) compares the updated measurement of $m_W$ to the SM prediction from the global EW fit~\cite{deBlas:2017wmn} and measurements from other experiments. The new ATLAS $m_W$ measurement has moved even closer to the SM prediction.

The EW theory also precisely predicts the $W$ decay width $\Gamma_W$, as the sum of the partial decay widths into SM particles~\cite{Denner1990,deBlas:2017wmn}. ATLAS uses the same input distributions and fit methods that are employed to extract $m_W$ to derive the first measurement of $\Gamma_W$ at the LHC, resulting in: $\Gamma_W$ = 2202 $\pm$  32 (stat) $\pm$ 34 (syst)~\MeV $= 2202 \pm  47$~\MeV. Figure~\ref{fig:mW}(b), compares the ATLAS measurement of $\Gamma_W$  with the SM prediction and measurements from other experiments. The measurement agrees with the SM prediction within two standard deviations.

\begin{figure}[b!]
\begin{center}
\subfloat[]{\includegraphics[width=0.475\textwidth]{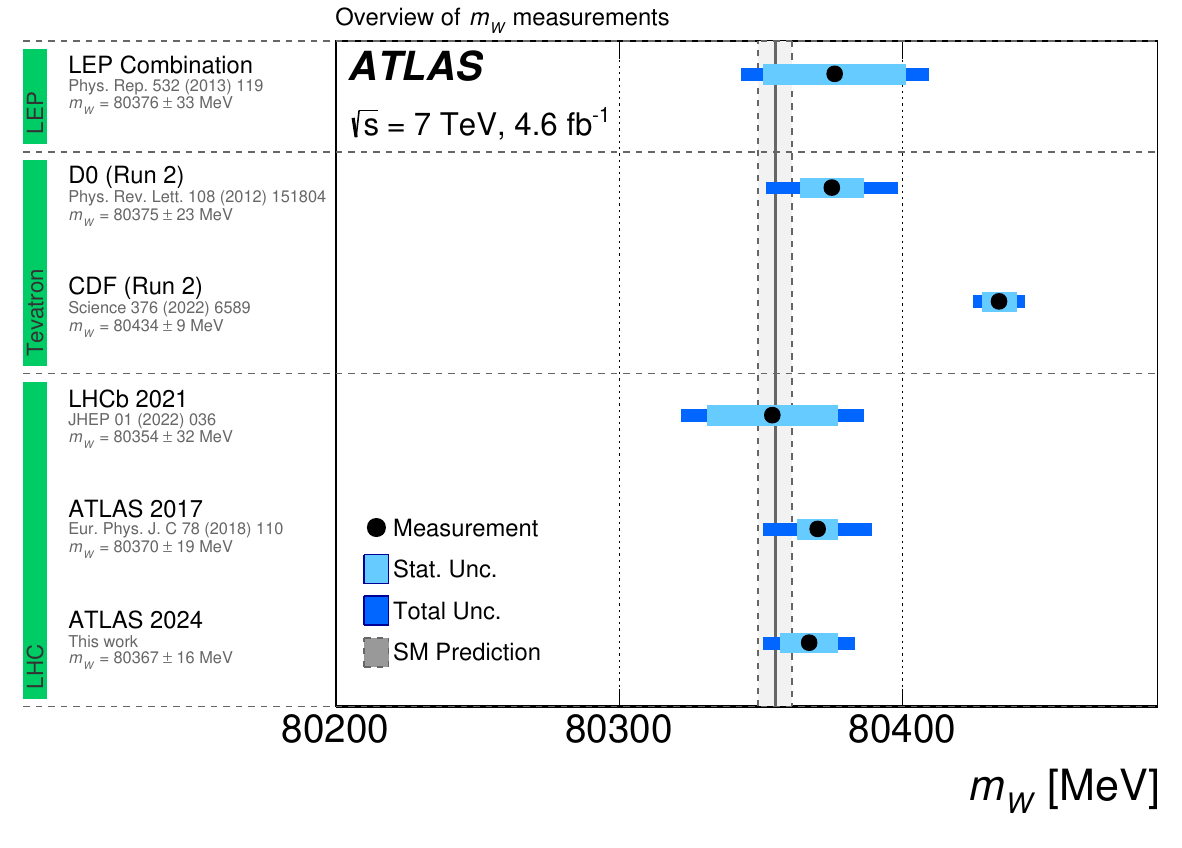}}
\hspace{0.3cm}
\subfloat[]{\includegraphics[width=0.475\textwidth]{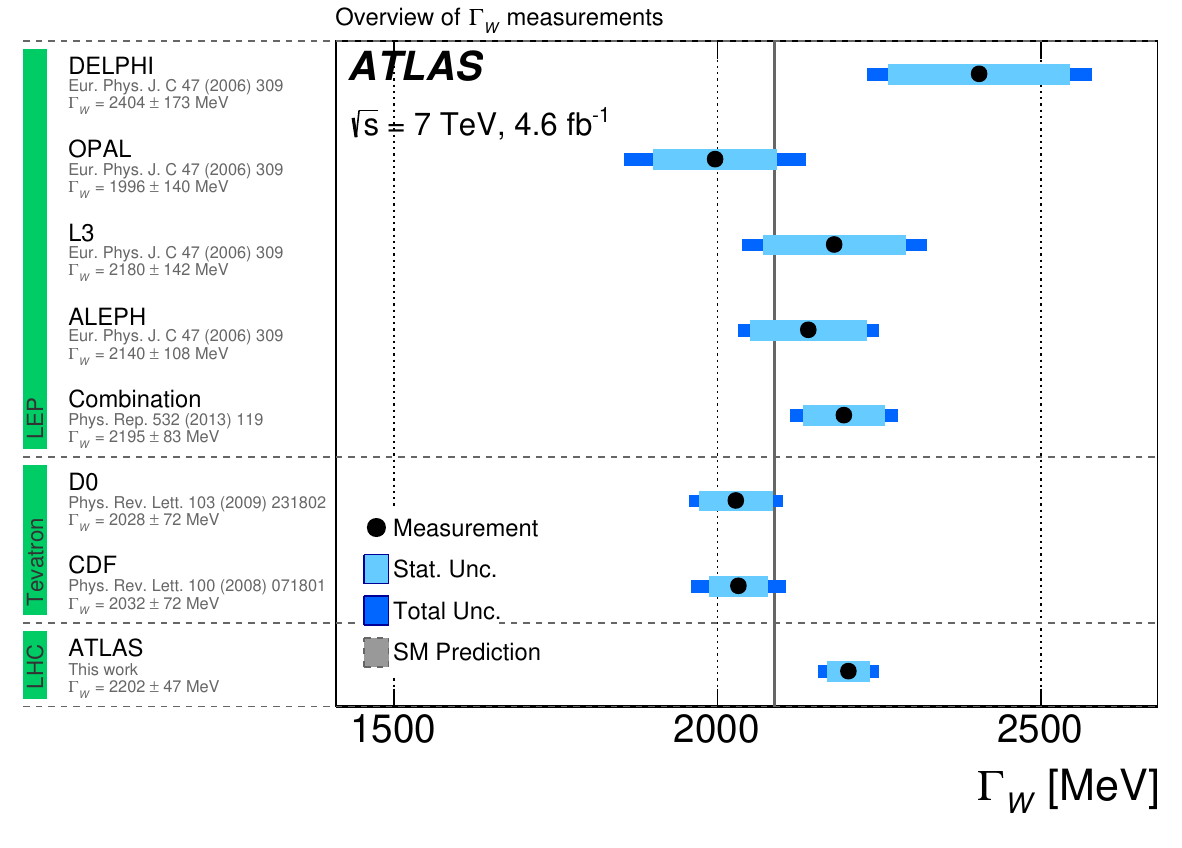}}
\end{center}
\caption{The measured values of (a) $m_W$ and (b) $\Gamma_W$ compared with the SM prediction from the global EW fit and measurements from other experiments~\cite{STDM-2019-24}.}
\label{fig:mW}
\end{figure}

\subsection{Determination of $\alphas$ from $Z$ boson \pt}

The strong coupling $\alphas$, measured at a reference energy scale, is the least precisely determined fundamental coupling constant~\cite{PDG2022}. While the precision of the ATLAS TEEC based $\alphas$ measurement in jet events, detailed in  Section~\ref{sec:evshapes}, has significantly improved, it is still limited by the residual uncertainty in the NNLO theory prediction. Recently, ATLAS performed a novel measurement of $\alphas$ in Drell--Yan events, which exceeds the precision of the jet-based measurements:

In LHC Drell--Yan $Z$ production, QCD initial-state radiation leads to the recoil of the $Z$ boson which acquires non-zero transverse momentum. The ATLAS Run~1 $\sqrt{s} = 8$~\TeV data sample is used to determine $\alphas$  from the low-momentum Sudakov region~\cite{Sudakov:1954sw}  of the \pt\ distribution of $Z$ bosons~\cite{STDM-2023-01} (see Figure~\ref{fig:alphas}(a)) as measured in~\cite{STDM-2018-05} (see Section~\ref{sec:DY2D}). Determining the cross-sections in the full phase space~\cite{STDM-2018-05}, allows comparison to a prediction at N$^3$LO and  approximate N$^4$LL accuracy in QCD calculated with DYTurbo~\cite{Camarda_2020, Camarda_2021}, using the approximate N$^3$LO MSHT20 PDF set~\cite{McGowan:2022nag}. QED ISR corrections are evaluated at LL accuracy with \pythiaeight\ using the AZ tune. The parameter $\alphas$ is extracted via a $\chi^2$ fit to the measured double-differential $\pt$--$y$ distribution of the $Z$ boson, using 72 bins in $|y|<3.6$ and $\pt < 29$~GeV. The fit directly includes only experimental and MSHT20aN$^3$LO PDF uncertainties (applying Hessian profiling) and two non-perturbative form factors. All other theory uncertainties are conservatively assessed via sequential fits with varied inputs. They include the QCD scale choice, matching to fixed order, the non-perturbative model, higher-order QED ISR, the approximate 4-loop calculation and the effects of heavy-quark masses and thresholds.
Figure~\ref{fig:alphas}(a), shows the post-fit ratios of the double-differential cross-sections to the predictions. The resulting value is $\alphas(m_Z) = 0.1183 \pm  0.0009$, where the largest contributions to the uncertainty are from experimental effects, PDFs, scale choices and heavy quarks. The result demonstrates the running of $\alphas$ in a single analysis, in contrast to almost all other $\alphas$
measurements which target one particular scale.
A conservative estimate of the residual PDF model dependence is derived by repeating the $\alphas$ extraction at a lower QCD order
using different NNLO PDF sets, resulting in a spread of $\alphas$ comparable to the total uncertainty on the original measurement.
Figure~\ref{fig:alphas}(b) presents the new $\alphas$  measurement together with other determinations of $\alphas$. This result is the most precise experimental determination of $\alphas(m_Z)$  achieved so far.

\begin{figure}[t!]
\begin{center}
\subfloat[]{\includegraphics[width=0.49\textwidth]{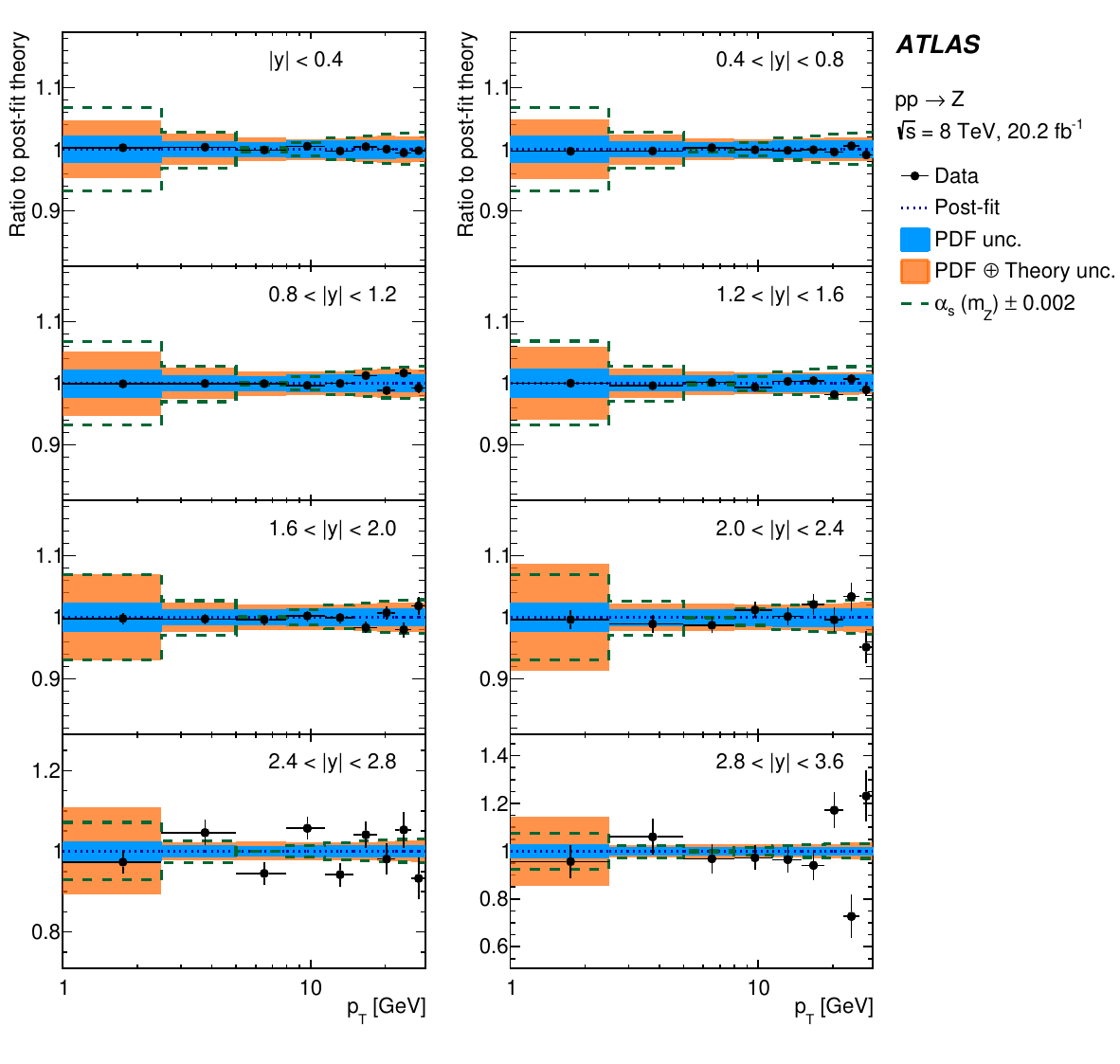}}
\subfloat[]{\includegraphics[width=0.49\textwidth]{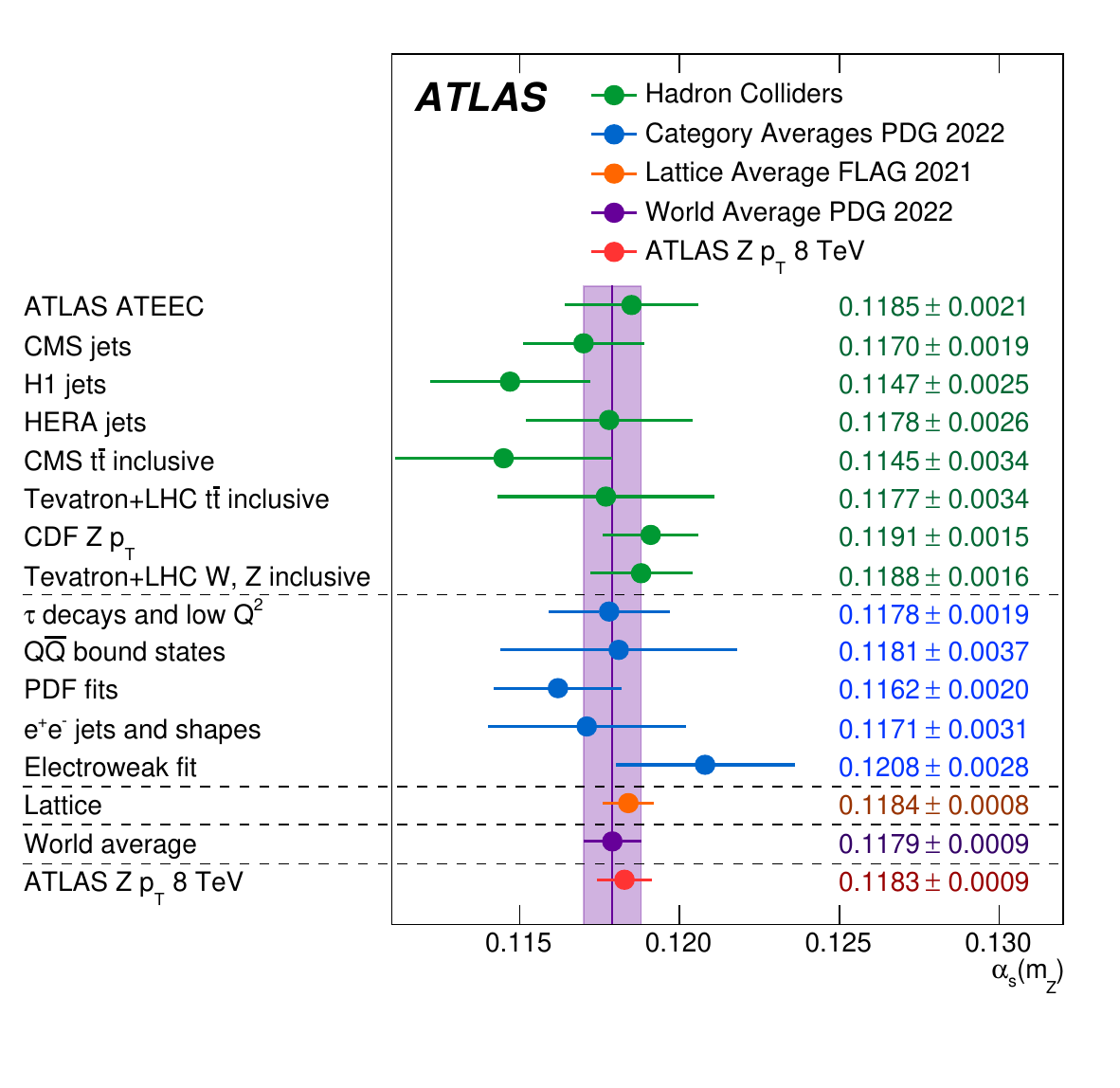}}
\end{center}
\caption{(a) Ratios of the measured double-differential cross-sections to the post-fit predictions, both as functions of the transverse momentum and rapidity of the $Z$ boson. The dependency on $\alphas$ is indicated. (b) Comparison of the determination of $\alphas(m_Z)$ from the $Z$ boson transverse-momentum distribution with other determinations at hadron colliders, the PDG category averages, the lattice QCD determination and with the PDG world average~\cite{STDM-2023-01}.}\label{fig:alphas}
\end{figure}

\subsection{Measurement of the $Z$ boson invisible width}

Part of the Run~2 data sample is used to perform a direct measurement of the invisible $Z$ width $\Gamma(Z\to \textrm{inv})$~\cite{STDM-2019-01}  using the ratio of
$Z(\to \nu\nu)$ + jets to $Z(\to \ell\ell)$ + jets cross-sections, defined as
\begin{equation}
R^\textrm{miss}(p_{\textrm T, Z})
= \frac{ \frac{\mathrm{d}\sigma(Z+\textrm{jets})\times B(Z\to \nu\nu)}{\mathrm{d}(p_{\textrm T, Z})}}{\frac{\mathrm{d}\sigma(Z+\textrm{jets})\times B(Z\to \ell\ell)}{\mathrm{d}(p_{\textrm T, Z})}}
\end{equation}
in a common phase space with  $p_{\text{T},Z} > 130$~\GeV and a jet with $\pt> 110$~\GeV. After bin-wise correction for detector effects and an additional correction of the $Z\to\ell\ell$ component for the $\mll$ requirement and for the $\gamma^*$ contributions, $R^\textrm{miss}$  is independent of $p_{\textrm T,Z}$ (see Figure~\ref{fig:gammainv}(a)). $\Gamma(Z\to \textrm{inv})$ is then extracted from the result $\widehat{R}^\textrm{miss}$ of a fit to $R^\textrm{miss}(p_{\textrm T,Z})$  as $\Gamma(Z\to \textrm{inv})  =  \widehat{R}^\textrm{miss} \Gamma(Z\to\ell\ell)$ using the well-constrained $e^+e^-$ measurement of $\Gamma(Z\to\ell\ell)$.
The invisible width is determined with 2.5$\%$ uncertainty as $\Gamma(Z\to \textrm{inv})  = 506 \pm 13$~\MeV. This is in good agreement with the lineshape-based measurement at LEP and the most precise experimental result to date for a measurement based on recoil final states (see Figure~\ref{fig:gammainv}(b)).

\begin{figure}[h!]
\begin{center}
\subfloat[]{\includegraphics[width=0.47\textwidth]{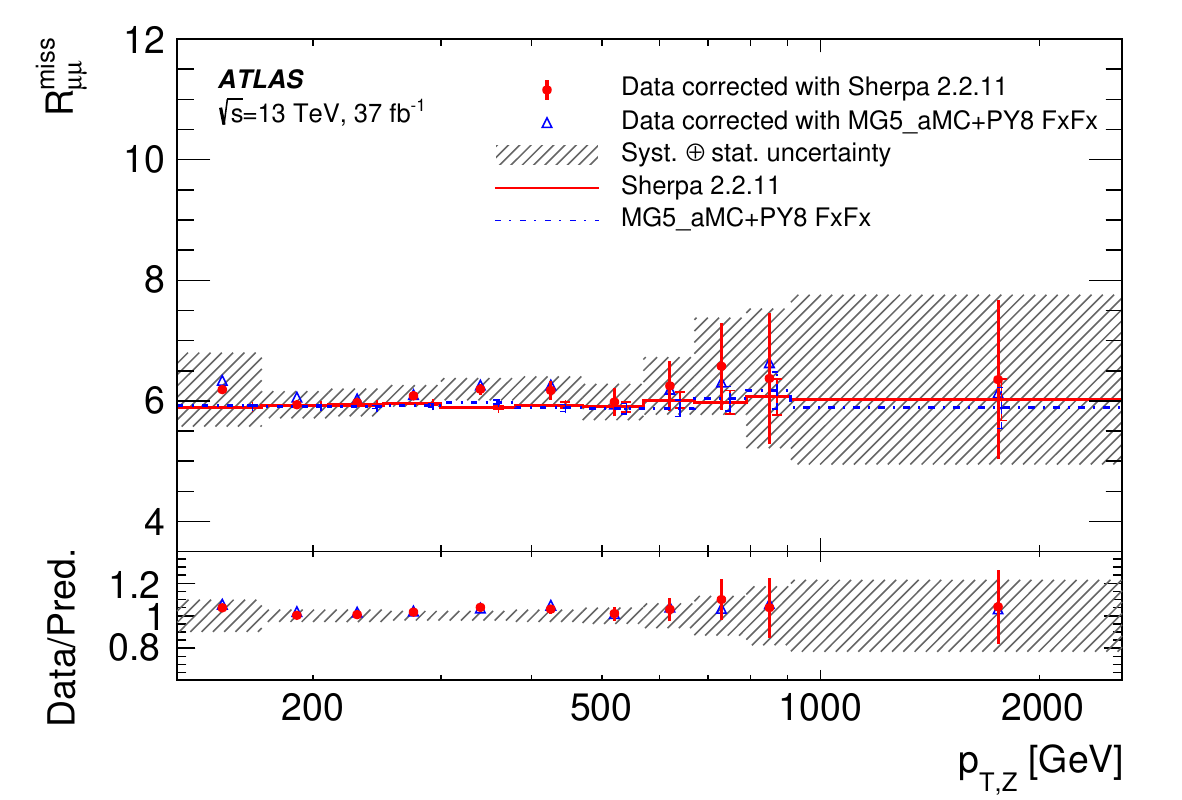}}
\subfloat[]{\includegraphics[width=0.51\textwidth]{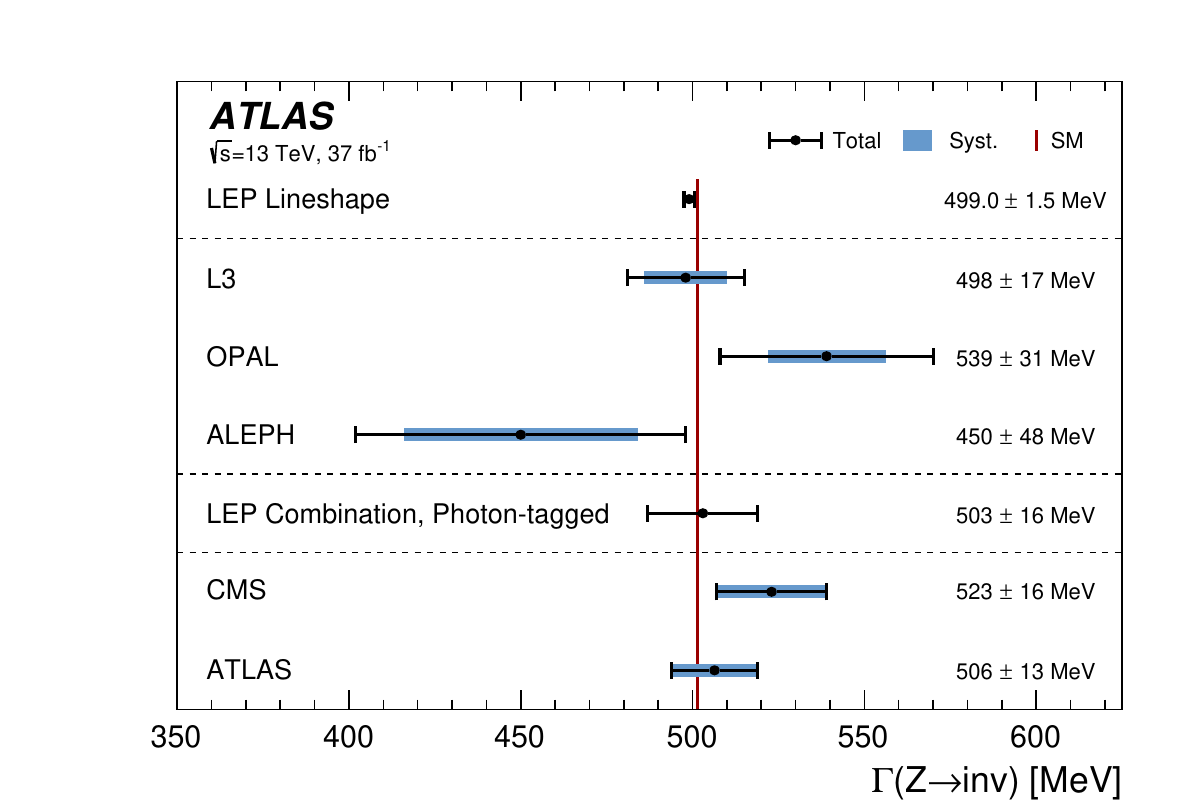}}
\end{center}
\caption{(a) Measured $R^\textrm{miss}$ in the muon channel as a function $p_{\textrm T,Z}$ in the common phase space and comparison with predictions~\cite{PMGR-2021-01,Bothmann:2019yzt,Frederix:2012ps} and  (b) comparison of the ATLAS  $\Gamma(Z\to \textrm{inv})$ measurement to direct measurements by other experiments~\cite{STDM-2019-01}.}\label{fig:gammainv}
\end{figure}


\section{Precision measurements of $b$-hadron decays in searches for contributions from new physics}
In addition to direct searches for new physics and new particles, a very promising direction of indirect searches proceeds via precision studies of low-energy phenomena. The detailed studies of the $b$-quark plays a special role in testing the flavour structure of the SM and searching for BSM physics~\cite{Artuso:2022ijh}.
This section summarises three such studies. The first study concerns the CP violation arising from an interference between mixing and decay amplitudes of the $B_s$ meson. Secondly, the search for the rare decays of $B$ mesons into a pair of oppositely charged muons is discussed. Finally, the lifetime of the $B_s$ meson is measured in the rare dimuon decay channel.

\label{sec:bphys}
\subsection{ CP violation with $B_s \rightarrow J/\psi \phi$ }

In the presence of BSM phenomena, new sources of \CP\ violation in $b$-hadron decays can arise in addition to those  predicted by the  SM~\cite{BOTELLA20071} and ~\cite{Lenz:2019lvd}.
In the  \Bst\ decay, \CP\  violation occurs due to interference between the \BsaBs mixing and the  \Bst\ decay. %
The \CP-violating phase \phis\ is defined as the weak phase difference between the \BsaBs\ mixing amplitude and the $b \rightarrow c \overline{c} s$ decay amplitude. In the SM, the phase \phis\ is small
and is  related to the Cabibbo--Kobayashi--Maskawa (CKM) quark mixing matrix elements via the relation $\phis \simeq -2 \beta_{s}$, with $\beta_{s} = \mathrm{arg} [- (V_{ts} V^{*}_{tb})/(V_{cs} V_{cb}^{*}) ]$.
A value of   $-2 \beta_{s} =  - 0.0368 \pm 0.0010$ rad is predicted by the UTfit Collaboration~\cite{UTFit2006}. While large enhancements are excluded by the precise  measurement of the oscillation frequency~\cite{Aaij:2013Mix}, any new physics couplings involved in the mixing may  still increase the size of the observed \CP\  violation  by enhancing the mixing phase $\phis $ relative to the SM value.

Using \ilumi\ of integrated luminosity collected  from \CoMEnergy\ proton--proton collisions at the LHC, combined with data from  \ilumiRunone\ of \CoMEnergySeven\ and \CoMEnergyEight,
ATLAS measures   the \Bst\ decay parameters in the channel  \Bsto\,  including the \CP-violating phase \phis,  the width difference \DGs\  between the \Bs\ meson mass eigenstates  and   the average decay width \Gs~\cite{BPHY-2018-01}.

The ATLAS result is presented in the form of the two-dimensional likelihood contours in the \phis--\DGs\ plane and is compared with the results up to 2021 from CMS~\cite{CMS-BPH-20-001} and LHCb~\cite{Aaij2019, Aaij:2016psitwoS, Aaij:2014Ds,Aaij:2014dka,Aaij:2019mhf} in Figure~\ref{fig:2DContour},  prepared by the HFLAV Collaboration~\cite{HFLAV:2022esi}.
The combination of experimental results  is performed with the \DGs\ errors scaled by a factor of 1.78 because of a tension in current experimental results. The SM prediction~\cite{Lenz:2019lvd} is shown in the same Figure~\ref{fig:2DContour}.
Older results from CDF~\cite{PhysRevLett.109.171802} and D0~\cite{PhysRevD.85.032006} are also shown.
So far all results are consistent  with the SM prediction~\cite{Lenz:2019lvd}. However the current experimental uncertainties on the CP violation phase  $\phis$ are too large in comparison with the SM prediction uncertainty, so there is still a place for BSM contributions. By including data from Run~3 and HL-LHC~\cite{CERNYellow:2019},  the experimental sensitivity will increase to the level allowing to exclude or confirm the SM prediction.

\begin{figure}[!t]
\begin{center}
\includegraphics[width=0.6\textwidth]{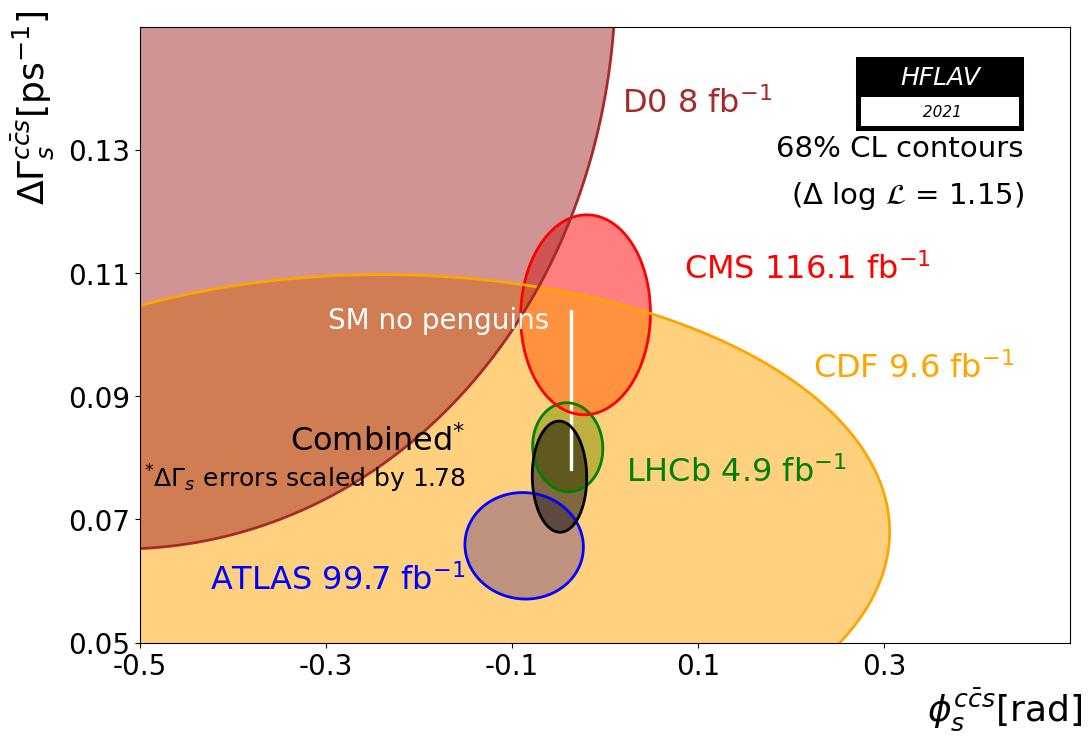}
\end{center}
\caption{  Contours of 68\% confidence level in the \phis--\DGs\ plane, including results from CMS and LHCb using all  \Bs \  channels, prepared by the HFLAV Collaboration~\cite{HFLAV:2022esi}. The  blue contour shows the ATLAS result for  \CoMEnergy\ combined  with \CoMEnergySeven\ and \CoMEnergyEight.  The LHC combination is shown in black. Older results from CDF and D0 are also shown.  In  all contours  the statistical and systematic uncertainties are combined in quadrature. The SM prediction neglecting penguin contributions is shown as a very thin white rectangle~\cite{Lenz:2019lvd}.}
\label{fig:2DContour}
\end{figure}

\subsection{Rare $B_{(s)}^0\rightarrow \mu^+\mu^- $  decays: measurement of branching fractions }
Flavour-changing neutral-current processes are highly suppressed
in the SM\@.
The branching fractions of the decays $B_{(s)}^0\rightarrow \mu^+\mu^- $\
are, in addition, helicity suppressed in the SM, and are predicted to be
$\BR$($B_{s}^0\rightarrow \mu^+\mu^- $) = (3.65 $\pm$ 0.23) $\times$ 10$^{-9}$ and
$\BR$($B_{d}^0\rightarrow \mu^+\mu^- $) = (1.06 $\pm$ 0.09) $\times$ 10$^{-10}$~\cite{Bobeth:2013uxa}.
The small values and the high precision of these predictions provide a favourable environment to search for contributions
from BSM physics.  Significant deviations from SM predictions could arise in models involving
non-SM heavy particles, such as those predicted in the
minimal supersymmetric SM~\cite{Huang:1998vb,
Hamzaoui:1998nu,Choudhury:1998ze,Babu:1999hn,Choudhury:2005rz} and
in extensions such as minimal flavour violation~\cite{DAmbrosio:2002ex, Buras:2003td},
two-Higgs-doublet models~\cite{Choudhury:2005rz},
and others~\cite{Davidson:2010uu,Guadagnoli:2013mru}.  The branching fractions of the decay $B_{(s)}^0\rightarrow \mu^+\mu^- $\  is measured by the LHCb~\cite{LHCb_2022} and CMS~\cite{CMS-BPH-21-006} Collaborations.

Using $pp$   LHC data at $13$ \TeV\ corresponding to an integrated luminosity of $26.3$~fb$^{-1}$  (collected in 2015 and 2016)~\cite{BPHY-2018-09},
the \Bs  branching fraction is measured as $\BR (B_{s}^0\rightarrow \mu^+\mu^- ) = \left( 3.2^{+1.1}_{-1.0} \right) \times 10^{-9}$,
where the uncertainty includes both the statistical and systematic contributions.
For the $B_{d}^0$ an upper limit $\BR (B_{d}^0\rightarrow \mu^+\mu^- ) < 4.3 \times 10^{-10}$ is placed at 95\% CL. %
Combining with the Run~1 data sample that used $25.0$~fb$^{-1}$ of  $7/8$ \TeV\ data~\cite{BPHY-2012-01},  ATLAS  obtains
$\BR (B_{s}^0\rightarrow \mu^+\mu^- ) =  \left(2.8^{+0.8}_{-0.7}\right) \times 10^{-9}$ and $\BR (B_{d}^0\rightarrow \mu^+\mu^- ) < 2.1 \times 10^{-10}$.
All the results are compatible with the branching fractions predicted by the SM and with currently available results from other experiments.
Figure~\ref{app:fig:2DLH-run1comb} shows the likelihood contours for the  combined Run~1 and Run~2 result for
$\BR(B_{s}^0\rightarrow \mu^+\mu^- )$\ and $\BR(B_{d}^0\rightarrow \mu^+\mu^- )$.  The LHC combination of the branching fractions of the decays $B_{(s)}^0\rightarrow \mu^+\mu^- $\  are published in \cite{ATLAS-CONF-2020-049}.
\begin{figure}[!t]
\begin{center}
\hspace{-1.2cm}
\includegraphics[width=0.6\textwidth]{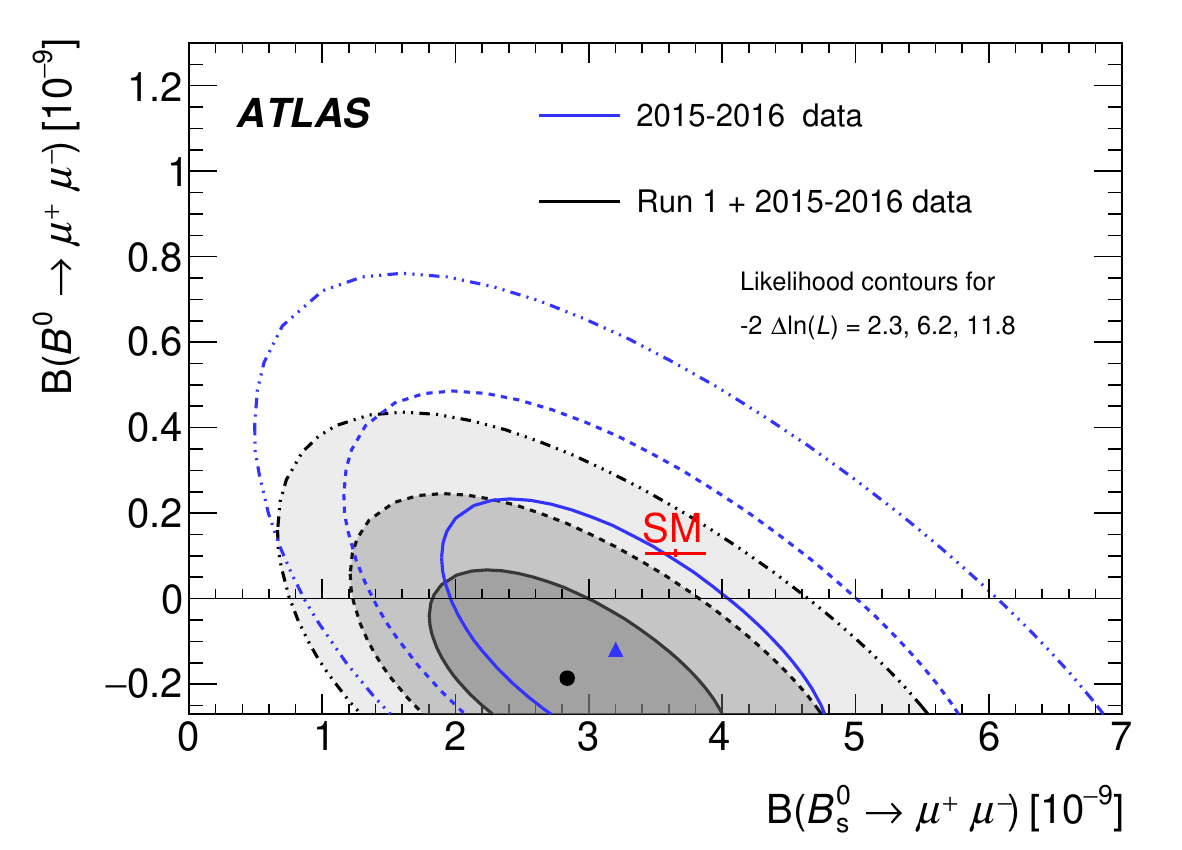}
\caption{Likelihood contours for the combination of the Run~1 and 2015--2016 Run~2 ATLAS results (shaded areas) on $\BR (B^0\rightarrow \mu^+\mu^-)$ and $\BR(B_{s}^0\rightarrow \mu^+\mu^- )$	~\cite{BPHY-2018-09}.
The contours are obtained
from the combined likelihoods of the two analyses,
for values of $-2\Delta \ln\left(\mathcal{L}\right)$ equal to 2.3, 6.2 and 11.8.
The empty contours represent the result from 2015--2016 Run~2 data alone. The SM prediction~\cite{Bobeth:2013uxa} with uncertainties is indicated.
}
\label{app:fig:2DLH-run1comb}
\end{center}
\end{figure}

\subsection{Measurement of the $B_s\to\mu\mu$ effective lifetime}
The SM predicts that only the CP-odd heavy-mass eigenstate in the $B_s$--$\bar{B}_s$ pair decays into a dimuon final state~\cite{Beneke_Power_enhance_SM_Bmumu, new_phys_via_eff_lifetime}. This statement does not generally hold when considering BSM contributions, such as, for instance, minimal supersymmetric SM extensions~\cite{np_in_flavour}, which can potentially alter the effective lifetime in $B_s\to\mu\mu$ decays. These perturbations can be significant, even in the absence of measurable BSM effects, on the $B_s\to\mu\mu$ branching fraction. The effective $B_s\to\mu\mu$ lifetime is defined as $\taumumu={\int_0^\infty t \Gamma\left(B_s\left(t\right)\to\mu\mu\right) dt }/{\int_0^\infty \Gamma\left(B_s\left(t\right)\to\mu\mu\right) dt}$, where $t$ is the proper decay time of the $B^0_s$ and $\bar{B}^0_s$ mesons and  $\Gamma\left(B_s\left(t\right)\to\mu\mu\right)=\Gamma\left(B^0_s\left(t\right)\to\mu\mu\right)+\Gamma\left(\bar{B}^0_s\left(t\right)\to\mu\mu\right)$.  In the SM,  $\taumumu$ coincides with the lifetime of the heavy $B_s$ eigenstate.
The experimental average produced by the HFLAV Collaboration ~\cite{HFLAV:2022esi} yields the SM prediction $\taumumu^\mathrm{SM}=\left( 1.624 \pm 0.009\right)\text{ ps}$. The $B_s\to\mu\mu$ effective lifetime was measured  by LHCb~\cite{LHCb_2022} and CMS~\cite{CMS-BPH-21-006}. The  combined  value of LHCb and CMS  $B_s\to\mu\mu$ effective lifetimes  are published in \cite{ATLAS-CONF-2020-049}.

The ATLAS measurement of the $\taumumu$ is based on 26.3 fb$^{-1}$ of 13~\TeV LHC $pp$ collisions,  collected in 2015--2016~\cite{BPHY-2020-07}.
The proper decay-time distribution of $58\pm13$ background-subtracted signal candidates is fitted with  simulated signal templates, parameterised
as a function of the $B_s$ effective lifetime (see Figure~\ref{fig:BsmumuTimefit}(a)).
The measured value  of $\taumumu$ is extracted by minimising the binned $\chi^2$ between the data histogram and the signal MC template series, generated for different lifetimes (see Figure~\Ref{fig:BsmumuTimefit}(b)).  The  statistical uncertainties are extracted through a Neyman construction~\cite{neyman}.
A small bias in the analysis of $0.082$~ps is determined in pseudo-data MC simulations and corrected.
Systematic uncertainties in \taumumu\ are currently subdominant and arise from fit-procedure assumptions,  discrepancies between data and the MC simulation
and from neglected backgrounds.
The final result is \taumumuobs\  =  $0.99^{+0.42}_{-0.07} \, (\text{stat.})\pm 0.17 \text{ (syst.)}\,\mathrm{ps}$.
The Figure ~\ref{fig:BsmumuTimefit}(c) shows  a comparison  of the ATLAS  $B_s\to\mu\mu$ effective lifetime ~\cite{BPHY-2020-07}  with the results from the LHCb based on 2011-2018 data LHCb~\cite{LHCb_2022} and 2011-2016 data \cite{LHCb_2017}, as well as the CMS  results on 2011- 2016 data \cite{CMS-BPH-16-004} and 2016-2018 data ~\cite{CMS-BPH-21-006}.  It is clear that  all experimental results are consistent with each other and  with  the SM prediction  ~\cite{HFLAV:2022esi}.

\begin{figure}[!t]
\centering
\subfloat[]{\includegraphics[width=0.45\textwidth]{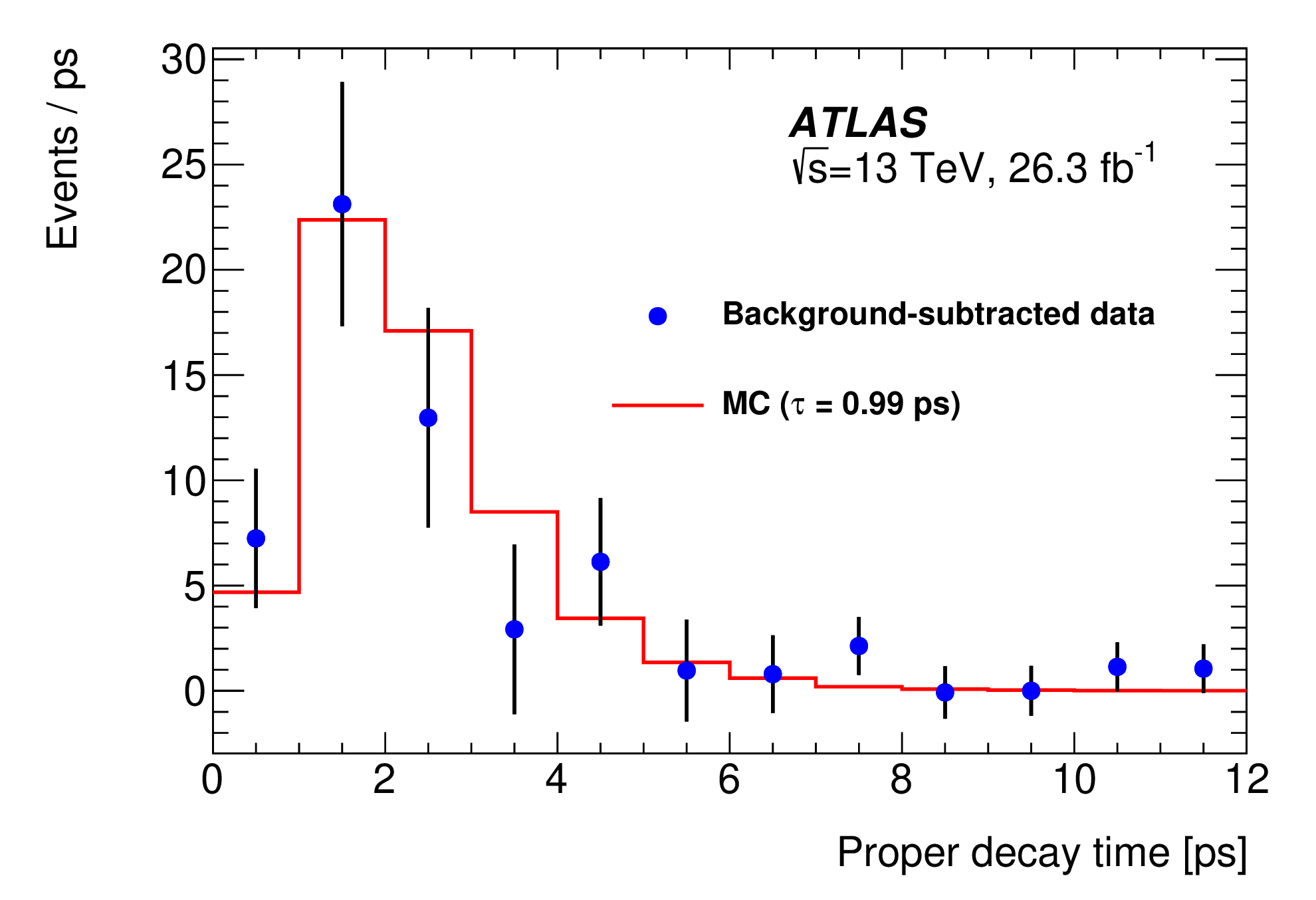}}
\subfloat[]{\includegraphics[width=0.45\textwidth]{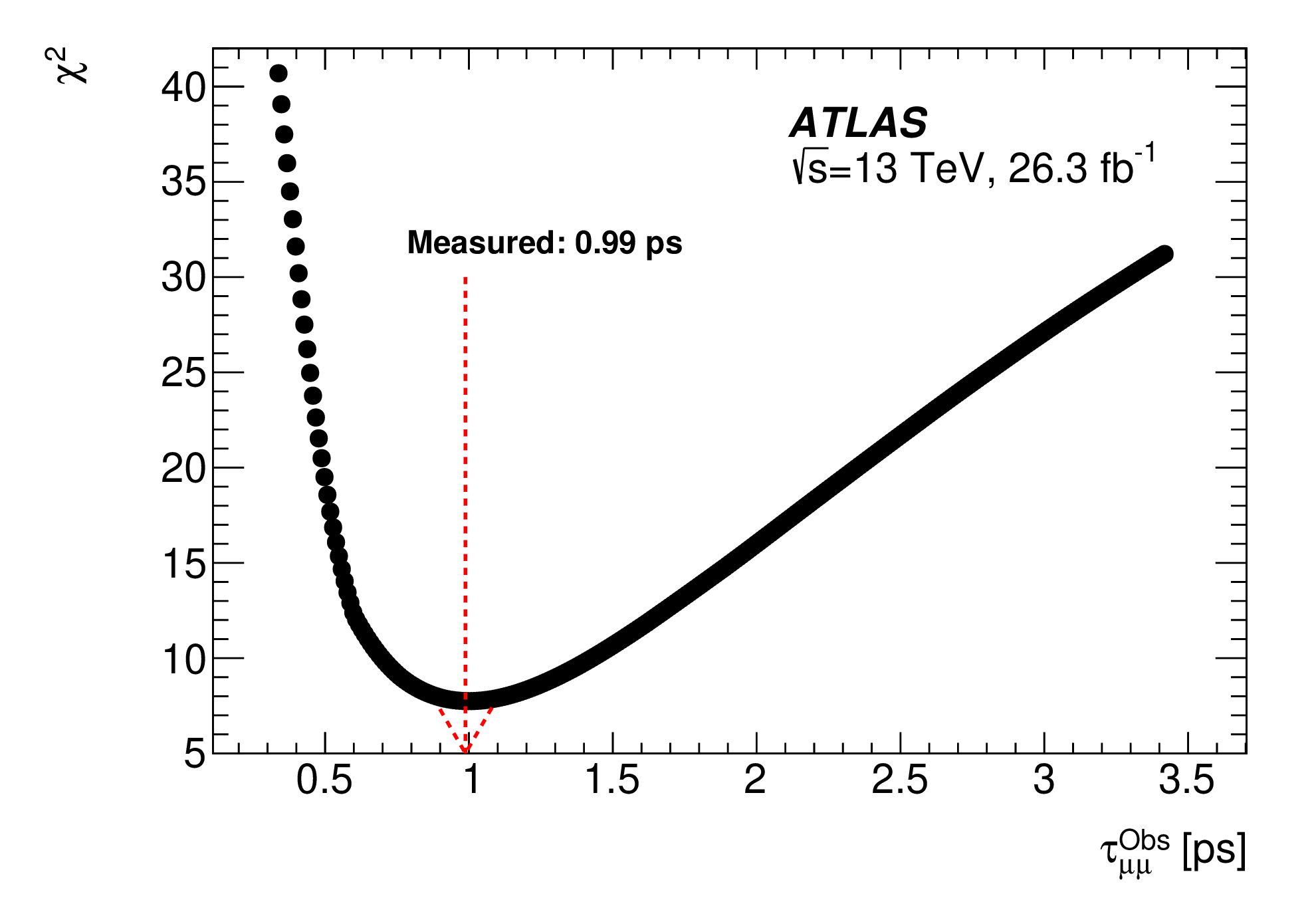} } \\
\centering
\subfloat[]{\includegraphics[width=0.45\textwidth]{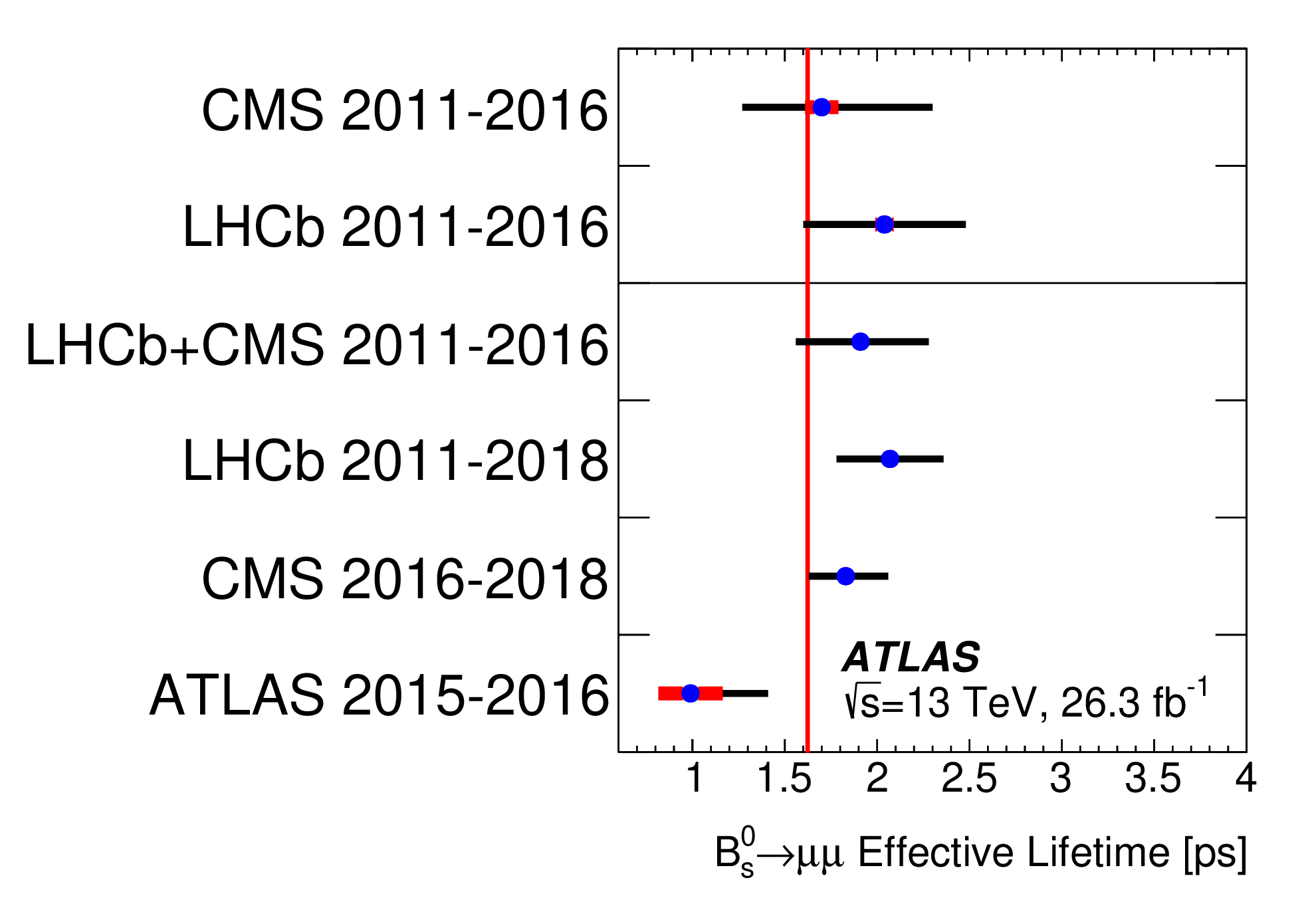} }
\caption{ATLAS results~\cite{BPHY-2020-07} on (a) the signal proper decay time distribution extracted with the \textit{sPlot} background subtraction procedure applied to the dimuon invariant mass fit. The superimposed signal MC template is the result of the   lifetime fit procedure. Uncertainties in the data points are calculated as Poisson fluctuations centred on the MC yield prediction (continuous histogram) in the corresponding bin. (b) $\chi^2$ scan versus the lifetime used in the MC template. The minimum of the scan ($\chi^2/\mathrm{ndf}=7.7/11$), located   at 0.99 ps, is indicated by the vertical dashed arrow.
(c) Comparison of ATLAS results~\cite{BPHY-2020-07}  $B_s\to\mu\mu$ effective lifetime effective lifetime  with the results from the LHCb collaboration based on 2011-2018 data LHCb~\cite{LHCb_2022} and 2011-2016 data \cite{LHCb_2017}, as well as the CMS collaboration results on 2011- 2016 data \cite{CMS-BPH-16-004} and 2016-2018 data ~\cite{CMS-BPH-21-006}. In all measurements, the horizontal bars represent the statistical (thinner) and systematic (thicker) uncertainties. The published combination \cite{ATLAS-CONF-2020-049}, on 2011-2016 data by LHCb and CMS collaborations is included as well ( LHCb + CMS 2011-2016), only reporting the total uncertainty. The SM prediction and its uncertainty ~\cite{HFLAV:2022esi}  are represented by the vertical line and its thickness, respectively. }
\label{fig:BsmumuTimefit}
\end{figure}


\section{Probing QCD with heavy-flavour hadrons}
\label{sec:hf}
\subsection{Precision measurement of \BcJpsiDsOrStar decays}
The \Bc meson represents a unique system comprised of the two heavy quarks, $b$ and $c$.
This makes studying production, decays, and spectroscopy of the $B_c$ family a powerful probe
of different QCD calculation approaches.
The \BcJpsiDsOrStar decays occur via the $\bar{b}\to \bar{c}c\bar{s}$ transition at quark level.
The decay processes can be divided into contributions involving a weak decay of the $b$- or $\bar c$-quark,
with the other one acting as a spectator, and the $b\bar c$ weak annihilation.
Corresponding diagrams are shown in Figure~\ref{fig:sturchikhin:Bc:diagrams}.
Beside the $\bar b\to \bar c$ tree diagrams, the annihilation topology can also contribute, although it is not expected
to have a large effect and is therefore often neglected~\cite{Colangelo:1999zn}.

\begin{figure}[t!]
\centering
\subfloat[]{\includegraphics[width=0.3\textwidth]{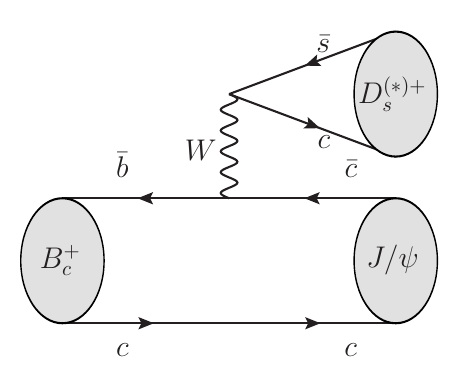}}
\subfloat[]{\includegraphics[width=0.3\textwidth]{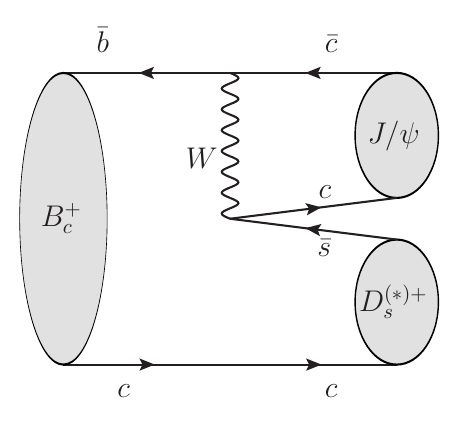}}
\subfloat[]{\includegraphics[width=0.3\textwidth]{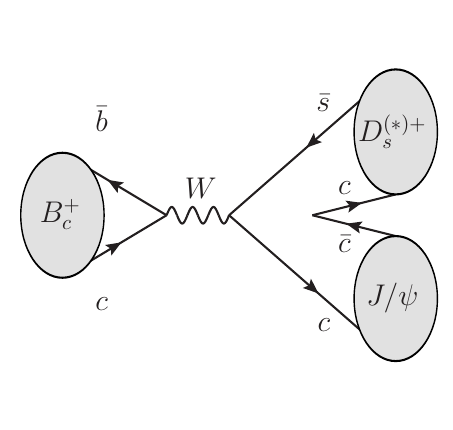}}
\caption{Quark diagrams for $\BcJpsiDsOrStar$ decays:
(a) colour-favoured,
(b) colour-suppressed
$b\to c$ tree and
(c) annihilation topology.
\label{fig:sturchikhin:Bc:diagrams}}
\end{figure}

These decays were first observed by the LHCb Collaboration~\cite{LHCb-PAPER-2013-010} and
later by ATLAS~\cite{BPHY-2014-04} using Run~1 data. Despite the lack of identification of kaons and pions,
the ATLAS measurement achieved competitive precision, especially for the polarisation in
the \BcJpsiDsStar decay, thanks to a more sophisticated signal fit strategy.

The ATLAS Run~2 study of these decays~\cite{BPHY-2018-08} benefits from larger numbers of events and improved
selection techniques. It aims to measure the branching fractions, relative to that of the reference decay
\BcJpsiPi. The following ratios are measured: $\RDsPi = \Br(\BcJpsiDs)/\Br(\BcJpsiPi)$,
$\RDsStarPi = \Br(\BcJpsiDsStar)/\Br(\BcJpsiPi)$,
and $\RDsStarDs = \Br(\BcJpsiDsStar)/\Br(\BcJpsiDs)$.

As the \BcJpsiDsStar decay is a transition of a pseudoscalar to two vector states, its decay products are polarised.
The decay can be described in terms of three helicity amplitudes, $A_{00}$, $A_{++}$, and $A_{--}$, where the indices
denote the helicities of the $\Jpsi$ and $\DsStar$ mesons. The $A_{00}$ amplitude corresponds to longitudinal polarisation and the other two refer to the transverse polarisations.
Although the soft photon from the $\DsStar\to\Dsp\gamma$ decay is not reconstructed in the analysis, the invariant mass
of the reconstructed \Bc decay products and angular shapes allow the fraction of transverse polarisation
\Gpp to be measured.

Figure~\ref{fig:sturchikhin:Bc:ValComp} shows the comparison of the Run~2 measurement results with those of Run~1 ATLAS and LHCb
measurements together with the results of various model calculations~\cite{Colangelo:1999zn,Kiselev:2002vz,Dubnicka:2017job,Dhir:2008hh,Ke:2013yka,Rui:2014tpa,Kar:2013fna, Nayak:2022qaq,Mohammadi:2018jcp}.
The new measurement achieves the best precision to date.
Overall the best description of all the ratios of branching fraction is given by the predictions
of a QCD relativistic potential model~\cite{Colangelo:1999zn}.
Several other predictions tend to underestimate the \RDsPi ratio, while still describing the \RDsStarPi well.
The measured value of \Gpp clearly agrees with a naive spin-counting expectation of $2/3$, being larger than
the values predicted by the dedicated calculations, which are below 0.5.

\begin{figure}[htbp]
\centering
\includegraphics[width=0.9\textwidth]{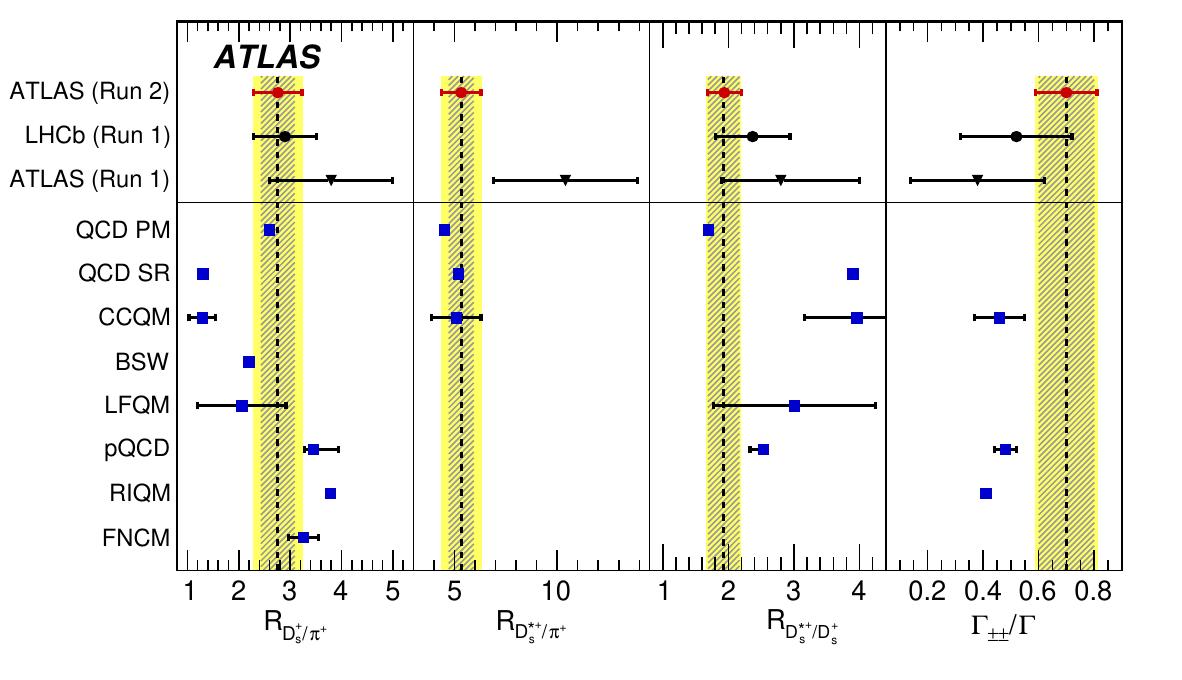}
\caption{Comparison of the results of the ATLAS \RunTwo \BcJpsiDsOrStar decay measurements~\cite{BPHY-2018-08} with those of
ATLAS \RunOne~\cite{BPHY-2014-04},
LHCb~\cite{LHCb-PAPER-2013-010} and theoretical predictions based
on a QCD relativistic potential model (QCD PM)~\cite{Colangelo:1999zn},
QCD sum rules (QCD SR)~\cite{Kiselev:2002vz},
covariant confined quark model (CCQM)~\cite{Dubnicka:2017job},
Bauer--Stech--Wirbel relativistic quark model (BSW)~\cite{Dhir:2008hh},
light-front quark model (LFQM)~\cite{Ke:2013yka},
perturbative QCD (pQCD)~\cite{Rui:2014tpa},
relativistic independent quark model (RIQM)~\cite{Kar:2013fna, Nayak:2022qaq},
and calculations in the QCD factorisation approach (FNCM)~\cite{Mohammadi:2018jcp}.
Hatched areas show the statistical uncertainties of this measurement and the wider bands correspond to the total uncertainties.
The uncertainties in the theoretical predictions are shown only if explicitly quoted in the corresponding papers.
\label{fig:sturchikhin:Bc:ValComp}
}
\end{figure}

Another interesting comparison can be made between the measured ratios of branching fractions and the
transverse polarisation fraction for \Bc decays to those for lighter $B$ mesons that occur predominantly
via either colour-favoured or
colour-suppressed tree diagrams.
Colour-favoured decays of $B^+$, $B^0$, or $B^0_s$ can be obtained by replacing the \Jpsi in the \Bc decay final state
with $\bar D^{*0}$, $D^{*-}$, or $D_s^{*-}$, while colour-suppressed modes are obtained
by replacing the \DsOrStar with $K^{(*)+}$, $K^{(*)0}$, or $\phi$, respectively.

These comparisons are presented in Figure~\ref{fig:sturchikhin:Bc:ValComp_AUX}.
The \RDsStarDs value agrees with the corresponding ratio calculated
for both the $B^0$ and $B^+$ decays into $D$ mesons and is larger than that obtained for
their decays into \Jpsi and kaons.
The measured value of \Gpp lies between the transverse polarisation
fraction values in the $B^0\to D^{*-} D_s^{*}$ and $B^0_s\to D_s^{*-} D_s^{*}$ decays
and is larger than those in
the considered $B$ decays occurring via the colour-suppressed tree diagram.
These results support the assumption that the colour-favoured tree diagram dominates
the \BcJpsiDsOrStar decay amplitudes.

\begin{figure}[htbp]
\centering
\includegraphics[width=0.9\textwidth]{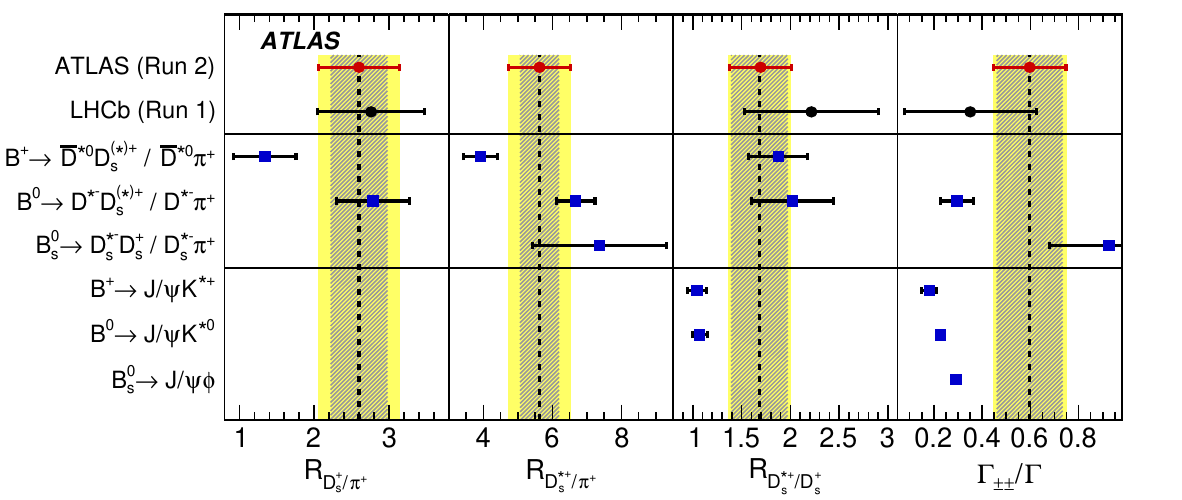}
\caption{Comparison of the measured ratios of the \Bc decay branching fractions
and the transverse polarisation fraction~\cite{BPHY-2018-08}
to the corresponding values for $B^+$, $B^0$, and $B^0_s$
decays~\cite{pdg2020} occurring predominantly via the colour-favoured or colour-suppressed spectator diagrams (see text).
No phase-space corrections are applied to the ratios of branching fractions
and the quoted uncertainties are the ones propagated from the world average uncertainties
of the individual decay branching fractions.
Hatched areas show the statistical uncertainties in these measurements and the wider bands correspond to the total uncertainties.
\label{fig:sturchikhin:Bc:ValComp_AUX}
}
\end{figure}

\subsection{Charmonium production measurements}
Despite a long history of studying heavy quarkonium production in hadronic collisions, these processes still present
a significant challenge to both theory and experiment. Two mechanisms play a role in production of charmonium states:
prompt production from `direct' QCD processes and non-prompt production from decays of $b$-hadrons. While the latter
can be described reasonably well within perturbative QCD~\cite{Cacciari:2001td, Cacciari:2012ny}, a satisfactory
understanding of prompt production is still to be achieved. The conventional approach to describe the prompt process is based on
non-relativistic QCD (NRQCD) and introduces a number of phenomenological parameters, namely long-distance matrix elements (LDMEs), that need to be extracted from fits to experimental data. Attempts to build a universal set of LDMEs
able to provide a consistently good description of charmonium polarisation, associated production,  and
photo- and electro-production have not been successful so far.

A wide range of experimental measurements of charmonium production characteristics have been provided by the LHC experiments
during the past decade. One path to add information useful for building theoretical models, is to extend the kinematic
reach of these measurements. With the full Run~2 data sample ATLAS performed a measurement
of \Jpsi and \psip production~\cite{BPHY-2019-08}
using their dimuon decay channels over the largest transverse momentum range ever achieved to date:
from 8 to 360\,\GeV for \Jpsi and up to 140\,\GeV for \psip. This was achieved by using a combination of two types of
triggers: dimuon triggers to cover the lower \pT range up to about 100\,\GeV, and single-muon triggers with a threshold of $\pT > 50\,\GeV$ above, where dimuon triggers are inefficient because of the small angular separation between the muon.

The signal extraction is performed by a simultaneous fit to the dimuon invariant mass and pseudo-proper lifetime distributions. Peaks
of \Jpsi and \psip are clearly separated in the mass spectrum,
while the lifetime distribution in the fit allows the prompt and non-prompt production to
be distinguished.
Double-differential production measurements of both
charmonium states are performed for prompt and non-prompt mechanisms.

Figure~\ref{fig:sturchikhin:CharmoniumProd:prompt} shows the prompt \Jpsi production cross-sections and
the comparison of the prompt production measurement results
with various theory predictions: NLO NRCDQ calculations~\cite{Butenschoen:2010rq} using pre-defined
LDMEs~\cite{Butenschoen:2011yh, Butenschoen:2022qka}, $k_\mathrm{T}$-factorisation model calculations made with
the \textsc{PEGASUS} generator~\cite{Lipatov:2019oxs} and a different set of LDMEs~\cite{Baranov:2019lhm}, and the `improved
colour evaporation model' (ICEM)~\cite{Cheung:2021epq} predictions. Overall, all approaches tend to predict
harder \pT spectra for both \Jpsi and \psip, while the ICEM also underestimates the total \psip production.

\begin{figure}[htbp]
\begin{center}
\subfloat[]{\includegraphics[width=0.32\textwidth]{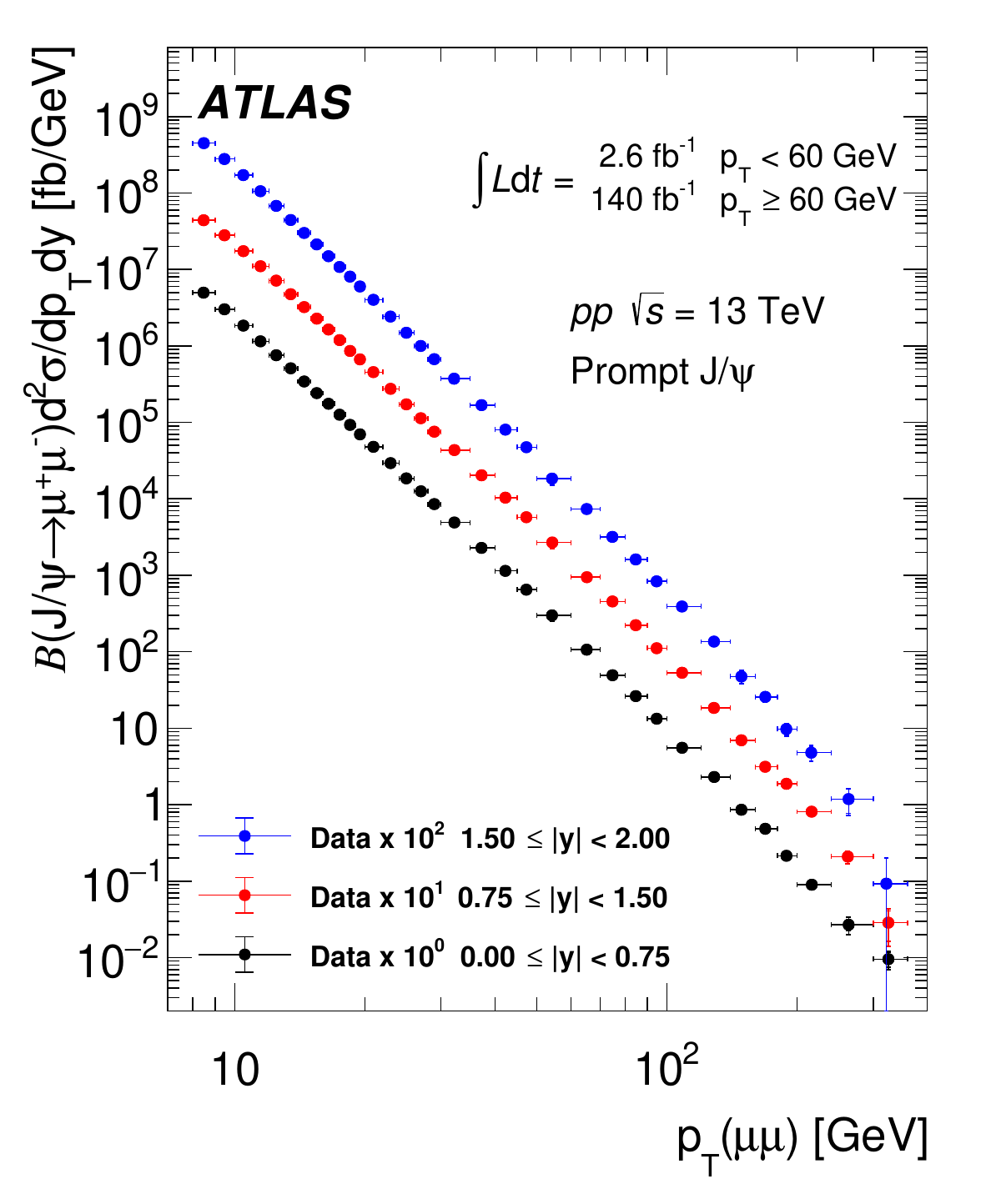}}
\subfloat[]{\includegraphics[width=0.32\textwidth]{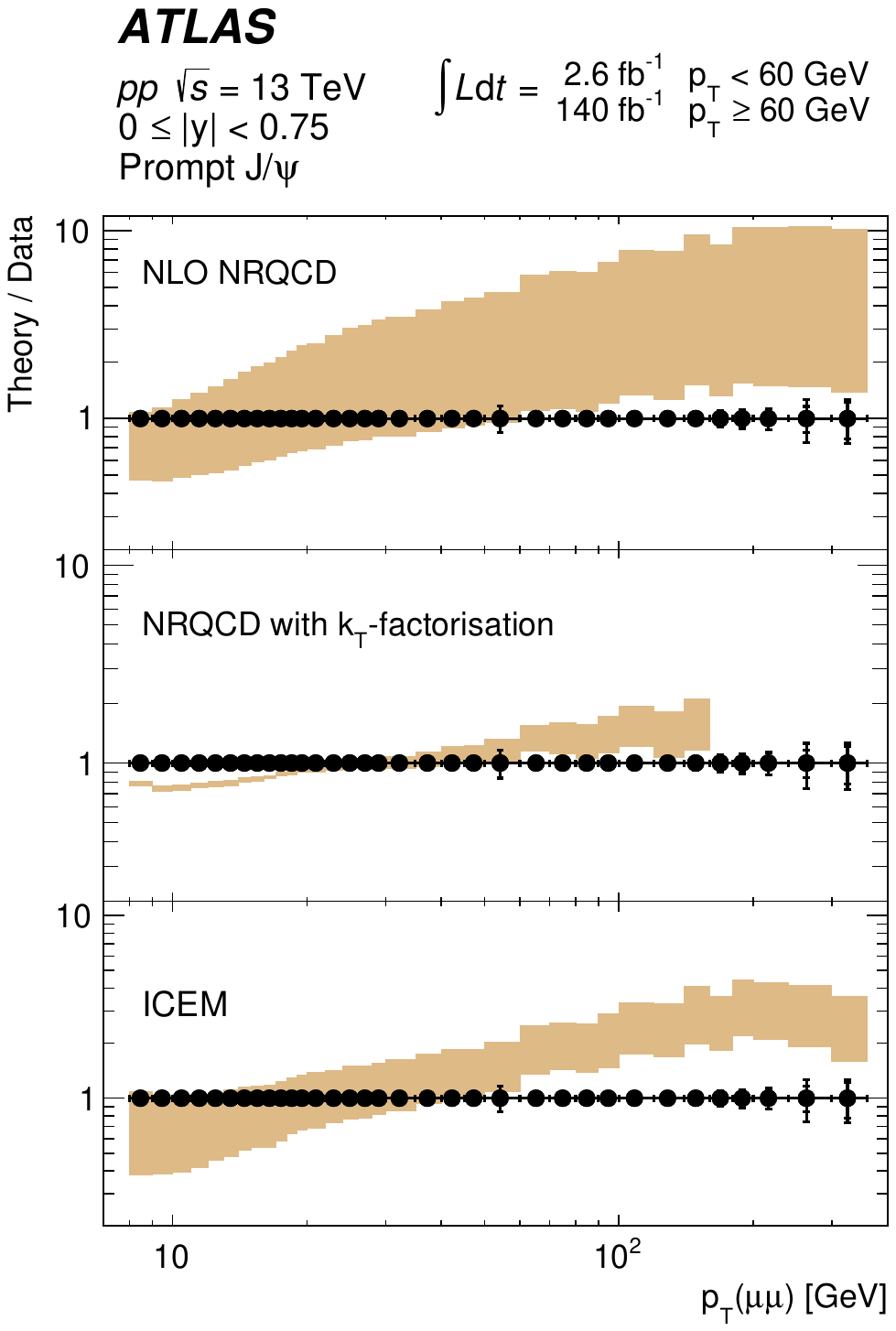}}
\subfloat[]{\includegraphics[width=0.32\textwidth]{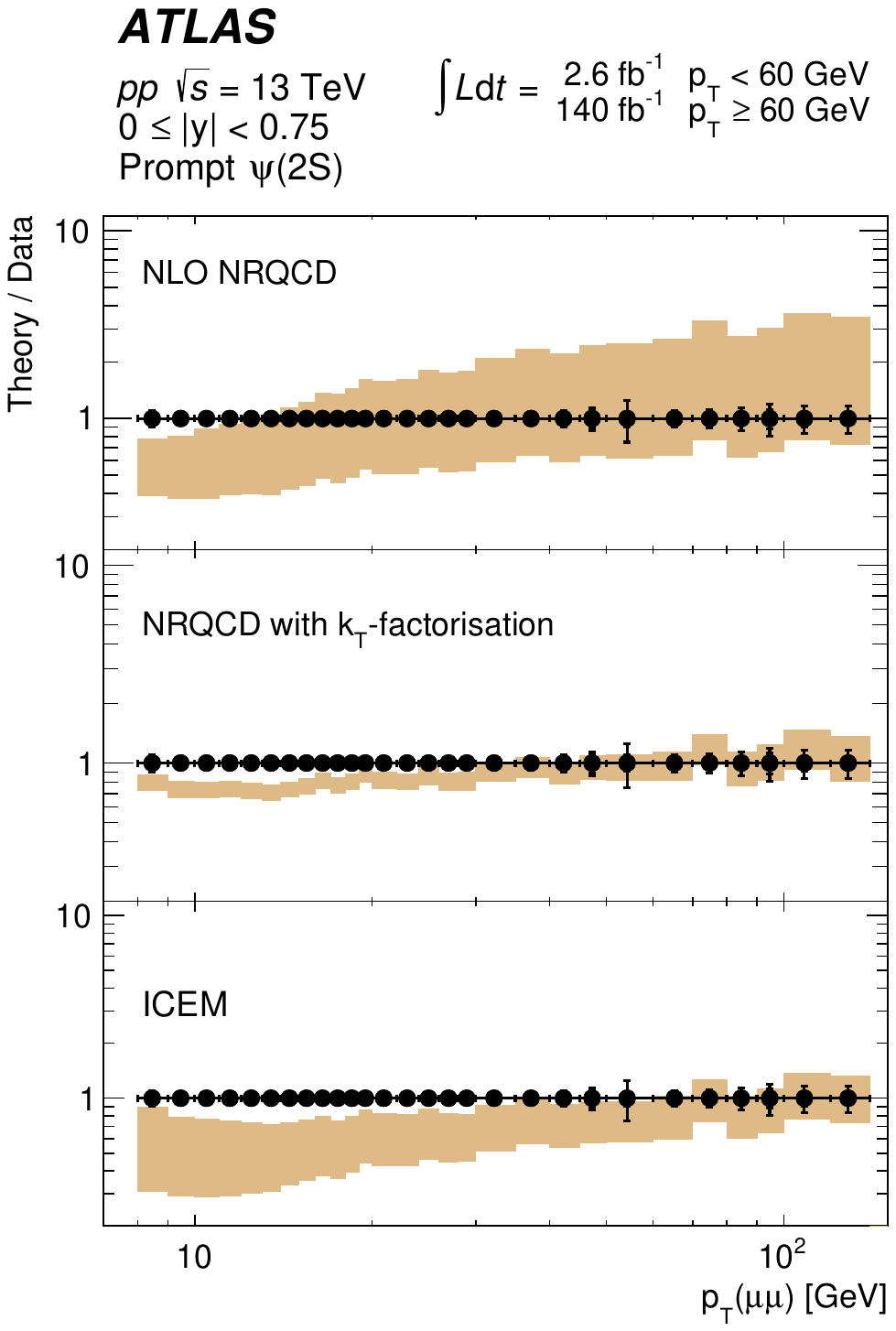}}
\end{center}
\caption{(a) Differential cross-sections of prompt \Jpsi production,
comparison of (b) prompt  \Jpsi and (c) \psip production measurement results with various
theoretical predictions (see text)~\cite{BPHY-2019-08}.
\label{fig:sturchikhin:CharmoniumProd:prompt}}
\end{figure}

These measurements reach an unprecedentedly wide kinematic range of charmonium production, challenge the existing models, and provide unique input for their further tuning.
Figure~\ref{fig:sturchikhin:CharmoniumProd:nonPrompt} shows the non-prompt \Jpsi production fraction
and compares the measured non-prompt production
with calculations: the traditional fixed-order-next-to-leading-log (FONLL)
approach~\cite{Cacciari:2001td, Cacciari:2012ny} predictions, those based on general-mass-variable-flavour-number scheme
(GM-VFNS)~\cite{Bolzoni:2013tca}, and $k_\mathrm{T}$-factorisation-based calculations~\cite{Lipatov:2019oxs, Baranov:2018cmp}.
None of these models is able to describe the data over the full \pT range, while the general trend in all of them is
the slower decrease of cross-section with \pT. This can be related to insufficient account of parton distribution
function evolution or to possible dependence of LDMEs on transverse momentum.

\begin{figure}[htbp]
\begin{center}
\subfloat[]{\raisebox{-0.5\height}{\includegraphics[width=0.32\textwidth]{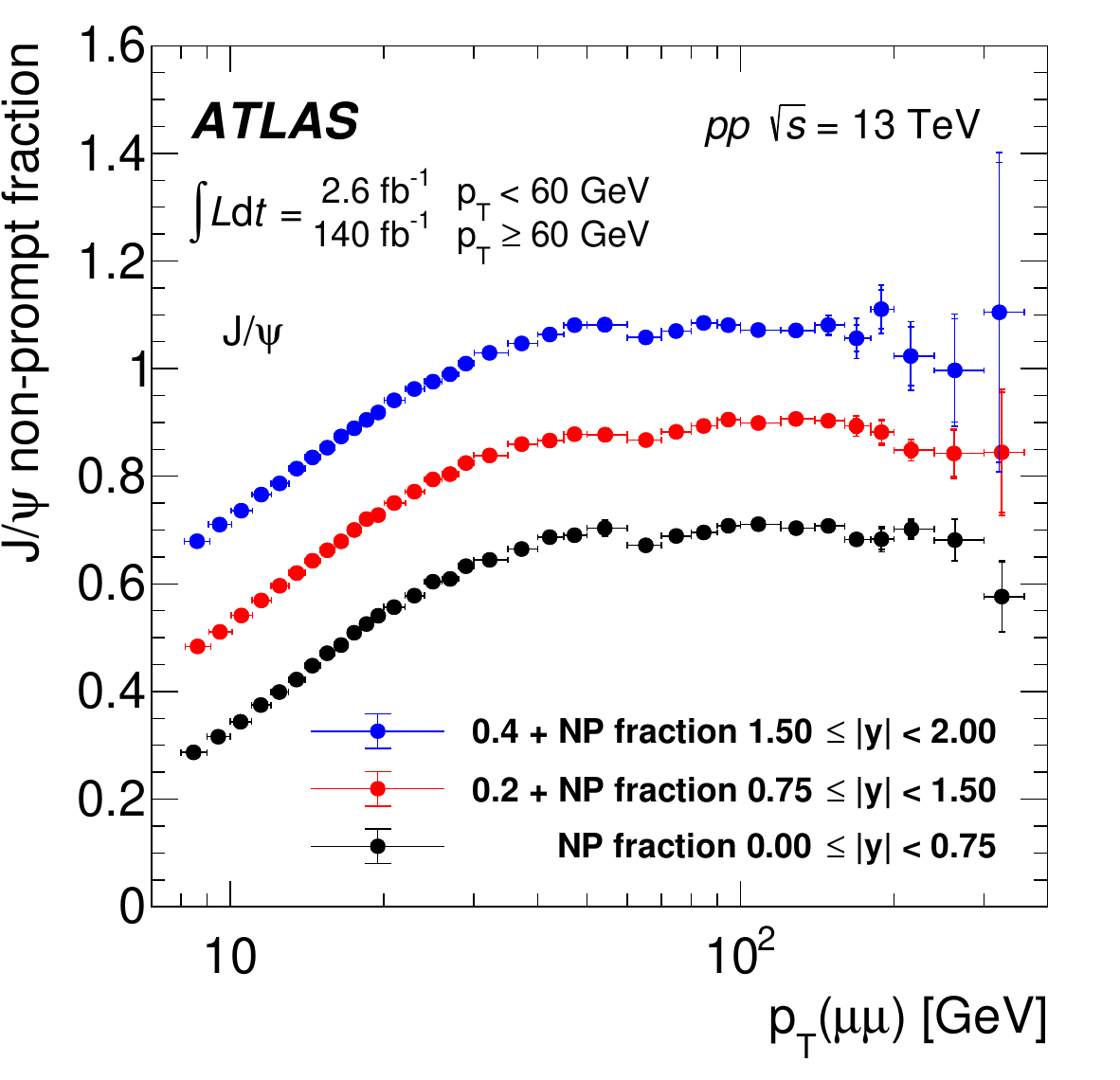}}}
\subfloat[]{\raisebox{-0.5\height}{\includegraphics[width=0.32\textwidth]{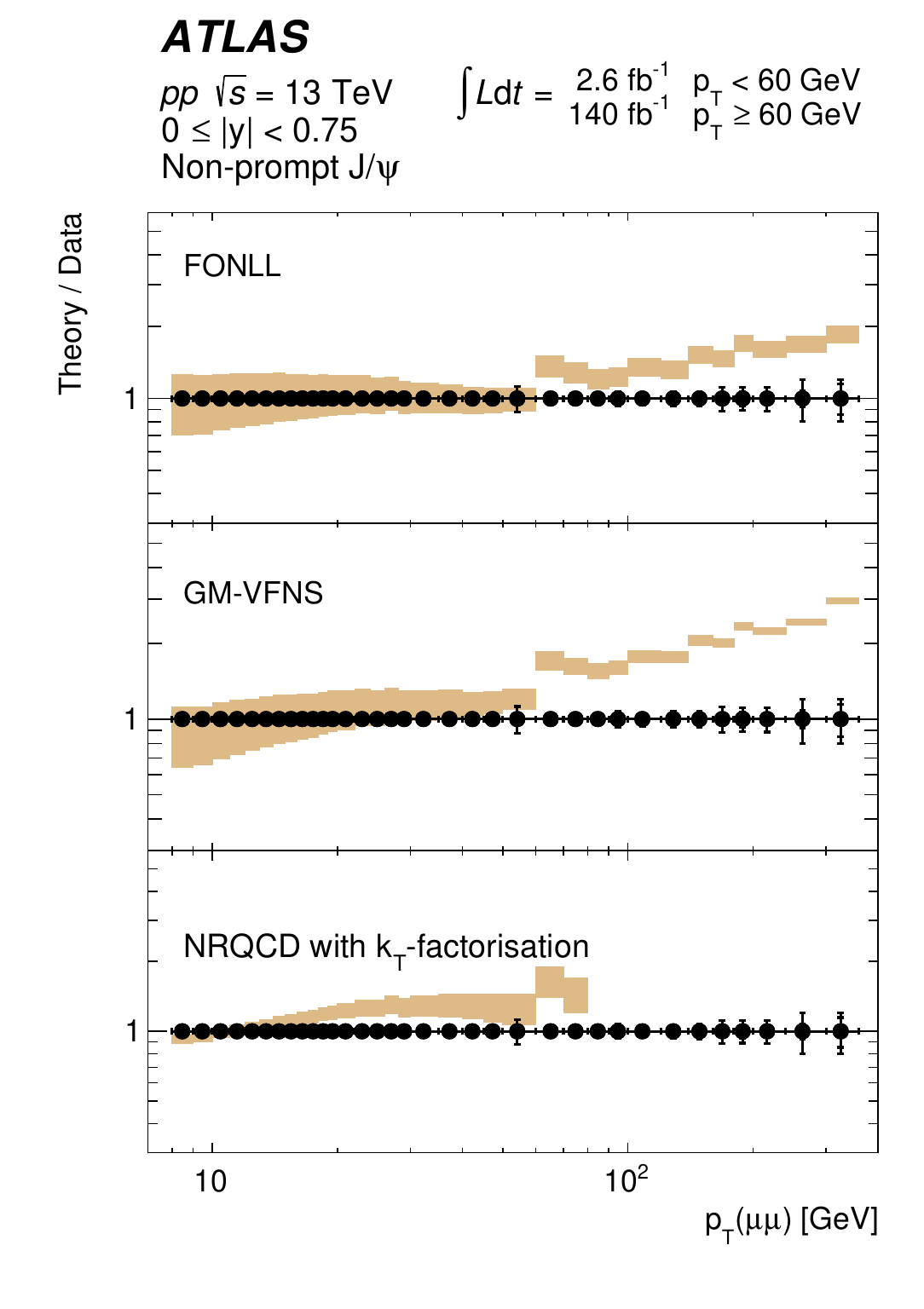}}}
\subfloat[]{\raisebox{-0.5\height}{\includegraphics[width=0.32\textwidth]{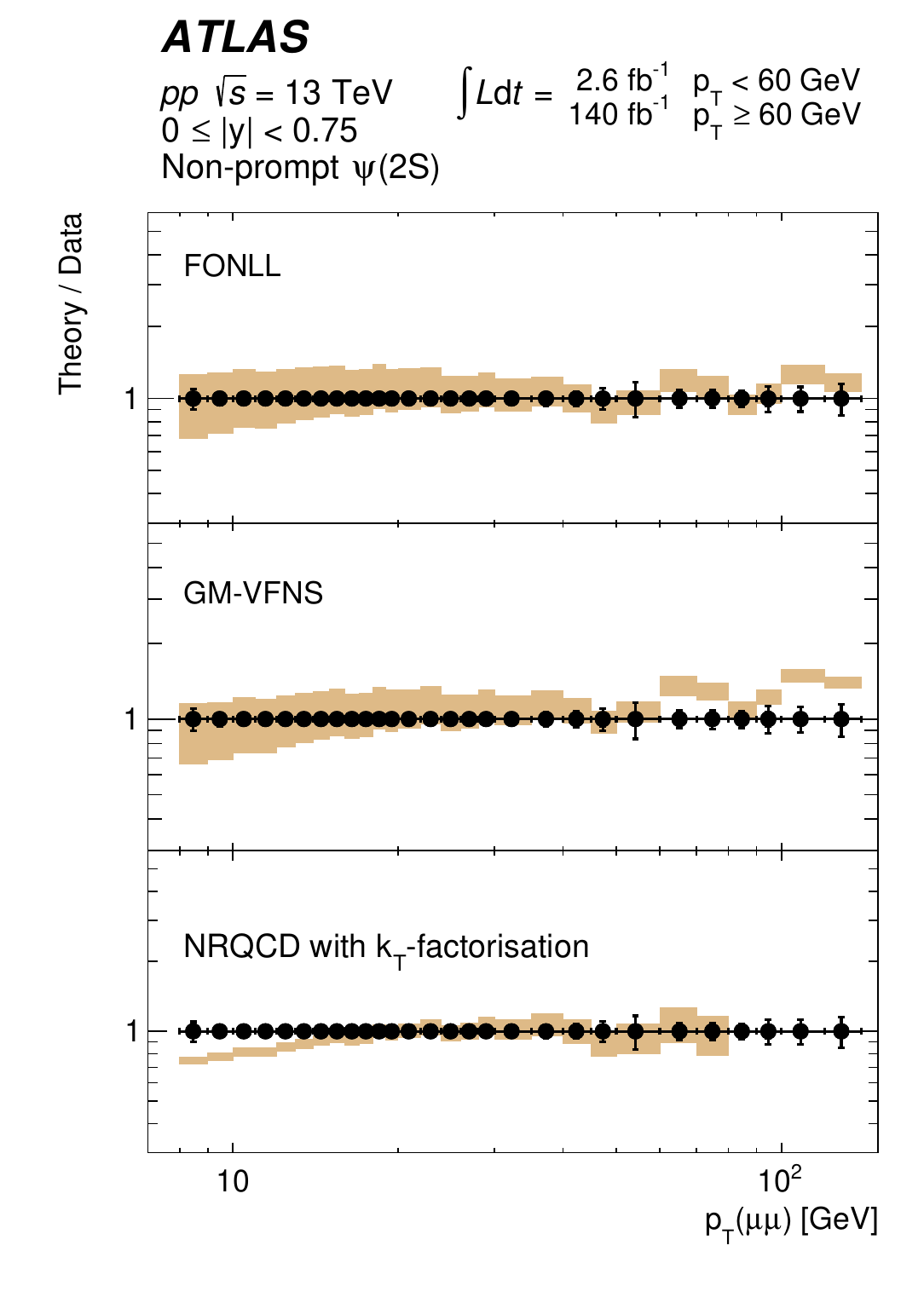}}}
\end{center}
\caption{(a) Non-prompt production fraction of \Jpsi,
comparison of (b) non-prompt  \Jpsi and (c) \psip production measurement results with various
theoretical predictions (see text)~\cite{BPHY-2019-08}.
\label{fig:sturchikhin:CharmoniumProd:nonPrompt}}
\end{figure}

\subsection{Studies of exotic hadron states}
Beside the conventional hadrons comprised of three quarks ($qqq$) or a quark and an antiquark ($q\bar q$),
QCD allows the existence of more complex
systems such as pentaquarks ($qqqq\bar q$) and tetraquarks ($qq\bar q\bar q$). A number of such states were discovered
in the last couple of decades~\cite{Olsen:2017bmm}.
One of them was observed by LHCb as a narrow structure in the di-\Jpsi channel at a mass of  6.9\,\GeV, along with
an enhancement in the mass spectrum closer to the di-\Jpsi threshold at about 6.2\,\GeV~\cite{LHCb-PAPER-2020-011}.
That structure could be interpreted as
a tetraquark composed of four charm quarks.

ATLAS performed a search for such states~\cite{BPHY-2022-01} in both di-\Jpsi and $\Jpsi+\psip$ channels
using the four-muon final state for both.
Figure~\ref{fig:sturchikhin:4mu:massPlots} shows the results of the fits to the corresponding invariant mass
distributions.
In the di-\Jpsi channel, two models are used for the fit. In the first one
(Figure~\ref{fig:sturchikhin:4mu:massPlots:diJpsi_A}), the signal probability density function consists of three
interfering S-wave Breit--Wigner resonances multiplied by a phase-space factor and convolved with a mass resolution
function. In the second model (Figure~\ref{fig:sturchikhin:4mu:massPlots:diJpsi_B}),
only two resonances are considered, one of which interferes with the amplitude of the
background \Jpsi pair production via single parton scattering (SPS), and the other is standalone.
Both models describe well the enhancement near the mass threshold and the enhancement at 6.9\,\GeV, attributed to a $X(6900)$ resonance. The significance of the resonance far exceeds five standard deviations and its mass and width agree with those measured
by LHCb~\cite{LHCb-PAPER-2020-011}. However, the broad structure at the lower mass could result
from many physical effects, such as feed-down from higher di-charmonium resonances, e.g.,
$\Tcccc\to\chi_{c1}\chi_{c1}\to\Jpsi\gamma\Jpsi\gamma$.

In the $\Jpsi+\psip$ channel fit, two models are also used. The first one
(Figure~\ref{fig:sturchikhin:4mu:massPlots:JpsiPsi2S_a}) assumes that the same three interfering
resonances from the first model of the di-\Jpsi fit can also decay into $\Jpsi+\psip$, in addition to a fourth
standalone resonance exclusively decaying into this channel.
Parameters of the first three resonances, contributing to the enhancement
just above the $\Jpsi+\psip$ threshold, are fixed to their values from the di-\Jpsi fit.
The second model (Figure~\ref{fig:sturchikhin:4mu:massPlots:JpsiPsi2S_b}) assumes only a single resonance in this
channel. The signal significance of the fit results with the two models is 4.7 and 4.3 standard deviations respectively.
In the fit
to the first model, the significance of the additional resonant structure near 7.2\,\GeV\ alone
is three standard deviations.

\begin{figure}[htbp]
\centering
\subfloat[]{
\includegraphics[width=0.49\textwidth]{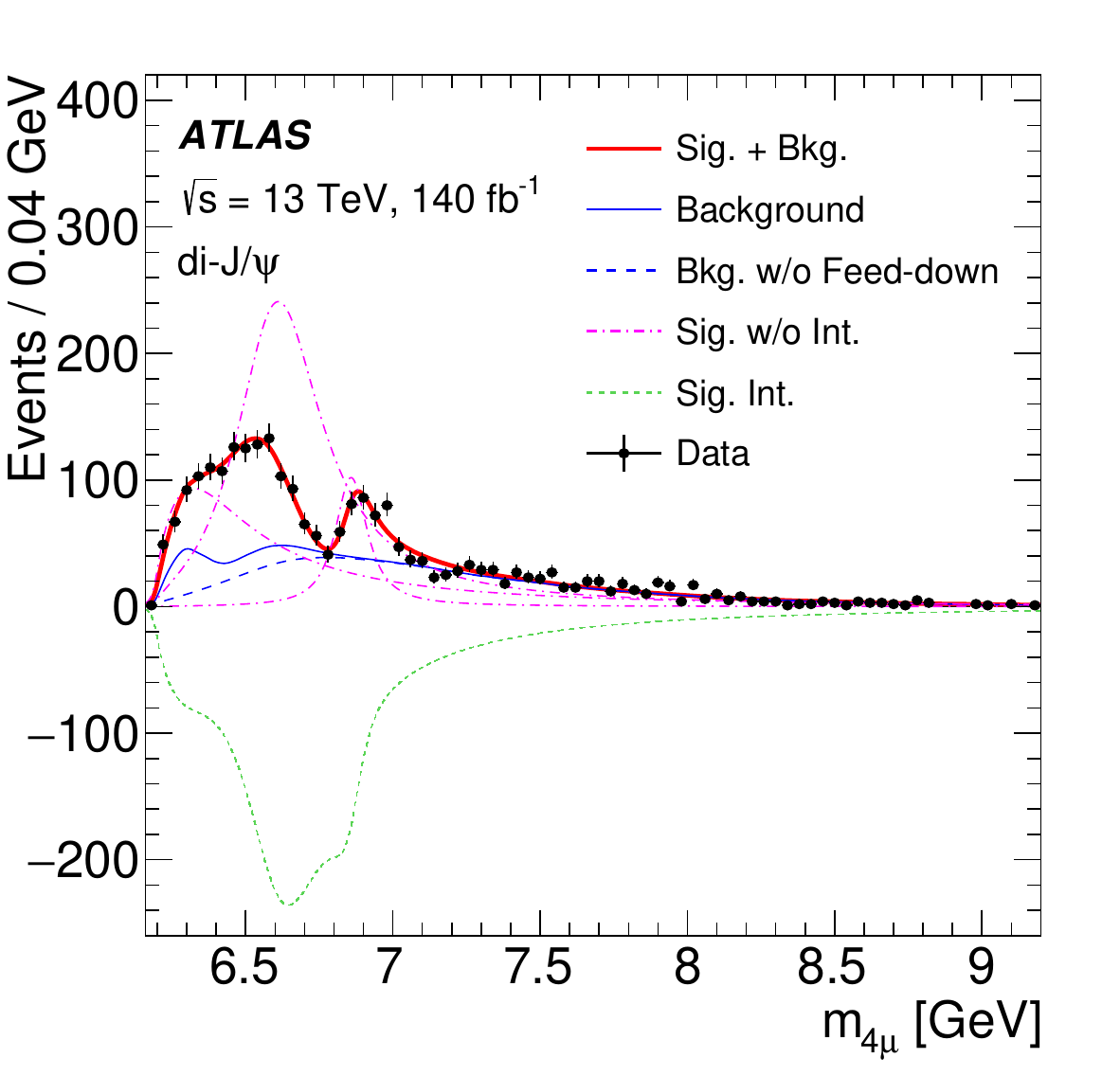}
\label{fig:sturchikhin:4mu:massPlots:diJpsi_A}
}
\subfloat[]{
\includegraphics[width=0.49\textwidth]{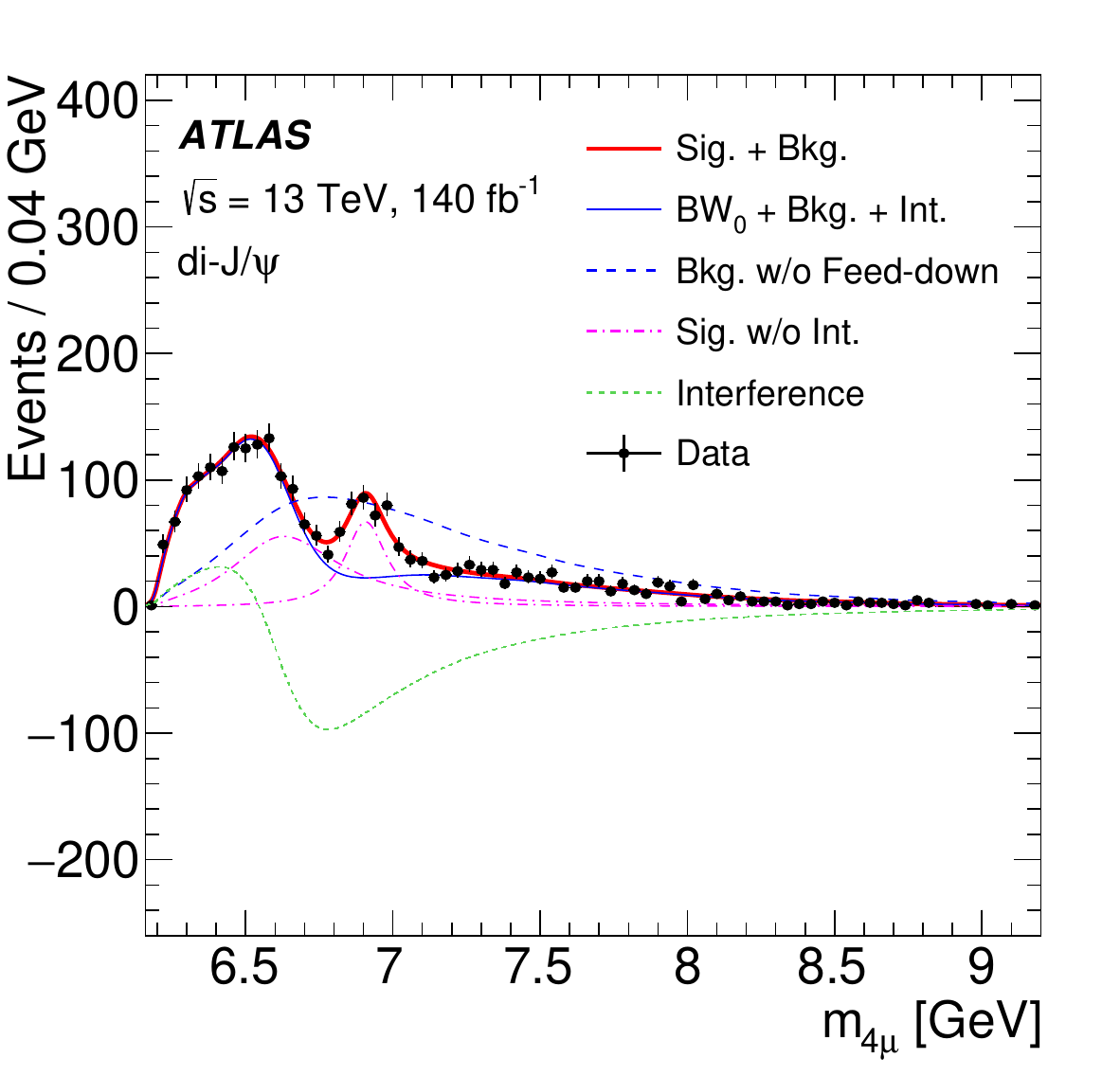}
\label{fig:sturchikhin:4mu:massPlots:diJpsi_B}
}\\
\subfloat[]{
\includegraphics[width=0.49\textwidth]{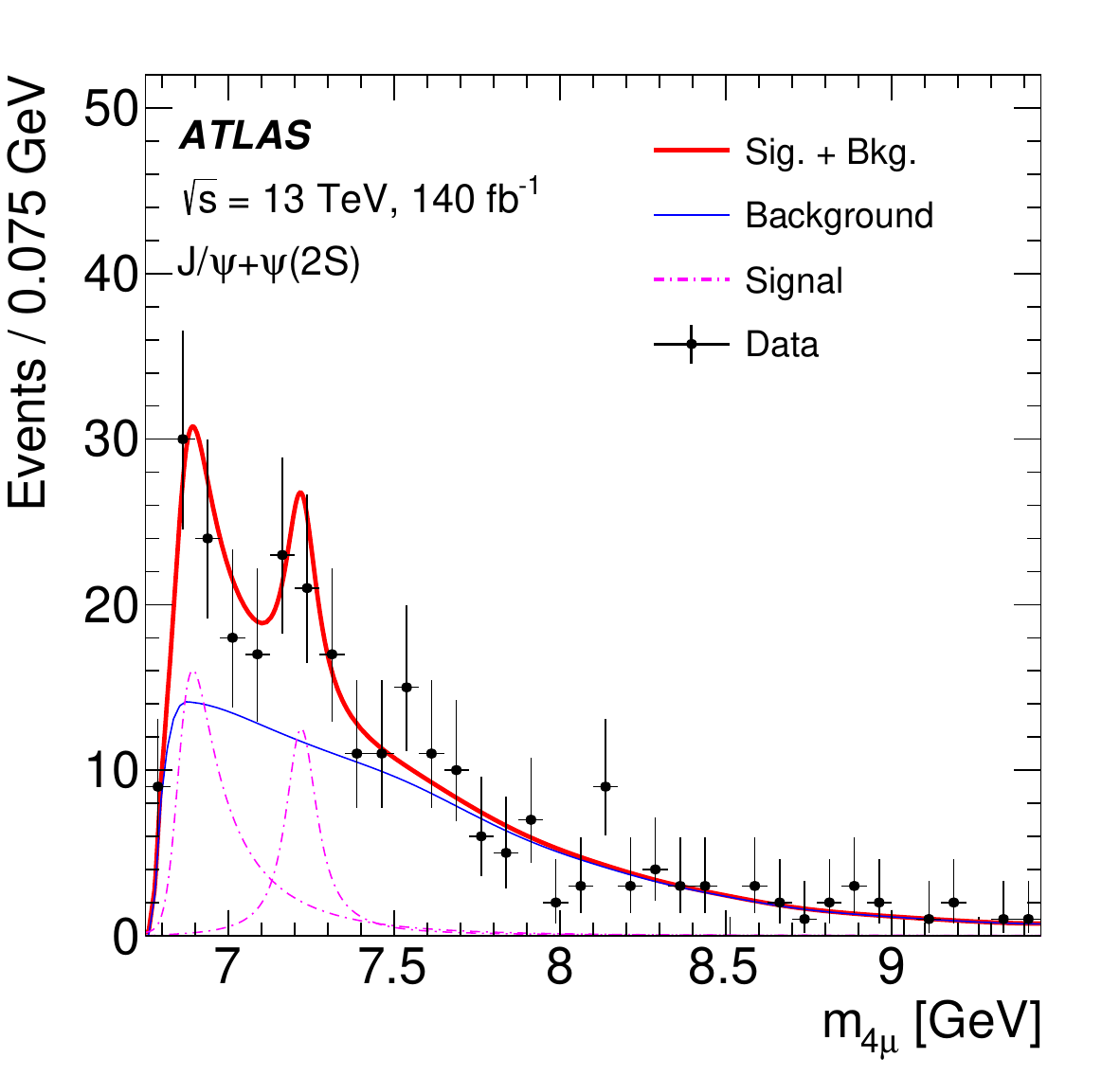}
\label{fig:sturchikhin:4mu:massPlots:JpsiPsi2S_a}
}
\subfloat[]{
\includegraphics[width=0.49\textwidth]{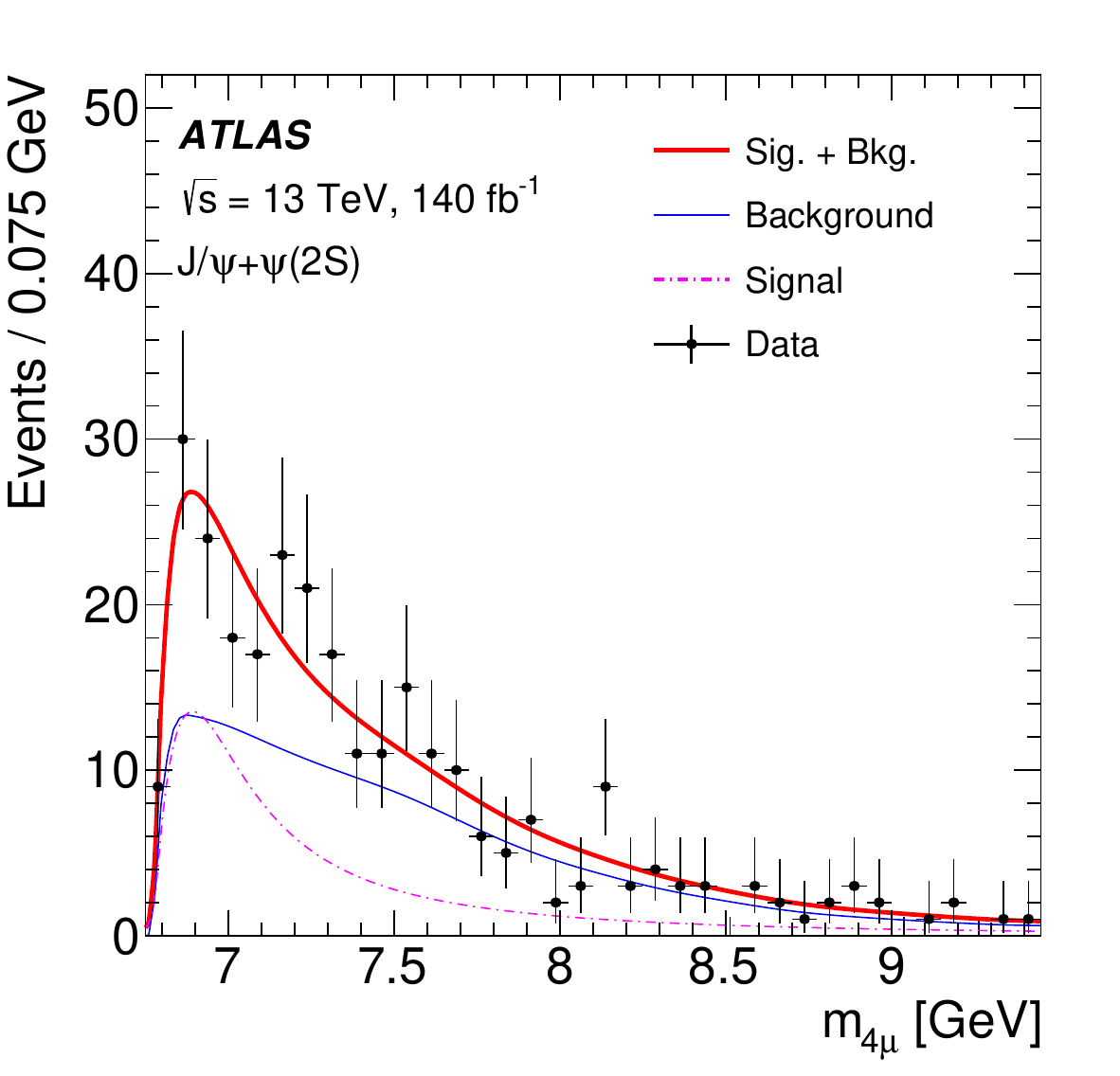}
\label{fig:sturchikhin:4mu:massPlots:JpsiPsi2S_b}
}
\caption{Results of the fits to di-\Jpsi mass spectra using models
\protect\subref{fig:sturchikhin:4mu:massPlots:diJpsi_A} with three interfering resonances and
\protect\subref{fig:sturchikhin:4mu:massPlots:diJpsi_B} with two resonances,
and to $\Jpsi+\psip$ spectra using models
\protect\subref{fig:sturchikhin:4mu:massPlots:JpsiPsi2S_a} with the three di-\Jpsi resonances and an additional standalone one, and
\protect\subref{fig:sturchikhin:4mu:massPlots:JpsiPsi2S_b} with a single resonance~\cite{BPHY-2022-01}.
\label{fig:sturchikhin:4mu:massPlots}}
\end{figure}

\FloatBarrier


%

%
%
%

%
\section{Conclusion}
\label{sec:conclusion}

This report summarises precision electroweak and QCD measurements performed by the ATLAS experiment during Run~2
of the Large Hadron Collider from 2015 to 2018. Most results are based on data taken at $\sqrt{s} = 13$~TeV corresponding to up to $140~\ifb$ but
selected recent precision measurements on Run~1 data at $\sqrt{s} = 7$~TeV and $\sqrt{s} = 8$~TeV are also reported.
The excellent performance of the upgraded ATLAS detector and significant progress in the performance of object reconstruction and identification,
together with an increased centre-of-mass energy and a large data sample, allows a large range of novel high-quality measurements and the
observation of rare processes.
The review covers measurements published until spring 2024, with several further Run~2 measurements still expected to be published.

QCD in its non-perturbative regime is tested via the measurements of the total ($\sigma_{\textrm{tot}}$), elastic ($\sigma_{\textrm{el}}$) and inelastic ($\sigma_{\textrm{inel}}$) $pp$ cross-sections, and via the production of charged particles in $pp$ collisions.
In particular, the ATLAS measurements of $\sigma_{\textrm{tot}}$, $\sigma_{\textrm{el}}$ and $\sigma_{\textrm{inel}}$ reach the best experimental precision among the existing LHC measurements, allowing for a detailed test of the energy evolution for $\sigma_{\textrm{tot}}$.

Perturbative QCD tests include the measurements of inclusive jets and isolated photons, but also jets in association with single EW bosons or EW boson pairs and the measurement of transverse momentum and other kinematic variables of single EW bosons and boson pairs.
These high-precision multi-differential measurements directly probe the higher-order QCD corrections, and are used to constrain parton distribution functions of the proton. The production of EW bosons with heavy-flavour jets, allows for tests of pQCD, flavour and mass schemes and of the $s$, $c$ and $b$ content of the proton.

In a series of measurements, ATLAS also studies the internal structure of jets.
These novel measurements are sensitive to both perturbative and non-perturbative QCD effects.

EW interactions are tested by measurements targeting triple and quartic EW boson interactions in vector-boson fusion and vector-boson scattering processes and in the production of three gauge bosons. The EW production of two gauge bosons ($W^\pm W^\pm$, $W^\pm W^\mp$, $W^\pm Z$, $ZZ$ and $Z\gamma$)  and the production of several triboson combinations that include heavy EW bosons ($WWW$, $WZ\gamma$ and $W\gamma\gamma$) are observed for the first time in Run~2.
Unique tests of EW interactions are also performed using measurements of photon--photon interactions in dilepton, diphoton, and $WW$ final-states, exploring both $pp$ and Pb+Pb collision systems.
This leads to the first direct observations of $\gamma\gamma \to \gamma\gamma$ and $\gamma\gamma \to WW$ scattering  processes.

Fundamental  parameters of the SM are extracted with unprecedented precision, based on novel techniques: the mass and width of the $W$ boson, the strong coupling constant and the invisible decay width of the $Z$ boson.

This report also covers studies of heavy-flavour hadrons, including charmonium and exotic states.
In CP-violating and rare $b$-hadron decays, a large data sample allows
the sensitivity of searches for new physics effects to be substantially improved, but more data is necessary to obtain conclusive results.
The double-heavy $B_c$ meson is studied including new decay modes and with unprecedented precision.
The extended kinematic reach of charmonium production measurements allows QCD calculations to be tested in a range never explored before.
Studies of recently discovered exotic resonances help to further establish their status, motivating the development of underlying theories.


%

%
\section*{Acknowledgements}
%

%
%

%
%

We thank CERN for the very successful operation of the LHC and its injectors, as well as the support staff at
CERN and at our institutions worldwide without whom ATLAS could not be operated efficiently.

The crucial computing support from all WLCG partners is acknowledged gratefully, in particular from CERN, the ATLAS Tier-1 facilities at TRIUMF/SFU (Canada), NDGF (Denmark, Norway, Sweden), CC-IN2P3 (France), KIT/GridKA (Germany), INFN-CNAF (Italy), NL-T1 (Netherlands), PIC (Spain), RAL (UK) and BNL (USA), the Tier-2 facilities worldwide and large non-WLCG resource providers. Major contributors of computing resources are listed in Ref.~\cite{ATL-SOFT-PUB-2023-001}.

We gratefully acknowledge the support of ANPCyT, Argentina; YerPhI, Armenia; ARC, Australia; BMWFW and FWF, Austria; ANAS, Azerbaijan; CNPq and FAPESP, Brazil; NSERC, NRC and CFI, Canada; CERN; ANID, Chile; CAS, MOST and NSFC, China; Minciencias, Colombia; MEYS CR, Czech Republic; DNRF and DNSRC, Denmark; IN2P3-CNRS and CEA-DRF/IRFU, France; SRNSFG, Georgia; BMBF, HGF and MPG, Germany; GSRI, Greece; RGC and Hong Kong SAR, China; ISF and Benoziyo Center, Israel; INFN, Italy; MEXT and JSPS, Japan; CNRST, Morocco; NWO, Netherlands; RCN, Norway; MNiSW, Poland; FCT, Portugal; MNE/IFA, Romania; MESTD, Serbia; MSSR, Slovakia; ARIS and MVZI, Slovenia; DSI/NRF, South Africa; MICIU/AEI, Spain; SRC and Wallenberg Foundation, Sweden; SERI, SNSF and Cantons of Bern and Geneva, Switzerland; NSTC, Taipei; TENMAK, T\"urkiye; STFC/UKRI, United Kingdom; DOE and NSF, United States of America.

Individual groups and members have received support from BCKDF, CANARIE, CRC and DRAC, Canada; PRIMUS 21/SCI/017, CERN-CZ and FORTE, Czech Republic; COST, ERC, ERDF, Horizon 2020, ICSC-NextGenerationEU and Marie Sk{\l}odowska-Curie Actions, European Union; Investissements d'Avenir Labex, Investissements d'Avenir Idex and ANR, France; DFG and AvH Foundation, Germany; Herakleitos, Thales and Aristeia programmes co-financed by EU-ESF and the Greek NSRF, Greece; BSF-NSF and MINERVA, Israel; Norwegian Financial Mechanism 2014-2021, Norway; NCN and NAWA, Poland; La Caixa Banking Foundation, CERCA Programme Generalitat de Catalunya and PROMETEO and GenT Programmes Generalitat Valenciana, Spain; G\"{o}ran Gustafssons Stiftelse, Sweden; The Royal Society and Leverhulme Trust, United Kingdom.

In addition, individual members wish to acknowledge support from CERN: European Organization for Nuclear Research (CERN PJAS); Chile: Agencia Nacional de Investigaci\'on y Desarrollo (FONDECYT 1190886, FONDECYT 1210400, FONDECYT 1230812, FONDECYT 1230987); China: Chinese Ministry of Science and Technology (MOST-2023YFA1605700, MOST-2023YFA1609300), National Natural Science Foundation of China (NSFC - 12175119, NSFC 12275265, NSFC-12075060); Czech Republic: Czech Science Foundation (GACR - 24-11373S), Ministry of Education Youth and Sports (FORTE CZ.02.01.01/00/22\_008/0004632), PRIMUS Research Programme (PRIMUS/21/SCI/017); EU: H2020 European Research Council (ERC - 101002463); European Union: European Research Council (ERC - 948254, ERC 101089007), Horizon 2020 Framework Programme (MUCCA - CHIST-ERA-19-XAI-00), European Union, Future Artificial Intelligence Research (FAIR-NextGenerationEU PE00000013), Italian Center for High Performance Computing, Big Data and Quantum Computing (ICSC, NextGenerationEU); France: Agence Nationale de la Recherche (ANR-20-CE31-0013, ANR-21-CE31-0013, ANR-21-CE31-0022, ANR-22-EDIR-0002), Investissements d'Avenir Labex (ANR-11-LABX-0012); Germany: Baden-Württemberg Stiftung (BW Stiftung-Postdoc Eliteprogramme), Deutsche Forschungsgemeinschaft (DFG - 469666862, DFG - CR 312/5-2); Italy: Istituto Nazionale di Fisica Nucleare (ICSC, NextGenerationEU), Ministero dell'Università e della Ricerca (PRIN - 20223N7F8K - PNRR M4.C2.1.1); Japan: Japan Society for the Promotion of Science (JSPS KAKENHI JP21H05085, JSPS KAKENHI JP22H01227, JSPS KAKENHI JP22H04944, JSPS KAKENHI JP22KK0227); Netherlands: Netherlands Organisation for Scientific Research (NWO Veni 2020 - VI.Veni.202.179); Norway: Research Council of Norway (RCN-314472); Poland: Ministry of Science and Higher Education (IDUB AGH, POB8, D4 no 9722), Polish National Agency for Academic Exchange (PPN/PPO/2020/1/00002/U/00001), Polish National Science Centre (NCN 2021/42/E/ST2/00350, NCN OPUS nr 2022/47/B/ST2/03059, NCN UMO-2019/34/E/ST2/00393, NCN \& H2020 MSCA 945339, UMO-2020/37/B/ST2/01043, UMO-2021/40/C/ST2/00187, UMO-2022/47/O/ST2/00148, UMO-2023/49/B/ST2/04085); Slovenia: Slovenian Research Agency (ARIS grant J1-3010); Spain: Generalitat Valenciana (Artemisa, FEDER, IDIFEDER/2018/048), Ministry of Science and Innovation (MCIN \& NextGenEU PCI2022-135018-2, MICIN \& FEDER PID2021-125273NB, RYC2019-028510-I, RYC2020-030254-I, RYC2021-031273-I, RYC2022-038164-I), PROMETEO and GenT Programmes Generalitat Valenciana (CIDEGENT/2019/023, CIDEGENT/2019/027); Sweden: Carl Trygger Foundation (Carl Trygger Foundation CTS 22:2312), Swedish Research Council (Swedish Research Council 2023-04654, VR 2018-00482, VR 2022-03845, VR 2022-04683, VR 2023-03403, VR grant 2021-03651), Knut and Alice Wallenberg Foundation (KAW 2018.0157, KAW 2018.0458, KAW 2019.0447, KAW 2022.0358); Switzerland: Swiss National Science Foundation (SNSF - PCEFP2\_194658); United Kingdom: Leverhulme Trust (Leverhulme Trust RPG-2020-004), Royal Society (NIF-R1-231091); United States of America: U.S. Department of Energy (ECA DE-AC02-76SF00515), Neubauer Family Foundation.

%
%


%
%
%
%
%
%

%
%

%
%
%

\printbibliography

@article{Baur_1998,
   author = "Baur, U. and Keller, S. and Sakumoto, W. K.",
    title = "{QED radiative corrections to $Z$ boson production and the forward backward asymmetry at hadron colliders}",
    eprint = "hep-ph/9707301",
    archivePrefix = "arXiv",
    reportNumber = "UB-HET-97-02, FERMILAB-PUB-97-221-T",
    doi = "10.1103/PhysRevD.57.199",
    journal = "Phys. Rev. D",
    volume = "57",
    pages = "199--215",
    year = "1998"
}

@article{Dittmaier_2010,
   author = "Dittmaier, Stefan and Huber, Max",
    title = "{Radiative corrections to the neutral-current Drell-Yan process in the Standard Model and its minimal supersymmetric extension}",
    eprint = "0911.2329",
    archivePrefix = "arXiv",
    primaryClass = "hep-ph",
    reportNumber = "MPP-2009-164",
    doi = "10.1007/JHEP01(2010)060",
    journal = "JHEP",
    volume = "01",
    pages = "060",
    year = "2010"
}

@article{Arbuzov_2006,
   author = "Arbuzov, A. and Bardin, D. and Bondarenko, S. and Christova, P. and Kalinovskaya, L. and Nanava, G. and Sadykov, R.",
    title = "{One-loop corrections to the Drell-Yan process in SANC (I). The charged current case}",
    eprint = "hep-ph/0506110",
    archivePrefix = "arXiv",
    doi = "10.1140/epjc/s2006-02505-y",
    journal = "Eur. Phys. J. C",
    volume = "46",
    pages = "407--412",
    year = "2006",
    note = "[Erratum: Eur.Phys.J.C 50, 505 (2007)]"
}

@article{Carloni_Calame_2017,
   author = "Carloni Calame, Carlo Michel and Chiesa, Mauro and Martinez, Homero and Montagna, Guido and Nicrosini, Oreste and Piccinini, Fulvio and Vicini, Alessandro",
    title = "{Precision Measurement of the W-Boson Mass: Theoretical Contributions and Uncertainties}",
    eprint = "1612.02841",
    archivePrefix = "arXiv",
    primaryClass = "hep-ph",
    reportNumber = "TIF-UNIMI-2016-10",
    doi = "10.1103/PhysRevD.96.093005",
    journal = "Phys. Rev. D",
    volume = "96",
    number = "9",
    pages = "093005",
    year = "2017"
}

@article{Dittmaier_2002,
%   title={Electroweak radiative corrections to W-boson production at hadron colliders},
%   volume={65},
%   DOI={10.1103/physrevd.65.073007},
%   number={7},
%   journal={Phys. Rev. D},
%   author={Dittmaier, Stefan and Krämer, Michael},
%   year={2002},
%   pages = {073007}}

@article{Denner1990,
  author = "Denner, Ansgar and Sack, Thomas",
    title = "{The W-boson width}",
    reportNumber = "ITP-SB-89-52",
    doi = "10.1007/BF01560267",
    journal = "Z. Phys. C",
    volume = "46",
    pages = "653--663",
    year = "1990"
}

@article{Denner_2020,
%   title={NLO QCD and EW corrections to vector-boson scattering into ZZ at the LHC},
%   volume={2020},
%   DOI={10.1007/jhep11(2020)110},
%   number={11},
%   journal={JHEP},
%   author={Denner, Ansgar and Franken, Robert and Pellen, Mathieu and Schmidt, Timo},
%   year={2020},
%   pages = {110}}

@article{Denner_2019,
   author = {Denner, Ansgar and Dittmaier, Stefan and Maierh\"ofer, Philipp and Pellen, Mathieu and Schwan, Christopher},
    title = "{QCD and electroweak corrections to WZ scattering at the LHC}",
    eprint = "1904.00882",
    archivePrefix = "arXiv",
    primaryClass = "hep-ph",
    reportNumber = "FR-PHENO-2019-001, Cavendish-HEP-19/05, TIF-UNIMI-2019-1,
  VBSCAN-PUB-02-19, TIF-UNIMI-2019-1, VBSCAN-PUB-02-19, Cavendish-HEP-19-05, TIF-UNIMI-2019-1",
    doi = "10.1007/JHEP06(2019)067",
    journal = "JHEP",
    volume = "06",
    pages = "067",
    year = "2019"
}

@article{Kallweit_2016,
%   title={NLO QCD+EW predictions for V + jets including off-shell vector-boson decays and multijet merging},
%   volume={2016},
%   DOI={10.1007/jhep04(2016)021},
%   number={4},
%   journal={JHEP},
%   author={Kallweit, S. and Lindert, J. M. and Maierhöfer, P. and Pozzorini, S. and Schönherr, M.},
%   year={2016},
%   month=apr,
%   pages={021}
%   }

@article{Jadach_2019,
   author = "Jadach, S. and Ward, B. F. L. and Was, Z. A. and Yost, S. A.",
    title = "{Systematic Studies of Exact ${\cal O}(\alpha^2L)$ CEEX EW Corrections in a Hadronic MC for Precision $Z/\gamma^*$ Physics at LHC Energies}",
    eprint = "1707.06502",
    archivePrefix = "arXiv",
    primaryClass = "hep-ph",
    reportNumber = "BU-HEPP-17-01, IFJPAN-IV-2017-15",
    doi = "10.1103/PhysRevD.99.076016",
    journal = "Phys. Rev. D",
    volume = "99",
    number = "7",
    pages = "076016",
    year = "2019"
}

@article{Banfi_2011,
   author = "Banfi, A. and Redford, S. and Vesterinen, M. and Waller, P. and Wyatt, T. R.",
    title = "{Optimisation of variables for studying dilepton transverse momentum distributions at hadron colliders}",
    eprint = "1009.1580",
    archivePrefix = "arXiv",
    primaryClass = "hep-ex",
    reportNumber = "MAN-HEP-2010-12",
    doi = "10.1140/epjc/s10052-011-1600-y",
    journal = "Eur. Phys. J. C",
    volume = "71",
    pages = "1600",
    year = "2011"
}

@article{Khoze:2004ak,
    author = "Khoze, V. A. and Martin, A. D. and Ryskin, M. G. and Stirling, W. J.",
    title = "{Diffractive gamma-gamma production at hadron colliders}",
    eprint = "hep-ph/0409037",
    archivePrefix = "arXiv",
    reportNumber = "IPPP-04-55, DCPT-04-110",
    doi = "10.1140/epjc/s2004-02059-0",
    journal = "Eur. Phys. J. C",
    volume = "38",
    pages = "475--482",
    year = "2005"
}

@article{Jaeger_2014,
    author = {J\"ager, Barbara and Karlberg, Alexander and Zanderighi, Giulia},
    title = "{Electroweak $ZZjj$ production in the Standard Model and beyond in the POWHEG-BOX V2}",
    eprint = "1312.3252",
    archivePrefix = "arXiv",
    primaryClass = "hep-ph",
    doi = "10.1007/JHEP03(2014)141",
    journal = "JHEP",
    volume = "03",
    pages = "141",
    year = "2014"
}

@article{J_ger_2011,
    author = "Jager, Barbara and Zanderighi, Giulia",
    title = "{NLO corrections to electroweak and QCD production of W+W+ plus two jets in the POWHEGBOX}",
    eprint = "1108.0864",
    archivePrefix = "arXiv",
    primaryClass = "hep-ph",
    reportNumber = "MZ-TH-11-20, OUTP-11-45P",
    doi = "10.1007/JHEP11(2011)055",
    journal = "JHEP",
    volume = "11",
    pages = "055",
    year = "2011"
}

@article{Kauer_2012,
    author = "Kauer, Nikolas and Passarino, Giampiero",
    title = "{Inadequacy of zero-width approximation for a light Higgs boson signal}",
    eprint = "1206.4803",
    archivePrefix = "arXiv",
    primaryClass = "hep-ph",
    doi = "10.1007/JHEP08(2012)116",
    journal = "JHEP",
    volume = "08",
    pages = "116",
    year = "2012"
}

@article{Kauer_2013,
    author = "Kauer, Nikolas",
    title = "{Interference effects for H $\to$ WW/ZZ $\to \ell\bar{\nu}_\ell\bar{\ell}\nu_\ell$ searches in gluon fusion at the LHC}",
    eprint = "1310.7011",
    archivePrefix = "arXiv",
    primaryClass = "hep-ph",
    doi = "10.1007/JHEP12(2013)082",
    journal = "JHEP",
    volume = "12",
    pages = "082",
    year = "2013"
}

@article{Hamilton_2016,
    author = "Hamilton, Keith and Melia, Tom and Monni, Pier Francesco and Re, Emanuele and Zanderighi, Giulia",
    title = "{Merging WW and WW+jet with MINLO}",
    eprint = "1606.07062",
    archivePrefix = "arXiv",
    primaryClass = "hep-ph",
    reportNumber = "CERN-TH-2016-146, LAPTH-031-16, OUTP-16-17P",
    doi = "10.1007/JHEP09(2016)057",
    journal = "JHEP",
    volume = "09",
    pages = "057",
    year = "2016"
}

@article{Monni_2020,
    author = "Monni, Pier Francesco and Nason, Paolo and Re, Emanuele and Wiesemann, Marius and Zanderighi, Giulia",
    title = "{MiNNLO$_{PS}$: a new method to match NNLO QCD to parton showers}",
    eprint = "1908.06987",
    archivePrefix = "arXiv",
    primaryClass = "hep-ph",
    reportNumber = "CERN-TH-2019-117, LAPTH-042/19, MPP-2019-177",
    doi = "10.1007/JHEP05(2020)143",
    journal = "JHEP",
    volume = "05",
    pages = "143",
    year = "2020",
    note = "[Erratum: JHEP 02, 031 (2022)]"
}

@Article{Hou_2021,
    author = "Hou, Tie-Jiun and others",
    title = "{New CTEQ global analysis of quantum chromodynamics with high-precision data from the LHC}",
    eprint = "1912.10053",
    archivePrefix = "arXiv",
    primaryClass = "hep-ph",
    reportNumber = "MSUHEP-19-025, PITT-PACC-1911, SMU-HEP-19-03",
    doi = "10.1103/PhysRevD.103.014013",
    journal = "Phys. Rev. D",
    volume = "103",
    number = "1",
    pages = "014013",
    year = "2021"
}

@Article{Alwall_2014,
    author = "Alwall, J. and Frederix, R. and Frixione, S. and Hirschi, V. and Maltoni, F. and Mattelaer, O. and Shao, H. -S. and Stelzer, T. and Torrielli, P. and Zaro, M.",
    title = "{The automated computation of tree-level and next-to-leading order differential cross sections, and their matching to parton shower simulations}",
    eprint = "1405.0301",
    archivePrefix = "arXiv",
    primaryClass = "hep-ph",
    reportNumber = "CERN-PH-TH-2014-064, CP3-14-18, LPN14-066, MCNET-14-09, ZU-TH-14-14",
    doi = "10.1007/JHEP07(2014)079",
    journal = "JHEP",
    volume = "07",
    pages = "079",
    year = "2014"
}

@Article{Camarda_2020,
    author = "Camarda, Stefano and others",
    title = "{DYTurbo: Fast predictions for Drell-Yan processes}",
    eprint = "1910.07049",
    archivePrefix = "arXiv",
    primaryClass = "hep-ph",
    doi = "10.1140/epjc/s10052-020-7757-5",
    journal = "Eur. Phys. J. C",
    volume = "80",
    number = "3",
    pages = "251",
    year = "2020",
    note = "[Erratum: Eur.Phys.J.C 80, 440 (2020)]"
}

@Article{Camarda_2021,
    author = "Camarda, Stefano and Cieri, Leandro and Ferrera, Giancarlo",
    title = "{Drell\textendash{}Yan lepton-pair production: q$_T$ resummation at N$^3$LL accuracy and fiducial cross sections at N$^3$LO}",
    eprint = "2103.04974",
    archivePrefix = "arXiv",
    primaryClass = "hep-ph",
    doi = "10.1103/PhysRevD.104.L111503",
    journal = "Phys. Rev. D",
    volume = "104",
    number = "11",
    pages = "L111503",
    year = "2021"
}

@Article{Camarda_2022,
    author = "Camarda, Stefano and Cieri, Leandro and Ferrera, Giancarlo",
    title = "{Fiducial perturbative power corrections within the $\mathbf{q}_T$ subtraction formalism}",
    eprint = "2111.14509",
    archivePrefix = "arXiv",
    primaryClass = "hep-ph",
    reportNumber = "IFIC/21-51, FTUV-21-1129.5289",
    doi = "10.1140/epjc/s10052-022-10510-x",
    journal = "Eur. Phys. J. C",
    volume = "82",
    number = "6",
    pages = "575",
    year = "2022"
}

@Article{Camarda_2023,
    author = "Camarda, Stefano and Cieri, Leandro and Ferrera, Giancarlo",
    title = "{Drell\textendash{}Yan lepton-pair production: q$_T$ resummation at N$^4$LL accuracy}",
    eprint = "2303.12781",
    archivePrefix = "arXiv",
    primaryClass = "hep-ph",
    doi = "10.1016/j.physletb.2023.138125",
    journal = "Phys. Lett. B",
    volume = "845",
    pages = "138125",
    year = "2023"
}

@article{Bondarenko_2023,
    author = "Bondarenko, Serge and Dydyshka, Yahor and Kalinovskaya, Lidia and Sadykov, Renat and Yermolchyk, Vitaly",
    title = "{Hadron-hadron collision mode in ReneSANCe-v1.3.0}",
    eprint = "2207.04332",
    archivePrefix = "arXiv",
    primaryClass = "hep-ph",
    doi = "10.1016/j.cpc.2022.108646",
    journal = "Comput. Phys. Commun.",
    volume = "285",
    pages = "108646",
    year = "2023"
}

@article{Baur_2004,
    author = "Baur, U. and Wackeroth, D.",
    title = "{Electroweak radiative corrections to $p \bar{p} \to W^\pm \to \ell^\pm \nu$ beyond the pole approximation}",
    eprint = "hep-ph/0405191",
    archivePrefix = "arXiv",
    reportNumber = "UB-HET-04-01, NSF-KITP-04-45",
    doi = "10.1103/PhysRevD.70.073015",
    journal = "Phys. Rev. D",
    volume = "70",
    pages = "073015",
    year = "2004"
}

@article{Calame_2006,
    author = "Carloni Calame, C. M. and Montagna, G. and Nicrosini, O. and Vicini, A.",
    title = "{Precision electroweak calculation of the charged current Drell-Yan process}",
    eprint = "hep-ph/0609170",
    archivePrefix = "arXiv",
    reportNumber = "FNT-T-2006-08, IFUM-874-FT",
    doi = "10.1088/1126-6708/2006/12/016",
    journal = "JHEP",
    volume = "12",
    pages = "016",
    year = "2006"
}

@article{Arbuzov_2008,
    author = "Arbuzov, A. and Bardin, D. and Bondarenko, S. and Christova, P. and Kalinovskaya, L. and Nanava, G. and Sadykov, R.",
    title = "{One-loop corrections to the Drell--Yan process in SANC. (II). The Neutral current case}",
    eprint = "0711.0625",
    archivePrefix = "arXiv",
    primaryClass = "hep-ph",
    doi = "10.1140/epjc/s10052-008-0531-8",
    journal = "Eur. Phys. J. C",
    volume = "54",
    pages = "451--460",
    year = "2008"
}

@article{Baur_2002,
    author = "Baur, U. and Brein, O. and Hollik, W. and Schappacher, C. and Wackeroth, D.",
    title = "{Electroweak radiative corrections to neutral current Drell-Yan processes at hadron colliders}",
    eprint = "hep-ph/0108274",
    archivePrefix = "arXiv",
    reportNumber = "KA-TP-26-2001, UB-HET-01-04, UR-1642",
    doi = "10.1103/PhysRevD.65.033007",
    journal = "Phys. Rev. D",
    volume = "65",
    pages = "033007",
    year = "2002"
}

@article{Dittmaier_2014,
    author = "Dittmaier, Stefan and Huss, Alexander and Schwinn, Christian",
    title = "{Mixed QCD-electroweak $\mathcal{O}(\alpha_s\alpha)$ corrections to Drell-Yan processes in the resonance region: pole approximation and non-factorizable corrections}",
    eprint = "1403.3216",
    archivePrefix = "arXiv",
    primaryClass = "hep-ph",
    reportNumber = "FR-PHENO-2013-014",
    doi = "10.1016/j.nuclphysb.2014.05.027",
    journal = "Nucl. Phys. B",
    volume = "885",
    pages = "318--372",
    year = "2014"
}

@article{LesHouches_2021,
    author = "Huss, Alexander and Huston, Joey and Jones, Stephen and Pellen, Mathieu",
    title = "{Les Houches 2021\textemdash{}physics at TeV colliders: report on the standard model precision wishlist}",
    eprint = "2207.02122",
    archivePrefix = "arXiv",
    primaryClass = "hep-ph",
    doi = "10.1088/1361-6471/acbaec",
    journal = "J. Phys. G",
    volume = "50",
    number = "4",
    pages = "043001",
    year = "2023"
}

@article{LesHouches_2023,
      title={Les Houches 2023: Physics at TeV Colliders: Standard Model Working Group Report}, 
      author={Andersen, J. and Others},
      year={2024},
      eprint={2406.00708},
      archivePrefix={arXiv},
      primaryClass={hep-ph},
      %url={https://arxiv.org/abs/2406.00708}
}

@Article{Biedermann_2017,
    author = "Biedermann, Benedikt and Denner, Ansgar and Pellen, Mathieu",
    title = "{Complete NLO corrections to W$^{+}$W$^{+}$ scattering and its irreducible background at the LHC}",
    eprint = "1708.00268",
    archivePrefix = "arXiv",
    primaryClass = "hep-ph",
    doi = "10.1007/JHEP10(2017)124",
    journal = "JHEP",
    volume = "10",
    pages = "124",
    year = "2017"
}

@article{Dittmaier_2023,
   author = {Dittmaier, Stefan and Maierh\"ofer, Philipp and Schwan, Christopher and Winterhalder, Ramon},
    title = "{Like-sign W-boson scattering at the LHC \textemdash{} approximations and full next-to-leading-order predictions}",
    eprint = "2308.16716",
    archivePrefix = "arXiv",
    primaryClass = "hep-ph",
    reportNumber = "FR-PHENO-2023-09, IRMP-CP3-23-42",
    doi = "10.1007/JHEP11(2023)022",
    journal = "JHEP",
    volume = "11",
    pages = "022",
    year = "2023"
}

@Article{Haller_2018,
    author = {Haller, Johannes and Hoecker, Andreas and Kogler, Roman and M\"onig, Klaus and Peiffer, Thomas and Stelzer, J\"org},
    title = "{Update of the global electroweak fit and constraints on two-Higgs-doublet models}",
    eprint = "1803.01853",
    archivePrefix = "arXiv",
    primaryClass = "hep-ph",
    doi = "10.1140/epjc/s10052-018-6131-3",
    journal = "Eur. Phys. J. C",
    volume = "78",
    number = "8",
    pages = "675",
    year = "2018"
}

@Article{mtop_2024,
   title={Combination of Measurements of the Top Quark Mass from Data Collected by the ATLAS and CMS Experiments at $\sqrt{s} = $~7~TeV and 8~TeV},
   author="{ATLAS and CMS collaborations}",
    eprint = "2402.08713",
    archivePrefix = "arXiv",
    primaryClass = "hep-ex",
    reportNumber = "CMS-TOP-22-001, ATLAS-TOPQ-2019-13, CERN-EP-2024-020",
    doi = "10.1103/PhysRevLett.132.261902",
    journal = "Phys. Rev. Lett.",
    volume = "132",
    number = "26",
    pages = "261902",
    year = "2024"
}

@Article{ATLAS_mH_2023,
   title={Combined Measurement of the Higgs Boson Mass from the $H\to\gamma\gamma$ and $H\to ZZ^*\to 4\ell$ decay channels with the ATLAS detector using $\sqrt{s} = $ 7,8 and 13~TeV pp Collision data},
   author="{ATLAS Collaboration}",
    eprint = "2308.04775",
    archivePrefix = "arXiv",
    primaryClass = "hep-ex",
    reportNumber = "CERN-EP-2023-156",
    doi = "10.1103/PhysRevLett.131.251802",
    journal = "Phys. Rev. Lett.",
    volume = "131",
    number = "25",
    pages = "251802",
    year = "2023"
}

@Article{ParticleDataGroup:2024cfk,
    author = "{Particle Data Group Collaboration}",
    collaboration = "Particle Data Group",
    title = "{Review of particle physics}",
    doi = "10.1103/PhysRevD.110.030001",
    journal = "Phys. Rev. D",
    volume = "110",
    number = "3",
    pages = "030001",
    year = "2024"
}

@article{Ciuchini_2013,
    author = "Ciuchini, Marco and Franco, Enrico and Mishima, Satoshi and Silvestrini, Luca",
    title = "{Electroweak Precision Observables, New Physics and the Nature of a 126 GeV Higgs Boson}",
    eprint = "1306.4644",
    archivePrefix = "arXiv",
    primaryClass = "hep-ph",
    doi = "10.1007/JHEP08(2013)106",
    journal = "JHEP",
    volume = "08",
    pages = "106",
    year = "2013"
}

@Article{Awramik_2004,
    author = "Awramik, M. and Czakon, M. and Freitas, A. and Weiglein, G.",
    title = "{Precise prediction for the W boson mass in the standard model}",
    eprint = "hep-ph/0311148",
    archivePrefix = "arXiv",
    reportNumber = "DCPT-03-146, DESY-03-184, FERMILAB-PUB-03-239-T, IPPP-03-73, DCPT/03/146, DESY 03-184, FERMILAB-Pub-03/239-T, IPPP/03/73",
    doi = "10.1103/PhysRevD.69.053006",
    journal = "Phys. Rev. D",
    volume = "69",
    pages = "053006",
    year = "2004"
}

@Article{Sirlin_1980,
  title = {Radiative corrections in the $\mathrm{SU}{(2)}_{L}\times\mathrm{U}(1)$ theory: A simple renormalization framework},
    author = "Sirlin, A.",
    reportNumber = "PRINT-80-0267 (IAS,PRINCETON)",
    doi = "10.1103/PhysRevD.22.971",
    journal = "Phys. Rev. D",
    volume = "22",
    pages = "971--981",
    year = "1980"
}

@Article{Luisoni:2013cuh,
%    author = "Luisoni, Gionata and Nason, Paolo and Oleari, Carlo and Tramontano, Francesco",
%    title = "{$HW^{\pm}$/HZ + 0 and 1 jet at NLO with the POWHEG BOX interfaced to GoSam and their merging within MiNLO}",
%    eprint = "1306.2542",
%    archivePrefix = "arXiv",
%    primaryClass = "hep-ph",
%    doi = "10.1007/JHEP10(2013)083",
%    journal = "JHEP",
%    volume = "10",
%    pages = "083",
%    year = "2013"
%

@article{Nason11,
    author = {Melia, Tom and Nason, Paolo and R\"ontsch, Raoul and Zanderighi, Giulia},
    title = "{$W^+W^-$, $WZ$ and $ZZ$  production in the POWHEG BOX}",
    eprint = "1107.5051",
    archivePrefix = "arXiv",
    primaryClass = "hep-ph",
    doi = "10.1007/JHEP11(2011)078",
    journal = "JHEP",
    volume = "11",
    pages = "078",
    year = "2011"
}

@article{Nason14,
    author = "Nason, Paolo and Zanderighi, Giulia",
    title = "{$W^+ W^-$ , $W Z$ and $Z Z$ production in the POWHEG-BOX-V2}",
    eprint = "1311.1365",
    archivePrefix = "arXiv",
    primaryClass = "hep-ph",
    doi = "10.1140/epjc/s10052-013-2702-5",
    journal = "Eur. Phys. J. C",
    volume = "74",
    number = "1",
    pages = "2702",
    year = "2014"
}

@Article{Krauss_2014,
    author = "Krauss, Frank and Petrov, Petar and Schoenherr, Marek and Spannowsky, Michael",
    title = "{Measuring collinear W emissions inside jets}",
    eprint = "1403.4788",
    archivePrefix = "arXiv",
    primaryClass = "hep-ph",
    reportNumber = "IPPP-14-04, DCPT-14-08, MCNET-14-06, LPN14-054",
    doi = "10.1103/PhysRevD.89.114006",
    journal = "Phys. Rev. D",
    volume = "89",
    number = "11",
    pages = "114006",
    year = "2014"
}

@Article{Cascioli_2014,
    author = {Cascioli, F. and H\"oche, S. and Krauss, F. and Maierh\"ofer, P. and Pozzorini, S. and Siegert, F.},
    title = "{Precise Higgs-background predictions: merging NLO QCD and squared quark-loop corrections to four-lepton + 0,1 jet production}",
    eprint = "1309.0500",
    archivePrefix = "arXiv",
    primaryClass = "hep-ph",
    reportNumber = "IPPP-13-66, DCPT-13-132, MCNET-13-12, SLAC-PUB-15714, ZU-TH-15-13, LPN13-056, FR-PHENO-2013-007",
    doi = "10.1007/JHEP01(2014)046",
    journal = "JHEP",
    volume = "01",
    pages = "046",
    year = "2014"
}

@article{Krause:2017nxq,
    author = "Krause, Johannes and Siegert, Frank",
    title = "{NLO QCD predictions for $Z+\gamma$ + jets production with Sherpa}",
    eprint = "1708.06283",
    archivePrefix = "arXiv",
    primaryClass = "hep-ph",
    doi = "10.1140/epjc/s10052-018-5627-1",
    journal = "Eur. Phys. J. C",
    volume = "78",
    number = "2",
    pages = "161",
    year = "2018"
}

@article{Hoeche:2014rya,
    author = "Hoeche, S. and Krauss, F. and Pozzorini, S. and Schoenherr, M. and Thompson, J. M. and Zapp, K. C.",
    title = "{Triple vector boson production through Higgs-Strahlung with NLO multijet merging}",
    eprint = "1403.7516",
    archivePrefix = "arXiv",
    primaryClass = "hep-ph",
    reportNumber = "DCP-14-46, IPPP-14-23, MCNET-14-007, SLAC-PUB-15933, ZU-13-14, CERN-PH-TH-2014-051",
    doi = "10.1103/PhysRevD.89.093015",
    journal = "Phys. Rev. D",
    volume = "89",
    number = "9",
    pages = "093015",
    year = "2014"
}

@article{WZjjRun2,
    author = "{ATLAS collaboration}",
    title = "{Measurements of electroweak $W^{\pm}Z$ boson pair production in association with two jets in pp collisions at $\sqrt{s} = 13$~TeV with the ATLAS detector}",
    journal = "JHEP",
    year = "2024",
    volume = "06",
    pages = "192",
    doi = "10.1007/JHEP06(2024)19",
    eprint = "2403.15296",
    archivePrefix = "arXiv",
    primaryClass = "hep-ex"
}

@article{Hoeche:2009xc,
    author = "Hoeche, Stefan and Schumann, Steffen and Siegert, Frank",
    title = "{Hard photon production and matrix-element parton-shower merging}",
    eprint = "0912.3501",
    archivePrefix = "arXiv",
    primaryClass = "hep-ph",
    reportNumber = "ZU-TH-19-09, IPPP-09-94, DCPT-09-188, HD-THEP-09-28, MCNET-09-18",
    doi = "10.1103/PhysRevD.81.034026",
    journal = "Phys. Rev. D",
    volume = "81",
    pages = "034026",
    year = "2010"
}

@article{Chahal:2022rid,
    author = "Chahal, Gurpreet Singh and Krauss, Frank",
    title = "{Cluster Hadronisation in Sherpa}",
    eprint = "2203.11385",
    archivePrefix = "arXiv",
    primaryClass = "hep-ph",
    reportNumber = "IPPP/22/14",
    doi = "10.21468/SciPostPhys.13.2.019",
    journal = "SciPost Phys.",
    volume = "13",
    number = "2",
    pages = "019",
    year = "2022"
}

@article{ATLAS:2023dns,
    author = "{ATLAS Collaboration}",
    collaboration = "ATLAS",
    title = "{The ATLAS experiment at the CERN Large Hadron Collider: a description of the detector configuration for Run~3}",
    eprint = "2305.16623",
    archivePrefix = "arXiv",
    primaryClass = "physics.ins-det",
    reportNumber = "CERN-EP-2022-259",
    doi = "10.1088/1748-0221/19/05/P05063",
    journal = "JINST",
    volume = "19",
    number = "05",
    pages = "P05063",
    year = "2024"
}

@article{AH:2023kor,
    author = "A H, Ajjath and Chaubey, Ekta and Fraaije, Mathijs and Hirschi, Valentin and Shao, Hua-Sheng",
    title = "{Light-by-light scattering at next-to-leading order in QCD and QED}",
    eprint = "2312.16956",
    archivePrefix = "arXiv",
    primaryClass = "hep-ph",
    doi = "10.1016/j.physletb.2024.138555",
    journal = "Phys. Lett. B",
    volume = "851",
    pages = "138555",
    year = "2024"
}

@article{Frixione:1998jh,
%    author = "Frixione, Stefano",
%    title = "{Isolated photons in perturbative QCD}",
%    eprint = "hep-ph/9801442",
%    archivePrefix = "arXiv",
%    reportNumber = "ETH-TH-97-40",
%    doi = "10.1016/S0370-2693(98)00454-7",
%    journal = "Phys. Lett. B",
%    volume = "429",
%    pages = "369--374",
%    year = "1998"
%}

@article{Pancheri:2016yel,
    author = "Pancheri, Giulia and Srivastava, Y. N.",
    title = "{Introduction to the physics of the total cross-section at LHC}: {A Review of Data and Models}",
    eprint = "1610.10038",
    archivePrefix = "arXiv",
    primaryClass = "hep-ph",
    reportNumber = "MIT-CTP-4828, INFN-16-13-LNF",
    doi = "10.1140/epjc/s10052-016-4585-8",
    journal = "Eur. Phys. J. C",
    volume = "77",
    number = "3",
    pages = "150",
    year = "2017"
}

@article{Currie:2016bfm,
    author = "Currie, J and Glover, E. W. N. and Pires, J",
    title = "{Next-to-Next-to Leading Order QCD Predictions for Single Jet Inclusive Production at the LHC}",
    eprint = "1611.01460",
    archivePrefix = "arXiv",
    primaryClass = "hep-ph",
    reportNumber = "IPPP-16-110, MPP-2016-322",
    doi = "10.1103/PhysRevLett.118.072002",
    journal = "Phys. Rev. Lett.",
    volume = "118",
    number = "7",
    pages = "072002",
    year = "2017"
}

@article{Dittmaier:2012kx,
    author = "Dittmaier, Stefan and Huss, Alexander and Speckner, Christian",
    title = "{Weak radiative corrections to dijet production at hadron colliders}",
    eprint = "1210.0438",
    archivePrefix = "arXiv",
    primaryClass = "hep-ph",
    reportNumber = "FR-PHENO-2012-024",
    doi = "10.1007/JHEP11(2012)095",
    journal = "JHEP",
    volume = "11",
    pages = "095",
    year = "2012"
}

@article{Nagy:2003tz,
    author = "Nagy, Zoltan",
    title = "{Next-to-leading order calculation of three jet observables in hadron hadron collision}",
    eprint = "hep-ph/0307268",
    archivePrefix = "arXiv",
    doi = "10.1103/PhysRevD.68.094002",
    journal = "Phys. Rev. D",
    volume = "68",
    pages = "094002",
    year = "2003"
}

@article{Sjostrand:2014zea,
%    author = {Sj\"ostrand, Torbj\"orn and Ask, Stefan and Christiansen, Jesper R. and Corke, Richard and Desai, Nishita and Ilten, Philip and Mrenna, Stephen and Prestel, Stefan and Rasmussen, Christine O. and Skands, Peter Z.},
%    title = "{An introduction to PYTHIA 8.2}",
%    eprint = "1410.3012",
%    archivePrefix = "arXiv",
%    primaryClass = "hep-ph",
%    reportNumber = "LU-TP-14-36, MCNET-14-22, CERN-PH-TH-2014-190, FERMILAB-PUB-14-316-CD, DESY-14-178, SLAC-PUB-16122",
%    doi = "10.1016/j.cpc.2015.01.024",
%    journal = "Comput. Phys. Commun.",
%    volume = "191",
%    pages = "159--177",
%    year = "2015"
%}

@article{Capella:1999ms,
    author = "Capella, A. and Dremin, I. M. and Gary, J. W. and Nechitailo, V. A. and Tran Thanh Van, J.",
    title = "{Evolution of average multiplicities of quark and gluon jets}",
    eprint = "hep-ph/9910226",
    archivePrefix = "arXiv",
    reportNumber = "FIAN-TD-22-99, LPT-99-74, UCRHEP-E264",
    doi = "10.1103/PhysRevD.61.074009",
    journal = "Phys. Rev. D",
    volume = "61",
    pages = "074009",
    year = "2000"
}

@article{Gehrmann:2018szu,
    author = "Gehrmann, Thomas and others",
    title = "{Jet cross sections and transverse momentum distributions with NNLOJET}",
    eprint = "1801.06415",
    archivePrefix = "arXiv",
    primaryClass = "hep-ph",
    reportNumber = "CFTP-18-001",
    doi = "10.22323/1.290.0074",
    journal = "PoS",
    volume = "RADCOR2017",
    pages = "074",
    year = "2018"
}

@article{Binoth:1999qq,
    author = "Binoth, T. and Guillet, J. P. and Pilon, E. and Werlen, M.",
    title = "{A Full next-to-leading order study of direct photon pair production in hadronic collisions}",
    eprint = "hep-ph/9911340",
    archivePrefix = "arXiv",
    reportNumber = "LAPTH-760-99",
    doi = "10.1007/s100520050024",
    journal = "Eur. Phys. J. C",
    volume = "16",
    pages = "311--330",
    year = "2000"
}

@article{Chen:2022gpk,
    author = {Chen, X. and Gehrmann, T. and Glover, E. W. N. and H\"ofer, M. and Huss, A. and Sch\"urmann, R.},
    title = "{Single photon production at hadron colliders at NNLO QCD with realistic photon isolation}",
    eprint = "2205.01516",
    archivePrefix = "arXiv",
    primaryClass = "hep-ph",
    reportNumber = "ZU-TH 16/22, KA-TP-12-2022, P3H-22-044, IPPP/22/30, IPPP/22/30,
  CERN-TH-2022-072, LMU-ASC 17/22, CERN-TH-2022-072",
    doi = "10.1007/JHEP08(2022)094",
    journal = "JHEP",
    volume = "08",
    pages = "094",
    year = "2022"
}

@article{Aurenche:2006vj,
    author = "Aurenche, Patrick and Fontannaz, Michel and Guillet, Jean-Philippe and Pilon, Eric and Werlen, Monique",
    title = "{A New critical study of photon production in hadronic collisions}",
    eprint = "hep-ph/0602133",
    archivePrefix = "arXiv",
    reportNumber = "LAPTH-1140-06, LPT-ORSAY-05-75",
    doi = "10.1103/PhysRevD.73.094007",
    journal = "Phys. Rev. D",
    volume = "73",
    pages = "094007",
    year = "2006"
}

@article{Catani:2002ny,
    author = "Catani, S. and Fontannaz, M. and Guillet, J. P. and Pilon, E.",
    title = "{Cross-section of isolated prompt photons in hadron hadron collisions}",
    eprint = "hep-ph/0204023",
    archivePrefix = "arXiv",
    reportNumber = "CERN-TH-2002-017, LPT-ORSAY-02-24, LAPTH-907-02",
    doi = "10.1088/1126-6708/2002/05/028",
    journal = "JHEP",
    volume = "05",
    pages = "028",
    year = "2002"
}

@article{Dyndal:2014yea,
    author = "Dyndal, Mateusz and Schoeffel, Laurent",
    title = "{The role of finite-size effects on the spectrum of equivalent photons in proton\textendash{}proton collisions at the LHC}",
    eprint = "1410.2983",
    archivePrefix = "arXiv",
    primaryClass = "hep-ph",
    doi = "10.1016/j.physletb.2014.12.019",
    journal = "Phys. Lett. B",
    volume = "741",
    pages = "66--70",
    year = "2015"
}

@article{Frixione_2007,
    author = "Frixione, Stefano and Nason, Paolo and Oleari, Carlo",
    title = "{Matching NLO QCD computations with Parton Shower simulations: the POWHEG method}",
    eprint = "0709.2092",
    archivePrefix = "arXiv",
    primaryClass = "hep-ph",
    reportNumber = "BICOCCA-FT-07-9, GEF-TH-21-2007",
    doi = "10.1088/1126-6708/2007/11/070",
    journal = "JHEP",
    volume = "11",
    pages = "070",
    year = "2007"
}

@article{Alioli_2010,
    author = "Alioli, Simone and Nason, Paolo and Oleari, Carlo and Re, Emanuele",
    title = "{A general framework for implementing NLO calculations in shower Monte Carlo programs: the POWHEG BOX}",
    eprint = "1002.2581",
    archivePrefix = "arXiv",
    primaryClass = "hep-ph",
    reportNumber = "DESY-10-018, SFB-CPP-10-22, IPPP-10-11, DCPT-10-22",
    doi = "10.1007/JHEP06(2010)043",
    journal = "JHEP",
    volume = "06",
    pages = "043",
    year = "2010"
}

@article{Nason_2004,
    author = "Nason, Paolo",
    title = "{A New method for combining NLO QCD with shower Monte Carlo algorithms}",
    eprint = "hep-ph/0409146",
    archivePrefix = "arXiv",
    reportNumber = "BICOCCA-FT-04-11",
    doi = "10.1088/1126-6708/2004/11/040",
    journal = "JHEP",
    volume = "11",
    pages = "040",
    year = "2004"
}

@article{Gleisberg:2008ta,
    author = "Gleisberg, T. and Hoeche, Stefan. and Krauss, F. and Schonherr, M. and Schumann, S. and Siegert, F. and Winter, J.",
    title = "{Event generation with SHERPA 1.1}",
    eprint = "0811.4622",
    archivePrefix = "arXiv",
    primaryClass = "hep-ph",
    reportNumber = "FERMILAB-PUB-08-477-T, SLAC-PUB-13420, ZU-TH-17-08, DCPT-08-138, IPPP-08-69, EDINBURGH-2008-30, MCNET-08-14",
    doi = "10.1088/1126-6708/2009/02/007",
    journal = "JHEP",
    volume = "02",
    pages = "007",
    year = "2009"
}

@article{Cacciari:2011ma,
    author = "Cacciari, Matteo and Salam, Gavin P. and Soyez, Gregory",
    title = "{FastJet User Manual}",
    eprint = "1111.6097",
    archivePrefix = "arXiv",
    primaryClass = "hep-ph",
    reportNumber = "CERN-PH-TH-2011-297",
    doi = "10.1140/epjc/s10052-012-1896-2",
    journal = "Eur. Phys. J. C",
    volume = "72",
    pages = "1896",
    year = "2012"
}

@article{STDM-2023-07,
      author         = "{ATLAS Collaboration}",
      title = "{Measurements of Lund subjet multiplicities in 13 TeV proton-proton collisions with the ATLAS detector}",
      eprint = "2402.13052",
      archivePrefix = "arXiv",
      primaryClass = "hep-ex",
      reportNumber = "CERN-EP-2024-029",
      month = "2",
      year = "2024"
      }

@article{FRIXIONE19923,
title = {Strong corrections to WZ production at hadron colliders},
journal = {Nucl. Phys. B},
volume = {383},
number = {1},
pages = {3-44},
year = {1992},
doi = {https://doi.org/10.1016/0550-3213(92)90668-2},
author = {S. Frixione and P. Nason and G. Ridolfi}
}

@article{Baur_1994,
   title={Amplitude zeros in $W^\pm Z$ production},
   volume={72},
   DOI={10.1103/physrevlett.72.3941},
   number={25},
   journal={Phys. Rev. Lett.},
   publisher={American Physical Society (APS)},
   author={Baur, U. and Han, T. and Ohnemus, J.},
   year={1994},
   month=jun, pages={3941–3944} }

@article{PDG2022,
    author = "{Particle Data Group Collaboration}",
    collaboration = "Particle Data Group",
    title = "{Review of Particle Physics}",
    doi = "10.1093/ptep/ptac097",
    journal = "PTEP",
    volume = "2022",
    pages = "083C01",
    year = "2022"
}

@article{CDF:2022hxs,
    author = "{CDF Collaboration}",
    collaboration = "CDF",
    title = "{High-precision measurement of the $W$ boson mass with the CDF II detector}",
    reportNumber = "FERMILAB-PUB-22-254-PPD",
    doi = "10.1126/science.abk1781",
    journal = "Science",
    volume = "376",
    number = "6589",
    pages = "170--176",
    year = "2022"
}

@misc{haller2022status,
      title={Status of the global electroweak fit with Gfitter in the light of new precision measurements}, 
      author={Johannes Haller and Andreas Hoecker and Roman Kogler and Klaus Mönig and Jörg Stelzer},
      year={2022},
      eprint={2211.07665},
      archivePrefix={arXiv},
      primaryClass={hep-ph}
}

@article{STDM-2020-01,
      author         = "{ATLAS Collaboration}",
      title          = "{Studies of the energy dependence of diboson polarization fractions and the Radiation Amplitude Zero effect in $WZ$ production with the ATLAS detector}",
      year           = "2024",
      eprint         = "2402.16365",
      archivePrefix  = "arXiv",
      primaryClass   = "hep-ex"
      }

@article{STDM-2018-31,
      author         = "{ATLAS Collaboration}",
      title          = "{Fiducial and differential cross-section measurements of electroweak $W\gamma jj$ production in pp collisions at $\sqrt{s} = 13$~TeV with the ATLAS detector}",
      year           = "2024",
      eprint         = "2403.02809",
      archivePrefix  = "arXiv",
      primaryClass   = "hep-ex"
      }

@article{STDM-2022-06,
      author         = "{ATLAS Collaboration}",
      title          = "{Observation of electroweak production of \Wp\Wm\ in association with jets in proton-proton collisions at $\sqrt{s} = 13$~TeV with the ATLAS Detector}",
      year           = "2024",
      eprint         = "2403.04869",
      archivePrefix  = "arXiv",
      primaryClass   = "hep-ex"
      }

@article{deBlas:2017wmn,
    author = "de Blas, Jorge and Ciuchini, Marco and Franco, Enrico and Mishima, Satoshi and Pierini, Maurizio and Reina, Laura and Silvestrini, Luca",
    title = "{The Global Electroweak and Higgs Fits in the LHC era}",
    eprint = "1710.05402",
    archivePrefix = "arXiv",
    primaryClass = "hep-ph",
    doi = "10.22323/1.314.0467",
    journal = "PoS EPS-HEP2017",
    pages = "467",
    year = "2017"
}

@article{STDM-2018-43,
      author         = "{ATLAS Collaboration}",
      title          = "{Measurements of the production cross-section for a $Z$~boson in association with $b$- or $c$-jets in proton--proton collisions at $\sqrt{s} = 13$~\TeV with the ATLAS detector}",
      year           = "2024",
      eprint         = "2403.15093",
      archivePrefix  = "arXiv",
      primaryClass   = "hep-ex"
      }

@article{STDM-2019-24,
      author         = "{ATLAS Collaboration}",
      title          = "{Measurement of the W-boson mass and width with the ATLAS detector using proton-proton collisions at  $\sqrt{s} = 7$~TeV}",
      year           = "2024",
      eprint         = "2403.15085",
      archivePrefix  = "arXiv",
      primaryClass   = "hep-ex"
      }

@article{Frederix:2012ps,
%      author         = "Frederix, Rikkert and Frixione, Stefano",
%      title          = "{Merging meets matching in MC@NLO}",
%      journal        = "JHEP",
%      volume         = "12",
%      year           = "2012",
%      pages          = "061",
%      doi            = "10.1007/JHEP12(2012)061",
%      eprint         = "1209.6215",
%      archivePrefix  = "arXiv",
%      primaryClass   = "hep-ph",
%      reportNumber   = "CERN-PH-TH-2012-247, ZU-TH-21-12"
%}

@article{Gauld:2022lem,
    author = "Gauld, Rhorry and Huss, Alexander and Stagnitto, Giovanni",
    title = "{Flavor Identification of Reconstructed Hadronic Jets}",
    eprint = "2208.11138",
    archivePrefix = "arXiv",
    primaryClass = "hep-ph",
    reportNumber = "BONN-TH-2022-17, CERN-TH-2022-121, ZU-TH 31/22",
    doi = "10.1103/PhysRevLett.130.161901",
    journal = "Phys. Rev. Lett.",
    volume = "130",
    number = "16",
    pages = "161901",
    year = "2023"
}

@article{Brodsky:1980pb,
    author = "Brodsky, S.J. and Hoyer, P. and Peterson, C. and Sakai, N.",
    title = "{The Intrinsic Charm of the Proton}",
    reportNumber = "NORDITA-80-18",
    doi = "10.1016/0370-2693(80)90364-0",
    journal = "Phys. Lett. B",
    volume = "93",
    pages = "451--455",
    year = "1980"
}

@article{STDM-2018-17,
   author         = "{ATLAS Collaboration}",
   title          = "{Precise measurements of $W$ and $Z$ transverse momentum spectra with the ATLAS detector at $\sqrt{s} = 5.02$~TeV and 13~TeV}",
   howpublished   = "CERN-EP-2024-080",
   year           = "2024",
   eprint         = "2404.06204",
   archivePrefix  = "arXiv",
   primaryClass   = "hep-ex"
}

@article{ALFA,
    author = "Abdel Khalek, S. and others",
    title = "{The ALFA Roman Pot Detectors of ATLAS}",
    eprint = "1609.00249",
    archivePrefix = "arXiv",
    primaryClass = "physics.ins-det",
    doi = "10.1088/1748-0221/11/11/P11013",
    journal = "JINST",
    volume = "11",
    number = "11",
    pages = "P11013",
    year = "2016"
}

@Booklet{ATL-PHYS-PUB-2023-039,
    author         = "{ATLAS Collaboration}",
    title          = "{Standard Model Summary Plots October 2023}",
    howpublished   = "{ATL-PHYS-PUB-2023-039}",
    url            = "https://cds.cern.ch/record/2882448",
    year           = "2023",
}

@article{Campbell:2022qmc,
    author = "Campbell, J. M. and others",
    title = "{Event Generators for High-Energy Physics Experiments}",
    booktitle = "{Snowmass 2021}",
    eprint = "2203.11110",
    archivePrefix = "arXiv",
    primaryClass = "hep-ph",
    reportNumber = "CP3-22-12, DESY-22-042, FERMILAB-PUB-22-116-SCD-T, IPPP/21/51,
  JLAB-PHY-22-3576, KA-TP-04-2022, LA-UR-22-22126, LU-TP-22-12, MCNET-22-04,
  OUTP-22-03P, P3H-22-024, PITT-PACC 2207, UCI-TR-2022-02",
    month = "3",
    year = "2022"
}

@article{Frederix_2020,
    author = "Frederix, R. and Frixione, S. and Prestel, S. and Torrielli, P.",
    title = "{On the reduction of negative weights in MC@NLO-type matching procedures}",
    eprint = "2002.12716",
    archivePrefix = "arXiv",
    primaryClass = "hep-ph",
    reportNumber = "LU-TP 20-09, MCNET-20-08",
    doi = "10.1007/JHEP07(2020)238",
    journal = "JHEP",
    volume = "07",
    pages = "238",
    year = "2020"
}

@article{Grazzini_2020,
    author = "Grazzini, Massimiliano and Kallweit, S. and Lindert, Jonas M. and Pozzorini, Stefano and Wiesemann, Marius",
    title = "{NNLO QCD + NLO EW with Matrix+OpenLoops: precise predictions for vector-boson pair production}",
    eprint = "1912.00068",
    archivePrefix = "arXiv",
    primaryClass = "hep-ph",
    reportNumber = "MPP-2019-175, ZU-TH 49/19",
    doi = "10.1007/JHEP02(2020)087",
    journal = "JHEP",
    volume = "02",
    pages = "087",
    year = "2020"
}

@article{Denner_2022,
    author = "Denner, Ansgar and Franken, Robert and Schmidt, Timo and Schwan, Christopher",
    title = "{NLO QCD and EW corrections to vector-boson scattering into W$^{+}$W$^{-}$ at the LHC}",
    eprint = "2202.10844",
    archivePrefix = "arXiv",
    primaryClass = "hep-ph",
    doi = "10.1007/JHEP06(2022)098",
    journal = "JHEP",
    volume = "06",
    pages = "098",
    year = "2022"
}

@article{Alvarez_2023,
    author = "Alvarez, Manuel and Cantero, Josu and Czakon, Michal and Llorente, Javier and Mitov, Alexander and Poncelet, Rene",
    title = "{NNLO QCD corrections to event shapes at the LHC}",
    eprint = "2301.01086",
    archivePrefix = "arXiv",
    primaryClass = "hep-ph",
    reportNumber = "Cavendish-HEP-22/11, P3H-22-129, TTK-22-49",
    doi = "10.1007/JHEP03(2023)129",
    journal = "JHEP",
    volume = "03",
    pages = "129",
    year = "2023"
}

@article{Dasgupta_2018,
    author = "Dasgupta, Mrinal and Dreyer, Fr\'ed\'eric A. and Hamilton, Keith and Monni, Pier Francesco and Salam, Gavin P.",
    title = "{Logarithmic accuracy of parton showers: a fixed-order study}",
    eprint = "1805.09327",
    archivePrefix = "arXiv",
    primaryClass = "hep-ph",
    reportNumber = "CERN-TH-2018-113",
    doi = "10.1007/JHEP09(2018)033",
    journal = "JHEP",
    volume = "09",
    pages = "033",
    year = "2018",
    related        = "Dasgupta_2018-err", 
    relatedstring  = "Erratum:", 
}

@article{Czakon_2023,
    author = "Czakon, Michal and Mitov, Alexander and Poncelet, Rene",
    title = "{Infrared-safe flavoured anti-k$_{T}$ jets}",
    eprint = "2205.11879",
    archivePrefix = "arXiv",
    primaryClass = "hep-ph",
    reportNumber = "Cavendish-HEP-22/06, P3H-22-056, TTK-22-17",
    doi = "10.1007/JHEP04(2023)138",
    journal = "JHEP",
    volume = "04",
    pages = "138",
    year = "2023"
}

@article{Caletti_2022,
    author = "Caletti, Simone and Larkoski, Andrew J. and Marzani, Simone and Reichelt, Daniel",
    title = "{Practical jet flavour through NNLO}",
    eprint = "2205.01109",
    archivePrefix = "arXiv",
    primaryClass = "hep-ph",
    reportNumber = "SLAC-PUB-17674, IPPP/22/27",
    doi = "10.1140/epjc/s10052-022-10568-7",
    journal = "Eur. Phys. J. C",
    volume = "82",
    number = "7",
    pages = "632",
    year = "2022"
}

@article{Alekhin:2014irh,
    author = "Alekhin, S. and others",
    title = "{HERAFitter}",
    eprint = "1410.4412",
    archivePrefix = "arXiv",
    primaryClass = "hep-ph",
    reportNumber = "DESY-14-188, DESY-REPORT-14-188, FERMILAB-PUB-14-603-CMS",
    doi = "10.1140/epjc/s10052-015-3480-z",
    journal = "Eur. Phys. J. C",
    volume = "75",
    number = "7",
    pages = "304",
    year = "2015"
}

@article{McGowan:2022nag,
    author = "McGowan, J. and Cridge, T. and Harland-Lang, L. A. and Thorne, R. S.",
    title = "{Approximate N$^{3}$LO parton distribution functions with theoretical uncertainties: MSHT20aN$^3$LO PDFs}",
    eprint = "2207.04739",
    archivePrefix = "arXiv",
    primaryClass = "hep-ph",
    doi = "10.1140/epjc/s10052-023-11236-0",
    journal = "Eur. Phys. J. C",
    volume = "83",
    number = "3",
    pages = "185",
    year = "2023",
    related        = "McGowan:2022nag-err", 
    relatedstring  = "Erratum:", 
}

@article{Mirkes:1992hu,
    author = "Mirkes, E.",
    title = "{Angular decay distribution of leptons from W-bosons at NLO in hadronic collisions}",
    reportNumber = "TTP-92-12",
    doi = "10.1016/0550-3213(92)90046-E",
    journal = "Nucl. Phys. B",
    volume = "387",
    pages = "3--85",
    year = "1992"
}

@article{Collins:1977iv,
    author = "Collins, John C. and Soper, Davison E.",
    title = "{Angular Distribution of Dileptons in High-Energy Hadron Collisions}",
    reportNumber = "Print-77-0288 (PRINCETON)",
    doi = "10.1103/PhysRevD.16.2219",
    journal = "Phys. Rev. D",
    volume = "16",
    pages = "2219",
    year = "1977"
}

@article{Sudakov:1954sw,
    author = "Sudakov, V. V.",
    title = "{Vertex parts at very high-energies in quantum electrodynamics}",
    journal = "Sov. Phys. JETP",
    volume = "3",
    pages = "65--71",
    year = "1956"
}

@article{Georgi:1985nv,
    author = "Georgi, Howard and Machacek, Marie",
    title = "{Doubly charged Higgs bosons}",
    reportNumber = "HUTP-85/A051",
    doi = "10.1016/0550-3213(85)90325-6",
    journal = "Nucl. Phys. B",
    volume = "262",
    pages = "463--477",
    year = "1985"
}

@article{Brivio:2017vri,
    author = "Brivio, Ilaria and Trott, Michael",
    title = "{The Standard Model as an Effective Field Theory}",
    eprint = "1706.08945",
    archivePrefix = "arXiv",
    primaryClass = "hep-ph",
    doi = "10.1016/j.physrep.2018.11.002",
    journal = "Phys. Rept.",
    volume = "793",
    pages = "1--98",
    year = "2019"
}

@Booklet{ATL-PHYS-PUB-2021-022,
%    author         = "{ATLAS Collaboration}",
%    title          = "{Combined effective field theory interpretation of differential cross-sections measurements of WW, WZ, 4l, and Z-plus-two-jets production using ATLAS data}",
%    howpublished   = "{ATL-PHYS-PUB-2021-022}",
%    url            = "https://cds.cern.ch/record/2776648",
%    year           = "2021",
%}

@article{Lukaszuk:1973nt,
    author = "Lukaszuk, L. and Nicolescu, B.",
    title = "{A Possible interpretation of pp rising total cross-sections}",
    doi = "10.1007/BF02824484",
    journal = "Lett. Nuovo Cim.",
    volume = "8",
    pages = "405--413",
    year = "1973"
}

@article{Harland-Lang:2015cta,
    author = "Harland-Lang, L. A. and Khoze, V. A. and Ryskin, M. G.",
    title = "{Exclusive physics at the LHC with SuperChic 2}",
    eprint = "1508.02718",
    archivePrefix = "arXiv",
    primaryClass = "hep-ph",
    reportNumber = "IPPP-15-34, DCPT-15-68",
    doi = "10.1140/epjc/s10052-015-3832-8",
    journal = "Eur. Phys. J. C",
    volume = "76",
    number = "1",
    pages = "9",
    year = "2016"
}

@article{STDM-2018-36,
%  author = "{ATLAS Collaboration}",
%  title = "Measurement of the cross-sections of the electroweak and total production of a $Z\gamma$ pair in association with two jets in pp collisions at $\sqrt{s} = 13\,\text{TeV}$ with the ATLAS detector",
%  year = "2023",
%  archivePrefix = "arXiv",
%  eprint = "2305.19142",
%  primaryClass = "hep-ex",
%  doi = "10.1016/j.physletb.2023.138222",
%  journal = "Phys. Lett. B",
%  volume = "846",
%  pages = "138222"
%}

@article{STDM-2018-33,
%  author = "{ATLAS Collaboration}",
%  title = "{Observation of $W\gamma\gamma$  triboson production in proton-proton collisions at $\sqrt{s} = 13\,\text{TeV}$ with the ATLAS detector}",
%  year = "2024",
%  archivePrefix = "arXiv",
%  eprint = "2308.03041",
%  primaryClass = "hep-ex",
%  doi = "10.1016/j.physletb.2023.138400",
%  journal = "Phys. Lett. B",
%  volume = "848",
%  pages = "138400"
%}

@article{STDM-2019-17,
%  author = "{ATLAS Collaboration}",
%  title = "Observation of $WZ\gamma$ production in pp collisions at $\sqrt{s} = 13\,\text{TeV}$ with the ATLAS detector",
%  year = "2024",
%  archivePrefix = "arXiv",
%  eprint = "arXiv:2305.16994",
%  primaryClass = "hep-ex",
%  doi = "10.1103/PhysRevLett.132.021802",
%  journal = "Phys. Rev. Lett.",
%  volume = "132",
%  pages = "021802"
%}

@article{Campbell11,
    author = "Campbell, John M. and Ellis, R. Keith and Williams, Ciaran",
    title = "{Vector Boson Pair Production at the LHC}",
    eprint = "1105.0020",
    archivePrefix = "arXiv",
    primaryClass = "hep-ph",
    reportNumber = "FERMILAB-PUB-11-182-T",
    doi = "10.1007/JHEP07(2011)018",
    journal = "JHEP",
    volume = "07",
    pages = "018",
    year = "2011"
}

@article{Campbell17,
    author = "Campbell, John M. and Neumann, Tobias and Williams, Ciaran",
    title = "{$Z\gamma$ Production at NNLO Including Anomalous Couplings}",
    eprint = "1708.02925",
    archivePrefix = "arXiv",
    primaryClass = "hep-ph",
    reportNumber = "FERMILAB-PUB-17-303-T",
    doi = "10.1007/JHEP11(2017)150",
    journal = "JHEP",
    volume = "11",
    pages = "150",
    year = "2017"
}

@article{Denner20,
    author = "Denner, Ansgar and Pelliccioli, Giovanni",
    title = "{NLO QCD predictions for doubly-polarized WZ production at the LHC}",
    eprint = "2010.07149",
    archivePrefix = "arXiv",
    primaryClass = "hep-ph",
    doi = "10.1016/j.physletb.2021.136107",
    journal = "Phys. Lett. B",
    volume = "814",
    pages = "136107",
    year = "2021"
}

@article{herwig1,
    author = {B\"ahr, M. and others},
    title = "{Herwig++ Physics and Manual}",
    eprint = "0803.0883",
    archivePrefix = "arXiv",
    primaryClass = "hep-ph",
    reportNumber = "CERN-PH-TH-2008-038, CAVENDISH-HEP-08-03, KA-TP-05-2008, DCPT-08-22, IPPP-08-11, CP3-08-05",
    doi = "10.1140/epjc/s10052-008-0798-9",
    journal = "Eur. Phys. J. C",
    volume = "58",
    pages = "639--707",
    year = "2008"
}

@article{herwig2,
    author = "Bellm, Johannes and others",
    title = "{Herwig 7.0/Herwig++ 3.0 release note}",
    eprint = "1512.01178",
    archivePrefix = "arXiv",
    primaryClass = "hep-ph",
    reportNumber = "CERN-PH-TH-2015-289, MAN-HEP-2015-15, IFJPAN-IV-2015-13, KA-TP-18-2015, DCPT-15-142, MCNET-15-28, IPPP-15-71, HERWIG-2015-01",
    doi = "10.1140/epjc/s10052-016-4018-8",
    journal = "Eur. Phys. J. C",
    volume = "76",
    number = "4",
    pages = "196",
    year = "2016"
}

@article{vbfnlo,
    author = "Arnold, K. and others",
    title = "{VBFNLO: A Parton level Monte Carlo for processes with electroweak bosons}",
    eprint = "0811.4559",
    archivePrefix = "arXiv",
    primaryClass = "hep-ph",
    reportNumber = "KA-TP-31-2008, SFB-CPP-08-95",
    doi = "10.1016/j.cpc.2009.03.006",
    journal = "Comput. Phys. Commun.",
    volume = "180",
    pages = "1661--1670",
    year = "2009"
}

@article{Fabres09,
    author = "F. Cordero, Fernando and Reina, L. and Wackeroth, D.",
    title = "{W- and Z-boson production with a massive bottom-quark pair at the Large Hadron Collider}",
    eprint = "0906.1923",
    archivePrefix = "arXiv",
    primaryClass = "hep-ph",
    reportNumber = "FSU-HEP-2009-0314, UCLA-09-TEP-49",
    doi = "10.1103/PhysRevD.80.034015",
    journal = "Phys. Rev. D",
    volume = "80",
    pages = "034015",
    year = "2009"
}

@article{Campbell04,
    author = "Campbell, John and Ellis, R. Keith and Maltoni, F. and Willenbrock, S.",
    title = "{Associated production of a $Z$ boson and a single heavy quark jet}",
    eprint = "hep-ph/0312024",
    archivePrefix = "arXiv",
    reportNumber = "ANL-HEP-PR-03-104, FERMILAB-PUB-03-378-T, RM3-TH-03-17, ILL-TH-03-10, NSF-KITP-03-98",
    doi = "10.1103/PhysRevD.69.074021",
    journal = "Phys. Rev. D",
    volume = "69",
    pages = "074021",
    year = "2004"
}

@article{Maltoni12,
    author = "Maltoni, Fabio and Ridolfi, Giovanni and Ubiali, Maria",
    title = "{b-initiated processes at the LHC: a reappraisal}",
    eprint = "1203.6393",
    archivePrefix = "arXiv",
    primaryClass = "hep-ph",
    reportNumber = "CP3-12-15, TTK-12-11",
    doi = "10.1007/JHEP04(2013)095",
    journal = "JHEP",
    volume = "07",
    pages = "022",
    year = "2012",
    note = "[Erratum: \href{https://doi.org/10.1007/JHEP04(2013)095}{JHEP 04, 095 (2013)}]"
}

@article{Ubiali20,
    author = "Faura, Ferran and Iranipour, Shayan and Nocera, Emanuele R. and Rojo, Juan and Ubiali, Maria",
    title = "{The Strangest Proton?}",
    eprint = "2009.00014",
    archivePrefix = "arXiv",
    primaryClass = "hep-ph",
    doi = "10.1140/epjc/s10052-020-08749-3",
    journal = "Eur. Phys. J. C",
    volume = "80",
    number = "12",
    pages = "1168",
    year = "2020"
}

@article{Azatov17,
    author = "Azatov, A. and Elias-Mir\'o, J. and Reyimuaji, Y. and Venturini, E.",
    title = "{Novel measurements of anomalous triple gauge couplings for the LHC}",
    eprint = "1707.08060",
    archivePrefix = "arXiv",
    primaryClass = "hep-ph",
    reportNumber = "SISSA-35-2017-FISI",
    doi = "10.1007/JHEP10(2017)027",
    journal = "JHEP",
    volume = "10",
    pages = "027",
    year = "2017"
}

@article{Panico18,
    author = "Panico, Giuliano and Riva, Francesco and Wulzer, Andrea",
    title = "{Diboson interference resurrection}",
    eprint = "1708.07823",
    archivePrefix = "arXiv",
    primaryClass = "hep-ph",
    reportNumber = "CERN-TH-2017-185",
    doi = "10.1016/j.physletb.2017.11.068",
    journal = "Phys. Lett. B",
    volume = "776",
    pages = "473--480",
    year = "2018"
}

@article{Grazzini16,
    author = "Grazzini, Massimiliano and Kallweit, Stefan and Rathlev, Dirk and Wiesemann, Marius",
    title = "{$W^{\pm}Z$ production at hadron colliders in NNLO QCD}",
    eprint = "1604.08576",
    archivePrefix = "arXiv",
    primaryClass = "hep-ph",
    reportNumber = "ZU-TH-15-16, MITP-16-037, NSF-KITP-16-046, DESY-16-074",
    doi = "10.1016/j.physletb.2016.08.017",
    journal = "Phys. Lett. B",
    volume = "761",
    pages = "179--183",
    year = "2016"
}

@article{Grazzini_2020b,
    author = "Grazzini, Massimiliano and Kallweit, Stefan and Wiesemann, Marius and Yook, Jeong Yeon",
    title = "{$W^+W^-$ production at the LHC: NLO QCD corrections to the loop-induced gluon fusion channel}",
    eprint = "2002.01877",
    archivePrefix = "arXiv",
    primaryClass = "hep-ph",
    reportNumber = "MPP-2019-176, ZU-TH 04/20",
    doi = "10.1016/j.physletb.2020.135399",
    journal = "Phys. Lett. B",
    volume = "804",
    pages = "135399",
    year = "2020"
}

@misc{Kallweit_2015,
    author = {Kallweit, Stefan and Lindert, Jonas M. and Maierh\"ofer, Philipp and Pozzorini, Stefano and Sch\"onherr, Marek},
    title = "{NLO electroweak automation and precise predictions for W+multijet production at the LHC}",
    eprint = "1412.5157",
    archivePrefix = "arXiv",
    primaryClass = "hep-ph",
    reportNumber = "LPN14-127, IPPP-14-107, DCPT-14-214, MCNET-14-26, ZU-TH-42-14, MITP-14-102",
    doi = "10.1007/JHEP04(2015)012",
    journal = "JHEP",
    volume = "04",
    pages = "012",
    year = "2015"
}

@article{Chanowitz88,
    author = "Chanowitz, Michael S. and Golden, Mitchell",
    title = "{Like-Charged Gauge-Boson Pairs as a Probe of Electroweak Symmetry Breaking}",
    reportNumber = "LBL-25441",
    doi = "10.1103/PhysRevLett.61.1053",
    journal = "Phys. Rev. Lett.",
    volume = "61",
    pages = "1053",
    year = "1988",
    related        = "Chanowitz88-err", 
    relatedstring  = "Erratum:", 
}

@article{Belanger92,
    author = "B\'elanger, G. and Boudjema, F.",
    title = "{Probing quartic couplings of weak bosons through three vector production at a 500 GeV NLC}",
    reportNumber = "ENSLAPP-A-363-92, UDEM-LPN-TH-82",
    doi = "10.1016/0370-2693(92)91978-I",
    journal = "Phys. Lett. B",
    volume = "288",
    pages = "201--209",
    year = "1992"
}

@article{Grazzini15a,
    author = "Grazzini, Massimiliano and Kallweit, Stefan and Rathlev, Dirk",
    title = "{ZZ production at the LHC: fiducial cross sections and distributions in NNLO QCD}",
    eprint = "1507.06257",
    archivePrefix = "arXiv",
    primaryClass = "hep-ph",
    doi = "10.1016/j.physletb.2015.09.055",
    journal = "Phys. Lett. B",
    volume = "750",
    pages = "407--410",
    year = "2015"
}

@article{Grazzini15b,
    author = "Grazzini, Massimiliano and Kallweit, Stefan and Rathlev, Dirk",
    title = "{$W\gamma$ and $Z\gamma$ production at the LHC in NNLO QCD}",
    eprint = "1504.01330",
    archivePrefix = "arXiv",
    primaryClass = "hep-ph",
    reportNumber = "ZU-TH-04-15, MITP-15-021",
    doi = "10.1007/JHEP07(2015)085",
    journal = "JHEP",
    volume = "07",
    pages = "085",
    year = "2015"
}

@article{Grazzini18,
    author = "Grazzini, Massimiliano and Kallweit, Stefan and Wiesemann, Marius",
    title = "{Fully differential NNLO computations with MATRIX}",
    eprint = "1711.06631",
    archivePrefix = "arXiv",
    primaryClass = "hep-ph",
    reportNumber = "ZU-TH-30-17, CERN-TH-2017-232, ZU-TH 30/17",
    doi = "10.1140/epjc/s10052-018-5771-7",
    journal = "Eur. Phys. J. C",
    volume = "78",
    number = "7",
    pages = "537",
    year = "2018"
}

@article{Jaeger09,
    author = {J\"ager, B. and Oleari, C. and Zeppenfeld, D.},
    title = "{Next-to-leading order QCD corrections to W$^+$W$^+$ jj and W$^-$W$^-$ jj production via weak-boson fusion}",
    eprint = "0907.0580",
    archivePrefix = "arXiv",
    primaryClass = "hep-ph",
    reportNumber = "KA-TP-09-2009",
    doi = "10.1103/PhysRevD.80.034022",
    journal = "Phys. Rev. D",
    volume = "80",
    pages = "034022",
    year = "2009"
}

@article{Hoeche13,
    author = {H\"oche, Stefan and Krauss, Frank and Sch\"onherr, Marek and Siegert, Frank},
    title = "{QCD matrix elements + parton showers. The NLO case}",
    eprint = "1207.5030",
    archivePrefix = "arXiv",
    primaryClass = "hep-ph",
    reportNumber = "SLAC-PUB-15191, IPPP-12-52, DCPT-12-104, LPN12-081, MCNET-12-09, FR-PHENO-2012-017",
    doi = "10.1007/JHEP04(2013)027",
    journal = "JHEP",
    volume = "04",
    pages = "027",
    year = "2013"
}

@article{Rubin10,
    author = "Rubin, Mathieu and Salam, Gavin P. and Sapeta, Sebastian",
    title = "{Giant QCD K-factors beyond NLO}",
    eprint = "1006.2144",
    archivePrefix = "arXiv",
    primaryClass = "hep-ph",
    doi = "10.1007/JHEP09(2010)084",
    journal = "JHEP",
    volume = "09",
    pages = "084",
    year = "2010"
}

@article{Prestel16,
    author = "Christiansen, Jesper Roy and Prestel, Stefan",
    title = "{Merging weak and QCD showers with matrix elements}",
    eprint = "1510.01517",
    archivePrefix = "arXiv",
    primaryClass = "hep-ph",
    doi = "10.1140/epjc/s10052-015-3871-1",
    journal = "Eur. Phys. J. C",
    volume = "76",
    number = "1",
    pages = "39",
    year = "2016"
}

@article{Boughezal15,
    author = "Boughezal, Radja and Focke, Christfried and Liu, Xiaohui",
    title = "{Jet vetoes versus giant K factors in the exclusive Z+1-jet cross section}",
    eprint = "1501.01059",
    archivePrefix = "arXiv",
    primaryClass = "hep-ph",
    doi = "10.1103/PhysRevD.92.094002",
    journal = "Phys. Rev. D",
    volume = "92",
    number = "9",
    pages = "094002",
    year = "2015"
}

@article{Sjoestrand14,
    author = {Christiansen, Jesper Roy and Sj\"ostrand, Torbj\"orn},
    title = "{Weak Gauge Boson Radiation in Parton Showers}",
    eprint = "1401.5238",
    archivePrefix = "arXiv",
    primaryClass = "hep-ph",
    reportNumber = "LU-TP-14-02, MCNET-14-01",
    doi = "10.1007/JHEP04(2014)115",
    journal = "JHEP",
    volume = "04",
    pages = "115",
    year = "2014"
}

@article{nnlojet1,
    author = "Gehrmann-De Ridder, A. and Gehrmann, T. and Glover, E. W. N. and Huss, A. and Morgan, T. A.",
    title = "{Precise QCD predictions for the production of a Z boson in association with a hadronic jet}",
    eprint = "1507.02850",
    archivePrefix = "arXiv",
    primaryClass = "hep-ph",
    reportNumber = "IPPP-15-44, ZU-TH-23-15",
    doi = "10.1103/PhysRevLett.117.022001",
    journal = "Phys. Rev. Lett.",
    volume = "117",
    number = "2",
    pages = "022001",
    year = "2016"
}

@article{nnlojet2,
    author = "Gehrmann-De Ridder, Aude and Gehrmann, T. and Glover, E. W. N. and Huss, A. and Morgan, T. A.",
    title = "{The NNLO QCD corrections to Z boson production at large transverse momentum}",
    eprint = "1605.04295",
    archivePrefix = "arXiv",
    primaryClass = "hep-ph",
    reportNumber = "IPPP-16-39, NSF-KITP-16-067, ZU-TH-18-16",
    doi = "10.1007/JHEP07(2016)133",
    journal = "JHEP",
    volume = "07",
    pages = "133",
    year = "2016"
}

@article{Komiske:2019fks,
    author = "Komiske, Patrick T. and Metodiev, Eric M. and Thaler, Jesse",
    title = "Metric Space of Collider Events",
    eprint = "1902.02346",
    archivePrefix = "arXiv",
    primaryClass = "hep-ph",
    reportNumber = "MIT-CTP 5102",
    doi = "10.1103/PhysRevLett.123.041801",
    journal = "Phys. Rev. Lett.",
    volume = "123",
    number = "4",
    pages = "041801",
    year = "2019"
}

@article{Cesarotti:2020hwb,
    author = "Cesarotti, Cari and Thaler, Jesse",
    title = "A Robust Measure of Event Isotropy at Colliders",
    eprint = "2004.06125",
    archivePrefix = "arXiv",
    primaryClass = "hep-ph",
    reportNumber = "MIT-CTP 5195",
    doi = "10.1007/JHEP08(2020)084",
    journal = "JHEP",
    volume = "08",
    pages = "084",
    year = "2020"
}

@article{wasserstein1969markov,
  title="Markov processes over denumerable products of spaces describing large systems of automata",
  author={Wasserstein, Leonid N},
  journal={Prob. Inf. Transm.},
  volume={5},
  number={3},
  pages={47--52},
  year={1969}
}

@Article{STDM-2020-14,
%    author         = "{ATLAS Collaboration}",
%    title          = "{Measurements of $Z\gamma$+jets differential cross sections in pp collisions at $\sqrt{s} = 13$~TeV with the ATLAS detector}",
%    year           = "2022",
%    reportNumber   = "CERN-EP-2022-240",
%    eprint         = "2212.07184",
%    archivePrefix  = "arXiv",
%    primaryClass   = "hep-ex",
%}

@Article{STDM-2021-09,
%    author         = "{ATLAS Collaboration}",
%    title          = "{Measurement of $Z\gamma\gamma$ production in pp collisions at $\sqrt{s} = 13$~TeV with the ATLAS detector}",
%    year            = "2022",
%    reportNumber   = "CERN-EP-2022-192",
%    eprint         = "2211.14171",
%    archivePrefix  = "arXiv",
%    primaryClass   = "hep-ex",
%}

@article{NNLOjet2016,
    author = "Gehrmann-De Ridder, A. and Gehrmann, T. and Glover, E. W. N. and Huss, A. and Morgan, T. A.",
    title = "{NNLO QCD corrections for Drell-Yan $p_T^Z$ and $\phi_{\eta}^*$ observables at the LHC}",
    eprint = "1610.01843",
    archivePrefix = "arXiv",
    primaryClass = "hep-ph",
    reportNumber = "IPPP-16-74, ZU-TH-36-16, IPPP/16/74, ZU-TH 36/16",
    doi = "10.1007/JHEP11(2016)094",
    journal = "JHEP",
    volume = "11",
    pages = "094",
    year = "2016",
    related        = "NNLOjet2016-err", 
    relatedstring  = "Erratum:", 
}

@article{Bizon2019,
    author = "Bizon, Wojciech and Gehrmann-De Ridder, Aude and Gehrmann, Thomas and Glover, Nigel and Huss, Alexander and Monni, Pier Francesco and Re, Emanuele and Rottoli, Luca and Walker, Duncan M.",
    title = "{The transverse momentum spectrum of weak gauge bosons at N ${}^3$ LL + NNLO}",
    eprint = "1905.05171",
    archivePrefix = "arXiv",
    primaryClass = "hep-ph",
    reportNumber = "ZU-TH 21/1, CERN-TH-2019-050, LAPTH-026/19, IPPP/19/38",
    doi = "10.1140/epjc/s10052-019-7324-0",
    journal = "Eur. Phys. J. C",
    volume = "79",
    number = "10",
    pages = "868",
    year = "2019"
}

@article{Bizon2018,
    author = "Bizo\'n, Wojciech and Chen, Xuan and Gehrmann-De Ridder, Aude and Gehrmann, Thomas and Glover, Nigel and Huss, Alexander and Monni, Pier Francesco and Re, Emanuele and Rottoli, Luca and Torrielli, Paolo",
    title = "{Fiducial distributions in Higgs and Drell-Yan production at N$^{3}$LL+NNLO}",
    eprint = "1805.05916",
    archivePrefix = "arXiv",
    primaryClass = "hep-ph",
    reportNumber = "CERN-TH-2018-105, IPPP/18/34, LAPTH-015/18, OUTP-17-19P, ZU-TH 17/18, IPPP-18-34, LAPTH-015-18, ZU-TH-17-18",
    doi = "10.1007/JHEP12(2018)132",
    journal = "JHEP",
    volume = "12",
    pages = "132",
    year = "2018"
}

@article{Avoni:2018iuv,
      author         = "Avoni, G. and others",
      title          = "{The new LUCID-2 detector for luminosity measurement and
                        monitoring in ATLAS}",
      journal        = "JINST",
      volume         = "13",
      year           = "2018",
      number         = "07",
      pages          = "P07017",
      doi            = "10.1088/1748-0221/13/07/P07017",
      SLACcitation   = "%%CITATION = JINST,13,P07017;%%"
}

@article{Sjostrand:2007gs,
    author = "Sjostrand, Torbjorn and Mrenna, Stephen and Skands, Peter",
    title = "{A Brief Introduction to PYTHIA 8.1}",
    eprint = "0710.3820",
    archivePrefix = "arXiv",
    primaryClass = "hep-ph",
    reportNumber = "CERN-LCGAPP-2007-04, LU-TP-07-28, FERMILAB-PUB-07-512-CD-T",
    doi = "10.1016/j.cpc.2008.01.036",
    journal = "Comput. Phys. Commun.",
    volume = "178",
    pages = "852--867",
    year = "2008"
}

@article{Ostapchenko:2010vb,
    author = "Ostapchenko, Sergey",
    title = "{Monte Carlo treatment of hadronic interactions in enhanced Pomeron scheme: QGSJET-II model }",
    eprint = "1010.1869",
    archivePrefix = "arXiv",
    primaryClass = "hep-ph",
    doi = "10.1103/PhysRevD.83.014018",
    journal = "Phys. Rev. D",
    volume = "83",
    pages = "014018",
    year = "2011"
}

@article{Pierog:2013ria,
    author = "Pierog, T. and Karpenko, Iu. and Katzy, J. M. and Yatsenko, E. and Werner, K.",
    title = "{EPOS LHC: Test of collective hadronization with data measured at the CERN Large Hadron Collider}",
    eprint = "1306.0121",
    archivePrefix = "arXiv",
    primaryClass = "hep-ph",
    reportNumber = "DESY-13-125",
    doi = "10.1103/PhysRevC.92.034906",
    journal = "Phys. Rev. C",
    volume = "92",
    number = "3",
    pages = "034906",
    year = "2015"
}

@article{ALICE:2022rdg,
    author = "{ALICE Collaboration}",
    title = "{Measurement of the angle between jet axes in pp collisions at $ \sqrt{s} $ = 5.02 TeV}",
    eprint = "2211.08928",
    archivePrefix = "arXiv",
    primaryClass = "nucl-ex",
    reportNumber = "CERN-EP-2022-242",
    doi = "10.1007/JHEP07(2023)201",
    journal = "JHEP",
    volume = "07",
    pages = "201",
    year = "2023"
}

@article{STAR:2020ejj,
    author = "{STAR Collaboration}",
    title = "{Measurement of groomed jet substructure observables in p+p collisions at $\sqrt {s}$ =200 GeV with STAR}",
    eprint = "2003.02114",
    archivePrefix = "arXiv",
    primaryClass = "hep-ex",
    doi = "10.1016/j.physletb.2020.135846",
    journal = "Phys. Lett. B",
    volume = "811",
    pages = "135846",
    year = "2020"
}

@article{ALICE:2012cor,
    author = "{ALICE Collaboration}",
    title = "{Transverse sphericity of primary charged particles in minimum bias proton-proton collisions at $\sqrt{s}=0.9$, 2.76 and 7 TeV}",
    eprint = "1205.3963",
    archivePrefix = "arXiv",
    primaryClass = "hep-ex",
    reportNumber = "CERN-PH-EP-2012-136",
    doi = "10.1140/epjc/s10052-012-2124-9",
    journal = "Eur. Phys. J. C",
    volume = "72",
    pages = "2124",
    year = "2012"
}

@article{Harland-Lang:2018iur,
    author = "Harland-Lang, L. A. and Khoze, V. A. and Ryskin, M. G.",
    title = "{Exclusive LHC physics with heavy ions: SuperChic 3}",
    eprint = "1810.06567",
    archivePrefix = "arXiv",
    primaryClass = "hep-ph",
    reportNumber = "IPPP/18/90",
    doi = "10.1140/epjc/s10052-018-6530-5",
    journal = "Eur. Phys. J. C",
    volume = "79",
    number = "1",
    pages = "39",
    year = "2019"
}

@article{Klein:2016yzr,
    author = "Klein, Spencer R. and Nystrand, Joakim and Seger, Janet and Gorbunov, Yuri and Butterworth, Joey",
    title = "{STARlight: A Monte Carlo simulation program for ultra-peripheral collisions of relativistic ions}",
    eprint = "1607.03838",
    archivePrefix = "arXiv",
    primaryClass = "hep-ph",
    doi = "10.1016/j.cpc.2016.10.016",
    journal = "Comput. Phys. Commun.",
    volume = "212",
    pages = "258--268",
    year = "2017"
}

@article{Klein:2020fmr,
    author = "Klein, Spencer and Steinberg, Peter",
    title = "{Photonuclear and Two-photon Interactions at High-Energy Nuclear Colliders}",
    eprint = "2005.01872",
    archivePrefix = "arXiv",
    primaryClass = "nucl-ex",
    doi = "10.1146/annurev-nucl-030320-033923",
    journal = "Ann. Rev. Nucl. Part. Sci.",
    volume = "70",
    pages = "323--354",
    year = "2020"
}

@article{Baltz:2007kq,
    author = "Baltz, A. J. and others",
    title = "{The Physics of Ultraperipheral Collisions at the LHC}",
    eprint = "0706.3356",
    archivePrefix = "arXiv",
    primaryClass = "nucl-ex",
    doi = "10.1016/j.physrep.2007.12.001",
    journal = "Phys. Rept.",
    volume = "458",
    pages = "1--171",
    year = "2008"
}

@article{TOTEM:2017asr,
    author = "{TOTEM Collaboration}",
    collaboration = "TOTEM",
    title = "{First measurement of elastic, inelastic and total cross-section at $\sqrt{s}=13$ TeV by TOTEM and overview of cross-section data at LHC energies}",
    eprint = "1712.06153",
    archivePrefix = "arXiv",
    primaryClass = "hep-ex",
    reportNumber = "CERN-EP-2017-321, CERN-EP-2017-321-V2",
    doi = "10.1140/epjc/s10052-019-6567-0",
    journal = "Eur. Phys. J. C",
    volume = "79",
    number = "2",
    pages = "103",
    year = "2019"
}

@article{LHCb:2018ehw,
    author = "{LHCb Collaboration}",
    collaboration = "LHCb",
    title = "{Measurement of the inelastic $pp$ cross-section at a centre-of-mass energy of 13 TeV}",
    eprint = "1803.10974",
    archivePrefix = "arXiv",
    primaryClass = "hep-ex",
    reportNumber = "LHCb-PAPER-2018-003, CERN-EP-2018-044, LHCB-PAPER-2018-003",
    doi = "10.1007/JHEP06(2018)100",
    journal = "JHEP",
    volume = "06",
    pages = "100",
    year = "2018"
}

@article{mitov2022,
    title={Next-to-Next-to-Leading Order Study of Three-Jet Production at the LHC},
    author={Michal Czakon and Alexander Mitov and Rene Poncelet},
    journal={Phys. Rev. Lett.},
    volume={127},
    year={2021},
    pages={152001},
    doi={10.1103/PhysRevLett.127.152001},
    eprint={2106.05331},
    archivePrefix={arXiv},
    primaryClass={hep-ph}
}

@article{Wilson:1953zz,
  author =	 "Wilson, Robert R.",
  title =	 "{Scattering of 1.33 MeV Gamma-Rays by an Electric
                  Field}",
  journal =	 "Phys. Rev.",
  volume =	 "90",
  year =	 "1953",
  pages =	 "720-721",
  doi =		 "10.1103/PhysRev.90.720",
  SLACcitation = "%%CITATION = PHRVA,90,720;%%"
}

@article{Akhmadaliev:2001ik,
      author         = "Akhmadaliev, Sh. Zh. and Kezerashvili, G. Ya. and Klimenko, S. G. and Lee, R. N. and Malyshev, V. M. and Maslennikov, A. L. and Milov, A. M. and Milstein, A. I. and Muchnoi, N. Yu. and Naumenkov, A. I. and Panin, V. S. and Peleganchuk, S. V. and Pospelov, G. E. and Protopopov, I. Ya. and Romanov, L. V. and Shamov, A. G. and Shatilov, D. N. and Simonov, E. A. and Strakhovenko, V. M. and Tikhonov, Yu. A.",
      title          = "{Experimental Investigation of High-Energy Photon Splitting in Atomic Fields}",
      journal        = "Phys. Rev. Lett.",
      volume         = "89",
      year           = "2002",
      pages          = "061802",
      doi            = "10.1103/PhysRevLett.89.061802",
      eprint         = "hep-ex/0111084",
      archivePrefix  = "arXiv",
      reportNumber   = "BUDKER-INP-2001-80",
      SLACcitation   = "%%CITATION = HEP-EX/0111084;%%"
}

@article{Hanneke:2008tm,
      author         = "Hanneke, D. and Fogwell, S. and Gabrielse, G.",
      title          = "{New Measurement of the Electron Magnetic Moment and the
                        Fine Structure Constant}",
      journal        = "Phys. Rev. Lett.",
      volume         = "100",
      year           = "2008",
      pages          = "120801",
      doi            = "10.1103/PhysRevLett.100.120801",
      eprint         = "0801.1134",
      archivePrefix  = "arXiv",
      primaryClass   = "physics.atom-ph",
      SLACcitation   = "%%CITATION = ARXIV:0801.1134;%%"
}

@article{Muong:2021ojo,
    author = "{Muon g-2 Collaboration}",
    collaboration = "Muon g-2",
    title = "{Measurement of the Positive Muon Anomalous Magnetic Moment to 0.46 ppm}",
    eprint = "2104.03281",
    archivePrefix = "arXiv",
    primaryClass = "hep-ex",
    reportNumber = "FERMILAB-PUB-21-132-E",
    doi = "10.1103/PhysRevLett.126.141801",
    journal = "Phys. Rev. Lett.",
    volume = "126",
    number = "14",
    pages = "141801",
    year = "2021"
}

@article{Enterria:2013yra,
      author         = "d'Enterria, David and da Silveira, Gustavo G.",
      title          = "{Observing Light-by-Light Scattering at the Large Hadron Collider}",
      journal        = "Phys. Rev. Lett.",
      volume         = "111",
      year           = "2013",
      pages          = "080405",
      doi            = "10.1103/PhysRevLett.111.080405",
      eprint         = "1305.7142",
      archivePrefix  = "arXiv",
      primaryClass   = "hep-ph",
    related        = "Enterria:2013yra-err", 
    relatedstring  = "Erratum:", 
}

@article{Klusek-Gawenda:2016euz,
      author         = {K\l{}usek-Gawenda, Mariola and Lebiedowicz, Piotr and
                        Szczurek, Antoni},
      title          = "{Light-by-light scattering in ultraperipheral Pb-Pb
                        collisions at energies available at the CERN Large Hadron
                        Collider}",
      journal        = "Phys. Rev. C",
      volume         = "93",
      year           = "2016",
      number         = "4",
      pages          = "044907",
      doi            = "10.1103/PhysRevC.93.044907",
      eprint         = "1601.07001",
      archivePrefix  = "arXiv",
      primaryClass   = "nucl-th",
      SLACcitation   = "%%CITATION = ARXIV:1601.07001;%%"
}

@techreport{Jenni:1009649,
      author        = "{ATLAS Collaboration}",
      title         = "{Zero Degree Calorimeters for ATLAS}",
      reportNumber  = "CERN-LHCC-2007-001, LHCC-I-016",
      year          = "2007",
      url           = "https://cds.cern.ch/record/1009649",
}

@article{DELPHI:2003nah,
    author = "{DELPHI Collaboration}",
    title = "{Study of tau-pair production in photon-photon collisions at LEP and limits on the anomalous electromagnetic moments of the tau lepton}",
    eprint = "hep-ex/0406010",
    archivePrefix = "arXiv",
    reportNumber = "CERN-EP-2003-058",
    doi = "10.1140/epjc/s2004-01852-y",
    journal = "Eur. Phys. J. C",
    volume = "35",
    pages = "159--170",
    year = "2004"
}

@article{Chen:2019zmr,
    author = {Chen, Xuan and Gehrmann, Thomas and Glover, Nigel and H\"ofer, Marius and Huss, Alexander},
    title = "{Isolated photon and photon+jet production at NNLO QCD accuracy}",
    eprint = "1904.01044",
    archivePrefix = "arXiv",
    primaryClass = "hep-ph",
    reportNumber = "IPPP-19-23, ZU-TH-13-19, IPPP/19/23, ZU-TH 13/19, CERN-TH-2019-034",
    doi = "10.1007/JHEP04(2020)166",
    journal = "JHEP",
    volume = "04",
    pages = "166",
    year = "2020"
}

@article{Cacciari:2008gp,
    author = "Cacciari, Matteo and Salam, Gavin P. and Soyez, Gregory",
    title = "{The anti-$k_t$ jet clustering algorithm}",
    eprint = "0802.1189",
    archivePrefix = "arXiv",
    primaryClass = "hep-ph",
    reportNumber = "LPTHE-07-03",
    doi = "10.1088/1126-6708/2008/04/063",
    journal = "JHEP",
    volume = "04",
    pages = "063",
    year = "2008"
}

@article{Metodiev:2018ftz,
    author = "Metodiev, Eric M. and Thaler, Jesse",
    title = "{Jet Topics: Disentangling Quarks and Gluons at Colliders}",
    eprint = "1802.00008",
    archivePrefix = "arXiv",
    primaryClass = "hep-ph",
    reportNumber = "MIT-CTP-4979",
    doi = "10.1103/PhysRevLett.120.241602",
    journal = "Phys. Rev. Lett.",
    volume = "120",
    number = "24",
    pages = "241602",
    year = "2018"
}

@article{Andersson:1988gp,
    author = {Andersson, Bo and Gustafson, Gosta and L\"onnblad, Leif and Pettersson, Ulf},
    title = "{Coherence Effects in Deep Inelastic Scattering}",
    reportNumber = "LU-TP-88-14",
    doi = "10.1007/BF01550942",
    journal = "Z. Phys. C",
    volume = "43",
    pages = "625",
    year = "1989"
}

@article{Larkoski:2014wba,
    author = "Larkoski, Andrew J. and Marzani, Simone and Soyez, Gregory and Thaler, Jesse",
    title = "{Soft Drop}",
    eprint = "1402.2657",
    archivePrefix = "arXiv",
    primaryClass = "hep-ph",
    reportNumber = "MIT-CTP-4531, DCPT-14-24, IPPP-14-12",
    doi = "10.1007/JHEP05(2014)146",
    journal = "JHEP",
    volume = "05",
    pages = "146",
    year = "2014"
}

@Article{compete,
      author         = "{COMPETE Collaboration} and Cudell, J. R. and others",
      title          = "{Benchmarks for the Forward Observables at RHIC, the
                        Tevatron-Run II, and the LHC}",
      collaboration  = "COMPETE Collaboration",
      doi = "10.1103/PhysRevLett.89.201801",
      journal        = "Phys. Rev. Lett.",
      volume         = "89",
      pages          = "201801",
      year           = "2002",
      eprint         = "hep-ph/0206172",
      archivePrefix  = "arXiv",
      primaryClass   = "hep-ph",
      SLACcitation   = "%%CITATION = HEP-PH/0206172;%%",
}

@article{TOTEM_2p5km,
    author = {{TOTEM Collaboration}},
    collaboration = "TOTEM",
    title = "{First determination of the ${\rho }$ parameter at ${\sqrt{s} = 13}$ TeV: probing the existence of a colourless C-odd three-gluon compound state}",
    eprint = "1812.04732",
    archivePrefix = "arXiv",
    primaryClass = "hep-ex",
    reportNumber = "CERN-EP-2017-335, CERN-EP-2017-335-v3",
    doi = "10.1140/epjc/s10052-019-7223-4",
    journal = "Eur. Phys. J. C",
    volume = "79",
    number = "9",
    pages = "785",
    year = "2019"
}

@article{TOTEM_D0,
    author = "{D0 and TOTEM Collaborations}",
    xcollaboration = "D0, TOTEM",
    title = "{Odderon Exchange from Elastic Scattering Differences between $pp$ and $p \bar{p}$ Data at 1.96~TeV and from $pp$ Forward Scattering Measurements}",
    eprint = "2012.03981",
    archivePrefix = "arXiv",
    primaryClass = "hep-ex",
    reportNumber = "FERMILAB-PUB-20-568-E, ~CERN-EP-2020-236",
    doi = "10.1103/PhysRevLett.127.062003",
    journal = "Phys. Rev. Lett.",
    volume = "127",
    number = "6",
    pages = "062003",
    year = "2021"
    }

@article{Colangelo:1999zn,
      author         = "Colangelo, Pietro and De Fazio, Fulvia",
      title          = "{Using heavy quark spin symmetry in semileptonic $B_c$
                        decays}",
      journal        = "Phys. Rev. D",
      volume         = "61",
      pages          = "034012",
      doi            = "10.1103/PhysRevD.61.034012",
      year           = "2000",
      eprint         = "hep-ph/9909423",
      archivePrefix  = "arXiv",
      primaryClass   = "hep-ph",
      reportNumber   = "BARI-TH-99-351, UGVA-DPT-1999-09-1051",
      SLACcitation   = "%%CITATION = HEP-PH/9909423;%%",
}

@booklet{Kiselev:2002vz,
      author         = "Kiselev, V. V.",
      title          = "{Exclusive decays and lifetime of $B_c$ meson in QCD sum
                        rules}",
      year           = "2002",
      eprint         = "hep-ph/0211021",
      archivePrefix  = "arXiv",
      primaryClass   = "hep-ph",
      SLACcitation   = "%%CITATION = HEP-PH/0211021;%%",
}

@article{Dubnicka:2017job,
    author = {Dubni\v{c}ka, Stanislav and Dubni\v{c}kov\'a, Anna Z. and Issadykov, Aidos and Ivanov, Mikhail A. and Liptaj, Andrej},
    title = "{Study of $B_c$ decays into charmonia and $D$ mesons}",
    eprint = "1708.09607",
    archivePrefix = "arXiv",
    primaryClass = "hep-ph",
    doi = "10.1103/PhysRevD.96.076017",
    journal = "Phys. Rev. D",
    volume = "96",
    number = "7",
    pages = "076017",
    year = "2017"
}

@article{Dhir:2008hh,
      author         = "Dhir, Rohit and Verma, R. C.",
      title          = "{$B_c$ meson form factors and $B_c \to PV$ decays
                        involving flavor dependence of transverse quark momentum}",
      journal        = "Phys. Rev. D",
      volume         = "79",
      pages          = "034004",
      doi            = "10.1103/PhysRevD.79.034004",
      year           = "2009",
      eprint         = "0810.4284",
      archivePrefix  = "arXiv",
      primaryClass   = "hep-ph",
      SLACcitation   = "%%CITATION = ARXIV:0810.4284;%%",
}

@article{Ke:2013yka,
      author         = "Ke, Hong-Wei and Liu, Tan and Li, Xue-Qian",
      title          = "{Transitions of $B_c\rightarrow \psi(1S,2S)$ and the 
                         modified harmonic oscillator wave function in the light front quark model}",
      journal        = "Phys. Rev. D",
      volume         = "89",
      pages          = "017501",
      doi            = "10.1103/PhysRevD.89.017501",
      year           = "2014",
      eprint         = "1307.5925",
      archivePrefix  = "arXiv",
      primaryClass   = "hep-ph",
      SLACcitation   = "%%CITATION = ARXIV:1307.5925;%%",
}

@article{Rui:2014tpa,
      author         = "Rui, Zhou and Zou, Zhi-Tian",
      title          = "{S-wave ground state charmonium decays of $B_c$ mesons in
                        the perturbative QCD approach}",
      journal        = "Phys. Rev. D",
      volume         = "90",
      pages          = "114030",
      doi            = "10.1103/PhysRevD.90.114030",
      year           = "2014",
      eprint         = "1407.5550",
      archivePrefix  = "arXiv",
      primaryClass   = "hep-ph",
      SLACcitation   = "%%CITATION = ARXIV:1407.5550;%%",
}

@article{Kar:2013fna,
      author         = "Kar, Susmita and Dash, P.C. and Priyadarsini, M. and
                        Naimuddin, Sk. and Barik, N.",
      title          = "{Nonleptonic $B_c \to VV$ decays}",
      journal        = "Phys. Rev. D",
      volume         = "88",
      pages          = "094014",
      doi            = "10.1103/PhysRevD.88.094014",
      year           = "2013",
      SLACcitation   = "%%CITATION = PHRVA,D88,094014;%%",
}

@article{Nayak:2022qaq,
    author = "Nayak, Lopamudra and Dash, P. C. and Kar, Susmita and Barik, N.",
    title = "{Exclusive nonleptonic $B_c$-meson decays to S-wave charmonium states}",
    eprint = "2202.01167",
    archivePrefix = "arXiv",
    primaryClass = "hep-ph",
    doi = "10.1103/PhysRevD.105.053007",
    journal = "Phys. Rev. D",
    volume = "105",
    number = "5",
    pages = "053007",
    year = "2022"
}

@article{Mohammadi:2018jcp,
    author = "Mohammadi, Behnam",
    title = "{The branching fraction calculations of $B_c^+ \to \psi(2S)\pi^+, B_c^+ \to J/\psi K^+$ and $B_c^+ \to J/\psi D_s^+$ decays relative to that of the $B_c^+ \to J/\psi \pi^+$ mode}",
    doi = "10.1142/S0217751X18500446",
    journal = "Int. J. Mod. Phys. A",
    volume = "33",
    number = "08",
    pages = "1850044",
    year = "2018",
    related        = "Mohammadi:2018jcp-err", 
    relatedstring  = "Erratum:", 
}

@article{pdg2020,
    author = "{Particle Data Group Collaboration}",
    collaboration = "Particle Data Group",
    title = "{Review of Particle Physics}",
    doi = "10.1093/ptep/ptaa104",
    journal = "PTEP",
    volume = "2020",
    number = "8",
    pages = "083C01",
    year = "2020"
}

@article{HFLAV:2022esi,
author = "{Heavy Flavor Averaging Group Collaboration,  Y. S. Amhis et al.}",
collaboration = "HFLAV",
title = "{Averages of b-hadron, c-hadron, and \ensuremath{\tau}-lepton properties as of 2021}",
eprint = "2206.07501",
archivePrefix = "arXiv",
primaryClass = "hep-ex",
doi = "10.1103/PhysRevD.107.052008",
journal = "Phys. Rev. D",
volume = "107",
number = "5",
pages = "052008",
year = "2023"
}

@article{BOTELLA20071,
author = "Botella, Francisco J. and Branco, Gustavo C. and Nebot, Miguel",
    title = "{CP violation and limits on New Physics including recent $B_s$ measurements}",
    eprint = "hep-ph/0608100",
    archivePrefix = "arXiv",
    reportNumber = "IFIC-06-22, FTUV-06-0809",
    doi = "10.1016/j.nuclphysb.2006.12.022",
    journal = "Nucl. Phys. B",
    volume = "768",
    pages = "1--20",
    year = "2007"
}

@article{UTFit2006,
    author = "{UTfit Collaboration,  M. Bona et al.}",
    collaboration = "UTfit",
    title = "{The Unitarity Triangle Fit in the Standard Model and Hadronic Parameters from Lattice QCD: A Reappraisal after the Measurements of $\Delta m_s$ and BR($B \to \tau \nu_{\tau}$)}",
    eprint = "hep-ph/0606167",
    archivePrefix = "arXiv",
    doi = "10.1088/1126-6708/2006/10/081",
    journal = "JHEP",
    volume = "10",
    pages = "081",
    year = "2006"
}

@article{Aaij:2013Mix,
      author         = "{LHCb Collaboration}",
      title          = "{Precision measurement of the  $B^0_s - \bar{B}^0_s$ oscillation frequency with the decay 
                        $B^0_s \to D_s^- \pi^+$ }",
      collaboration  = "LHCb",
      journal        = "New J. Phys.",
      volume         = "15",
      pages          = "053021",
      doi = {10.1088/1367-2630/15/5/053021},
      year           = "2013",
      eprint         = "1304.4741",
      archivePrefix  = "arXiv",
      primaryClass   = "hep-ex",
      SLACcitation   = "%%CITATION = ARXIV:1304.4741;%%",
}

@article{Aaij2019,
      author         = "{LHCb Collaboration}",
      title          = "{Updated measurement of time-dependent CP-violating
                        observables in $B^{0}_{s}\to J/\psi K^+ K^-$ decays}",
      collaboration  = "LHCb",
      journal        = "Eur. Phys. J. C",
      volume         = "79",
      year           = "2019",
      number         = "8",
      pages          = "706",
      doi            = "10.1140/epjc/s10052-019-7159-8",
      eprint         = "1906.08356",
      archivePrefix  = "arXiv",
      primaryClass   = "hep-ex",
      reportNumber   = "LHCb-PAPER-2019-013, CERN-EP-2019-108",
      SLACcitation   = "%%CITATION = ARXIV:1906.08356;%%",
      related        = "Aaij2019-err",
      relatedstring  = "Erratum:"
}

@article{Aaij:2016psitwoS,
      author         = "{LHCb Collaboration}",
      title          = "{First study of the CP -violating phase and decay-width
                        difference in $B_s^0\to\psi(2S)\phi$ decays}",
      collaboration  = "LHCb",
      journal        = "Phys. Lett. B",
      volume         = "762",
      year           = "2016",
      pages          = "253-262",
      doi            = "10.1016/j.physletb.2016.09.028",
      eprint         = "1608.04855",
      archivePrefix  = "arXiv",
      primaryClass   = "hep-ex",
      reportNumber   = "CERN-EP-2016-192, LHCB-PAPER-2016-027",
      SLACcitation   = "%%CITATION = ARXIV:1608.04855;%%"
}

@article{Aaij:2014Ds,
      author        = "{LHCb Collaboration}",
      title         = "{Measurement of the CP-Violating Phase $\phi_s$ in
                       $\bar{B}^{0}_{s}\to D_{s}^{+}D_{s}^{-}$ decays}",
      journal       = "Phys. Rev. Lett.",
      collaboration = "LHCb collaboration",
      number        = "LHCB-PAPER-2014-051. LHCB-PAPER-2014-051.
                       CERN-PH-EP-2014-223",
      volume        = "113",
      pages         = "211801,",
      month         = "9",
      year          = "2014",
      reportNumber  = "LHCB-PAPER-2014-051",
%      url           = "https://cds.cern.ch/record/1756036",
      doi           = "10.1103/PhysRevLett.113.211801",
      eprint         = "1409.4619",
      archivePrefix  = "arXiv",
      primaryClass   = "hep-ex",
      SLACcitation   = "%%CITATION = ARXIV:1409.4619;%%"

}

@article{Aaij:2014dka,
      author         = "{LHCb Collaboration}",
      title          = "{Measurement of the CP-violating phase $\phi_s$ in
                        $\overline{B}^0_s\rightarrow J/\psi \pi^+\pi^-$ decays}",
      collaboration  = "LHCb",
      journal        = "Phys. Lett. B",
      volume         = "736",
      year           = "2014",
      pages          = "186-195",
      doi            = "10.1016/j.physletb.2014.06.079",
      eprint         = "1405.4140",
      archivePrefix  = "arXiv",
      primaryClass   = "hep-ex",
      reportNumber   = "CERN-PH-EP-2014-086, LHCB-PAPER-2014-019",
      SLACcitation   = "%%CITATION = ARXIV:1405.4140;%%"
}

@article{Aaij:2019mhf,
      author         = "{LHCb Collaboration}",
      title          = "{Measurement of the $CP$-violating phase $\phi_s$ from
                        $B_{s}^{0}\to J/\psi\pi^+\pi^-$ decays in 13 TeV $pp$
                        collisions}",
      collaboration  = "LHCb",
      journal        = "Phys. Lett. B",
      volume         = "797",
      year           = "2019",
      pages          = "134789",
      doi            = "10.1016/j.physletb.2019.07.036",
      eprint         = "1903.05530",
      archivePrefix  = "arXiv",
      primaryClass   = "hep-ex",
      reportNumber   = "LHCb-PAPER-2019-003; CERN-EP-2019-037, CERN-EP-2019-037,
                        LHCb-PAPER-2019-003",
      SLACcitation   = "%%CITATION = ARXIV:1903.05530;%%"
}

@article{PhysRevD.85.032006,
author         = "{D0 Collaboration}",
  title = {Measurement of the $CP$-violating phase ${\ensuremath{\phi}}_{s}^{J/\ensuremath{\psi}\ensuremath{\phi}}$ using the flavor-tagged decay ${B}_{s}^{0}\ensuremath{\rightarrow}J/\ensuremath{\psi}\ensuremath{\phi}$ in $8\text{ }\text{ }{\mathrm{fb}}^{\ensuremath{-}1}$ of $p\overline{p}$ collisions},
collaboration = {The D0 Collaboration},
  journal = {Phys. Rev. D},
  volume = {85},
  pages = {032006},
  numpages = {24},
  year = {2012},
  publisher = {American Physical Society},
  doi = {10.1103/PhysRevD.85.032006},
      eprint = "1109.3166",
    archivePrefix = "arXiv",
    primaryClass = "hep-ex",
}

@article{PhysRevLett.109.171802,
author         = "{CDF Collaboration}",
  title = {Measurement of the Bottom-Strange Meson Mixing Phase in the Full CDF Data Set},
 collaboration = {CDF Collaboration},
  journal = {Phys. Rev. Lett.},
  volume = {109},
  pages = {171802},
  numpages = {8},
  year = {2012},
  publisher = {American Physical Society},
  doi = {10.1103/PhysRevLett.109.171802},
      eprint = "1208.2967",
    archivePrefix = "arXiv",
    primaryClass = "hep-ex",
}

@book{Artuso:2022ijh,
    author = "Artuso, Marina and Isidori, Gino and Stone, Sheldon",
    title = "{New Physics in b Decays}",
    doi = "10.1142/12696",
    isbn = "978-981-12-5129-0",
    %isbn = "978-981-12-5129-0, 978-981-12-5131-3",
    publisher = "World Scientific",
    month = "5",
    year = "2022"
}

@article{Lenz:2019lvd,
    author = "Lenz, Alexander and Tetlalmatzi-Xolocotzi, Gilberto",
    title = "{Model-independent bounds on new physics effects in non-leptonic tree-level decays of B-mesons}",
    eprint = "1912.07621",
    archivePrefix = "arXiv",
    primaryClass = "hep-ph",
    reportNumber = "IPPP/19/49, Nikhef-2019-054, SI-HEP-2019, P3H-19-044",
    doi = "10.1007/JHEP07(2020)177",
    journal = "JHEP",
    volume = "07",
    pages = "177",
    year = "2020"
}

@book{CERNYellow:2019,
    editor = "Dainese, Andrea and Mangano, Michelangelo and Meyer, Andreas B. and Nisati, Aleandro and Salam, Gavin and Vesterinen, Mika Anton",
    title = "{Report on the Physics at the HL-LHC, and Perspectives for the HE-LHC}",
    reportNumber = "CERN-2019-007",
    doi = "10.23731/CYRM-2019-007",
    isbn = "978-92-9083-549-3",
    publisher = "CERN",
    address = "Geneva, Switzerland",
    series = "CERN Yellow Reports: Monographs",
    volume = "7/2019",
    year = "2019"
}

@article{Bobeth:2013uxa,
      author         = "{C. Bobeth et al.}",
      title          = "$B_{s,d} \to \ell^+ \ell^-$ in the Standard Model with Reduced
                        Theoretical Uncertainty",
      journal        = "Phys. Rev. Lett.",
      volume         = "112",
      pages          = "101801",
      doi            = "10.1103/PhysRevLett.112.101801",
      year           = "2014",
      eprint         = "1311.0903",
      archivePrefix  = "arXiv",
      primaryClass   = "hep-ph",
      reportNumber   = "FLAVOUR(267104)-ERC-53, LTH-990, SFB-CPP-13-82,
                        TTP13-033",
      SLACcitation   = "%%CITATION = ARXIV:1311.0903;%%"
}

@article{Huang:1998vb,
  title = "{Promising process to distinguish supersymmetric models with large tan \ensuremath{\beta} from the standard model:
  $B \to X_{s} \mu^+ \mu^-$}",
  author = "{C.-S. Huang, W. Liao, and Q.-S. Yan}",
  journal = "{Phys. Rev. D}",
  volume = "{59}",
 % issue = "{1}",
  pages = "{011701}",
  numpages = "{4}",
  year = {1998},
  publisher = "{American Physical Society}",
  doi = "10.1103/PhysRevD.59.011701",
  eprint         = "hep-ph/9803460",
  archivePrefix  = "arXiv",
  primaryClass   = "hep-ph"
}

@article{Hamzaoui:1998nu,
  title = "{Higgs-boson-mediated FCNC in supersymmetric models with large tan \ensuremath{\beta}}",
  author = "{C. Hamzaoui, M. Pospelov and M. Toharia}",
  journal = "{Phys. Rev. D}",
  volume = "{59}",
 % issue = "{9}",
  pages = "{095005}",
  numpages = "{8}",
  year = {1999},
  publisher = "{American Physical Society}",
  doi = "10.1103/PhysRevD.59.095005",
  eprint         = "hep-ph/9807350",
  archivePrefix  = "arXiv",
  primaryClass   = "hep-ph"
%  url = {https://link.aps.org/doi/10.1103/PhysRevD.59.095005}
}

@article{Choudhury:1998ze,
      author         = "{S. Rai Choudhury and N. Gaur}",
      title          = "{Dileptonic decay of $B_s$ meson in SUSY models with large tan\ensuremath{\beta}}",
      journal        = "Phys. Lett. B",
      volume         = "451",
      year           = "1999",
      pages          = "86-92",
      doi            = "10.1016/S0370-2693(99)00203-8",
      eprint         = "hep-ph/9810307",
      archivePrefix  = "arXiv",
      primaryClass   = "hep-ph",
      SLACcitation   = "%%CITATION = HEP-PH/9810307;%%"
}

@article{Babu:1999hn,
      author         = "Babu, K. S. and Kolda, Christopher",
      title          = "{Higgs-Mediated $B^0 \to \mu^{+} \mu^{-}$ in Minimal
                        Supersymmetry}",
      journal        = "Phys. Rev. Lett.",
      volume         = "84",
      year           = "2000",
      pages          = "228-231",
      doi            = "10.1103/PhysRevLett.84.228",
      eprint         = "hep-ph/9909476",
      archivePrefix  = "arXiv",
      primaryClass   = "hep-ph",
      reportNumber   = "OSU-HEP-99-10, LBNL-44284, UCB-PTH-99-43, LBL-44284",
      SLACcitation   = "%%CITATION = HEP-PH/9909476;%%"
}

@article{Choudhury:2005rz,
      author         = "Rai Choudhury, S. and Cornell, Alan S. and Gaur, Naveen
                        and Joshi, Girish C.",
      title          = "{Signatures of new physics in dileptonic B-decays}",
      journal        = "Int. J. Mod. Phys. A",
      volume         = "21",
      year           = "2006",
      pages          = "2617-2634",
      doi            = "10.1142/S0217751X06029491",
      eprint         = "hep-ph/0504193",
      archivePrefix  = "arXiv",
      primaryClass   = "hep-ph",
      reportNumber   = "YITP-05-18",
      SLACcitation   = "%%CITATION = HEP-PH/0504193;%%"
}

@article{DAmbrosio:2002ex,
      author         = "D'Ambrosio, G. and Giudice, G. F. and Isidori, G. and
                        Strumia, A.",
      title          = "{Minimal flavour violation: an effective field theory
                        approach}",
      journal        = "Nucl. Phys. B",
      volume         = "645",
      year           = "2002",
      pages          = "155-187",
      doi            = "10.1016/S0550-3213(02)00836-2",
      eprint         = "hep-ph/0207036",
      archivePrefix  = "arXiv",
      primaryClass   = "hep-ph",
      reportNumber   = "CERN-TH-2002-147, IFUP-TH-2002-17",
      SLACcitation   = "%%CITATION = HEP-PH/0207036;%%"
}

@article{Buras:2003td,
      author         = "Buras, Andrzej J.",
      title          = "{Relations between $\Delta M_{s,d}$ and $B_{s,d} \to \mu^{+} \mu^{-}$
                        in models with minimal flavour violation}",
      journal        = "Phys. Lett. B",
      volume         = "566",
      year           = "2003",
      pages          = "115-119",
      doi            = "10.1016/S0370-2693(03)00561-6",
      eprint         = "hep-ph/0303060",
      archivePrefix  = "arXiv",
      primaryClass   = "hep-ph",
      reportNumber   = "TUM-HEP-502-03",
      SLACcitation   = "%%CITATION = HEP-PH/0303060;%%"
}

@Article{LHCb_2022,
	author = "{LHCb Collaboration}",
	title = "{Analysis of Neutral $B$-Meson Decays into Two Muons}",
	journal = "Phys. Rev. Lett.",
	volume = "128",
	year = "2022",
	pages = "041801",
	doi = "10.1103/physrevlett.128.041801",
	eprint         = "2108.09284",
     	archivePrefix  = "arXiv",
     	primaryClass   = "hep-ex",
}

@Article{LHCb_2017,
	author = "{LHCb Collaboration}",
	title = "{Measurement of the \(B^0_{s} \to \mu^+\mu^-\) branching fraction and effective lifetime and search for  \(B^0_{s} \to \mu^+\mu^-\) decays}",
	journal = "Phys. Rev. Lett.",
	volume = "118",
	year = "2017",
	pages = "191801",
	doi = "10.1103/physrevlett.128.041801", 
	eprint         = "1703.05747",
     	archivePrefix  = "arXiv",
     	primaryClass   = "hep-ex",
}

@Article{CMS-BPH-21-006,
%    author         = "{CMS Collaboration}",
%    title          = "{Measurement of the \(B^0_{s} \to \mu^+\mu^-\) decay properties and search for the \(B^0 \to \mu^+\mu^-\) decay in proton--proton collisions at \(\sqrt{s} = 13\,\text{TeV}\)}",
%    journal        = "Phys. Lett. B",
%    volume         = "842",
%    year           = "2023",
%    pages          = "137955",
%    doi            = "10.1016/j.physletb.2023.137955",
%    reportNumber   = "CERN-EP-2022-270",
%    eprint         = "2212.10311",
%    archivePrefix  = "arXiv",
%    primaryClass   = "hep-ex",
%}

@Article{CMS-BPH-16-004,
%    author         = "{CMS Collaboration}",
%    title          = "{Measurement of properties of  \(B^0_{s} \to \mu^+\mu^-\) decays and search for the \(B^0 \to \mu^+\mu^-\) with CMS experiment}",
%    journal        = "JHEP 04",
%    volume         = "04",
%    year           = "2022",
%    pages          = "188",
%    doi            = "10.1007/HEP04(2020)188",  	
%    reportNumber   = "CERN-EP-2019-215",
%    eprint         = "1910.12127",
%    archivePrefix  = "arXiv",
%    primaryClass   = "hep-ex",
%}

@Booklet{ATLAS-CONF-2020-049,
%    author         = "{ATLAS, CMS and LHCb Collaborations}",
%    title          = "{Combination of the ATLAS, CMS and LHCb results on the \(B^0_{(s)}\to\mu^+\mu^-\) decays.}",
%    howpublished   = "{ATLAS-CONF-2020-049}",
%    url            = "https://cds.cern.ch/record/2727216",
%    year           = "2020",
}

@article{Davidson:2010uu,
      author         = "Davidson, Sacha and Descotes-Genon, Sebastien",
      title          = "{"Minimal Flavour Violation" for Leptoquarks}",
      journal        = "JHEP",
      volume         = "11",
      year           = "2010",
      pages          = "073",
      doi            = "10.1007/JHEP11(2010)073",
      eprint         = "1009.1998",
      archivePrefix  = "arXiv",
      primaryClass   = "hep-ph",
      reportNumber   = "LPT-ORSAY-10-72",
      SLACcitation   = "%%CITATION = ARXIV:1009.1998;%%"
}

@article{Guadagnoli:2013mru,
      author         = "Guadagnoli, Diego and Isidori, Gino",
      title          = "{$B(B_s \to \mu^{+} \mu^{-})$ as an electroweak precision
                        test}",
      journal        = "Phys. Lett. B",
      volume         = "724",
      year           = "2013",
      pages          = "63-67",
      doi            = "10.1016/j.physletb.2013.05.054",
      eprint         = "1302.3909",
      archivePrefix  = "arXiv",
      primaryClass   = "hep-ph",
      reportNumber   = "LAPTH-010-13, CERN-PH-TH-2013-028",
      SLACcitation   = "%%CITATION = ARXIV:1302.3909;%%"
}

@article{Olsen:2017bmm,
    author = "Olsen, Stephen Lars and Skwarnicki, Tomasz and Zieminska, Daria",
    title = "{Nonstandard heavy mesons and baryons: Experimental evidence}",
    eprint = "1708.04012",
    archivePrefix = "arXiv",
    primaryClass = "hep-ph",
    doi = "10.1103/RevModPhys.90.015003",
    journal = "Rev. Mod. Phys.",
    volume = "90",
    number = "1",
    pages = "015003",
    year = "2018"
}

@article{new_phys_via_eff_lifetime,
    author = "De Bruyn, Kristof and Fleischer, Robert and Knegjens, Robert and Koppenburg, Patrick and Merk, Marcel and Pellegrino, Antonio and Tuning, Niels",
    title = "{Probing New Physics via the $B^0_s\to \mu^+\mu^-$ Effective Lifetime}",
    eprint = "1204.1737",
    archivePrefix = "arXiv",
    primaryClass = "hep-ph",
    reportNumber = "NIKHEF-2012-006",
    doi = "10.1103/PhysRevLett.109.041801",
    journal = "Phys. Rev. Lett.",
    volume = "109",
    pages = "041801",
    year = "2012"
}

@article{Beneke_Power_enhance_SM_Bmumu,
    author = "Beneke, Martin and Bobeth, Christoph and Szafron, Robert",
    title = "{Power-enhanced leading-logarithmic QED corrections to $B_q \to \mu^+\mu^-$}",
    eprint = "1908.07011",
    archivePrefix = "arXiv",
    primaryClass = "hep-ph",
    reportNumber = "TUM-HEP-1212/19",
    doi = "10.1007/JHEP10(2019)232",
    journal = "JHEP",
    volume = "10",
    pages = "232",
    year = "2019",
    related        = "Beneke_Power_enhance_SM_Bmumu-err", 
    relatedstring  = "Erratum:", 
}

@article{np_in_flavour,
    author = "Straub, David M.",
    title = "{New Physics Searches in Flavour Physics}",
    eprint = "1107.0266",
    archivePrefix = "arXiv",
    primaryClass = "hep-ph",
    doi = "10.1393/ncc/i2012-11132-x",
    journal = "Nuovo Cim. C",
    volume = "035N1",
    pages = "249--256",
    year = "2012"
}

@article{neyman,
    author = "Neyman, J.",
    title = "{Outline of a Theory of Statistical Estimation Based on the Classical Theory of Probability}",
    doi = "10.1098/rsta.1937.0005",
    journal = "Phil. Trans. Roy. Soc. Lond. A",
    volume = "236",
    number = "767",
    pages = "333--380",
    year = "1937"
}

@article{Cacciari:2001td,
    author = "Cacciari, Matteo and Frixione, Stefano and Nason, Paolo",
    title = "{The $p_T$ spectrum in heavy flavor photoproduction}",
    eprint = "hep-ph/0102134",
    archivePrefix = "arXiv",
    reportNumber = "BICOCCA-FT-01-01, GEF-TH-2-01, YITP-SB-01-01",
    doi = "10.1088/1126-6708/2001/03/006",
    journal = "JHEP",
    volume = "03",
    pages = "006",
    year = "2001"
}

@article{Cacciari:2012ny,
    author = "Cacciari, Matteo and Frixione, Stefano and Houdeau, Nicolas and Mangano, Michelangelo L. and Nason, Paolo and Ridolfi, Giovanni",
    title = "{Theoretical predictions for charm and bottom production at the LHC}",
    eprint = "1205.6344",
    archivePrefix = "arXiv",
    primaryClass = "hep-ph",
    reportNumber = "CERN-PH-TH-2011-227",
    doi = "10.1007/JHEP10(2012)137",
    journal = "JHEP",
    volume = "10",
    pages = "137",
    year = "2012"
}

@article{Butenschoen:2010rq,
    author = {Butensch\"on, Mathias and Kniehl, Bernd A.},
    title = "{Reconciling $J/\psi$ Production at HERA, RHIC, Tevatron, and LHC with Nonrelativistic QCD Factorization at Next-to-Leading Order}",
    eprint = "1009.5662",
    archivePrefix = "arXiv",
    primaryClass = "hep-ph",
    reportNumber = "DESY-10-101",
    doi = "10.1103/PhysRevLett.106.022003",
    journal = "Phys. Rev. Lett.",
    volume = "106",
    pages = "022003",
    year = "2011"
}

@article{Butenschoen:2011yh,
    author = {Butensch\"on, Mathias and Kniehl, Bernd A.},
    title = "{World data of J/$\psi$ production consolidate nonrelativistic QCD factorization at next-to-leading order}",
    eprint = "1105.0820",
    archivePrefix = "arXiv",
    primaryClass = "hep-ph",
    reportNumber = "DESY-11-046",
    doi = "10.1103/PhysRevD.84.051501",
    journal = "Phys. Rev. D",
    volume = "84",
    pages = "051501",
    year = "2011"
}

@article{Butenschoen:2022qka,
    author = {Butensch\"on, Mathias and Kniehl, Bernd A.},
    title = "{Global analysis of \ensuremath{\psi}(2S) inclusive hadroproduction at next-to-leading order in nonrelativistic-QCD factorization}",
    eprint = "2207.09346",
    archivePrefix = "arXiv",
    primaryClass = "hep-ph",
    reportNumber = "DESY-22-119",
    doi = "10.1103/PhysRevD.107.034003",
    journal = "Phys. Rev. D",
    volume = "107",
    number = "3",
    pages = "034003",
    year = "2023"
}

@article{Lipatov:2019oxs,
    author = "Lipatov, A. V. and Malyshev, M. A. and Baranov, S. P.",
    title = "{Particle Event Generator: A Simple-in-Use System PEGASUS version 1.0}",
    eprint = "1912.04204",
    archivePrefix = "arXiv",
    primaryClass = "hep-ph",
    doi = "10.1140/epjc/s10052-020-7898-6",
    journal = "Eur. Phys. J. C",
    volume = "80",
    number = "4",
    pages = "330",
    year = "2020"
}

@article{Baranov:2019lhm,
    author = "Baranov, S. P. and Lipatov, A. V.",
    title = "{Are there any challenges in the charmonia production and polarization at the LHC?}",
    eprint = "1906.07182",
    archivePrefix = "arXiv",
    primaryClass = "hep-ph",
    doi = "10.1103/PhysRevD.100.114021",
    journal = "Phys. Rev. D",
    volume = "100",
    number = "11",
    pages = "114021",
    year = "2019"
}

@article{Cheung:2021epq,
    author = "Cheung, Vincent and Vogt, Ramona",
    title = "{Production and polarization of direct J/\ensuremath{\psi} to O(\ensuremath{\alpha_s^3}) in the improved color evaporation model in collinear factorization}",
    eprint = "2102.09118",
    archivePrefix = "arXiv",
    primaryClass = "hep-ph",
    reportNumber = "LLNL-JRNL-819612",
    doi = "10.1103/PhysRevD.104.094026",
    journal = "Phys. Rev. D",
    volume = "104",
    number = "9",
    pages = "094026",
    year = "2021"
}

@article{Bolzoni:2013tca,
    author = "Bolzoni, Paolo and Kniehl, Bernd A. and Kramer, Gustav",
    title = "{Inclusive J/$\psi$ and $\psi$(2S) production from b-hadron decay in p$\bar{p}$ and pp collisions}",
    eprint = "1309.3389",
    archivePrefix = "arXiv",
    primaryClass = "hep-ph",
    reportNumber = "DESY-13-159",
    doi = "10.1103/PhysRevD.88.074035",
    journal = "Phys. Rev. D",
    volume = "88",
    number = "7",
    pages = "074035",
    year = "2013"
}

@article{Baranov:2018cmp,
    author = "Baranov, S. P. and Lipatov, A. V. and Malyshev, M. A.",
    title = "{Associated non-prompt $J/\psi + \mu$ and $J/\psi + J/\psi$ production at LHC as a test for TMD gluon density}",
    eprint = "1808.06233",
    archivePrefix = "arXiv",
    primaryClass = "hep-ph",
    reportNumber = "DESY-18-138, DESY 18-138",
    doi = "10.1140/epjc/s10052-018-6297-8",
    journal = "Eur. Phys. J. C",
    volume = "78",
    number = "10",
    pages = "820",
    year = "2018"
}

@article{H1:2015ubc,
    xauthor = "Abramowicz, H. and others",
    author = "{H1 and ZEUS Collaborations}",
    title = "{Combination of measurements of inclusive deep inelastic ${e^{\pm }p}$ scattering cross sections and QCD analysis of HERA data}",
    eprint = "1506.06042",
    archivePrefix = "arXiv",
    primaryClass = "hep-ex",
    reportNumber = "DESY-15-039",
    doi = "10.1140/epjc/s10052-015-3710-4",
    journal = "Eur. Phys. J. C",
    volume = "75",
    number = "12",
    pages = "580",
    year = "2015"
}

@Article{PERF-2007-01,
    author         = "{ATLAS Collaboration}",
    title          = "{The ATLAS Experiment at the CERN Large Hadron Collider}",
    journal        = "JINST",
    volume         = "3",
    year           = "2008",
    pages          = "S08003",
    doi            = "10.1088/1748-0221/3/08/S08003",
    primaryClass   = "hep-ex",
}

@Article{STDM-2011-33,
    author         = "{ATLAS Collaboration}",
    title          = "{Measurement of event shapes at large momentum transfer with the ATLAS detector in \(pp\) collisions at \(\sqrt{s} = 7\,\text{TeV}\)}",
    journal        = "Eur. Phys. J. C",
    volume         = "72",
    year           = "2012",
    pages          = "2211",
    doi            = "10.1140/epjc/s10052-012-2211-y",
    reportNumber   = "CERN-PH-EP-2012-119",
    eprint         = "1206.2135",
    archivePrefix  = "arXiv",
    primaryClass   = "hep-ex",
}

@Article{BPHY-2012-01,
    author         = "{ATLAS Collaboration}",
    title          = "{Study of the rare decays of \(B^0_s\) and \(B^0\) into muon pairs from data collected during the LHC Run~1 with the ATLAS detector}",
    journal        = "Eur. Phys. J. C",
    volume         = "76",
    year           = "2016",
    pages          = "513",
    doi            = "10.1140/epjc/s10052-016-4338-8",
    reportNumber   = "CERN-EP-2016-064",
    eprint         = "1604.04263",
    archivePrefix  = "arXiv",
    primaryClass   = "hep-ex",
}

@Article{HIGG-2012-27,
    author         = "{ATLAS Collaboration}",
    title          = "{Observation of a new particle in the search for the Standard Model Higgs boson with the ATLAS detector at the LHC}",
    journal        = "Phys. Lett. B",
    volume         = "716",
    year           = "2012",
    pages          = "1",
    doi            = "10.1016/j.physletb.2012.08.020",
    reportNumber   = "CERN-PH-EP-2012-218",
    eprint         = "1207.7214",
    archivePrefix  = "arXiv",
    primaryClass   = "hep-ex",
}

@Article{PERF-2012-04,
    author         = "{ATLAS Collaboration}",
    title          = "{Performance of \(b\)-jet identification in the ATLAS experiment}",
    journal        = "JINST",
    volume         = "11",
    year           = "2016",
    pages          = "P04008",
    doi            = "10.1088/1748-0221/11/04/P04008",
    reportNumber   = "CERN-PH-EP-2015-216",
    eprint         = "1512.01094",
    archivePrefix  = "arXiv",
    primaryClass   = "hep-ex",
}

@Article{STDM-2012-20,
    author         = "{ATLAS Collaboration}",
    title          = "{Precision measurement and interpretation of inclusive \(W^+\), \(W^-\) and \(Z/\gamma^*\) production cross sections with the ATLAS detector}",
    journal        = "Eur. Phys. J. C",
    volume         = "77",
    year           = "2017",
    pages          = "367",
    doi            = "10.1140/epjc/s10052-017-4911-9",
    reportNumber   = "CERN-EP-2016-272",
    eprint         = "1612.03016",
    archivePrefix  = "arXiv",
    primaryClass   = "hep-ex",
}

@Article{STDM-2012-23,
    author         = "{ATLAS Collaboration}",
    title          = "{Measurement of the \(Z/\gamma^*\) boson transverse momentum distribution in \(pp\) collisions at \(\sqrt{s} = 7\,\text{TeV}\) with the ATLAS detector}",
    journal        = "JHEP",
    volume         = "09",
    year           = "2014",
    pages          = "145",
    doi            = "10.1007/JHEP09(2014)145",
    reportNumber   = "CERN-PH-EP-2014-075",
    eprint         = "1406.3660",
    archivePrefix  = "arXiv",
    primaryClass   = "hep-ex",
}

@Article{STDM-2013-06,
    author         = "{ATLAS Collaboration}",
    title          = "{Evidence for Electroweak Production of \(W^{\pm}W^{\pm}jj\) in \(pp\) Collisions at \(\sqrt{s} = 8\,\text{TeV}\) with the ATLAS Detector}",
    journal        = "Phys. Rev. Lett.",
    volume         = "113",
    year           = "2014",
    pages          = "141803",
    doi            = "10.1103/PhysRevLett.113.141803",
    reportNumber   = "CERN-PH-EP-2014-079",
    eprint         = "1405.6241",
    archivePrefix  = "arXiv",
    primaryClass   = "hep-ex",
}

@Article{STDM-2013-10,
    author         = "{ATLAS Collaboration}",
    title          = "{Measurement of the total cross section from elastic scattering in \(pp\) collisions at \(\sqrt{s} = 7\,\text{TeV}\) with the ATLAS detector}",
    journal        = "Nucl. Phys. B",
    volume         = "889",
    year           = "2014",
    pages          = "486",
    doi            = "10.1016/j.nuclphysb.2014.10.019",
    reportNumber   = "CERN-PH-EP-2014-177",
    eprint         = "1408.5778",
    archivePrefix  = "arXiv",
    primaryClass   = "hep-ex",
}

@Article{STDM-2013-11,
    author         = "{ATLAS Collaboration}",
    title          = "{Measurement of the inclusive jet cross-section in proton--proton collisions at \(\sqrt{s} = 7\,\text{TeV}\) using \(4.5\,\text{fb}^{-1}\) of data with the ATLAS detector}",
    journal        = "JHEP",
    volume         = "02",
    year           = "2015",
    pages          = "153",
    doi            = "10.1007/JHEP02(2015)153",
    reportNumber   = "CERN-PH-EP-2014-155",
    eprint         = "1410.8857",
    archivePrefix  = "arXiv",
    primaryClass   = "hep-ex",
    related        = "STDM-2013-11-err",
    relatedstring  = "Erratum:",
}

@Article{BPHY-2014-04,
    author         = "{ATLAS Collaboration}",
    title          = "{Study of the \(B_c^+ \rightarrow J/\psi D_s^+\) and \(B_c^+ \rightarrow J/\psi D_s^{*+}\) decays with the ATLAS detector}",
    journal        = "Eur. Phys. J. C",
    volume         = "76",
    year           = "2016",
    pages          = "4",
    doi            = "10.1140/epjc/s10052-015-3743-8",
    reportNumber   = "CERN-PH-EP-2015-163",
    eprint         = "1507.07099",
    archivePrefix  = "arXiv",
    primaryClass   = "hep-ex",
}

@Article{PERF-2014-07,
    author         = "{ATLAS Collaboration}",
    title          = "{Topological cell clustering in the ATLAS calorimeters and its performance in LHC Run~1}",
    journal        = "Eur. Phys. J. C",
    volume         = "77",
    year           = "2017",
    pages          = "490",
    doi            = "10.1140/epjc/s10052-017-5004-5",
    reportNumber   = "CERN-PH-EP-2015-304",
    eprint         = "1603.02934",
    archivePrefix  = "arXiv",
    primaryClass   = "hep-ex",
}

@Article{STDM-2014-01,
    author         = "{ATLAS Collaboration}",
    title          = "{Measurements of \(Z\gamma\) and \(Z\gamma\gamma\) production in \(pp\) collisions at \(\sqrt{s} = 8\,\text{TeV}\) with the ATLAS detector}",
    journal        = "Phys. Rev. D",
    volume         = "93",
    year           = "2016",
    pages          = "112002",
    doi            = "10.1103/PhysRevD.93.112002",
    reportNumber   = "CERN-EP-2016-049",
    eprint         = "1604.05232",
    archivePrefix  = "arXiv",
    primaryClass   = "hep-ex",
}

@Article{STDM-2014-03,
    author         = "{ATLAS Collaboration}",
    title          = "{Measurement of transverse energy--energy correlations in multi-jet events in \(pp\) collisions at \(\sqrt{s} = 7\,\text{TeV}\) using the ATLAS detector and determination of the strong coupling constant \(\alpha_{\text{s}}(m_Z)\)}",
    journal        = "Phys. Lett. B",
    volume         = "750",
    year           = "2015",
    pages          = "427",
    doi            = "10.1016/j.physletb.2015.09.050",
    reportNumber   = "CERN-PH-EP-2015-177",
    eprint         = "1508.01579",
    archivePrefix  = "arXiv",
    primaryClass   = "hep-ex",
}

@Article{STDM-2014-10,
    author         = "{ATLAS Collaboration}",
    title          = "{Measurement of the angular coefficients in \(Z\)-boson events using electron and muon pairs from data taken at \(\sqrt{s} = 8\,\text{TeV}\) with the ATLAS detector}",
    journal        = "JHEP",
    volume         = "08",
    year           = "2016",
    pages          = "159",
    doi            = "10.1007/JHEP08(2016)159",
    reportNumber   = "CERN-EP-2016-087",
    eprint         = "1606.00689",
    archivePrefix  = "arXiv",
    primaryClass   = "hep-ex",
}

@Article{STDM-2014-18,
    author         = "{ATLAS Collaboration}",
    title          = "{Measurement of the \(W\)-boson mass in \(pp\) collisions at \(\sqrt{s} = 7\,\text{TeV}\) with the ATLAS detector}",
    journal        = "Eur. Phys. J. C",
    volume         = "78",
    year           = "2018",
    pages          = "110",
    doi            = "10.1140/epjc/s10052-017-5475-4",
    reportNumber   = "CERN-EP-2016-305",
    eprint         = "1701.07240",
    archivePrefix  = "arXiv",
    primaryClass   = "hep-ex",
    related        = "STDM-2014-18-err",
    relatedstring  = "Erratum:",
}

@Article{PERF-2015-09,
    author         = "{ATLAS Collaboration}",
    title          = "{Jet reconstruction and performance using particle flow with the ATLAS Detector}",
    journal        = "Eur. Phys. J. C",
    volume         = "77",
    year           = "2017",
    pages          = "466",
    doi            = "10.1140/epjc/s10052-017-5031-2",
    reportNumber   = "CERN-EP-2017-024",
    eprint         = "1703.10485",
    archivePrefix  = "arXiv",
    primaryClass   = "hep-ex",
}

@Article{STDM-2015-01,
    author         = "{ATLAS Collaboration}",
    title          = "{Measurement of the inclusive jet cross-sections in proton--proton collisions at \(\sqrt{s} = 8\,\text{TeV}\) with the ATLAS detector}",
    journal        = "JHEP",
    volume         = "09",
    year           = "2017",
    pages          = "020",
    doi            = "10.1007/JHEP09(2017)020",
    reportNumber   = "CERN-EP-2017-043",
    eprint         = "1706.03192",
    archivePrefix  = "arXiv",
    primaryClass   = "hep-ex",
}

@Article{STDM-2015-02,
    author         = "{ATLAS Collaboration}",
    title          = "{Charged-particle distributions in \(\sqrt{s} = 13\,\text{TeV}\) \(pp\) interactions measured with the ATLAS detector at the LHC}",
    journal        = "Phys. Lett. B",
    volume         = "758",
    year           = "2016",
    pages          = "67",
    doi            = "10.1016/j.physletb.2016.04.050",
    reportNumber   = "CERN-EP-2016-014",
    eprint         = "1602.01633",
    archivePrefix  = "arXiv",
    primaryClass   = "hep-ex",
}

@Article{STDM-2015-03,
    author         = "{ATLAS Collaboration}",
    title          = "{Measurement of \(W^{\pm}\) and \(Z\)-boson production cross sections in \(pp\) collisions at \(\sqrt{s} = 13\,\text{TeV}\) with the ATLAS detector}",
    journal        = "Phys. Lett. B",
    volume         = "759",
    year           = "2016",
    pages          = "601",
    doi            = "10.1016/j.physletb.2016.06.023",
    reportNumber   = "CERN-EP-2016-069",
    eprint         = "1603.09222",
    archivePrefix  = "arXiv",
    primaryClass   = "hep-ex",
}

@Article{STDM-2015-05,
    author         = "{ATLAS Collaboration}",
    title          = "{Measurement of the Inelastic Proton--Proton Cross Section at \(\sqrt{s} = 13\,\text{TeV}\) with the ATLAS Detector at the LHC}",
    journal        = "Phys. Rev. Lett.",
    volume         = "117",
    year           = "2016",
    pages          = "182002",
    doi            = "10.1103/PhysRevLett.117.182002",
    reportNumber   = "CERN-EP-2016-140",
    eprint         = "1606.02625",
    archivePrefix  = "arXiv",
    primaryClass   = "hep-ex",
}

@Article{STDM-2015-10,
    author         = "{ATLAS Collaboration}",
    title          = "{Measurement of exclusive \(\gamma\gamma\rightarrow W^+W^-\) production and search for exclusive Higgs boson production in \(pp\) collisions at \(\sqrt{s} = 8\,\text{TeV}\) using the ATLAS detector}",
    journal        = "Phys. Rev. D",
    volume         = "94",
    year           = "2016",
    pages          = "032011",
    doi            = "10.1103/PhysRevD.94.032011",
    reportNumber   = "CERN-EP-2016-123",
    eprint         = "1607.03745",
    archivePrefix  = "arXiv",
    primaryClass   = "hep-ex",
}

@Article{STDM-2015-16,
    author         = "{ATLAS Collaboration}",
    title          = "{Measurement of \(W\) boson angular distributions in events with high transverse momentum jets at \(\sqrt{s} = 8\,\text{TeV}\) using the ATLAS detector}",
    journal        = "Phys. Lett. B",
    volume         = "765",
    year           = "2017",
    pages          = "132",
    doi            = "10.1016/j.physletb.2016.12.005",
    reportNumber   = "CERN-EP-2016-182",
    eprint         = "1609.07045",
    archivePrefix  = "arXiv",
    primaryClass   = "hep-ex",
}

@Article{STDM-2015-17,
    author         = "{ATLAS Collaboration}",
    title          = "{Charged-particle distributions at low transverse momentum in \(\sqrt{s} = 13\,\text{TeV}\) \(pp\) interactions measured with the ATLAS detector at the LHC}",
    journal        = "Eur. Phys. J. C",
    volume         = "76",
    year           = "2016",
    pages          = "502",
    doi            = "10.1140/epjc/s10052-016-4335-y",
    reportNumber   = "CERN-EP-2016-099",
    eprint         = "1606.01133",
    archivePrefix  = "arXiv",
    primaryClass   = "hep-ex",
}

@Article{STDM-2015-22,
    author         = "{ATLAS Collaboration}",
    title          = "{Measurement of the total cross section from elastic scattering in \(pp\) collisions at \(\sqrt{s} = 8\,\text{TeV}\) with the ATLAS detector}",
    journal        = "Phys. Lett. B",
    volume         = "761",
    year           = "2016",
    pages          = "158",
    doi            = "10.1016/j.physletb.2016.08.020",
    reportNumber   = "CERN-EP-2016-158",
    eprint         = "1607.06605",
    archivePrefix  = "arXiv",
    primaryClass   = "hep-ex",
}

@Article{TOPQ-2015-06,
    author         = "{ATLAS Collaboration}",
    title          = "{Measurements of top-quark pair differential cross-sections in the lepton+jets channel in \(pp\) collisions at \(\sqrt{s} = 8\,\text{TeV}\) using the ATLAS detector}",
    journal        = "Eur. Phys. J. C",
    volume         = "76",
    year           = "2016",
    pages          = "538",
    doi            = "10.1140/epjc/s10052-016-4366-4",
    reportNumber   = "CERN-PH-EP-2015-239",
    eprint         = "1511.04716",
    archivePrefix  = "arXiv",
    primaryClass   = "hep-ex",
}

@Article{TOPQ-2015-07,
    author         = "{ATLAS Collaboration}",
    title          = "{Measurement of top quark pair differential cross sections in the dilepton channel in \(pp\) collisions at \(\sqrt{s} = 7\) and \(8\,\text{TeV}\) with ATLAS}",
    journal        = "Phys. Rev. D",
    volume         = "94",
    year           = "2016",
    pages          = "092003",
    doi            = "10.1103/PhysRevD.94.092003",
    reportNumber   = "CERN-EP-2016-144",
    eprint         = "1607.07281",
    archivePrefix  = "arXiv",
    primaryClass   = "hep-ex",
}

@Article{HION-2016-02,
    author         = "{ATLAS Collaboration}",
    title          = "{Exclusive dimuon production in ultraperipheral Pb+Pb collisions at \(\sqrt{s_{\text{NN}}} = 5.02\,\text{TeV}\) with ATLAS}",
    journal        = "Phys. Rev. C",
    volume         = "104",
    year           = "2021",
    pages          = "024906",
    doi            = "10.1103/PhysRevC.104.024906",
    reportNumber   = "CERN-EP-2020-138",
    eprint         = "2011.12211",
    archivePrefix  = "arXiv",
    primaryClass   = "nucl-ex",
}

@Article{HION-2016-05,
    author         = "{ATLAS Collaboration}",
    title          = "{Evidence for light-by-light scattering in heavy-ion collisions with the ATLAS detector at the LHC}",
    journal        = "Nature Phys.",
    volume         = "13",
    year           = "2017",
    pages          = "852",
    doi            = "10.1038/nphys4208",
    reportNumber   = "CERN-EP-2016-316",
    eprint         = "1702.01625",
    archivePrefix  = "arXiv",
    primaryClass   = "hep-ex",
}

@Article{STDM-2016-01,
    author         = "{ATLAS Collaboration}",
    title          = "{Measurements of the production cross section of a \(Z\) boson in association with jets in \(pp\) collisions at \(\sqrt{s} = 13\,\text{TeV}\) with the ATLAS detector}",
    journal        = "Eur. Phys. J. C",
    volume         = "77",
    year           = "2017",
    pages          = "361",
    doi            = "10.1140/epjc/s10052-017-4900-z",
    reportNumber   = "CERN-EP-2016-297",
    eprint         = "1702.05725",
    archivePrefix  = "arXiv",
    primaryClass   = "hep-ex",
}

@Article{STDM-2016-03,
    author         = "{ATLAS Collaboration}",
    title          = "{Measurement of inclusive jet and dijet cross-sections in proton--proton collisions at \(\sqrt{s} = 13\,\text{TeV}\) with the ATLAS detector}",
    journal        = "JHEP",
    volume         = "05",
    year           = "2018",
    pages          = "195",
    doi            = "10.1007/JHEP05(2018)195",
    reportNumber   = "CERN-EP-2017-157",
    eprint         = "1711.02692",
    archivePrefix  = "arXiv",
    primaryClass   = "hep-ex",
}

@Article{STDM-2016-04,
    author         = "{ATLAS Collaboration}",
    title          = "{Measurement of the Drell--Yan triple-differential cross section in \(pp\) collisions at \(\sqrt{s} = 8\,\text{TeV}\)}",
    journal        = "JHEP",
    volume         = "12",
    year           = "2017",
    pages          = "059",
    doi            = "10.1007/JHEP12(2017)059",
    reportNumber   = "CERN-EP-2017-152",
    eprint         = "1710.05167",
    archivePrefix  = "arXiv",
    primaryClass   = "hep-ex",
}

@Article{STDM-2016-06,
    author         = "{ATLAS Collaboration}",
    title          = "{Measurement of the production cross section of three isolated photons in \(pp\) collisions at \(\sqrt{s} = 8\,\text{TeV}\) using the ATLAS detector}",
    journal        = "Phys. Lett. B",
    volume         = "781",
    year           = "2018",
    pages          = "55",
    doi            = "10.1016/j.physletb.2018.03.057",
    reportNumber   = "CERN-EP-2017-302",
    eprint         = "1712.07291",
    archivePrefix  = "arXiv",
    primaryClass   = "hep-ex",
}

@Article{STDM-2016-07,
    author         = "{ATLAS Collaboration}",
    title          = "{Measurement of charged-particle distributions sensitive to the underlying event in \(\sqrt{s} = 13\,\text{TeV}\) proton--proton collisions with the ATLAS detector at the LHC}",
    journal        = "JHEP",
    volume         = "03",
    year           = "2017",
    pages          = "157",
    doi            = "10.1007/JHEP03(2017)157",
    reportNumber   = "CERN-EP-2016-28",
    eprint         = "1701.05390",
    archivePrefix  = "arXiv",
    primaryClass   = "hep-ex",
}

@Article{STDM-2016-10,
    author         = "{ATLAS Collaboration}",
    title          = "{Determination of the strong coupling constant \(\alpha_{\text{s}}\) from transverse energy-energy correlations in multijet events at \(\sqrt{s} = 8\,\text{TeV}\) using the ATLAS detector}",
    journal        = "Eur. Phys. J. C",
    volume         = "77",
    year           = "2017",
    pages          = "872",
    doi            = "10.1140/epjc/s10052-017-5442-0",
    reportNumber   = "CERN-EP-2017-093",
    eprint         = "1707.02562",
    archivePrefix  = "arXiv",
    primaryClass   = "hep-ex",
}

@Article{STDM-2016-11,
    author         = "{ATLAS Collaboration}",
    title          = "{Measurement of the inclusive cross-section for the production of jets in association with a \(Z\) boson in proton--proton collisions at \(8\,\text{TeV}\) using the ATLAS detector}",
    journal        = "Eur. Phys. J. C",
    volume         = "79",
    year           = "2019",
    pages          = "847",
    doi            = "10.1140/epjc/s10052-019-7321-3",
    reportNumber   = "CERN-EP-2019-133",
    eprint         = "1907.06728",
    archivePrefix  = "arXiv",
    primaryClass   = "hep-ex",
}

@Article{STDM-2016-13,
    author         = "{ATLAS Collaboration}",
    title          = "{Measurement of the exclusive \(\gamma \gamma \rightarrow \mu^{+} \mu^{-}\) process in proton--proton collisions at \(\sqrt{s} = 13\,\text{TeV}\) with the ATLAS detector}",
    journal        = "Phys. Lett. B",
    volume         = "777",
    year           = "2018",
    pages          = "303",
    doi            = "10.1016/j.physletb.2017.12.043",
    reportNumber   = "CERN-EP-2017-151",
    eprint         = "1708.04053",
    archivePrefix  = "arXiv",
    primaryClass   = "hep-ex",
}

@Article{STDM-2016-14,
    author         = "{ATLAS Collaboration}",
    title          = "{Measurement of differential cross sections and \(W^{+}/W^{-}\) cross-section ratios for \(W\) boson production in association with jets at \(\sqrt{s} = 8\,\text{TeV}\) with the ATLAS detector}",
    journal        = "JHEP",
    volume         = "05",
    year           = "2018",
    pages          = "077",
    doi            = "10.1007/JHEP05(2018)077",
    reportNumber   = "CERN-EP-2017-213",
    eprint         = "1711.03296",
    archivePrefix  = "arXiv",
    primaryClass   = "hep-ex",
}

@Article{STDM-2016-15,
    author         = "{ATLAS Collaboration}",
    title          = "{\(ZZ \rightarrow \ell^{+}\ell^{-}\ell^{\prime +}\ell^{\prime -}\) cross-section measurements and search for anomalous triple gauge couplings in \(13\,\text{TeV}\) \(pp\) collisions with the ATLAS detector}",
    journal        = "Phys. Rev. D",
    volume         = "97",
    year           = "2018",
    pages          = "032005",
    doi            = "10.1103/PhysRevD.97.032005",
    reportNumber   = "CERN-EP-2017-163",
    eprint         = "1709.07703",
    archivePrefix  = "arXiv",
    primaryClass   = "hep-ex",
}

@Article{TRIG-2016-01,
    author         = "{ATLAS Collaboration}",
    title          = "{Performance of the ATLAS trigger system in 2015}",
    journal        = "Eur. Phys. J. C",
    volume         = "77",
    year           = "2017",
    pages          = "317",
    doi            = "10.1140/epjc/s10052-017-4852-3",
    reportNumber   = "CERN-EP-2016-241",
    eprint         = "1611.09661",
    archivePrefix  = "arXiv",
    primaryClass   = "hep-ex",
}

@Article{STDM-2017-01,
    author         = "{ATLAS Collaboration}",
    title          = "{Measurement of the cross section for isolated-photon plus jet production in \(pp\) collisions at \(\sqrt{s}=13\,\text{TeV}\) using the ATLAS detector}",
    journal        = "Phys. Lett. B",
    volume         = "780",
    year           = "2018",
    pages          = "578",
    doi            = "10.1016/j.physletb.2018.03.035",
    reportNumber   = "CERN-EP-2017-265",
    eprint         = "1801.00112",
    archivePrefix  = "arXiv",
    primaryClass   = "hep-ex",
}

@Article{STDM-2017-03,
    author         = "{ATLAS Collaboration}",
    title          = "{Measurement of \(ZZ\) production in the \(\ell\ell\nu\nu\) final state with the ATLAS detector in \(pp\) collisions at \(\sqrt{s} = 13\,\text{TeV}\)}",
    journal        = "JHEP",
    volume         = "10",
    year           = "2019",
    pages          = "127",
    doi            = "10.1007/JHEP10(2019)127",
    reportNumber   = "CERN-EP-2019-066",
    eprint         = "1905.07163",
    archivePrefix  = "arXiv",
    primaryClass   = "hep-ex",
}

@Article{STDM-2017-04,
    author         = "{ATLAS Collaboration}",
    title          = "{Measurement of the Soft-Drop Jet Mass in \(pp\) Collisions at \(\sqrt{s} = 13\,\text{TeV}\) with the ATLAS detector}",
    journal        = "Phys. Rev. Lett.",
    volume         = "121",
    year           = "2018",
    pages          = "092001",
    doi            = "10.1103/PhysRevLett.121.092001",
    reportNumber   = "CERN-EP-2017-231",
    eprint         = "1711.08341",
    archivePrefix  = "arXiv",
    primaryClass   = "hep-ex",
}

@Article{STDM-2017-06,
    author         = "{ATLAS Collaboration}",
    title          = "{Observation of Electroweak Production of a Same-Sign \(W\) Boson Pair in Association with Two Jets in \(pp\) Collisions at \(\sqrt{s} = 13\,\text{TeV}\) with the ATLAS Detector}",
    journal        = "Phys. Rev. Lett.",
    volume         = "123",
    year           = "2019",
    pages          = "161801",
    doi            = "10.1103/PhysRevLett.123.161801",
    reportNumber   = "CERN-EP-2019-008",
    eprint         = "1906.03203",
    archivePrefix  = "arXiv",
    primaryClass   = "hep-ex",
}

@Article{STDM-2017-12,
    author         = "{ATLAS Collaboration}",
    title          = "{Measurement of the ratio of cross sections for inclusive isolated-photon production in \(pp\) collisions at \(\sqrt{s} = 13\) and \(8\,\text{TeV}\) with the ATLAS detector}",
    journal        = "JHEP",
    volume         = "04",
    year           = "2019",
    pages          = "093",
    doi            = "10.1007/JHEP04(2019)093",
    reportNumber   = "CERN-EP-2018-340",
    eprint         = "1901.10075",
    archivePrefix  = "arXiv",
    primaryClass   = "hep-ex",
}

@Article{STDM-2017-13,
    author         = "{ATLAS Collaboration}",
    title          = "{Measurement of the cross-section and charge asymmetry of \(W\) bosons produced in proton--proton collisions at \(\sqrt{s} = 8\,\text{TeV}\) with the ATLAS detector}",
    journal        = "Eur. Phys. J. C",
    volume         = "79",
    year           = "2019",
    pages          = "760",
    doi            = "10.1140/epjc/s10052-019-7199-0",
    reportNumber   = "CERN-EP-2019-053",
    eprint         = "1904.05631",
    archivePrefix  = "arXiv",
    primaryClass   = "hep-ex",
}

@Article{STDM-2017-16,
    author         = "{ATLAS Collaboration}",
    title          = "{Properties of jet fragmentation using charged particles measured with the ATLAS detector in \(pp\) collisions at \(\sqrt{s} = 13\,\text{TeV}\)}",
    journal        = "Phys. Rev. D",
    volume         = "100",
    year           = "2019",
    pages          = "052011",
    doi            = "10.1103/PhysRevD.100.052011",
    reportNumber   = "CERN-EP-2019-090",
    eprint         = "1906.09254",
    archivePrefix  = "arXiv",
    primaryClass   = "hep-ex",
}

@Article{STDM-2017-18,
    author         = "{ATLAS Collaboration}",
    title          = "{Measurement of the \(Z\gamma \rightarrow \nu\bar{\nu}\gamma\) production cross section in \(pp\) collisions at \(\sqrt{s} = 13\,\text{TeV}\) with the ATLAS detector and limits on anomalous triple gauge-boson couplings}",
    journal        = "JHEP",
    volume         = "12",
    year           = "2018",
    pages          = "010",
    doi            = "10.1007/JHEP12(2018)010",
    reportNumber   = "CERN-EP-2018-220",
    eprint         = "1810.04995",
    archivePrefix  = "arXiv",
    primaryClass   = "hep-ex",
}

@Article{STDM-2017-19,
    author         = "{ATLAS Collaboration}",
    title          = "{Observation of electroweak production of two jets and a \(Z\)-boson pair}",
    journal        = "Nature Phys.",
    volume         = "19",
    year           = "2023",
    pages          = "237--253",
    doi            = "10.1038/s41567-022-01757-y",
    reportNumber   = "CERN-EP-2020-016",
    eprint         = "2004.10612",
    archivePrefix  = "arXiv",
    primaryClass   = "hep-ex",
}

@Article{STDM-2017-21,
    author         = "{ATLAS Collaboration}",
    title          = "{Observation of photon-induced \(W^+W^-\) production in \(pp\) collisions at \(\sqrt{s} = 13\,\text{TeV}\) using the ATLAS detector}",
    journal        = "Phys. Lett. B",
    volume         = "816",
    year           = "2021",
    pages          = "136190",
    doi            = "10.1016/j.physletb.2021.136190",
    reportNumber   = "CERN-EP-2020-165",
    eprint         = "2010.04019",
    archivePrefix  = "arXiv",
    primaryClass   = "hep-ex",
}

@Article{STDM-2017-22,
    author         = "{ATLAS Collaboration}",
    title          = "{Evidence for the production of three massive vector bosons with the ATLAS detector}",
    journal        = "Phys. Lett. B",
    volume         = "798",
    year           = "2019",
    pages          = "134913",
    doi            = "10.1016/j.physletb.2019.134913",
    reportNumber   = "CERN-EP-2019-041",
    eprint         = "1903.10415",
    archivePrefix  = "arXiv",
    primaryClass   = "hep-ex",
}

@Article{STDM-2017-23,
    author         = "{ATLAS Collaboration}",
    title          = "{Observation of electroweak \(W^{\pm}Z\) boson pair production in association with two jets in \(pp\) collisions at \(\sqrt{s} = 13\,\text{TeV}\) with the ATLAS detector}",
    journal        = "Phys. Lett. B",
    volume         = "793",
    year           = "2019",
    pages          = "469",
    doi            = "10.1016/j.physletb.2019.05.012",
    reportNumber   = "CERN-EP-2018-286",
    eprint         = "1812.09740",
    archivePrefix  = "arXiv",
    primaryClass   = "hep-ex",
}

@Article{STDM-2017-24,
    author         = "{ATLAS Collaboration}",
    title          = "{Measurement of fiducial and differential \(W^+W^-\) production cross-sections at \(\sqrt{s} = 13\,\text{TeV}\) with the ATLAS detector}",
    journal        = "Eur. Phys. J. C",
    volume         = "79",
    year           = "2019",
    pages          = "884",
    doi            = "10.1140/epjc/s10052-019-7371-6",
    reportNumber   = "CERN-EP-2019-055",
    eprint         = "1905.04242",
    archivePrefix  = "arXiv",
    primaryClass   = "hep-ex",
}

@Article{STDM-2017-27,
    author         = "{ATLAS Collaboration}",
    title          = "{Differential cross-section measurements for the electroweak production of dijets in association with a \(Z\) boson in proton--proton collisions at ATLAS}",
    journal        = "Eur. Phys. J. C",
    volume         = "81",
    year           = "2021",
    pages          = "163",
    doi            = "10.1140/epjc/s10052-020-08734-w",
    reportNumber   = "CERN-EP-2020-045",
    eprint         = "2006.15458",
    archivePrefix  = "arXiv",
    primaryClass   = "hep-ex",
}

@Article{STDM-2017-28,
    author         = "{ATLAS Collaboration}",
    title          = "{Measurement of distributions sensitive to the underlying event in inclusive \(Z\) boson production in \(pp\) collisions at \(\sqrt{s} = 13\,\text{TeV}\) with the ATLAS detector}",
    journal        = "Eur. Phys. J. C",
    volume         = "79",
    year           = "2019",
    pages          = "666",
    doi            = "10.1140/epjc/s10052-019-7162-0",
    reportNumber   = "CERN-EP-2019-064",
    eprint         = "1905.09752",
    archivePrefix  = "arXiv",
    primaryClass   = "hep-ex",
}

@Article{STDM-2017-29,
    author         = "{ATLAS Collaboration}",
    title          = "{Measurement of the inclusive isolated-photon cross section in \(pp\) collisions at \(\sqrt{s} = 13\,\text{TeV}\) using \(36\,\text{fb}^{-1}\) of ATLAS data}",
    journal        = "JHEP",
    volume         = "10",
    year           = "2019",
    pages          = "203",
    doi            = "10.1007/JHEP10(2019)203",
    reportNumber   = "CERN-EP-2019-136",
    eprint         = "1908.02746",
    archivePrefix  = "arXiv",
    primaryClass   = "hep-ex",
}

@Article{STDM-2017-30,
    author         = "{ATLAS Collaboration}",
    title          = "{Measurement of the production cross section of pairs of isolated photons in \(pp\) collisions at \(13\,\text{TeV}\) with the ATLAS detector}",
    journal        = "JHEP",
    volume         = "11",
    year           = "2021",
    pages          = "169",
    doi            = "10.1007/JHEP11(2021)169",
    reportNumber   = "CERN-EP-2021-105",
    eprint         = "2107.09330",
    archivePrefix  = "arXiv",
    primaryClass   = "hep-ex",
}

@Article{STDM-2017-32,
    author         = "{ATLAS Collaboration}",
    title          = "{Measurement of isolated-photon plus two-jet production in \(pp\) collisions at \(\sqrt{s} = 13\,\text{TeV}\) with the ATLAS detector}",
    journal        = "JHEP",
    volume         = "03",
    year           = "2020",
    pages          = "179",
    doi            = "10.1007/JHEP03(2020)179",
    reportNumber   = "CERN-EP-2019-210",
    eprint         = "1912.09866",
    archivePrefix  = "arXiv",
    primaryClass   = "hep-ex",
}

@Article{STDM-2017-33,
    author         = "{ATLAS Collaboration}",
    title          = "{Measurement of soft-drop jet observables in \(pp\) collisions with the ATLAS detector at \(\sqrt{s} = 13\,\text{TeV}\)}",
    journal        = "Phys. Rev. D",
    volume         = "101",
    year           = "2020",
    pages          = "052007",
    doi            = "10.1103/PhysRevD.101.052007",
    reportNumber   = "CERN-EP-2019-269",
    eprint         = "1912.09837",
    archivePrefix  = "arXiv",
    primaryClass   = "hep-ex",
}

@Article{STDM-2017-34,
    author         = "{ATLAS Collaboration}",
    title          = "{Measurement of jet-substructure observables in top quark, \(W\) boson and light jet production in proton--proton collisions at \(\sqrt{s} = 13\,\text{TeV}\) with the ATLAS detector}",
    journal        = "JHEP",
    volume         = "08",
    year           = "2019",
    pages          = "033",
    doi            = "10.1007/JHEP08(2019)033",
    reportNumber   = "CERN-EP-2019-011",
    eprint         = "1903.02942",
    archivePrefix  = "arXiv",
    primaryClass   = "hep-ex",
}

@Article{STDM-2017-37,
    author         = "{ATLAS Collaboration}",
    title          = "{Measurement of cross sections for production of a \(Z\) boson in association with a flavor-inclusive or doubly \(b\)-tagged large-radius jet in proton--proton collisions at \(\sqrt{s} = 13\,\text{TeV}\) with the ATLAS experiment}",
    journal        = "Phys. Rev. D",
    volume         = "108",
    year           = "2023",
    pages          = "012022",
    doi            = "10.1103/PhysRevD.108.012022",
    reportNumber   = "CERN-EP-2021-140",
    eprint         = "2204.12355",
    archivePrefix  = "arXiv",
    primaryClass   = "hep-ex",
}

@Article{TOPQ-2017-19,
    author         = "{ATLAS Collaboration}",
    title          = "{Measurements of jet observables sensitive to \(b\)-quark fragmentation in \(t\bar{t}\) events at the LHC with the ATLAS detector}",
    journal        = "Phys. Rev. D",
    volume         = "106",
    year           = "2022",
    pages          = "032008",
    doi            = "10.1103/PhysRevD.106.032008",
    reportNumber   = "CERN-EP-2021-116",
    eprint         = "2202.13901",
    archivePrefix  = "arXiv",
    primaryClass   = "hep-ex",
}

@Article{BPHY-2018-01,
    author         = "{ATLAS Collaboration}",
    title          = "{Measurement of the \(CP\)-violating phase \(\phi_s\) in \(B^0_s \to J/\psi\phi\) decays in ATLAS at \(13\,\text{TeV}\)}",
    journal        = "Eur. Phys. J. C",
    volume         = "81",
    year           = "2021",
    pages          = "342",
    doi            = "10.1140/epjc/s10052-021-09011-0",
    reportNumber   = "CERN-EP-2019-218",
    eprint         = "2001.07115",
    archivePrefix  = "arXiv",
    primaryClass   = "hep-ex",
}

@Article{BPHY-2018-08,
    author         = "{ATLAS Collaboration}",
    title          = "{Study of \(B_c^+ \to J/\psi D_s^+\) and \(B_c^+ \to J/\psi D_s^{*+}\) decays in \(pp\) collisions at \(\sqrt{s} = 13\,\text{TeV}\) with the ATLAS detector}",
    journal        = "JHEP",
    volume         = "08",
    year           = "2022",
    pages          = "087",
    doi            = "10.1007/JHEP08(2022)087",
    reportNumber   = "CERN-EP-2022-025",
    eprint         = "2203.01808",
    archivePrefix  = "arXiv",
    primaryClass   = "hep-ex",
}

@Article{BPHY-2018-09,
    author         = "{ATLAS Collaboration}",
    title          = "{Study of the rare decays of \(B^0_s\) and \(B^0\) mesons into muon pairs using data collected during 2015 and 2016 with the ATLAS detector}",
    journal        = "JHEP",
    volume         = "04",
    year           = "2019",
    pages          = "098",
    doi            = "10.1007/JHEP04(2019)098",
    reportNumber   = "CERN-EP-2018-291",
    eprint         = "1812.03017",
    archivePrefix  = "arXiv",
    primaryClass   = "hep-ex",
}

@Article{EGAM-2018-01,
    author         = "{ATLAS Collaboration}",
    title          = "{Electron and photon performance measurements with the ATLAS detector using the 2015--2017 LHC proton--proton collision data}",
    journal        = "JINST",
    volume         = "14",
    year           = "2019",
    pages          = "P12006",
    doi            = "10.1088/1748-0221/14/12/P12006",
    reportNumber   = "CERN-EP-2019-145",
    eprint         = "1908.00005",
    archivePrefix  = "arXiv",
    primaryClass   = "hep-ex",
}

@Article{HION-2018-02,
    author         = "{ATLAS Collaboration}",
    title          = "{Measurements of \(W\) and \(Z\) boson production in \(pp\) collisions at \(\sqrt{s} = 5.02\,\text{TeV}\) with the ATLAS detector}",
    journal        = "Eur. Phys. J. C",
    volume         = "79",
    year           = "2019",
    pages          = "128",
    doi            = "10.1140/epjc/s10052-019-6622-x",
    reportNumber   = "CERN-EP-2018-259",
    eprint         = "1810.08424",
    archivePrefix  = "arXiv",
    primaryClass   = "hep-ex",
    related        = "HION-2018-02-err",
    relatedstring  = "Erratum:",
}

@Article{HION-2018-19,
    author         = "{ATLAS Collaboration}",
    title          = "{Observation of Light-by-Light Scattering in Ultraperipheral Pb+Pb Collisions with the ATLAS Detector}",
    journal        = "Phys. Rev. Lett.",
    volume         = "123",
    year           = "2019",
    pages          = "052001",
    doi            = "10.1103/PhysRevLett.123.052001",
    reportNumber   = "CERN-EP-2019-051",
    eprint         = "1904.03536",
    archivePrefix  = "arXiv",
    primaryClass   = "hep-ex",
}

@Article{JETM-2018-02,
    author         = "{ATLAS Collaboration}",
    title          = "{In situ calibration of large-radius jet energy and mass in \(13\,\text{TeV}\) proton--proton collisions with the ATLAS detector}",
    journal        = "Eur. Phys. J. C",
    volume         = "79",
    year           = "2019",
    pages          = "135",
    doi            = "10.1140/epjc/s10052-019-6632-8",
    reportNumber   = "CERN-EP-2018-191",
    eprint         = "1807.09477",
    archivePrefix  = "arXiv",
    primaryClass   = "hep-ex",
}
\clearpage
 
\begin{flushleft}
\hypersetup{urlcolor=black}
{\Large The ATLAS Collaboration}

\bigskip

\AtlasOrcid[0000-0002-6665-4934]{G.~Aad}$^\textrm{\scriptsize 103}$,
\AtlasOrcid[0000-0001-7616-1554]{E.~Aakvaag}$^\textrm{\scriptsize 16}$,
\AtlasOrcid[0000-0002-5888-2734]{B.~Abbott}$^\textrm{\scriptsize 121}$,
\AtlasOrcid[0000-0002-0287-5869]{S.~Abdelhameed}$^\textrm{\scriptsize 117a}$,
\AtlasOrcid[0000-0002-1002-1652]{K.~Abeling}$^\textrm{\scriptsize 55}$,
\AtlasOrcid[0000-0001-5763-2760]{N.J.~Abicht}$^\textrm{\scriptsize 49}$,
\AtlasOrcid[0000-0002-8496-9294]{S.H.~Abidi}$^\textrm{\scriptsize 29}$,
\AtlasOrcid[0009-0003-6578-220X]{M.~Aboelela}$^\textrm{\scriptsize 44}$,
\AtlasOrcid[0000-0002-9987-2292]{A.~Aboulhorma}$^\textrm{\scriptsize 35e}$,
\AtlasOrcid[0000-0001-5329-6640]{H.~Abramowicz}$^\textrm{\scriptsize 153}$
% ...
\bigskip
\\

% ...
$^{16}$Department for Physics and Technology, University of Bergen, Bergen; Norway.\\
% ...
$^{29}$Physics Department, Brookhaven National Laboratory, Upton NY; United States of America.\\
% ...
$^{35}$$^{(a)}$Facult\'e des Sciences Ain Chock, Universit\'e Hassan II de Casablanca;$^{(b)}$Facult\'{e} des Sciences, Universit\'{e} Ibn-Tofail, K\'{e}nitra;$^{(c)}$Facult\'e des Sciences Semlalia, Universit\'e Cadi Ayyad, LPHEA-Marrakech;$^{(d)}$LPMR, Facult\'e des Sciences, Universit\'e Mohamed Premier, Oujda;$^{(e)}$Facult\'e des sciences, Universit\'e Mohammed V, Rabat;$^{(f)}$Institute of Applied Physics, Mohammed VI Polytechnic University, Ben Guerir; Morocco.\\
% ...
$^{44}$Physics Department, Southern Methodist University, Dallas TX; United States of America.\\
% ...
$^{49}$Fakult\"{a}t Physik , Technische Universit{\"a}t Dortmund, Dortmund; Germany.\\
% ...
$^{55}$II. Physikalisches Institut, Georg-August-Universit\"{a}t G\"ottingen, G\"ottingen; Germany.\\
% ...
$^{103}$CPPM, Aix-Marseille Universit\'e, CNRS/IN2P3, Marseille; France.\\
% ...
$^{117}$$^{(a)}$New York University Abu Dhabi, Abu Dhabi;$^{(b)}$United Arab Emirates University, Al Ain; United Arab Emirates.\\
% ...
$^{121}$Homer L. Dodge Department of Physics and Astronomy, University of Oklahoma, Norman OK; United States of America.\\
% ...
$^{153}$Raymond and Beverly Sackler School of Physics and Astronomy, Tel Aviv University, Tel Aviv; Israel.\\
% ...
$^{a}$ Also Affiliated with an institute covered by a cooperation agreement with CERN.\\
$^{b}$ Also at An-Najah National University, Nablus; Palestine.\\
$^{c}$ Also at Borough of Manhattan Community College, City University of New York, New York NY; United States of America.\\
$^{d}$ Also at Center for High Energy Physics, Peking University; China.\\
$^{e}$ Also at Center for Interdisciplinary Research and Innovation (CIRI-AUTH), Thessaloniki; Greece.\\
$^{f}$ Also at Centro Studi e Ricerche Enrico Fermi; Italy.\\
% ...

\end{flushleft}


%

%
%
%
%

%
%
%
%

\end{document}